\documentclass[12pt]{article}
\pdfoutput=1

\usepackage[T1]{fontenc}
\usepackage{caption}
\usepackage{subcaption}
\usepackage{jheppub}
\usepackage{enumitem}
\usepackage{framed}
\usepackage{booktabs}
\usepackage{subcaption}
\usepackage{tikz}
\usetikzlibrary{calc,arrows,decorations.markings,decorations.pathreplacing}
\usepackage{mathrsfs}
\allowdisplaybreaks[1]

\tikzset{
  big arrow/.style={
    decoration={markings,mark=at position 1 with {\arrow[scale=2,#1]{>}}},
    postaction={decorate},
    shorten >=0.4pt},
  big arrow/.default=black}

\usepackage{diagbox}

\newcommand\scalemath[2]{\scalebox{#1}{\mbox{\ensuremath{\displaystyle #2}}}}

\def\IZ {{\mathbb Z}}

\title{SCFTs, Holography, and Topological Strings} 

\author[a]{Hirotaka Hayashi\,}
\author[b]{Patrick Jefferson\,}
\author[b,d]{Hee-Cheol Kim\,}
\author[c]{Kantaro Ohmori\,}
\author[b]{and Cumrun Vafa\,}

\affiliation[a]{Department of Physics, School of Science, Tokai University, 4-1-1 Kitakaname, Hiratsuka-shi, Kanagawa 259-1292, Japan}
\affiliation[b]{Jefferson Physical Laboratory, Harvard University, Cambridge, MA 02138, USA}
\affiliation[c]{School of Natural Sciences, Institute for Advanced Study, Princeton, NJ 08540, USA}
\affiliation[d]{Department of Physics, POSTECH, Pohang 790-784, Korea}

\abstract
{We show that the non-gravitational sectors of certain 6d and 5d supergravity theories can be decomposed into superconformal field theories (SCFTs) which are coupled together by pairwise identifying and gauging mutual global symmetries.  In the case of 6d supergravity,
we consider F-theory on compact elliptic Calabi-Yau 3-folds with base $B=T^4/\IZ_n\times \IZ_m$ and we show in many examples that the non-gravitational field theory sectors can be described as configurations of coupled 6d $(1,0)$ SCFTs.  We also conjecture that the effective 2d $(0,4)$ SCFTs living on the self-dual strings of the 6d theories lead to holographically dual descriptions of type IIB string theory on $AdS_3\times S^3\times B$ and moreover that their elliptic genera can be used to compute the degeneracies of 5d spinning BPS black holes along with all-genus topological string amplitudes on the corresponding compact 3-fold. In the case of 5d supergravity, we consider M-theory on compact Calabi-Yau 3-folds and using similar ideas as in the 6d case
 we show the complete non-gravitational sector of 5d supergravity theories can be decomposed into coupled 5d $\mathcal N=1$ SCFTs.  Furthermore, using this picture we propose a generalized topological vertex formalism which, excluding some curve classes, seems to capture all-genus topological string amplitudes for the mirror quintic.}

\begin{document}
\maketitle

\section{Introduction}
There has been quite a bit of progress recently in understanding $\mathcal{N}=(1,0)$ supreconformal field theories (SCFTs) in six dimensions \cite{Heckman:2013pva,Heckman:2015bfa,Bhardwaj:2015xxa}.  6d $(1,0)$ SCFTs arise as the low energy effective descriptions of F-theory on non-compact elliptic Calabi-Yau (CY) 3-folds.  It is natural to ask whether these advances in our understanding of the 6d SCFTs lead to a deeper understanding of the compact case, namely F-theory on compact CY 3-folds.   One goal of this paper is to give an affirmative answer to this question at least for a special class of elliptic 3-folds where the base $B$ of F-theory is $B = T^4/\IZ_m\times \IZ_n$, extending the work in \cite{Gopakumar:1996mu}.  In particular, we are able to identify the full non-gravitational sector of the resulting 6d (1,0) supergravity theories in terms of 6d $(1,0)$ SCFTs---it turns out that the non-gravitational dynamics of these theories can be described in terms of collections of 6d SCFTs coupled together by gauging their global symmetries in a specific manner.
We confirm these results by checking gauge and gravitational anomaly cancellation for these models along with the fact that the lattice associated to the 6d self-dual strings is both of the correct signature $(1,T)$ (where $T$ is the number of tensor multiplets) and self-dual, as should be the case.  

One of the successes of recent work on 6d $(1,0)$ SCFTs is an improved understanding of the strings charged under the tensor multiplets.  These strings are described at low energy by $\mathcal{N}=(0,4)$ supersymmetric quantum field theories in 2d.   In particular for a large number of them, concrete gauge theories have been proposed and checked in multiple ways \cite{Haghighat:2013gba,Haghighat:2013tka,Kim:2014dza,Gadde:2015tra,Haghighat:2014vxa,Kim:2015fxa,Kim:2016foj}.  It is natural to ask what can be learned from these strings in the case of F-theory on a compact 3-fold, where the strings belong to the spectrum of a 6d (1,0) supergravity theory rather than an SCFT.  This question has been raised and discussed in \cite{Haghighat:2015ega} (see also \cite{Couzens:2017way}).
It was proposed that for any F-theory model on an elliptic 3-fold with compact base $B$, the corresponding strings are holographically dual to $AdS_3\times S^3\times B$.
Moreover the elliptic genus of the strings leads to a count of 5d black holes, obtained by compactifying the 6d theory on a circle and considering strings wrapped on the circle carrying Kaluza-Klein (KK) momenta.    However an explicit description of such strings in terms of concrete 2d $(0,4)$ supersymmetric field theories is in general unknown.    The central charge of these 2d systems is enough to capture the black hole entropy to leading order for large KK momentum.   For the examples we consider, where $B=T^4/\IZ_m\times \IZ_n$, we show from the associated anomaly polynomials that the resulting strings indeed lead to the correct central charges expected from the black hole entropy.  Furthermore, for the specific case of $B=T^4/\IZ_2\times \IZ_2$ we are able to identify a concrete 2d $(0,4)$ quiver gauge theory which has many, but not all, of the needed ingredients.  The fact that this quiver model correctly reproduces the anticipated central charge is a strong indication that the model is close to the correct one.  An important indication that some features are missing is the fact that the proposed gauge theory has more symmetries than expected from the corresponding strings.  This suggests that perhaps some suitable modification of these quiver gauge theories leads to the correct $(0,4)$ theory living on the corresponding strings.  This also implies that computing the corresponding spinning BPS black holes states, which are also captured by all-genus topological string amplitudes, is related to computing the elliptic genus of a suitably modified version of these 2d gauge theories.  In other words, the all-genus topological string amplitudes for 
the CY 3-fold $T^6/\IZ_2\times \IZ_2$ (with Hodge numbers $(h^{1,1},h^{2,1})=(51,3)$) can be computed using the elliptic genus of the proposed theory.  An important consistency check is that these elliptic genus computations reproduce the expected asymptotic degeneracy for spinning black holes.

The basic ingredient in the identification of the 6d field theory sector of our models is the observation that local elliptic 3-folds with base of the form $B=\mathbb C^2/{\IZ_n\times \IZ_m}$, which look like local patches of the compact base $T^4/\IZ_n\times \IZ_m$, can be related to 6d $(1,0)$ SCFTs \cite{DelZotto:2015rca}. This geometric observation has a very natural field theoretic interpretation: the non-gravitational sector of the corresponding supergravity theory can be obtained by ``stitching'' together these SCFTs, where the stitching corresponds to additional gauging of global symmetries.

Motivated by the power of 6d SCFTs in capturing compact topological string amplitudes, we find it is possible to use 5d SCFTs very much in the same spirit, and taking inspiration from some of the constructions in \cite{Hayashi:2017jze}, we introduce an approach which appears to capture a large part of the associated topological string amplitudes for a class of compact Calabi-Yau 3-folds. In particular, for the mirror of the quintic 3-fold we propose that the all-genus amplitudes for topological strings, excluding those associated to a finite list of curve classes, can be computed by viewing the mirror quintic as 10 copies of the 5d SCFT $T_5$ whose global symmetries are gauged by 10 $SU(5)$ gauge groups. 

Our computational methodology entails generalizing the standard notion of the topological vertex ($+$)
to include its mirror ($-$), and further generalizing these two types of vertices by appending an additional integer $N$, leading to the notions of $N^-$ vertices and speculatively, $N^+$ vertices (note that the $1^-$ vertex was already studied in \cite{Aganagic:2004js}, and the groundwork for the $N^{-}$ vertices was pursued in \cite{Hayashi:2017jze}.) However, it turns out that straightforward application of the $N^{-}$ vertices is not enough to describe the mirror quintic and that an additional modification of the vertex formalism is required. To circumvent this problem, we apply a modification of the generalized vertex formalism described above to a closely related geometry which in certain loci is essentially a complex structure deformation of the local geometry of the mirror quintic. Remarkably, this additional modification appears to compute all-genus topological string amplitudes for the mirror quintic with the exception of invariants associated to a finite list of curve classes. The invariants predicted by this modified formalism agree with independent mathematical computations of the same invariants for curve classes of low degree; furthermore, our modified formalism makes numerous predictions for invariants which have yet to be computed by other means. 

The organization of this paper is as follows. In Section \ref{sec:Ftheory} we review some basic features of F-theory on compact elliptic CY 3-folds in order to introduce the 6d perspective on compact 3-folds.  In Section \ref{sec:6dsugra.from.orbifolds} we present our orbifold models, including a discussion of the associated $AdS_3\times S^3\times B$ holography, as well as connections with black hole entropy and topological strings.  In Section \ref{sec:5d} we present a 5d perspective on compact CY 3-folds and apply this perspective to the computation of topological string amplitudes in the specific example of the mirror quintic. In Section \ref{sec:concl} we present our conclusions. In Appendix \ref{sec:mirrorquintic} we describe the geometry of the resolved mirror quintic. Appendix \ref{sec:SU5gauging} summarizes the structure of K$\ddot{\text{a}}$hler parameters and framing factors for a trivalent $SU(5)$ gauging in a particular phase. In Appendix~\ref{sec:prepotential} we compute the triple intersection numbers of the mirror quintic using its decomposition into local 3-folds associated to 5d field theories.


\section{F-theory review}
\label{sec:Ftheory}
We begin by reviewing some properties of F-theory compactifications on elliptically fibered Calabi-Yau 3-folds which are used in later sections.

\subsection{6d supergravity and compact elliptic 3-folds}

F-theory compactified on an elliptically fibered CY 3-fold $X$ yields a 6d supergravity theory with eight supercharges as a low energy effective field theory \cite{Vafa:1996xn, Morrison:1996na, Morrison:1996pp}. $X$ has an elliptic fibration over a complex surface $B$ and may be described by a Weierstrass equation\footnote{It is also possible to consider an F-theory compactification on $X$ which has a torus fibration without a section \cite{Braun:2014oya}.}
\begin{equation}
y^2 = x^3 + f(s, t)x + g(s ,t), \label{weierstrass}
\end{equation}
where $x, y, s, t$ are complex coordinates on a local patch. Globally, $f$ and $g$ are sections of line bundles $\mathcal{O}(-4K_B)$ and $\mathcal{O}(-6K_B)$ respectively where $K_B$ is the canonical divisor of the base $B$. The elliptic fiber degenerates along the discriminant locus where the dsicriminant $\Delta$ vanishes. The discriminant of the Weierstrass equation \eqref{weierstrass} is given by
\begin{equation}
\Delta = 4f^3 + 27g^2, \label{discriminant}
\end{equation}
which is a section of a line bundle $\mathcal{O}(-12K_B)$. The discriminant locus is a complex curve inside the base $B$. Physically, the discriminant locus is a location of 7-branes in type IIB string theory. 
When we do not consider 6d theories with enhanced supersymmetry, the base $B$ is either an Enriques surface, a blow up of $\mathbb{P}^2$, Hirzebruch surface $\mathbb{F}_n$ or a surface with orbifold singularities whose resolution gives one of the geometries above \cite{Morrison:1996pp, MR1109635, MR1217381}. 

A specific choice of the sections $f$ and $g$ of the base $B$ may lead to a configuration where several 7-branes are put on top of each other. In this case, the worldvolume theory on the 7-branes can support a non-abelian gauge algebra $\mathfrak{g}$. In terms of the geometry, the non-abelian gauge theory is realized by having singularities over a complex curve on which the 7-branes are wrapped. 
The type of the singularities characterizes the gauge algebra $\mathfrak{g}$ of the worldvolume theory on the 7-branes. The singularity type is classified by the Tate's algorithm using the Tate form of the above Weierstrass equation \cite{Bershadsky:1996nh, Katz:2011qp}
\begin{equation}
y^2 + a_1(s, t)xy + a_3(s, t)y = x^3 + a_2(s, t)x^2 + a_4(s, t)x + a_6(s, t). \label{tate}
\end{equation}
Suppose $N$ 7-branes are wrapped on a curve $C$ in $B$ and the defining equation of $C$ is given by $\sigma(s, t) = 0$ on a local patch. The discriminant \eqref{discriminant} on the local patch can be written as
\begin{equation}
\Delta = \sigma^N Y
\end{equation}
where $Y$ is a residual polynomial of the discriminant. The lowest order of $\sigma$ in $a_1, a_2, a_3, a_4, a_6$ may fix the singularity type or equivalently the non-abelian gauge algebra $\mathfrak{g}$ for the worldvolume theory of the 7-branes\footnote{When $\mathfrak{g}=\mathfrak{so}(4k+4), \;(k=1, 2, \cdots)$, we need an extra condition for the defining equation.}. On the other hand, an abelian gauge algebra requires an additional section for the elliptic fibration of $X$ \cite{Morrison:1996pp, Fukae:1999zs}. The number of abelian gauge factors is equal to the rank of the Mordell-Weil group.

Let us consider the case when 7-branes are wrapped on a smooth curve $C$ and there is a simply laced non-abelian gauge algbera $\mathfrak{g}$ on the 7-branes. First, we note that if $C$ is a genus $g$ curve, then we have $g$ hypermultiplets in the adjoint representation \cite{Witten:1996qb}. On the curve $C$, which carries singular fibers associated to the simply laced non-abelian gauge algebra $\mathfrak{g}$, there can be a special point where the singularity is enhanced. Physically, this special point is a point of intersection between the 7-branes wrapping $C$ and another configuration of 7-branes. Then there is charged matter localized at this intersection point. 
If the enhanced singularity is associated to a non-abelian gauge algebra $\mathfrak{g}' \supset \mathfrak{g} \oplus \mathfrak{h}$, then the representation of the charged matter can be determined from the embedding of $\mathfrak{g}$ into $\mathfrak{g}'$ \cite{Bershadsky:1996nh, Katz:1996xe}. Namely, we consider a decomposition of the adjoint representation of $\mathfrak{g}'$ under $\mathfrak{g}$
\begin{equation}
\text{adj}(\mathfrak{g}') = \text{adj}(\mathfrak{g}) + \text{dim}\left(\text{adj}(\mathfrak{h})\right){\bf 1} + \sum_i({\bf r}_i + \bar{{\bf r}}_i). \label{decomposition}
\end{equation}
Then ${\bf r}_i$ appearing in \eqref{decomposition} is the representation of the localized matter\footnote{This methods may not fix uniquely the matter representation when there are several embeddings of $\mathfrak{g}$ into $\mathfrak{g}'$. For example, there are two inequivalent embeddings of $\mathfrak{su}(8)$ into $\mathfrak{e}_8$.}. The number of the intersection points where the singularity is enhanced to the type $\mathfrak{g}'$ is related to the number of hypermultiplets in the representation ${\bf r}_i$. 
The charged matter may also acquire a vacuum expectation value (vev), which breaks the gauge symmetry associated to $\mathfrak{g}$. This corresponds to considering more generic polynomials for $a_1, a_2, a_3, a_4, a_6$ on a local patch of $B$, and the maximally Higgsed phase is realized by the most generic polynomials for $a_1, a_2, a_3, a_4, a_6$ for a given choice of base.

Charged matter for a non-simply laced gauge algebra arises in a subtler fashion. A non-simply laced gauge algebra $\mathfrak{g}$ is realized by acting with an outer automorphism on a degenerated fiber associated to a simply laced gauge algebra $\tilde{\mathfrak{g}}$ on $C$. Due to a decomposition of the adjoint representaion of $\tilde{\mathfrak{g}}$ under $\mathfrak{g}$
\begin{equation}
\text{adj}(\tilde{\mathfrak{g}}) = \text{adj}(\mathfrak{g}) +\sum_i (d-1){\bf r}_i,
\end{equation}
there may be hypermultiplets in the representation ${\bf r}_i$ of $\mathfrak{g}$. In order to see their number, one may consider a branched cover $\tilde{C}$ of $C$ with degree $d$. Then the number of hypermultiplets in the representation ${\bf r}_i$ is given by \cite{Grassi:2011hq}
\begin{equation}
n_{{\bf r}_i} = (d-1)(g-1) + \frac{1}{2}\text{deg}(R),
\end{equation}
where $g$ is the geometric genus of $C$ and $R$ is the ramification divisor of $\tilde{C}$.

Moreover if a curve $C$ on which 7-branes are wrapped is singular, there could be some additional matter with a novel representation from the singular points \cite{Sadov:1996zm, Morrison:2011mb}. For example, when $C$ supports an $\mathfrak{su}(N)$ gauge algebra and has an ordinary double point singularity, it may give rise to a symmetric and antisymmetric hypermultiplet. 

In a maximally Higgsed phase, there may be still some unbroken non-abelian gauge symmetry. In other words, some singular fibers may remain over a curve in $B$. Geometrically it is possible to resolve these singularities by introducing exceptional divisors via a blowup. The elliptically fibered 3-fold $X$ becomes a smooth manifold $\tilde{X}$ after the resolution. In fact, the resolved phase does not exist in F-theory on $X$. To see this, we consider the duality between F-theory and M-theory: F-theory compactified on $X \times S^1$ is dual to M-theory compactified on the same $X$ \cite{Vafa:1996xn}. It is indeed possible to consider an M-theory compactification on the resolved 3-fold $\tilde{X}$. However, the 6d limit is a decompactification limit of the $S^1$ which corresponds to a limit where the size of the elliptic fiber in $\tilde{X}$ of the M-theory compactification vanishes. Hence, the effect of the resolution disappears in the 6d limit. Nevertheless, the resolution is useful for determining matter representations and other physical data. 

So far we have seen how the singularity structure of $X$ determines the gauge algebra on 7-branes and also  charged matter localized to intersections between 7-branes. We can also see the relation between the number of multiplets in a 6d supergravity obtained by an F-theory compactification on $X$ and the number of moduli of the resolved 3-fold $\tilde{X}$. A 6d supergravity theory with eight supercharges has a gravity multiplet, tensor multiplets, vector multiplets, and hypermultiplets. Let the numbers of tensor multiplets, vector multiplets, and hypermultiplets be $T$, $V$, and $H$ respectively. On the other hand, an elliptically fibered 3-fold $\tilde{X}$ has K\"ahler and complex structure moduli. Their numbers are related to the Hodge numbers of $\tilde{X}$: the number of K\"ahler moduli is $h^{1, 1}(\tilde{X})$ and the number of complex structure moduli is $h^{2, 1}(\tilde{X})$. The Hodge numbers of $\tilde X$ can be expressed in terms of $T$, $V$, and $H$. 

Strings in a 6d supergravity theory couple to anti-self-dual two-forms in tensor multiplets and a self-dual two-form in a gravity multiplet. From the F-theory viewpoint, strings are given by D3-branes wrapped on two-cycles in $B$. The string charges satisfy a Dirac pairing and the signature $(1, T)$ string charge lattice is identified with $H_2(B, \mathbb{Z})$. It follows that the number of tensor multiplets is given by\footnote{When $B$ is either an Enriques surface, a blow up of $\mathbb{P}^2$, or Hirzebruch surface $\mathbb{F}_n$, we have $h^{2,0}(B) = 0$.}
\begin{equation}
T = h^{1,1}(B) - 1. \label{tensor}
\end{equation}
Geometrically, the subtracted contribution in the above expression corresponds to the K\"ahler modulus which controls the overall size of the base $B$.

In order to see other relations, we consider a $T^2$ compactification of the 6d theory. The dimensional reduction yields a 4d $\mathcal{N}=2$ theory. The dimensional reduction splits the 6d gravity multiplet into a 4d gravity multiplet and two 4d abelian vector multiplets. The 6d tensor multiplet reduces to a 4d abelian vector multiplet. The 6d vector multiplet and hypermultiplet become a 4d vector multiplet and hypermultiplet respectively. On the other hand, F-theory compactified on $X \times T^2$ is dual to type IIA string theory on $X$ and is described at low energy by 4d $\mathcal{N}=2$ supergravity. At a generic point of the Coulomb branch, there are only abelian gauge symmetries and the low energy effective field theory is given by type IIA supergravity theory on the resolved 3-fold $\tilde{X}$. In this case, dimensional reduction implies  the number of abelian vector multiplets is $h^{1,1}(\tilde{X})$ and the number of massless hypermultiplets neutral under the abelian gauge symmetries is $h^{2, 1}(\tilde{X}) + 1$. Let $r(V)$ be the number of 6d vector multiplets in the Cartan subalgebra and let $H_0$ be the number of 6d hypermultiplets neutral under the Cartan. The duality with type IIA string theory then yields relations
\begin{eqnarray}
r(V) + T + 2 &=& h^{1,1}(\tilde{X}), \label{vector.tensor}\\
H_0 &=& h^{2,1}(\tilde{X}) + 1.   \label{neutralhyper}
\end{eqnarray}
The additional contribution in \eqref{neutralhyper} geometrically corresponds to the K\"ahler modulus controlling the overall size of $B$ which is subtracted in \eqref{tensor}. Combining \eqref{vector.tensor} with \eqref{tensor} gives
\begin{eqnarray}
r(V) = h^{1,1}(\tilde{X}) - h^{1,1}(B) - 1.\label{abelianvector}
\end{eqnarray}
The numbers of 6d vector multiplets and hypermultiplets charged under the Cartan subalgebra can be determined from the singularity structure as explained before.

\subsection{Anomaly cancellation}
\label{sec:anomaly}
6d supergravity theories may have anomalies which are characterized by the anomaly polynomial 8-form $I_8(R, F)$ where $R$ and $F$ are the spacetime and the Yang-Mills curvatures respectively. The 6d gravitational, non-abelian gauge, and mixed gauge-gravitational anoamlies can be cancelled by the Green-Schwarz mechanism \cite{Green:1984sg, Green:1984bx, Nishino:1985xp} provided the anomaly polynomial factorizes as \cite{Sagnotti:1992qw, Sadov:1996zm, Kumar:2010ru}
\begin{equation}
I_8 = \frac{1}{2}\Omega^{\alpha\beta}X_{4, \alpha}X_{4, \beta}. \label{GS}
\end{equation}
In the above expression $\Omega^{\alpha\beta}$ is a symmetric bilinear form with a signature $(1, T)$ and the 4-form $X_{4, \alpha}$ is given by
\begin{equation}
X_{4, \alpha} = \frac{1}{2}a_{\alpha}\text{tr}R^2 + \sum_I b_{I, \alpha}\left(\frac{2}{\lambda_I}\text{tr}F_I^2\right),
\end{equation}
where $I$ labels simple algebras $\mathfrak{g}_I$ of the 6d gauge theory, $a, b_I$ are vectors in $\mathbb{R}^{1, T}$, and $\lambda_I$ are the normalization constants for the gauge algebras $\mathfrak{g}_I$ given in Table \ref{tb:lambdai}.
\begin{table}[t]
\centering
\begin{tabular}{|c||c|c|c|c|c|c|c|c|}
\hline
$\mathfrak{g}$& $\mathfrak{su}(N)$ & $\mathfrak{so}(N)$ & $\mathfrak{usp}(2N)$ & $\mathfrak{g}_2$ & $\mathfrak{f}_4$ & $\mathfrak{e}_6$ & $\mathfrak{e}_7$ & $\mathfrak{e}_8$\\
\hline
$\lambda$ & 1 & 2 & 1 & 2 & 6 & 6 & 12 & 60\\ 
\hline
\end{tabular}
\caption{The normalization constants $\lambda_I$ for simple algebras.}
\label{tb:lambdai}
\end{table}
Notice the number of $b_I$ does not necessarily agree with the dimension of the vector space $\mathbb{R}^{1,T}$. Anomaly cancellation via \eqref{GS} implies the following conditions:
\begin{eqnarray}
H - V &=& 273  - 29T \label{AC1}\\
0&=& B_{adj, I} - \sum_{R_I}n_{R_I}B_{R_I}, \label{AC2}\\
a \cdot a &=& 9 - T, \label{AC3}\\
a \cdot b_I&=& \frac{1}{6}\lambda_I\left(A_{adj, I} - \sum_{R_I}n_{R_I}A_{R_I}\right), \label{AC4}\\
b_I\cdot b_I&=&-\frac{1}{3}\lambda_I^2\left(C_{adj, I} - \sum_{R_I}n_{R_I}C_{R_I}\right), \label{AC5}\\
b_I \cdot b_J &=& \lambda_I\lambda_J\sum_{R_I, R'_J}n_{R_IR'_J}A_{R_I}A_{R'_J}, \quad I \neq J, \label{AC6}
\end{eqnarray}
where $\Omega^{\alpha\beta}$ is used for the inner products $a \cdot a, a \cdot b_I, b_I \cdot b_I, b_I \cdot b_J$. Here $A_R, B_R, C_R$ are defined through
\begin{eqnarray}
\text{tr}_RF^2 &=& A_R\text{tr}F^2,\\
\text{tr}_RF^4 &=& B_R\text{tr}F^4 + C_R\left(\text{tr}F^2\right)^2.
\end{eqnarray}
$n_{R_I}$ is the number of hypermultiplets in the representation $R_I$ of $\mathfrak{g}_I$ and $n_{R_IR'_J}$ is the number of hypermultiplets in the representation $(R_I, R'_J)$ of $\mathfrak{g}_I \oplus \mathfrak{g}'_J$. 
In fact the anomaly cancellation conditions \eqref{AC3}-\eqref{AC6} imply that these inner products are all integers\footnote{The integrality condition of \eqref{AC5} automatically implies the absence of global gauge anomalies for $\mathfrak{su}(2), \mathfrak{su}(3), \mathfrak{g}_2$ \cite{Bershadsky:1997sb}.} \cite{Kumar:2010ru} and hence the vectors $a$ and $b_I$ form an integral lattice $\Lambda$.

In the context of F-theory, $\Omega^{\alpha\beta}, a_{\alpha}, b_{\alpha}$ are given by geometric data \cite{Sadov:1996zm, Kumar:2009ac, Kumar:2010ru}. First note that the self-intersection number of the canonical divisor in $B$ is always given by
\begin{equation}
K_B^2 = 10 - h^{1,1}(B). \label{Ksquared}
\end{equation}
By using \eqref{tensor} we can write \eqref{Ksquared} as
\begin{equation}
K_B^2 = 9-T. \label{Ksquare2} 
\end{equation}
Since the right hand side of \eqref{Ksquare2} is exactly the same as the right hand side of \eqref{AC3}, we obtain the relation 
\begin{equation}
K_B^2 = a \cdot a. \label{asquared}
\end{equation}
Therefore, we can naturally identify the vector $a$ with the canonical divisor $K_B$, and the symmetric bilinear form $\Omega^{\alpha\beta}$ with the intersection form in $H_2(B, \mathbb{Z})$. From \eqref{AC6}, the inner product $b_I \cdot b_J$ is related to the number of charged hypermultiplets which arise at the intesrections between 7-branes. When 7-branes wrapped on a curve $C_I$ supports the gauge algebra $\mathfrak{g}_I$, it can be shown that \cite{Sadov:1996zm, Grassi:2000we, Grassi:2011hq}
\begin{equation}
 b_I \cdot b_J = C_I \cdot C_J. \label{csquared}
\end{equation}
Then it is natural to identify the vector $b_I$ with the divisor class $C_I$ in $H_2(B, \mathbb{Z})$ on which the 7-branes are wrapped. Since $I$ labels simple gauge factors, we associate $C_I$ to $b_I$ when there is some non-abelian gauge symmetry on 7-branes wrapping $C_I$. These results are also consistent with the fact that the string charge lattice is identified with $H_2(B, \mathbb{Z})$. The integrality of the inner products of vectors $a, b_I$ is automatic from the F-theory viewpoint since they are intersection numbers between divisors. The maps $a \to K_B, b_I \to C_I$ now induce a lattice embedding $\Lambda \hookrightarrow H_2(B, \mathbb{Z})$.  

Let us give an expression for the canonical divisor in terms of a basis for $H_2(B, \mathbb{Z})$. We denote a basis of $H_2(B, \mathbb{Z})$ by $\tilde{C}^{\alpha},\; (\alpha = 1, \cdots, 1+T)$. Then the canonical divisor can be expressed as
\begin{equation}
K_B = \sum_{\alpha} a_{\alpha}\tilde{C}^{\alpha}.
\end{equation}
The adjunction formula relates the genus $g^{\beta}$ of $C^{\beta} =\sum_{\alpha}\delta_{\alpha}^{\beta}\tilde{C}^{\alpha}$ with intersection numbers by
\begin{eqnarray}
2g^{\beta} - 2 &=& K_B\cdot C^{\beta} + \left(C^{\beta}\right)^2\nonumber\\
&=&\sum_{\alpha}\Omega^{\beta\alpha}a_{\alpha} + \left(C^{\beta}\right)^2,
\end{eqnarray}
where $\tilde{C}^{\alpha}\cdot \tilde{C}^{\beta} = \Omega^{\alpha\beta}$. Therefore $a_{\alpha}$ is given by
\begin{equation}
a_{\alpha} = \sum_{\beta}(\Omega^{-1})_{\alpha\beta}(2g^{\beta} - 2 - \left(C^{\beta}\right)^2). \label{afromselfint}
\end{equation}

\subsection{6d SCFTs and LSTs as non-compact limits of compact 3-folds}

So far we have focused on compactifications of F-theory on compact elliptically fibered CY 3-folds. It is possible to take a decompactification limit of the base $B$ with some curves in $B$ kept compact. In this case, gravity is decoupled from the resulting low energy effective field theory. After the decompactification limit, there may be several disconnected collections of curves, giving disconnected field theories. We will focus on one connected configuration which gives a single theory. As discussed in \cite{Bhardwaj:2015oru} the decompactification limit leads to two possible cases: one case is that the intersection matrix of curves in $H_2(B, \mathbb{Z})$ is negative definite and the other case is that the intersection matrix is negative semidefinite with a one-dimensional eigenspace with a zero eigenvalue. The former yields a 6d theory with a conformal fixed point and the latter gives a little string theory (LST). When the intersection matrix is negative definite, it is possible to contract all the curves. Physically, this corresponds to turning off all the vevs for scalars in tensor multiplets and hence the 6d theory reaches a conformal fixed point characterized by the appearance of tensionless strings (coming from D3-branes wrapped on shrinking two-cycles.) On the other hand, when the intersection matrix is negative semidefinite, a curve associated to the null direction corresponds to a non-contractible curve. In this case, the volume of the non-contractible curve is controlled by the vev of a scalar in a non-dynamical tensor multiplet. The vev is related to a dimensionful parameter of the LST, namely the tension of a little string.

Then, classifying 6d SCFTs or LSTs in a maximally Higgsed phase amounts to classifying possible sets of curves which yield a negative definite or a negative semidefinite intersection matrix \cite{Heckman:2013pva, Heckman:2015bfa, Bhardwaj:2015oru}. In fact, these configurations of curves can be obtained by gluing ``non-Higgsable clusters'' \cite{Morrison:2012np}. The list of the non-Higgsable clusters associated with an isolated curve is given in Table \ref{tb:NHC1}.
\begin{table}[t]
\centering
\begin{tabular}{|c||c|c|c|c|c|c|c|}
\hline
self-intersection& $-3$ & $-4$ & $-5$ & $-6$ & $-7$ & $-8$ & $-12$\\
\hline
gauge & $\mathfrak{su}(3)$ & $\mathfrak{so}(8)$ & $\mathfrak{f}_4$ & $\mathfrak{e}_6$ & $\mathfrak{e}_7$ & $\mathfrak{e}_7$ & $\mathfrak{e}_8$\\
\hline
hyper & - & - & - & - & $\frac{1}{2}{\bf 56}$ & -&  -\\
\hline
\end{tabular}
\caption{Non-Higgsable clusters with an isolated curve.}
\label{tb:NHC1}
\end{table}
The local geometry of the non-Higgsable cluters is a line bundle $\mathcal{O}(-n)$ over $\mathbb{P}^1$ with $n=3, 4, 5, 6, 7, 8, 12$. We refer to the theories as $\mathcal{O}(-n)$ theories. Note that some gauge symmetries are still unbroken in the maximally Higgsed phase. 
There are additional non-Higgsable clusters with more than one curve, as displayed in Table \ref{tb:NHC2}. The intersection numbers between curves with non-Abelian gauge algebras for the non-Higgsable clusters in Table \ref{tb:NHC1} and Table \ref{tb:NHC2} satisfy the gauge anomaly cancellation conditions \eqref{AC5} and \eqref{AC6}. 
\begin{table}[t]
\centering
\begin{tabular}{|c||c|c|c|}
\hline
self-intersection & $-3, -2$ & $-3, -2, -2$ & $-2, -3, -2$\\
\hline
gauge & $\mathfrak{g}_2 \oplus \mathfrak{su}(2)$ & $\mathfrak{g}_2 \oplus \mathfrak{sp}(1)$ & $\mathfrak{su}(2) \oplus \mathfrak{so}(7) \oplus \mathfrak{su}(2)$\\
\hline
hyper & $\frac{1}{2}({\bf 7} + {\bf 1} , {\bf 2})$ & $\frac{1}{2}({\bf 7} + {\bf 1}, {\bf 2})$ & $\frac{1}{2}({\bf 2}, {\bf 8}, {\bf 1}) \oplus \frac{1}{2}({\bf 1}, {\bf 8}, {\bf 2})$\\
\hline
\end{tabular}
\caption{Non-Higgsable clusters with two or three curves. In the second example, no gauge algebra is supported on the rightmost curve with self-intersection $-2$. }
\label{tb:NHC2}
\end{table} 
For gluing the non-Higgsable clusters we use $\mathbb{P}^1$'s with self-intersection $-1$. Namely, we insert $\mathbb{P}^1$'s with self-intersection $-1$ between sets of curves describing non-Higgsable clusters. The local geometry used for the gluing, which is a line bundle $\mathcal{O}(-1)$ over $\mathbb{P}^1$, also leads to a 6d SCFT called the E-string theory.

\subsection{Decomposing orbifolds of $T^6$ into non-compact 3-folds}

A number of elliptically fibered CY 3-folds can be realized as toroidal orbifolds, including non-compact 3-folds which realize some of the SCFTs and LSTs described in the previous subsection, as well as compact 3-folds of the form $T^6/ \Gamma$ which realize 6d $(1,0)$ supergravity theories. The geometric relationship between these various constructions is instrumental in developing the perspective that the non-gravitational sector of certain 6d $(1,0)$ supergravities is captured by configurations of 6d SCFTs. Since we study several examples of toroidal orbifolds in depth in Section \ref{sec:6dsugra.from.orbifolds}, we pause briefly here to summarize their basic characteristics. 

A six torus $T^6$ can be constructed from a lattice $L$ with an identification 
\begin{equation}
x \sim x + v, \quad v \in L,
\end{equation}
where $x \in \mathbb{R}^6$. Namely, $T^6$ is a quotient $\mathbb{R}^6/L$. We then consider a quotient of $T^6$ by a point group $\Gamma$ for obtaining an orbifold. The point group must be an automorphism of the lattice $L$. We focus on only abelian point groups. In order for the orbifold to be a CY 3-fold (with an $SU(3)$ holonomy), the point group $\Gamma$ should be a subgroup of $SU(3)$. Let the complex coordinates of $T^6$ be $(z_1, z_2, z_3)$. Then the orbifold action of $g \in \Gamma$ is given by
\begin{equation}
g: \quad (z_1, z_2, z_3) \quad \to \quad (e^{2\pi i v_1} z_1, e^{2\pi i v_2} z_2, e^{2\pi i v_3} z_3), 
\end{equation}
where $g^N = 1$ for some integer $N$. Since $\Gamma \subset SU(3)$, we require that 
\begin{equation}
v_1 + v_2 + v_3 = 0. \label{orbifoldSU3}
\end{equation}
The condition \eqref{orbifoldSU3} together with the requirement that the point group is a symmetry of the lattice imposes  stringent constraints on the possible choices of $\Gamma$. In fact, $\Gamma$ needs to be either $\mathbb{Z}_n$ with $n=3, 4, 6, 7, 8, 12$ or $\mathbb{Z}_m \times \mathbb{Z}_{n}$ where $n$ is a multiple of $m$ with $m = 2, 3, 4, 6$ \cite{Dixon:1985jw, Dixon:1986jc}. The orbifolds are not smooth 3-folds due to the presence of fixed points and fixed lines. In this paper we consider only a subset of all possible examples point groups $\mathbb{Z}_m \times \mathbb{Z}_n$ examples; their orbifold actions are summarized in Table \ref{tb:orbifold1}.
\begin{table}[t]
\centering
\begin{tabular}{|c||c|c|}
\hline
$\Gamma=\mathbb{Z}_m \times \mathbb{Z}_n$ & $(v_1, v_2, v_3)$ for $\mathbb{Z}_m$ &  $(v_1, v_2, v_3)$ for $\mathbb{Z}_n$\\
\hline
$\mathbb{Z}_2 \times \mathbb{Z}_2$ & $\frac{1}{2}(-1, 1, 0)$ & $\frac{1}{2}(-1, 0, 1)$\\
\hline
$\mathbb{Z}_2 \times \mathbb{Z}_4$ & $\frac{1}{2}(-1, 1, 0)$ & $\frac{1}{4}(-1, 0, 1)$\\
\hline
$\mathbb{Z}_2 \times \mathbb{Z}_6$ & $\frac{1}{2}(-1, 1, 0)$ & $\frac{1}{6}(-1, 0, 1)$\\
\hline
$\mathbb{Z}_3 \times \mathbb{Z}_3$ & $\frac{1}{3}(-1, 1, 0)$ & $\frac{1}{3}(-1, 0, 1)$\\
\hline
$\mathbb{Z}_3 \times \mathbb{Z}_6$ & $\frac{1}{3}(-1, 1, 0)$ & $\frac{1}{6}(-1, 0, 1)$\\
\hline
$\mathbb{Z}_4 \times \mathbb{Z}_4$ & $\frac{1}{4}(-1, 1, 0)$ & $\frac{1}{4}(-1, 0, 1)$\\
\hline
$\mathbb{Z}_6 \times \mathbb{Z}_6$ & $\frac{1}{6}(-1, 1, 0)$ & $\frac{1}{6}(-1, 0, 1)$\\
\hline
\end{tabular}
\caption{Generators of $\Gamma = \mathbb{Z}_m \times \mathbb{Z}_n$.}
\label{tb:orbifold1}
\end{table}

Having summarized the basic properties of compact toroidal orbifolds $T^6/\Gamma$, we now turn our attention to orbifold realizations of some examples of non-compact 3-folds, and we will shortly see how these non-compact orbifolds can naturally be ``stitched'' together to form local descriptions of the compact orbifolds. First, note that some of the non-Higgsable cluster theories have an orbifold realization. Consider the 3-fold $X = (T^2 \times \mathbb{C}^2)/\mathbb{Z}_n$ \cite{Witten:1996qb, DelZotto:2015rca}. The orbifold action is given by
\begin{equation}
\quad (z_1, z_2, z_3) \quad \to \quad  (\omega^2z_1, \omega^{-1}z_2, \omega^{-1}z_3), \label{OnZn}
\end{equation}
where $\omega^n = 1$. Note that $z_1$ is a complex coordinates of $T^2$ and $z_2, z_3$ are complex coordinates of $\mathbb{C}^2$. Since the orbifold action is an isometry of the torus $T^2$, $n$ is restricted to the values $n=2, 3, 4, 6, 8, 12$. The $n=2$ case yields 6d $\mathcal{N}=(2, 0)$ SCFT of $A_1$ type. On the other hand, $n=3, 4, 6, 8, 12$ corresponds to the $\mathcal{O}(-n)$ theories with the same $n$. 

\begin{figure}[t]
  \centering
  \includegraphics[width=.80\linewidth]{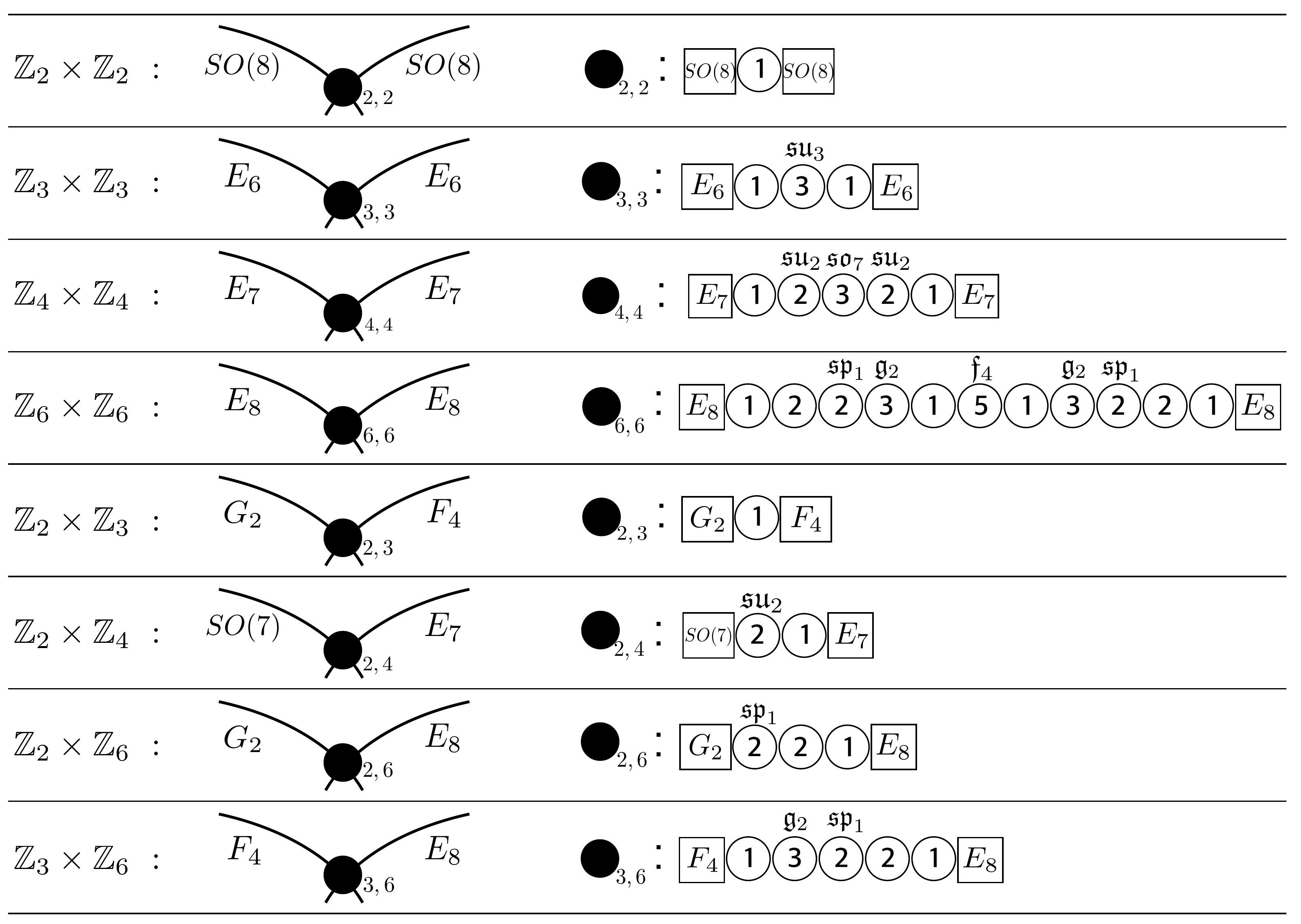}
  \caption{Conformal matter theories from elliptic CY 3-folds of $T^2\times \mathbb{C}^2/\mathbb{Z}_m\times \mathbb{Z}_n$.}
  \label{fig:orbifold}
\end{figure}

One may also consider 6d SCFTs realized by $X=T^2\times \mathbb{C}^2/\Gamma$ with point group $\Gamma=\mathbb{Z}_m\times \mathbb{Z}_n$. The orbifold action of $\Gamma$ is of the form 
\begin{equation}\label{eq:orbifold-action}
	g = (\alpha,\alpha^{-1},1) \ , \quad h=(\omega,1,\omega^{-1}) \ , \quad \alpha^m=1 \ , \quad \omega^n=1 \ .
\end{equation}
The orbifold generators $g$ and $h$ act trivially on one of two $\mathbb{C}$-planes in the base $B=\mathbb{C}\times \mathbb{C}$. It is known in \cite{Sen:1996vd,Dasgupta:1996ij} that each action gives rise to a line of singularities with gauge group $G=SO(8),E_6,E_7,E_8$ in the base for $m,n=2,3,4,6$ respectively. In F-theory, the gauge group $G$ is the symmetry on the 7-branes wrapped on the singular line. Moreover, since the trivial entries of $1$ in the actions of $g$ and $h$ are orthogonal to each other, the total orbifold action $\Gamma$ leads to two intersecting lines of singularities on the base $B$. Each singular line hosts an independent symmetry group $G$.
As studied in \cite{DelZotto:2015rca}, F-theory on this type of orbifold $X/\Gamma$ engineers 6d SCFTs called $(G,G')$ conformal matter theories \cite{DelZotto:2014hpa}.  The corresponding conformal matter theories, which depend on a choice of $\Gamma$, are given in Figure \ref{fig:orbifold}.
Other possible $\mathbb{Z}_m\times \mathbb{Z}_n$ actions are not compatible with the complex structure of $T^2$, so they cannot generate a consistent elliptic 3-fold. 
Note that when $m$ and $n$ are different, the symmetries on two singular lines become smaller from the naive expectation of $G=SO(8),E_6$ in certain cases.
This is because there is automorphism among the fixed points of an orbifold action, say $g$, induced by the other orbifold action, say $h$.
For the $\mathbb{Z}_2\times \mathbb{Z}_3$ case, the $SO(8)$ reduces to $G_2$ due to the $\mathbb{Z}_3$ outer automorphism by the $h$ action with $m=3$ and the $E_6$ reduces to $F_4$ due to the $\mathbb{Z}_2$ outer automorphism by the $g$ action with $n=2$. Similarly, the $SO(8)$ symmetry of the $\mathbb{Z}_2\times \mathbb{Z}_4$ orbifold becomes $SO(7)$ due to the $\mathbb{Z}_2$ outer automorphism induced by the $\mathbb{Z}_4$ orbifold action. The $SO(8)$ symmetry of the $\mathbb{Z}_2\times \mathbb{Z}_6$ orbifold and the $E_6$ symmetry of the $\mathbb{Z}_3\times \mathbb{Z}_6$ orbifold reduce to $G_2$ and $F_4$ (respectively) due to the $\mathbb{Z}_3$ and $\mathbb{Z}_2$ outer automorphisms induced by the $\mathbb{Z}_6$ action.

We next turn to 6d LSTs coming from $X=T^4\times \mathbb{C}/\mathbb{Z}_m\times \mathbb{Z}_n$.  LSTs can be realized by a simple generalization of the 6d SCFTs discussed above. Under the orbifold actions $g$ and $h$, the torus $T^2$ in the base $B=T^2\times \mathbb{C}$ will have a number of fixed points. The local geometry around each fixed point can be approximated as an elliptic 3-fold that gives rise to one of the 6d SCFTs in Figure \ref{fig:orbifold}. Therefore the full geometry of $X$ contains several copies of the 6d SCFTs in Figure \ref{fig:orbifold} localized to the fixed points of $T^2$ and these local SCFTs are glued together in an appropriate manner. Each fixed point theory has $G\times G'$ type global symmetry. One of $G\times G'$ symmetries generated by an orbifold acting trivially on the torus is localized to the compact $T^2$ in the base; this symmetry is gauged. Gauging of this global symmetry in all local 6d SCFTs gives rise to a 6d LST associated to the geometry $X$. For example, as we will discuss in more detail in the next section, the geometry $X=T^4\times \mathbb{C}/\mathbb{Z}_2\times \mathbb{Z}_2$ leads to the LST in Figure \ref{fig:T2-Z2Z2}. One of the $\mathbb{Z}_2$ actions has four fixed points on the torus and each fixed point is described by an E-string theory; these four E-string theories are glued by gauging the common $SO(8)$ symmetry carried by the $-4$ curve in the center. Many other examples of LSTs will arise as certain components in the construction of 6d supergravity models.

To summarize, in the above examples we explained how one can construct the field theory of an elliptic 3-fold by first identifying local theories around fixed points of the orbifold action on $T^2$ and then properly gluing these local theories together by gauging certain global symmetries. We can easily extend this to the construction of F-theory models on the compact orbifold $T^6/\Gamma$, where $\Gamma=\mathbb{Z}_m\times\mathbb{Z}_n$. In the next section, we will present explicit constructions of several 6d supergravity models arising from F-theory on $T^6/\Gamma$.

Before moving on, we describe some additional geometric aspects of $T^6/\Gamma$. Toroidal orbifolds $T^6/\Gamma$ have K\"ahler moduli and complex structure moduli as ordinary CY 3-folds. In each case, there are two types of geometric moduli: untwisted moduli and twisted moduli. The number of untwisted K\"ahler moduli is equal to the number of $(1, 1)$-forms $dz_i \wedge d\bar{z}_j, \; i, j=1, 2, 3$ which are invariant under the orbifold action. Since $dz_i \wedge d\bar{z}_i$ for $i=1, 2, 3$ are always invariant under the orbifold actions, these orbifolds all have at least three untwisted K\"ahler moduli. On the other hand, the number of untwisted complex structure moduli is equal to the number of $(2, 1)$-forms $dz_i \wedge dz_j \wedge d\bar{z}_k, \; i,j,k=1,2,3$ which are invariant under the orbifold actions. In order to see twisted moduli we need to resolve singularities of orbifolds. The singularities may be resolved by introducing exceptional divisors. The deformations of the K\"ahler forms associated to the exceptional divisors give rise to twisted K\"ahler moduli. In other words, the twisted K\"ahler moduli are set to be zero in the orbifold limit. Furthermore, there may be also twisted complex structure moduli. Twisted complex structure moduli arise when orbifolds contain fixed lines without any fixed points \cite{Lust:2006zg, Lust:2006zh}. 
For example, when there is a $\mathbb{Z}_N$ fixed line over a curve with $T^2$ topology then the resolution yields $N-1$ twisted complex structure moduli. The numbers of the geometric moduli for the $\mathbb{Z}_m \times \mathbb{Z}_n$ orbifolds which we will consider in this paper are summarized in Table \ref{tb:orbifold2}.
\begin{table}[t]
\centering
\begin{tabular}{|c||c|c|c|c|}
\hline
$\Gamma=\mathbb{Z}_m \times \mathbb{Z}_n$ & $h^{1,1}_{\text{untwisted}}$ &  $h^{2,1}_{\text{untwisted}}$ & $h^{1,1}_{\text{twisted}}$  & $h^{2,1}_{\text{twisted}}$ \\
\hline
$\mathbb{Z}_2 \times \mathbb{Z}_2$ & 3 & 3 & 48 & 0\\
\hline
$\mathbb{Z}_2 \times \mathbb{Z}_4$ & 3 & 1 & 58 & 0\\
\hline
$\mathbb{Z}_2 \times \mathbb{Z}_6$ & 3 & 1 & 48 & 2\\
\hline
$\mathbb{Z}_3 \times \mathbb{Z}_3$ &  3 & 0 & 81 & 0\\
\hline
$\mathbb{Z}_3 \times \mathbb{Z}_6$ &  3 & 0 & 70 & 1\\
\hline
$\mathbb{Z}_4 \times \mathbb{Z}_4$ &  3 & 0 & 87 & 0\\
\hline
$\mathbb{Z}_6 \times \mathbb{Z}_6$ &  3 & 0 & 81 & 0\\
\hline
\end{tabular}
\caption{The numbers of the geometric moduli for the $\mathbb{Z}_m \times \mathbb{Z}_n$ orbifolds.}
\label{tb:orbifold2}
\end{table}

\section{6d supergravity from F-theory on $T^6/\mathbb{Z}_m\times \mathbb{Z}_n$}
\label{sec:6dsugra.from.orbifolds}

In this section, we consider 6d $\mathcal{N}=1$ supergravity theories realized by elliptic CY 3-folds of orbifold type $T^6/\mathbb{Z}_m\times \mathbb{Z}_n$ in F-theory. We propose 6d field theories which fully capture the non-gravitational sector of the 6d supergravity theory associated to these orbifolds. We also investigate a connection between the elliptic genus of the self-dual strings in the 6d field theories and the topological string partition function.

\subsection{$T^6/\mathbb{Z}_m\times \mathbb{Z}_n$}
\label{sec:T6ZmZn}

We first focus on the geometry of the orbifolds $T^6/ \mathbb Z_m \times \mathbb Z_n$. We can interpret these compact 3-folds as being constructed by gluing local non-compact 3-folds around the fixed points of the orbifold action discussed in the previous section. One might then ask if the field theory sector of the corresponding supergravity theory can be fully constructed by gluing the SCFT models obtained from the local non-compact 3-folds in F-theory and coupling the resulting system to gravity. Following this local analysis, we now construct the glued field theory sectors and provide evidence that they fully capture the non-gravitational sector. 
As a byproduct of this analysis we construct 6d LSTs in F-theory compactified on $T^4\times \mathbb{C}/\mathbb{Z}_m\times \mathbb{Z}_n$.

Two important consistency checks we perform are: 1) that the 6d field content satisfies anomaly cancellation via the Green-Schwarz-West-Sagnotti mechanism, and 2) that the lattice of self-dual string charges is unimodular, self-dual, and of appropriate signature as is necessary \cite{Seiberg:2011dr} for the existence of a consistent 6d supergravity theory with eight supercharges. Performing these consistency checks requires  that we read off the 6d field content using the usual F-theory dictionary as well as identifying a basis for $H_2(B,\mathbb Z)$. 

The latter task is straightforward: we simply compute the endpoints of the bases $B = T^4 / \mathbb Z_m \times \mathbb Z_n$ by blowing down all $-1$ curves and then identifying the remaining independent complex 1-cycles. The set of exceptional curves together with the independent curves of the endpoint form a basis $\{ C^\alpha\}$ of $H_2(B,\mathbb Z)$ in terms of which all curves $C_I$ may be expressed. In all of the cases we study, the endpoints are of the form $\mathbb F_0 = \mathbb P^1 \times \mathbb P^1$, and hence apart from the $-1$ curves there are only two independent cycles $h,v$ corresponding to the ``horizontal'' and ``vertical'' $\mathbb P^1$'s of the ruling. We outline this procedure in the specific case of the model $T^6 /\mathbb Z_2 \times \mathbb Z_2$ in the following subsection, and note that the remaining models are treated in an analogous fashion.

\subsubsection{$T^6/\mathbb{Z}_2\times \mathbb{Z}_2$}

\begin{figure}
\begin{subfigure}{.5\textwidth}
  \centering
  \includegraphics[width=.6\linewidth]{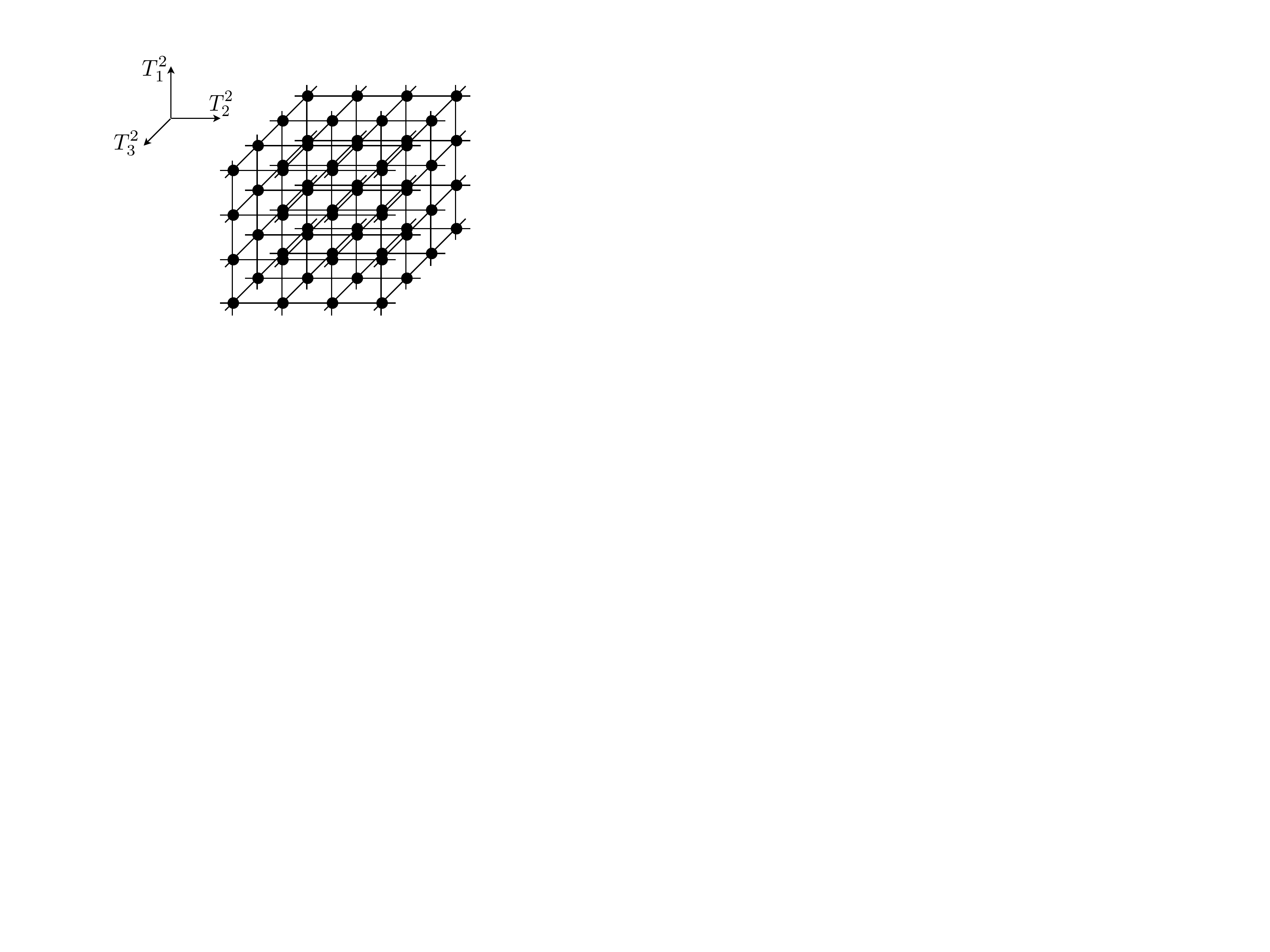}
  \caption{}
  \label{fig:T6-Z2Z2}
\end{subfigure}%
\begin{subfigure}{.5\textwidth}
  \centering
  \includegraphics[width=.6\linewidth]{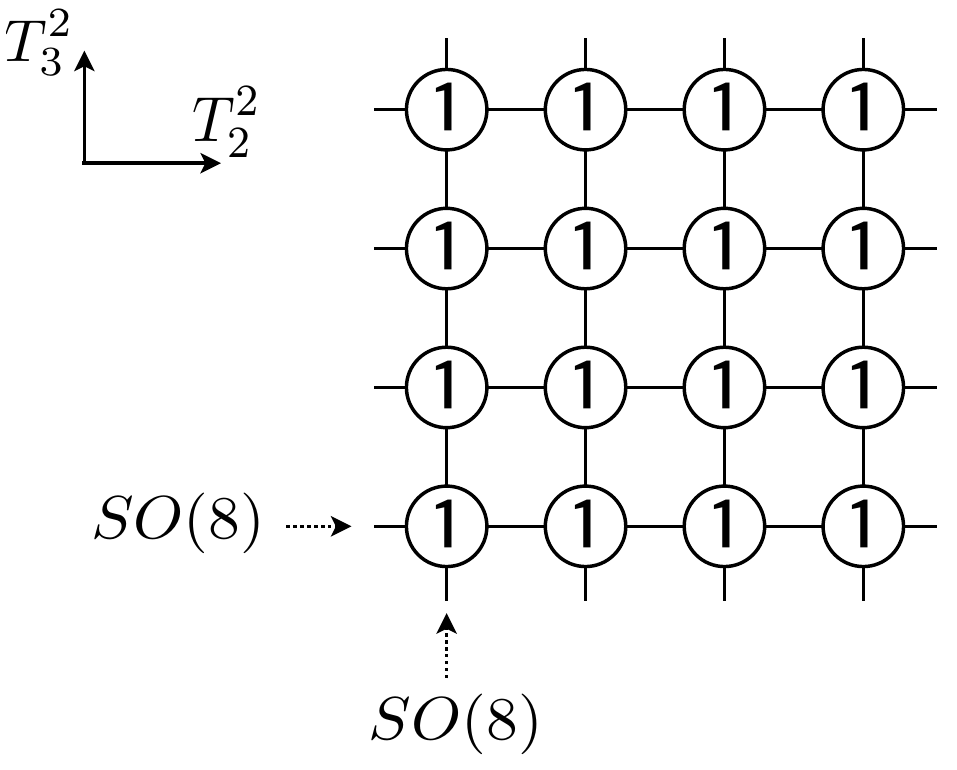}
  \caption{}
  \label{fig:T4-Z2Z2}
\end{subfigure}
\caption{$T^6/\mathbb{Z}_2\times \mathbb{Z}_2$ model where $T^6=T^2_1\times T^2_2\times T^2_3$. This geometry can be considered as a collection of local $\mathbb{C}^3/\mathbb{Z}_2\times \mathbb{Z}_2$ geometries around 64 orbifold fixed points like (a). Or this can also be considered to be gluing 16 local $\mathcal{O}(-1)$ minimal CFTs (or E-string theories) on the base $T^4=T^2_2\times T^2_3$ like (b).}
\label{fig:CY3-T6-Z2Z2}
\end{figure}

The simplest example is the $T^6/\mathbb{Z}_2\times \mathbb{Z}_2$ model. The orbifold action $\Gamma=\mathbb{Z}_2\times \mathbb{Z}_2$ on $T^6=T^2\times T^2\times T^2$ is generated by
\begin{equation}
	g = (-1,-1,1) \ , \quad h = (-1,1,-1) \ .
\end{equation}
This action has four fixed points on each torus $T^2$. The local geometry around each fixed point is $\mathbb{C}^3/\mathbb{Z}_2\times \mathbb{Z}_2$.
This implies that we have 64 such local geometries on $T^6$. Hence, the total compact 3-fold is obtained by gluing together these 64 local geometries in the manner depicted in Figure \ref{fig:T6-Z2Z2}.

In the previous section, we studied the non-compact 3-fold $T^2\times \mathbb{C}^2/\mathbb{Z}_2\times \mathbb{Z}_2$. In this case, we glue four copies of the local $\mathbb{C}^3/\mathbb{Z}_2\times \mathbb{Z}_2$ geometry at the fixed points of the elliptic fiber $T^2$. In F-theory, this gives rise to the 6d E-string theory or the $(D_4,D_4)$ conformal matter theory. In this section we replace $\mathbb{C}^2$ by $T^4=T^2\times T^2$.

The orbifold action  leads to 16 fixed points  on the base $B=T^4$. The local geometry near each fixed point is an elliptic CY 3-fold of $T^2\times \mathbb{C}^2/\mathbb{Z}_2\times \mathbb{Z}_2$ as depicted in Figure \ref{fig:T4-Z2Z2}. The horizontal and vertical lines in Figure \ref{fig:T4-Z2Z2} are $SO(8)$ seven-branes. Two $SO(8)$ seven-branes intersect at each fixed point. So the local theory at a fixed point has $SO(8)\times SO(8)$ global symmetry, which enhances to $E_8$ symmetry, coming from two $SO(8)$ flavor branes. However, the seven-branes are wrapping on one of two compact $T^2$'s. Hence, all the $SO(8)$ symmetries on the seven-branes become gauge symmetries when placed on the compact base $B$. Therefore, in the compact 3-fold, we have eight $SO(8)$ gauge symmetries gluing 16 $(D_4,D_4)$ conformal matter theories (or E-string theories) denoted by \raisebox{.5pt}{\textcircled{\raisebox{-.9pt} {1}}} in Figure \ref{fig:T4-Z2Z2}. Each seven-brane denoted by a horizontal or a vertical line intersects with four E-string theories and gauges a common $SO(8)$ global symmetry of these four E-string theories.
The theory living on the $SO(8)$ seven-brane in the compact model is also a 6d SCFT known as the $\mathcal{O}(-4)$ minimal SCFT. Therefore, the gravity theory of this compact model can be understood as a theory interacting with 16 $\mathcal{O}(-1)$ minimal CFTs glued by 8 $\mathcal{O}(-4)$ minimal CFTs. Each $\mathcal{O}(-1)$ theory intersects with two, a horizontal and a vertical, $\mathcal{O}(-4)$ theories, while each $\mathcal{O}(-4)$ theory intersects with four $\mathcal{O}(-1)$ theories. The global symmetry $SO(8)\times SO(8)$ in an $\mathcal{O}(-1)$ theory is gauged by two adjacent $\mathcal{O}(-4)$ theories.
We claim that this 6d field theory together with three neutral hypermultiplets when coupled to gravity describes the  massless degrees of freedom of 6d supergravity theory of $T^6/\mathbb{Z}_2\times \mathbb{Z}_2$.

\begin{figure}
  \centering
  \includegraphics[width=.25\linewidth]{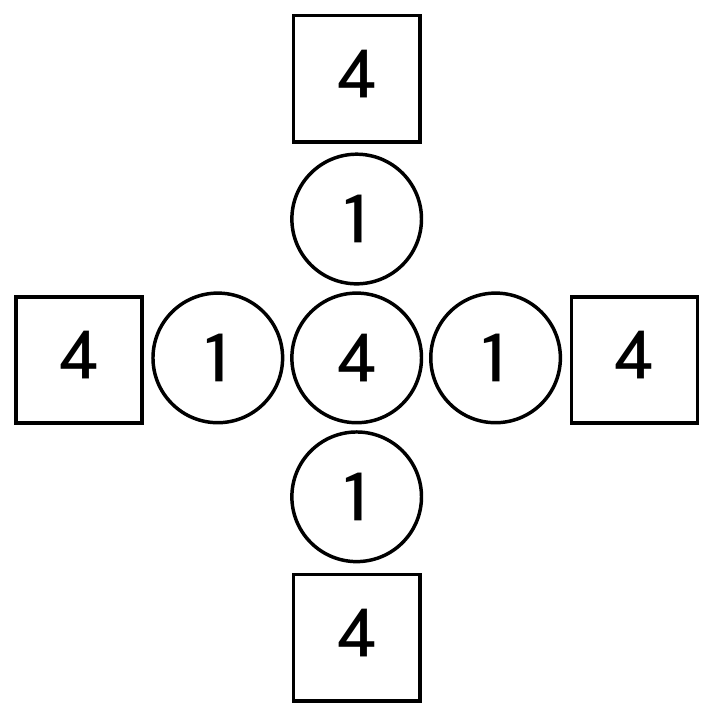}
  \caption{6d Little string theory of $T^4\!\times\!\mathbb{C}/\mathbb{Z}_2\!\times\!  \mathbb{Z}_2$. Each box with `4' denotes a $SO(8)$ flavor 7-brane.}
  \label{fig:T2-Z2Z2}
\end{figure}

We can also consider a doubly elliptic 3-fold $(T^4\times \mathbb{C})/\mathbb{Z}_2\times \mathbb{Z}_2$. This geometry in F-theory engineers a 6d LST. The orbifold action $\Gamma=\mathbb{Z}_2\times \mathbb{Z}_2$ leads to four fixed points in the base $B=T^2\times \mathbb{C}$. As we discussed above, each fixed point is associated to a 6d $\mathcal{O}(-1)$ minimal SCFT. There is an $SO(8)$ seven-brane, which wraps the torus $T^2$ in the base, intersecting with all four fixed points. Also four other $SO(8)$ flavor seven-branes each intersect with one of the four fixed points. The theory living on the seven-brane wrapping the torus on the base is the $\mathcal{O}(-4)$ minimal CFT. Therefore, the final theory is the 6d theory with four $\mathcal{O}(-1)$ theories and one $\mathcal{O}(-4)$ theory. One of the $SO(8)$ symmetries in each $\mathcal{O}(-1)$ theory is gauged by the $SO(8)$ symmetry of the $\mathcal{O}(-4)$ theory as drawn in Figure \ref{fig:T2-Z2Z2}. This theory has $SO(8)^4$ flavor symmetry coming from the four flavor 7-branes. 

There are five tensor nodes in this 6d theory as indicated in Figure \ref{fig:T2-Z2Z2}. The intersection matrix for them is
\begin{equation}
	\Omega^{LST}_{\mathbb{Z}_2\times\mathbb{Z}_2} = \left(\begin{array}{ccccc}
	-1 & 0 & 0 & 0 & 1 \\
	0 & -1 & 0 & 0 & 1 \\
	0 & 0 & -1 & 0 & 1 \\
	0 & 0 & 0 & -1 & 1 \\
	1 & 1 & 1 & 1 & -4 \\
	\end{array}\right) \ ,
\end{equation}
where the first four entries (row or column) correspond to the four $\mathcal{O}(-1)$ theories and the last entry corresponds to the $\mathcal{O}(-4)$ theory.
Among these five tensors, one combination, say a null tensor $T_{0}=(1,1,1,1,1)$, has zero eigenvalue with respect to this intersection matrix. The presence of one null tensor $T_0$ is a distinguished property of a 6d LST. This null tensor multiplet is non-dynamical as it has no kinetic term. The scalar field $\phi_0$ in this non-dynamical tensor multiplet sets the little string scale, $\frac{1}{2\pi \alpha'} = \phi_0$.

\begin{figure}
  \centering
  \includegraphics[width=.45\linewidth]{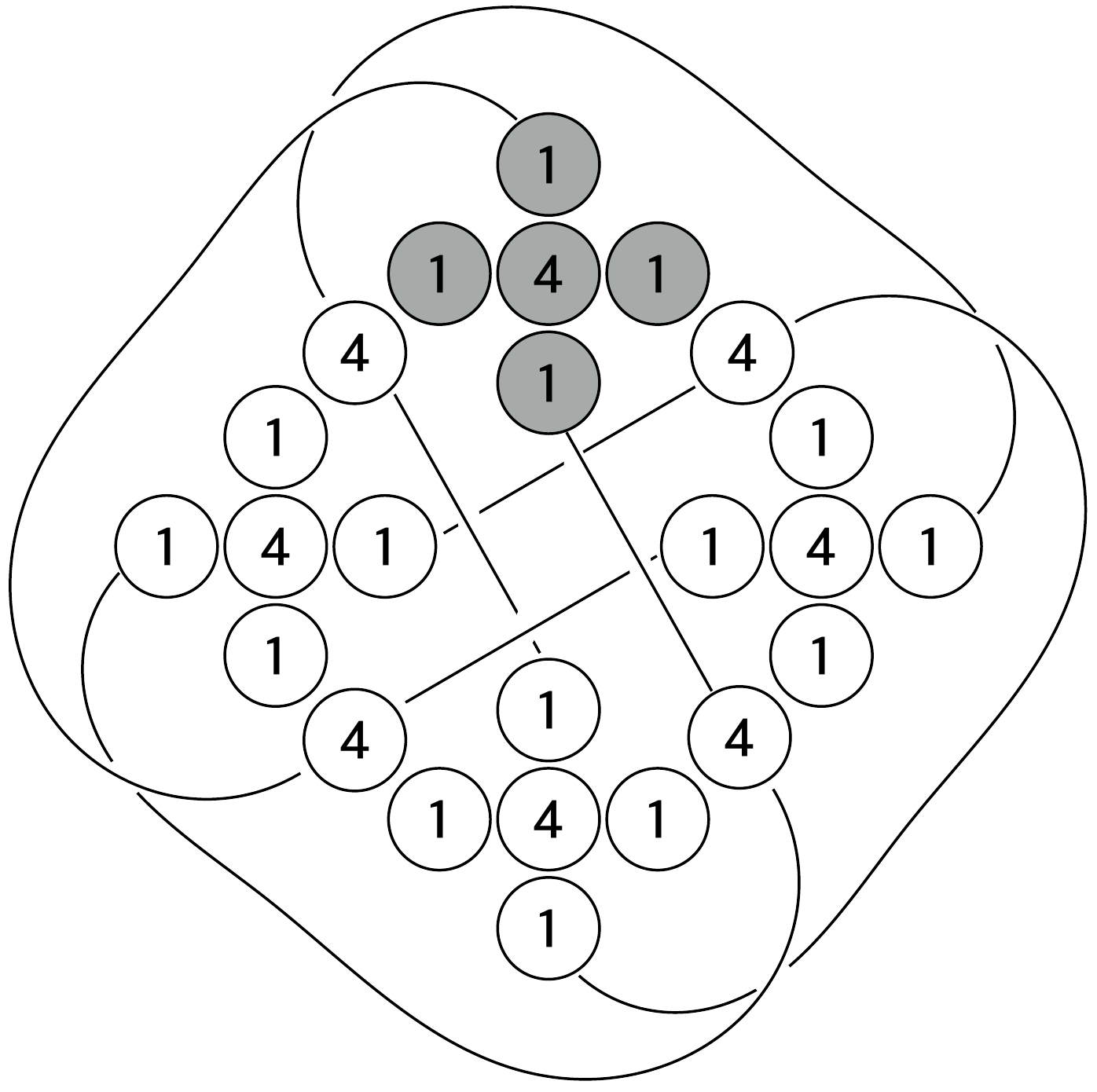}
  \caption{6d gravity theory of $T^6/\mathbb{Z}_2\times \mathbb{Z}_2$. It consists of 16 $\mathcal{O}(-1)$ nodes denoted by \raisebox{.5pt}{\textcircled{\raisebox{-.9pt} {1}}} and 8 $\mathcal{O}(-4)$ nodes denoted by \raisebox{.5pt}{\textcircled{\raisebox{-.9pt} {4}}}. A solid line glues a  $\mathcal{O}(-1)$ node to a $\mathcal{O}(-4)$ node by $SO(8)$ gauging. The 5 connected gray nodes form a 6d little string theory of $(T^4\times \mathbb{C})/\mathbb{Z}_2\times\mathbb{Z}_2$.}
  \label{fig:T6-Z2Z2nodes}
\end{figure}

Note that each horizontal or vertical line intersecting with the four $\mathcal{O}(-1)$ singularities in Figure \ref{fig:T4-Z2Z2} has the same local geometry as $(T^4\times \mathbb{C})/\mathbb{Z}_2\times \mathbb{Z}_2$. We find therefore that the compact 3-fold in Figure \ref{fig:T4-Z2Z2} can be considered as four copies of the local horizontal (or vertical) $(T^4\times \mathbb{C})/\mathbb{Z}_2\times \mathbb{Z}_2$ geometry glued by four vertical (or horizontal) $SO(8)$ 7-branes. This tells us that the 6d supergravity theory realized by F-theory in Figure \ref{fig:T6-Z2Z2} can also be constructed by (as displayed in Figure \ref{fig:T4-Z2Z2}) four LSTs (see Figure \ref{fig:T2-Z2Z2}) glued by four other $\mathcal{O}(-4)$ theories. The tensor quiver diagram in Figure \ref{fig:T6-Z2Z2nodes} shows this construction. The five shaded nodes denote the 6d LST realized by F-theory on $T^4\times \mathbb{C}/\mathbb{Z}_2\times \mathbb{Z}_2$; in the compact case we have four copies of these LSTs connected by four $\mathcal{O}(-4)$ nodes.
Thus we conclude the field theory sector of the 6d supergravity theory associated to $T^6/\mathbb Z_2 \times \mathbb Z_2$ is given by the quiver like diagram in Figure \ref{fig:T6-Z2Z2nodes}.

\subsubsection*{Gravitational anomaly}
Let us test our claim that the above model captures the non-gravitational field theory sector of the 6d supergravity theory associated to $T^6/\mathbb Z_2 \times \mathbb Z_2$. We consider a compactification of this 6d supergravity theory on a circle of radius $R$. This system in F-theory is equivalent to M-theory on an elliptic CY 3-fold by the duality between F-theory and M-theory. The volume of the elliptic fiber is identified with the inverse radius $1/R$.
The resulting theory is a 5d $\mathcal{N}=1$ supergravity theory coupled to Kaluza-Klein states along the circle.

Let us first count dynamical K\"ahler parameters of this 5d supergravity theory. Naively, we have $24$ real scalar fields in the 6d tensor multiplets from $16+8$ tensor nodes and $r(V)=32$ real scalars coming from $U(1)^{32}$ holonomies of $SO(8)^8$ gauge group where $r(V)$ denotes the rank of the gauge groups.
However, some of the tensor fields and their associated scalars turn out to be non-dynamical.
The intersection matrix $\Omega$ associated to our model has null vectors with zero eigenvalue, so the corresponding tensor fields have no interaction with other field theory degrees of freedom. Hence the actual number of dynamical tensors is smaller than the naive count. The naive intersection matrix $\Omega$ is a $24\times 24$ symmetric matrix with diagonal entries $\left\{(-1)^{16},(-4)^8\right\}$ for the 16 $\mathcal{O}(-1)$ nodes and 8 $\mathcal{O}(-4)$ nodes, and upper triangle elements given by
\begin{eqnarray}\label{omegaT6Z2Z2}
	\Omega^{ij} =  \left\{ \begin{array}{cc} 1 & \ \  {\rm if} \ j=16+m \,,\ j=20+n  \\
	0 & {\rm otherwise}
	\end{array} \right.
	\ , \quad i=4m+n-4 \ , \quad 1\le m,n \le 4 \ .
\end{eqnarray}
The off-diagonal elements in the intersection matrix reflect our geometric configuration in which each $-1$ curve intersects with two $-4$ curves, one horizontal and one vertical, and two adjacent curves always meet each other only once.
One can verify by direct computation that the signature of $\Omega$ is 
	\begin{align}
		(-,+,0) = (17,1,6),
	\end{align}
including $6$ null tensors with zero eigenvalue. This is in fact consistent with \eqref{tensor}, as the number of dynamical tensors $T$ is 
	\begin{align}
		h^{1,1}(B) = 24 -6 = 18 \ , \quad T=h^{1,1}(B)-1=17 \ ,
	\end{align}
where $h^{1,1}(B)$ corresponds to the K\"ahler classes in $B$. Here we subtract $1$ from $h^{1,1}(B)$  for $T$ since one of the K\"ahler classes controlling the overall size of $B$ becomes a hypermultiplet scalar in 6d supergravity.

We note the Hodge numbers of the elliptic 3-fold $X = T^6/\mathbb{Z}_2\times \mathbb{Z}_2$, given in Table \ref{tb:orbifold2}, are 
\begin{align}
		h^{1,1}(X) = 51,~~~ h^{2,1}(X) = 3 \ .
\end{align}
This agrees with our field theory result by using the relation \eqref{abelianvector} 
\begin{equation}
	h^{1,1}(X) = h^{1,1}(B) + r(V)+1 = 18 + 32 + 1 = 51 \ .
\end{equation}
The last contribution $+1$ in the this counting corresponds to the size of the elliptic fiber class which is proportional to $1/R$ where $R$ is radius of the M-theory compactification circle.
$h^{2,1}(X)$ corresponds to the number of neutral hypermultiplets which are not captured by our field theory model.

We next check gravitational anomaly cancellation. The number of vector multiplets obtained by summing all contributions from the 8 $\mathcal{O}(-4)$ theories is
	\begin{align}
		V=8 \times \text{dim } SO(8) = 224 \ .
	\end{align}
The tensor nodes of the $-1$ and $-4$ curves have no hypermultiplets except those coming from 
the overall size of $B$. Thus, the number of hypermultiplets given by \eqref{neutralhyper} is simply $H=h^{2,1}(X)+1=4$.
With these numbers, one can easily check that the gravitational anomaly is cancelled:
	\begin{align}
		H - V +29 T - 273 = 4 - 8 \cdot  28 + 29 \cdot 17 - 273 = 0 \ .
	\end{align}
This is a strong check that our field theory model can consistently couple to 6d gravity. Our field theory model can be naturally embedded in the 6d (or 5d) supergravity theory realized by F-theory (or M-theory) on $T^6/\mathbb{Z}_2\times \mathbb{Z}_2$.  This field theory content was already proposed in \cite{Gopakumar:1996mu}.

\subsubsection*{Gauge/gravity mixed anomaly}

Let us turn to the gauge/gravity mixed anomalies. In this case, we can confine our attention to the case $G = SO(8)$. 
The discussion of anomaly cancellation in Section \ref{sec:anomaly} implies that the following conditions have to be satisfied:
	\begin{align}
		a \cdot a &= K_B^2 =9 - T = -8\label{anomaly1T6Z2Z2}\\
		a \cdot b_I &= K_B \cdot C_I = \frac{\lambda_I}{6} \left( A_{\text{adj}}^I \right)\label{GSanomaly2}\\
		b_I \cdot b_I &= C_I^2 = - \frac{\lambda_I^2}{3} \left( C_{\text{adj}}^I \right), \label{GSanomaly3}
	\end{align}
where $I$ labels the eight $-4$ curves. For $SO(N)$ with $N \geq 5$, we have
	\begin{align}
		A_{\text{adj}} = N-2,~~~ B_{\text{adj}} = N - 8,~~~ C_{\text{adj}} = 3, \label{SON}
	\end{align}	
giving us (in the case $N =8$)
	\begin{align}
		a \cdot b_I &= \lambda_I =2,~~~ b_I \cdot b_I = - \lambda_I^2 = -4.\label{anomaly2T6Z2Z2}
	\end{align}
	
Let us see that the anomaly cancellation conditions \eqref{anomaly1T6Z2Z2}, \eqref{anomaly2T6Z2Z2} are satisfied by the 6d quiver model in Figure \ref{fig:T4-Z2Z2}. For this, we consider a reduced intersection matrix $\tilde{\Omega}$ obtained by choosing only the independent two-cycles in $B$ as a basis. We denote the $-4$ curves in the vertical and horizontal directions in Figure \ref{fig:T4-Z2Z2} by $C^{(-4)}_i$ with $i=1, 2, 3, 4$ from bottom to top and $C'^{(-4)}_i$ with $i=1, 2, 3, 4$ from left to right, respectively. We then denote a $-1$ curve which intersects with both $C^{(-4)}_i$ and $C'^{(-4)}_j$ by $C^{(SO(8), SO(8))}_{ij}$. As mentioned above, not all curves are independent. 
Since the intersection matrix \eqref{omegaT6Z2Z2} has six eigenvectors with their eigenvalues zero, the K$\ddot{\text{a}}$hler parameters for curves corresponding to the null directions are not dynamical parameters of the 6d model. We then identify two curves with each other if the difference between them is a curve class corresponding to the null direction with respect to the intersection matrix \eqref{omegaT6Z2Z2}. 
Hence we impose constraints
\begin{equation}
C \equiv C^{(-4)}_{i} + \sum_{j=1}^4C^{(SO(8), SO(8))}_{ij},~~  i = 1,\dots, 4, \label{condition1T6Z2Z2}
\end{equation}
and 
\begin{equation}
C' \equiv C'^{(-4)}_{i} + \sum_{j=1}^4C^{(SO(8), SO(8))}_{ji},~~ i =1,\dots, 4. 
\label{condition2T6Z2Z2}
\end{equation}
The constraints \eqref{condition1T6Z2Z2} are identifying with each other the curves which are null directions with respect to the intersection matrix of LSTs in the horizontal directions in Figure \ref{fig:T4-Z2Z2}. On the other hand, the constraints \eqref{condition2T6Z2Z2} are identifying the curves which are null directions with respect to the intersection matrix of LSTs in the vertical directions in Figure \ref{fig:T4-Z2Z2}. 

Note that the curves appearing in the constraints \eqref{condition1T6Z2Z2} have a clear geometric interpretation. This can be seen by blowing down all of the $-1$ curves in the base $T^4/\mathbb Z_2 \times \mathbb Z_2$, resulting in a configuration consisting of four horizontal and four vertical $\mathbb P^1$'s intersecting in $16$ points, each with self-intersection $0$. Comparing this information to the fact that $T^2/\mathbb Z_2 \cong \mathbb P^1$, we see there are two homologically-distinct $\mathbb P^1$ classes of self-intersection $0$, which we denote $C,C'$. Thus the configuration of curves should be viewed as four copies of $C \cong \mathbb P^1$ and four copies of $C' \cong \mathbb P^1$ in the compact surface $\mathbb F_0 = \mathbb P^1 \times \mathbb P^1$, mutually intersecting in 16 points. The constraints in \eqref{condition1T6Z2Z2} are then due to the fact that the four copies of the curves $C, C'$ are homologically equivalent. Now that we have precisely identified the geometry of the base, we can compute the vectors $a, b_i$ independently. For example, using the fact that the canonical class of a Hirzebruch surface $\mathbb F_n$ is given by $K_{\mathbb F_n} = -2C + (n-2)C'$, we find that the blowup of $\mathbb F_0$ at 16 points with exceptional divisors $C_{ij}^{(SO(8),SO(8))}$ where $i,j=1,\dots,4$ will have canonical class 
	\begin{align}
		K_{\text{Bl}_{16} \mathbb F_0} = K_{\mathbb F_0} + \sum_{i,j=1}^4 C_{ij}^{(SO(8),SO(8))} = -2 C - 2 C' +\sum_{i,j=1}^4 C_{ij}^{(SO(8),SO(8))}.
	\end{align}	

Thus, due to the conditions \eqref{condition1T6Z2Z2} and \eqref{condition2T6Z2Z2}, we can choose $C, C'$ and $C^{(SO(8), SO(8))}_{ij}$ for $i, j = 1, \dots, 4$ for our $18$ independent basis elements, so that a general curve class $v$ admits the following expansion:
\begin{equation}
v = v_1C + v_2C' + \sum_{i=1,j=1}^4v_{4i+j-2}C_{ij}^{(SO(8), SO(8))}.
\end{equation}
For example, from the condition \eqref{condition1T6Z2Z2}, the $C_{1}^{(-4)}$ curve is given by the vector 
\begin{equation}
b_1^{SO(8)} = (1, 0, -1, -1, -1, -1, 0, \cdots, 0). \label{bT6Z2Z2}
\end{equation}
The other $b_i^{SO(8)}$ for $C^{(-4)}_i$ and $b'^{SO(8)}_i$ for $C'^{(-4)}_i$ are determined similarly from the constraints \eqref{condition1T6Z2Z2} and \eqref{condition2T6Z2Z2}. 

In terms of the above basis, the reduced intersection matrix obtained from \eqref{omegaT6Z2Z2} is given by
\begin{eqnarray}
\tilde{\Omega} = \left(
\begin{array}{cc}
0 & 1\\
1 & 0
\end{array}
\right) \oplus \text{diag}\left(-1, -1, \cdots, -1\right), \label{redomegaT6Z2Z2}
\end{eqnarray}
where the diagonal matrix has $16$ components corresponding to the number of $-1$ curves. Since we set the curves with the zero eigenvalue with respect to the intersection matrix \eqref{omegaT6Z2Z2} to be zero, the reduced intersection matrix \eqref{redomegaT6Z2Z2} has no zero eigenvalue. Furthermore, either by computing the endpoint in the manner described above or by using (\ref{afromselfint}) combined with the adjunction formula, one can easily show the vector $a$ corresponding to the canonical class is 
\begin{equation}
a = \left(-2, -2, 1, \cdots, 1\right), \label{aT6Z2Z2}
\end{equation}
where in the above expression there are $16$ entries of $1$.

By using the reduced intersection matrix \eqref{redomegaT6Z2Z2}, we can compute the intersection numbers between $a$ and $b^{SO(8)}_i, b'^{SO(8)}_i$ and the result is given by 
\begin{align}
&a\cdot a = -8, \qquad a\cdot b^{SO(8)}_i = a\cdot b'^{SO(8)}_i =2, \cr
&b^{SO(8)}_i\cdot b^{SO(8)}_i = -4,\qquad b'^{SO(8)}_i\cdot b'^{SO(8)}_i = -4,\quad i=1, \cdots, 4,
\end{align}
which satisfies the anomaly cancellation conditions \eqref{anomaly1T6Z2Z2} and \eqref{anomaly2T6Z2Z2}. 
Note that this lattice is self-dual as it should be based on both geometric considerations as well as consistency conditions for ${\cal N}=(1,0)$ theories.




\subsubsection{$T^6/\mathbb{Z}_3\times \mathbb{Z}_3$}
Our next example is the supergravity theory realized by F-theory on $X=T^6/\mathbb{Z}_3\times \mathbb{Z}_3$. The elliptic 3-fold $X$ is constructed by means of the following orbifold action on $T^6$:
\begin{equation}
	g = (\alpha,\alpha^{-1},1)  \ , \quad h = (w,1,w^{-1}) \ , \quad \alpha^3=w^3=1 \ .
\end{equation}
This action generates 27 fixed points on $T^6$. The local geometry at each fixed point is $\mathbb{C}^3/\mathbb{Z}_3\times \mathbb{Z}_3$ and the total  geometry is a compact CY 3-fold formed by gluing 27 local patches of this non-compact geometry.

\begin{figure}
  \centering
  \includegraphics[width=.75\linewidth]{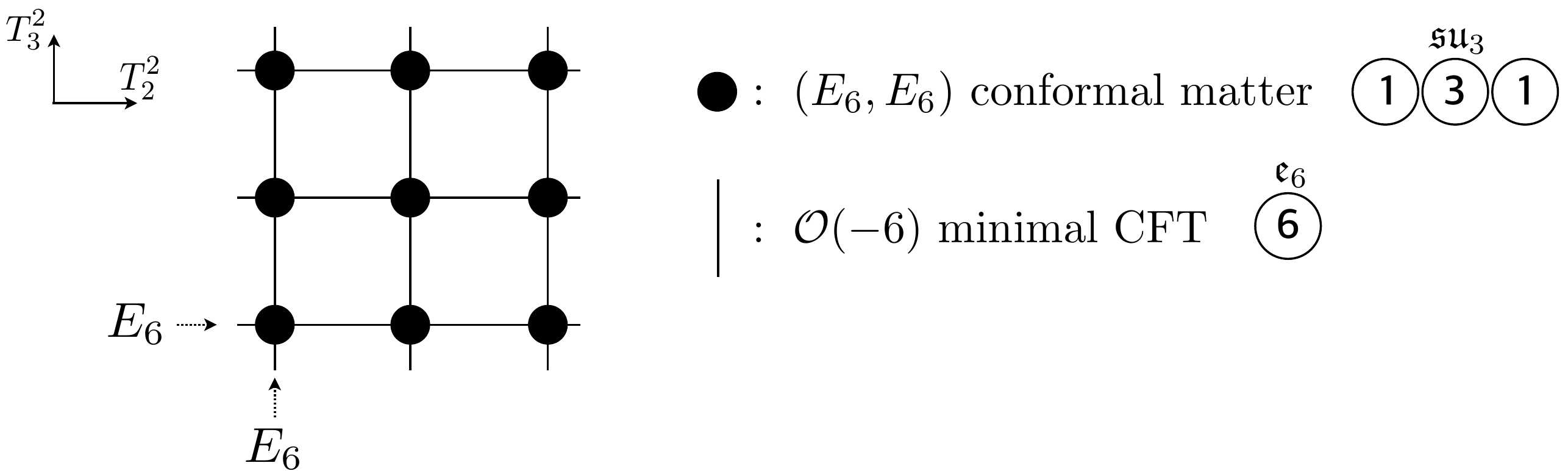}
  \caption{6d model of the supergravity theory realized by F-theory on $T^6/\mathbb{Z}_3\times \mathbb{Z}_3$ with base $B=T^4/\mathbb{Z}_3\times \mathbb{Z}_3$, where $T^4 = T_2^2 \times T_3^2$. In the above graph, the nine black dots are $(E_6,E_6)$ conformal matter theories, while the three horizontal and three vertical lines denote $\mathcal{O}(-6)$ minimal SCFTs.}
  \label{fig:T4-Z3Z3}
\end{figure}

On the base $B=T^4/\mathbb{Z}_3\times \mathbb{Z}_3$, we have 9 fixed points and each local geometry is an elliptic 3-fold with a non-compact base $\mathbb{C}^2/\mathbb{Z}_3\times \mathbb{Z}_3$. The corresponding local 6d theory is the $(E_6,E_6)$ conformal matter theory. The global symmetry $E_6\times E_6$ of this theory comes from two intersecting 7-branes with $E_6$ symmetry. In the compact base $B$, there are 6 such $E_6$ 7-branes and each $E_6$ 7-brane meets three $(E_6,E_6)$ conformal matter theories. Since they wrap compact $T^2$ in the base $B$, the 7-branes gauge $E_6$ global symmetries of adjacent conformal matter theories. The theory on an $E_6$ 7-brane is a non-Higgsable tensor theory with $E_6$ gauge group called the $\mathcal{O}(-6)$ minimal SCFT. Therefore, the full theory of the F-theory geometry $X$ is given by nine $(E_6,E_6)$ conformal matter theories glued together by six $\mathcal{O}(-6)$ minimal CFT theories as depicted in Figure \ref{fig:T4-Z3Z3}. We claim that this is the non-gravitational field theory sector of the 6d supergravity realized by F-theory on $T^6/\mathbb{Z}_3\times \mathbb{Z}_3$.

\subsubsection*{Gravitational anomaly}
A naive tensor counting gives $33 -1 = 32$ tensors. Howerver we can again expect that some tensors are non-dynamical. There are in fact four null tensors that can be computed from the intersection matrix $\Omega$. In this geometry, $\Omega$ is a $33\times 33$ symmetric matrix with diagonal entries $\text{diag}(\Omega)=-\left(\{1,3,1\}^9,6^6\right)$ and off-diagonal entries 
\begin{equation}\label{eq:intersect-matrix-off}
	\Omega^{ij} = \left \{ \begin{array}{cc} +1 & \ \ {\rm if } \ i{\text-}{\rm th \ and} \  j{\text-} {\rm th  \ curves \ intersect } \\
	0 & {\rm otherwise}
	\end{array} \right. \ .
\end{equation}
The signature of this matrix $\Omega$ is
\begin{equation}
	(-,+,0) = (28,1,4) \ .
\end{equation}
Thus we find four null tensor multiplets among 33 tensor multiplets.
We therefore compute
\begin{equation}
	h^{1,1}(B)=33-4=29 \ , \quad T = h^{1,1}(B)-1=28 \ .
\end{equation}
We can also compute the number of vector multiplets by summing over all non-abelian gauge factors, namely nine $SU(3)$'s and six $E_6$'s, to obtain
\begin{equation}
	r(V) = 9 \times 2 + 6 \times 6 = 54 \ , \quad V =  9 \times 8+ 6 \times 78  = 540 \ .
\end{equation}
Finally, there is only one hypermultiplet, $H=1$, associated to the overall size of $B$.  We can compare these  numbers with the topological numbers of the orbifold geometry $X=T^6/\mathbb{Z}_3\times \mathbb{Z}_3$ in Table \ref{tb:orbifold2}, for which
	\begin{align}
		h^{1,1}(X) =  84 \ , \quad h^{2,1}(X)=0 \ .
	\end{align}
Thus the geometric counting for $h^{1,1}(Y)$ agrees with our field theory counting $h^{1,1}(B)+r(V)+1=84$. Also the gravitational anomaly is cancelled:
\begin{equation}
	H-V+29T -273 = 1 - 540 +29\times 28 - 273 = 0 \ .
\end{equation}

\subsubsection*{Gauge/gravity mixed anomaly}
	
The anomaly cancellation conditions for this model are given by
	\begin{align}
		a \cdot a &= K_B^2 = 9 - T = -19\label{anomaly1T6Z3Z3}
	\end{align}
and also \eqref{GSanomaly2}, \eqref{GSanomaly3} for gauge groups $SU(3), E_6$.
The constants $\lambda$ can be found in Table \ref{tb:lambdai}. For $SU(m)$, $m=2,3$, we have 
	\begin{align}
		\text{tr}_{\text{adj}} F^2 = 2m\, \text{tr} F^2,~~~\text{tr}_{\text{adj}} F^4 = (m+6) (\text{tr} F^2)^2,\label{SUm} 
	\end{align}
and therefore
	\begin{align}
		A_\text{adj} = 6,~~~ B_\text{adj} = 0,~~~ C_\text{adj} = 9,~~~ \lambda_{SU(3)}=1, 
		\end{align}
	which gives us
		\begin{align}
		 a \cdot b_I^{SU(3)}=  1,~~~ b_I^{SU(3)} \cdot b_I^{SU(3)} = -3, \label{anomaly2T6Z3Z3}
	\end{align}
where $I$ labels the nine $-3$ curves supporting $SU(3)$ gauge symmetry. For $E_6$, we have 
	\begin{align}
		\text{tr}_{\text{adj}} F^2= 4 \text{tr} F^2,~~~ \text{tr}_{\text{adj}} F^4 = \frac{1}{2} (\text{tr} F^2)^2,
	\end{align}
and thus
	\begin{align}
		A_\text{adj} = 4,~~~ B_{\text{adj}} =0,~~~ C_\text{adj} = \frac{1}{2},~~~ \lambda_{E_6} = 6,
	\end{align}	
which gives us
	\begin{align}
		a \cdot b_{I}^{E_6} = 4,~~~ b_I^{E_6} \cdot b^{E_6}_{I} =  -6, \label{anomaly3T6Z3Z3}
	\end{align} 
where $I$ labels the six $-6$ curves supporting $E_6$ gauge symmetry.

Let us see how the anomaly cancellation conditions \eqref{anomaly1T6Z3Z3}, \eqref{anomaly2T6Z3Z3} and \eqref{anomaly3T6Z3Z3} follow from the intersection numbers between curves in the quiver model in Figure \ref{fig:T4-Z3Z3}. As we have done for the $T^6/\mathbb{Z}_2\times\mathbb{Z}_2$ model, we choose $29$ independent curves in the base $B$ from the $33$ curves in Figure \ref{fig:T4-Z3Z3} by setting the curves with zero eigenvalue with respect to the intersection matrix $\Omega$ to be zero. We denote the $-6$ curves in the horizontal and vertical directions in Figure \ref{fig:T4-Z3Z3} by $C^{(-6)}_i$ with $i=1, 2, 3$ from bottom to top and $C'^{(-6)}_i$ with $i=1,2,3$ from left to right, respectively. We also denote the three curves of each $(E_6, E_6)$ conformal matter which are sandwiched between $C_i^{(-6)}$ and $C'^{(-6)}_j$ by $C_{ijk}^{(E_6, E_6)}$ with $k=1, 2, 3$. In each $(E_6, E_6)$ conformal matter, $k=1$ corresponds to the $-1$ curve which intersects with $C^{(-6)}_i$, $k=2$ corresponds to the $-3$ curve, and $k=3$ corresponds to the $-1$ curve which intersects with $C'^{(-6)}_j$. The $29$ independent curves are obtained by the constraints
\begin{eqnarray}
C &\equiv & C_{i}^{(-6)} + \sum_{j=1}^3\left(2C_{ij1}^{(E_6, E_6)} + \sum_{k=2}^3C_{ijk}^{(E_6, E_6)} \right), 
\label{condition1T6Z3Z3}\\
C' &\equiv & C'^{(-6)}_i+ \sum_{j=1}^3\left(\sum_{k=1}^2C_{jik}^{(E_6, E_6)} + 2C_{ji3}^{(E_6, E_6)}\right), 
\label{condition2T6Z3Z3}
\end{eqnarray}
for all $i = 1, 2, 3$. With $C, C'$ defined by \eqref{condition1T6Z3Z3} and \eqref{condition2T6Z3Z3}, we can choose a basis $C, C', C_{ijk}^{(E_6, E_6)}$ for $i, j, k= 1, 2, 3$, consisting of $29$ independent classes. A vector $v$ then admits the following expansion in this basis:
\begin{equation}
v = v_1C + v_2C' + \sum_{i,j,k=1}^3v_{9i+3j+k-10}C_{ijk}^{(E_6, E_6)}.
\end{equation}
For example, the $-3$ curve corresponding to $C^{(E_6, E_6)}_{112}$ is described by
\begin{equation}
b^{SU(3)}_{112}= (0, 0, 0, 1, 0, 0, \cdots, 0) \label{m3bT6Z3Z3}
\end{equation}
and the $-6$ curve corresponding to $C_{1}^{(-6)}$ is given by
\begin{equation}
b_{1}^{E_6} = (1, 0, -2, -1, -1, -2, -1, -1, -2, -1, -1, 0, \cdots, 0) \label{m6bT6Z3Z3}
\end{equation}
from \eqref{condition1T6Z3Z3}. The reduced intersection matrix in this basis is then 
\begin{eqnarray}
\tilde{\Omega} = \left(
\begin{array}{cc}
0 & 1 \\
1 & 0
\end{array}
\right) \oplus \left(
\begin{array}{ccc}
-1 & 1 & 0\\
1 & -3 & 1\\
0 & 1 & -1
\end{array}
\right) \oplus \cdots \oplus \left(
\begin{array}{ccc}
-1 & 1 & 0\\
1 & -3 & 1\\
0 & 1 & -1
\end{array}
\right), \label{redomegaT6Z3Z3}
\end{eqnarray}
where the matrix
\begin{eqnarray}
 \left(
\begin{array}{ccc}
-1 & 1 & 0\\
1 & -3 & 1\\
0 & 1 & -1
\end{array}
\right)
\end{eqnarray}
is the matrix for the $(E_6, E_6)$ conformal matter and there are nine such matrices in \eqref{redomegaT6Z3Z3}. The matrix \eqref{redomegaT6Z3Z3} describes a self-dual lattice as expected.
The canonical class can be determined by using \eqref{afromselfint} and requiring that the genera of the $-1$, $-3$, and $-6$ curves are all zero. Then the vector $a$ is given by
\begin{equation}
a = \left(-2, -2, \{2, 1, 2\}^9\right).\label{aT6Z3Z3}
\end{equation}
In the above expression, the notation $\{ 2,1,2\}^9$ indicates that there are nine repeated sequences of entries $2,1,2$ in the vector $a$. 

We are now ready to check the anomaly cancellation conditions \eqref{anomaly1T6Z3Z3}, \eqref{anomaly2T6Z3Z3}, and \eqref{anomaly3T6Z3Z3}. First, \eqref{aT6Z3Z3} yields
\begin{equation}
a \cdot a = -19,
\end{equation}
which agrees with \eqref{anomaly1T6Z3Z3}. Furthermore, \eqref{m3bT6Z3Z3} and \eqref{m6bT6Z3Z3} give
\begin{equation}
a\cdot b_{11}^{SU(3)} = 1, \quad b_{112}^{SU(3)}\cdot b_{112}^{SU(3)} = -3, 
\end{equation}
and
\begin{equation}
a\cdot b_1^{E_6} = 4, \quad b_1^{E_6} \cdot b_1^{E_6} = -6, 
\end{equation}
agreeing with \eqref{anomaly2T6Z3Z3} and \eqref{anomaly3T6Z3Z3}. It is straghtforward to extend these computations to the remaining $-3$ and $-6$ curves. We find perfect agreement with \eqref{anomaly2T6Z3Z3} and \eqref{anomaly3T6Z3Z3}.


\subsubsection{$T^6/\mathbb{Z}_4\times \mathbb{Z}_4$}
\begin{figure}
  \centering
  \includegraphics[width=.7\linewidth]{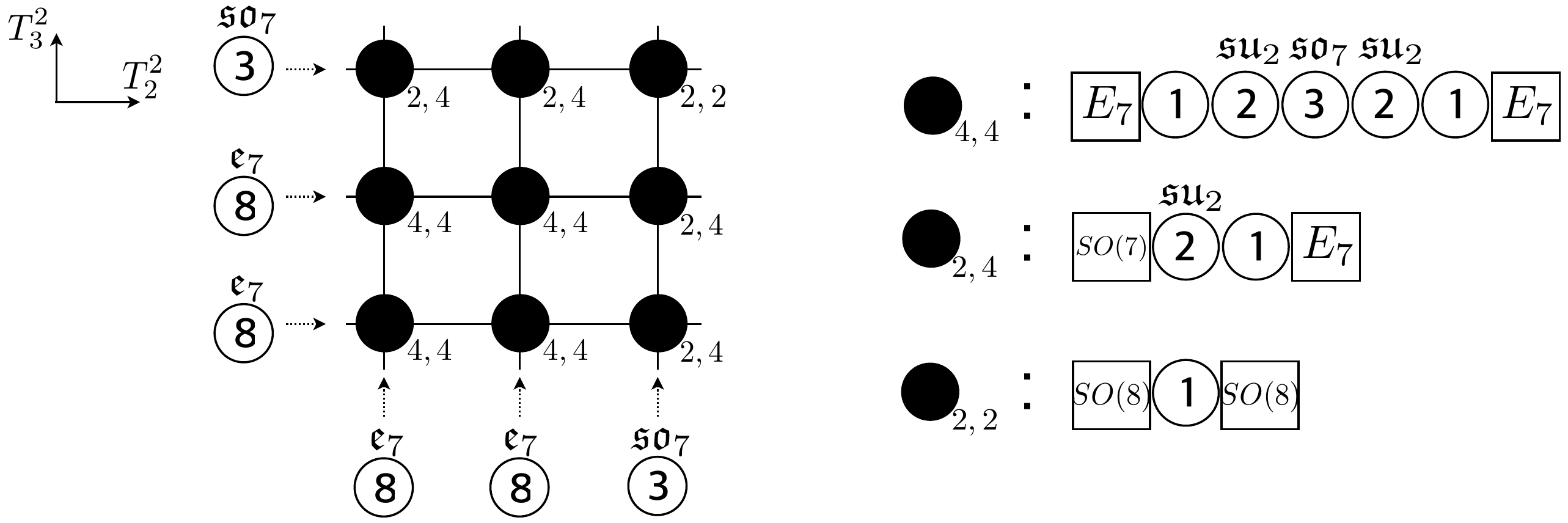}
  \caption{6d gravity model of $T^6/\mathbb{Z}_4\times \mathbb{Z}_4$ geometry. }
  \label{fig:T4-Z4Z4}
\end{figure}

We now construct a 6d model for the gravity theory realized by F-theory on $T^6/\mathbb{Z}_4\times \mathbb{Z}_4$. The orbifold action is generated by $g$ and $h$ in (\ref{eq:orbifold-action}) with $\alpha^4=1$ and $w^4=1$. The action $g$ has one $\mathbb{Z}_2$ fixed point and two identical $\mathbb{Z}_4$ fixed points on the first $T^2$ in the base, while the action $h$ has one $\mathbb{Z}_2$ fixed point and two identical $\mathbb{Z}_4$ fixed points on the second $T^2$. Thus there are nine fixed points on the base $B=T^4/\mathbb{Z}_4\times \mathbb{Z}_4$ as drawn in Figure \ref{fig:T4-Z4Z4}. The four fixed points denoted by black solid circles with subscript `$4,4$' have local geometry $\mathbb{C}^2/\mathbb{Z}_4\times \mathbb{Z}_4$; the four fixed points with subscript `$2,4$' have local geometry $\mathbb{C}^2/\mathbb{Z}_2\times \mathbb{Z}_4$; and the one fixed point with subscript `$2,2$' has local geometry $\mathbb{C}^2/\mathbb{Z}_2\times \mathbb{Z}_2$ on the base.
This tells us that the local 6d theories at the fixed points are conformal matter theories of type $(E_7,E_7)$, $(SO(7),E_7)$, and $(SO(8),SO(8))$ respectively. The full field theory is obtained by gluing these conformal matter theories by four $E_7$ and two $SO(7)$ gaugings as drawn in Figure \ref{fig:T4-Z4Z4}. We claim that this theory coupled to gravity describes the 6d gravity theory associated to $T^6/\mathbb{Z}_4\times \mathbb{Z}_4$.

\subsubsection*{Gravitational anomaly}
The intersection matrix of this geometry is a $35\times 35$ matrix whose diagonal elements are $\text{diag} \, \Omega=-\left(\{1,2,3,2,1\}^4,\{2,1\}^4,1,8^4,3^2\right)$ and off-diagonal elements are assigned by following the rule in \eqref{eq:intersect-matrix-off}. The signature of $\Omega$ is
\begin{equation}
	(-,+,0) = (30,1,4) \ .
\end{equation}
There are four null tensors, so this geometry has the following number of dynamical tensor multiplets:
\begin{equation}
	h^{1,1}(B) = 31 \ , \quad T=h^{1,1}(B)-1 = 30 \ .
\end{equation}
We now count the number of vector multiplets. There are eight $SU(2)$ and four $SO(7)$ gauge symmetries from the $(E_7,E_7)$ conformal matter theories, four $SU(2)$ gauge symmetries from the $(SO(7),E_7)$ conformal matter theories, and two $SO(7)$ and four $E_7$ gauge symmetries living on the gluing curves with self-intersection numbers $-3$ and $-8$ respectively. Therefore, we have
\begin{equation}
	r(V) = 12 +6\times 3+4\times 7= 58 \ , \quad V = 12\times 3 + 6 \times 21 + 4 \times 133 = 694 \ .
\end{equation}
The number of the moduli for the geometry $T^6/\mathbb{Z}_4\times \mathbb{Z}_4$ in Table \ref{tb:orbifold2} is
\begin{equation}
	h^{1,1}(X) = 90 \ , \quad h^{2,1}(X) = 0 \ ,
\end{equation}
which agrees with the field theory counting $h^{1,1}(X) = h^{1,1}(B) + r(V)+1 = 90$.

A single $(E_7,E_7)$ conformal matter has 16 hypermultiplets between $-3$ and $-2$ curves and an $(SO(7),E_7)$ conformal matter contains eight hypermultiplets on the $-2$ curve. In total, this geometry has the number of hypermultiplets:
\begin{equation} 
	H =4\times 16 + 4 \times 8 +1 = 97 \ ,
\end{equation}
where the last $+1$ corresponds to the overall size of $B$. Now one can easily check that gravitational anomaly cancellation is satisfied:
\begin{equation}
	H - V + 29 T - 273 = 97 - 694 +29\times 30 -273= 0 \ .
\end{equation}

\subsubsection*{Gauge/gravity mixed anomaly}

The gauge/gravity mixed anomaly cancellation conditions are given by
\begin{align}
                0 &= B_{\text{adj}} - \sum_{R_I}n_{R_I}B_{R_I}\label{anomaly0T6Z4Z4}\\
		a \cdot a &= K_B^2 = 9 - T = -21,\label{anomaly1T6Z4Z4}\\
		a \cdot b_I &= K_B \cdot C_I = \frac{\lambda_I}{6} \left( A_{\text{adj}}^I  - \sum_{R_I}n_{R_I}A_{R_I}\right), \label{GSanomaly4}\\
		b_I \cdot b_I &= C_I^2 = - \frac{\lambda_I^2}{3} \left( C_{\text{adj}}^I - \sum_{R_I}n_{R_I}C_{R_I}\right),\label{GSanomaly5}\\
		b_I \cdot b_J &= C_I \cdot C_J = \lambda_I\lambda_J\sum_{R_I, R'_I}n_{R_IR'_J}A_{R_I}A_{R'_J}, \label{bifundT6Z4Z4}
	\end{align}
for gauge groups $SU(2), SO(7)$, and $E_7$.
Note that we also have hypermultiplets in the representation of $\frac{1}{2}({\bf 2}, {\bf 8})$ under $SU(2) \times SO(7)$. For $SU(2)$, we have 
\begin{equation}
A_{\text{adj}} = 4, \quad B_{\text{adj}} = 0, \quad C_{\text{adj}} = 8, \quad \lambda_{SU(2)} = 1, \label{SU2.adj.numbers}
\end{equation}
from \eqref{SUm} for the adjoint representation. Since we have hypermultiplets in the fundamental representation of $SU(2)$, we also need
\begin{equation}
A_{\text{fund}} = 1, \quad B_{\text{fund}} = 0, \quad C_{\text{fund}} = \frac{1}{2}. \label{SU2.fund.numbers}
\end{equation}
Hence for curves on which we have $SU(2)$, the anomaly cancellation conditions become
\begin{align}
		a \cdot b_{I}^{SU(2)} = 0,~~~ b_I^{SU(2)} \cdot b_{I}^{SU(2)} =  -2, \label{anomalySU2T6Z4Z4}
\end{align} 
For $SO(7)$, we can use the result \eqref{SON} with $N=7$ which gives
\begin{equation}
A_{\text{adj}} = 5, \quad B_{\text{adj}} = -1, \quad C_{\text{adj}} = 3, \quad \lambda_{SO(7)} = 2,
\end{equation}
for the adjoint representaion. We also have hypermultiplets in the spinor representation 
\begin{equation}
A_{\text{spinor}} = 1, \quad B_{\text{spinor}} = -\frac{1}{2}, \quad C_{\text{spinor}} = \frac{3}{8}.
\end{equation}
Hence for curves on which we have $SO(7)$, the anomaly cancellation conditions are 
\begin{align}
		a \cdot b_{I}^{SO(7)} = 1,~~~ b^{SO(7)}_I \cdot b^{SO(7)}_{I} =  -3, \label{anomalySO7T6Z4Z4}
		\end{align} 
and \eqref{anomaly0T6Z4Z4} is satisfied. The condition \eqref{bifundT6Z4Z4} becomes 
\begin{align}
b_I^{SU(2)} \cdot b^{SO(7)}_J =  1. \label{anomalyBFT6Z4Z4}
\end{align}
Finaly, for $E_7$, we have only adjoint representations and the necessary information for anomaly cancellation is
	\begin{align}
		\text{tr}_{\text{adj}} F^2= 3 \text{tr} F^2,~~~ \text{tr}_{\text{adj}} F^4 = \frac{1}{6} (\text{tr} F^2)^2,
	\end{align}
thus
	\begin{align}
		A_\text{adj} = 3,~~~ B_{\text{adj}} =0,~~~ C_\text{adj} = \frac{1}{6},~~~ \lambda_{E_7} = 12,
	\end{align}	
which gives us
	\begin{align}
		a \cdot b^{E_7}_{I} = 6,~~~ b^{E_7}_I \cdot b^{E_7}_{I} = -8, \label{anomalyE7T6Z4Z4}
	\end{align} 
where $b^{E_7}_I$ corresponds to a $-8$ curve carrying $E_7$ gauge symmetry.

Let us then see the anomaly cancellation conditions \eqref{anomaly1T6Z4Z4}, \eqref{anomalySU2T6Z4Z4}, \eqref{anomalySO7T6Z4Z4}, \eqref{anomalyBFT6Z4Z4} and \eqref{anomalyE7T6Z4Z4} are reproduced from the quiver model in Figure \ref{fig:T4-Z4Z4}. For this purpose we use a reduced intersection matrix which only involves independent curves. As in the cases of the $T^6/\mathbb{Z}_2 \times \mathbb{Z}_2, T^6/\mathbb{Z}_3 \times \mathbb{Z}_3$ models, we first name the curves in Figure \ref{fig:T4-Z4Z4}, in the following way. We denote the $-n$ curves corresponding to horizontal lines by $C^{(-n)}_i, i=1, \cdots$ from bottom to top. The $-n$ curves corresponding to vertical lines are represented by $C'^{(-n)}_i, i=1, \cdots$ from left to right. The curves between $C^{(-n)}_i$ with a gauge group $G$ and $C'^{(-n')}_j$ with a gauge group $G'$ are denoted by $C^{(G, G')}_{ijk}, k=1, \cdots$ where $k$ is in order from the curve next to $C^{(-n)}_i$ to the curve next to $C'^{(-n')}_j$. We will use this notation for the remainder of Section \ref{sec:T6ZmZn}.   
Then, the constraints among the curve classes are 
\begin{eqnarray}
C &\equiv & C^{(-8)}_1 + \sum_{j=1}^2\left(3C_{1j1}^{(E_7, E_7)} + 2C_{1j2}^{(E_7, E_7)} + \sum_{k=3}^5C_{1jk}^{(E_7, E_7)}\right) + \left(2C^{(E_7, SO(7))}_{131} + C^{(E_7, SO(7))}_{132}\right)\nonumber\\
&=& C^{(-8)}_2 + \sum_{j=1}^2\left(3C_{2j1}^{(E_7, E_7)} + 2C_{2j2}^{(E_7, E_7)} + \sum_{k=3}^5C_{2jk}^{(E_7, E_7)}\right) + \left(2C^{(E_7, SO(7))}_{231} + C^{(E_7, SO(7))}_{232}\right)\nonumber\\
&=& C^{(-3)}_3 + \sum_{j=1, k=1}^2 C^{(SO(7), E_7)}_{3jk} + C^{(SO(7), SO(7))}_{331}
\end{eqnarray}
and 
\begin{eqnarray}
C' &\equiv & C'^{(-8)}_1 + \sum_{i=1}^2\left(\sum_{k=1}^3C_{i1k}^{(E_7, E_7)} + 2C_{i14}^{(E_7, E_7)} + 3C_{i15}^{(E_7, E_7)}\right) + \left(C^{(SO(7), E_7)}_{311} + 2C^{(SO(7), E_7)}_{312}\right) \nonumber\\
&=&C'^{(-8)}_2 + \sum_{i=1}^2\left(\sum_{k=1}^3C_{i2k}^{(E_7, E_7)} + 2C_{i24}^{(E_7, E_7)} + 3C_{i25}^{(E_7, E_7)}\right) + \left(C^{(SO(7), E_7)}_{321} + 2C^{(SO(7), E_7)}_{322}\right) \nonumber\\
&=& C'^{(-3)}_3 + \sum_{i=1, k=1}^2 C_{i3k}^{(E_7, SO(7))} + C^{(SO(7), SO(7))}_{331}.
\end{eqnarray}
We then choose the following basis:
\begin{align}
\big(C, C', \{C_{11k}^{(E_7, E_7)}\}, \{C_{12k}^{(E_7, E_7)}\}, \{C_{13k}^{(E_7, SO(7))}\}, &\{C_{21k}^{(E_7, E_7)}\}, \{C_{22k}^{(E_7, E_7)}\}, \{C_{23k}^{(E_7, SO(7))}\}, \nonumber\\ 
&\{C^{(SO(7), E_7)}_{31k}\}, \{C^{(SO(7), E_7)}_{32k}\}, C^{(SO(7), SO(7)}_{331}\big). \label{basesT4Z4Z4}
\end{align} 
The reduced intersection matrix in this basis is
\begin{eqnarray}
\tilde{\Omega} = \left(
\begin{array}{cc}
0 & 1 \\
1 & 0
\end{array}
\right) \oplus \left[\left\{\left(
\begin{array}{ccccc}
-1 & 1 & 0 & 0 & 0\\
1 & -2 & 1 & 0 & 0\\
0 & 1 & -3 & 1 & 0\\
0 & 0 & 1 & -2 & 1\\
0 & 0 & 0 & 1 & -1
\end{array}
\right)\right\}^2 \oplus  \left(
\begin{array}{cc}
-1 & 1 \\
1 & -2 \\
\end{array}
\right)\right]^2 \oplus \left\{\left(
\begin{array}{cc}
-2 & 1 \\
1 & -1 \\
\end{array}
\right)\right\}^2 \oplus\left( -1 \right).\nonumber\\ \label{redomegaT4Z4Z4}
\end{eqnarray}

Using \eqref{redomegaT4Z4Z4}, we check the intersection numbers \eqref{anomaly1T6Z4Z4}, \eqref{anomalySU2T6Z4Z4}, \eqref{anomalySO7T6Z4Z4}, \eqref{anomalyBFT6Z4Z4} and \eqref{anomalyE7T6Z4Z4}. First, we determine the canonical class from \eqref{afromselfint}, giving us
\begin{align}
a = (-2, -2, 3, 2, 1, 2, 3, 3, 2, 1, 2, 3, 2, 1, 3, 2, 1, 2, 3, 3, 2, 1, 2, 3, 2, 1, 1, 2, 1, 2, 1). \label{aT4Z4Z4}
\end{align}
We can see that \eqref{aT4Z4Z4} with \eqref{redomegaT4Z4Z4} satisfies \eqref{anomaly1T6Z4Z4}. 

$-2$ curves with an $SU(2)$ gauge group appear either inside the $(E_7, E_7)$ conformal matter or $(SO(7), E_7)$ conformal matter. For example the $-2$ curve of $C_{112}^{(E_7, E_7)}$ is
\begin{align}
b_{112}^{SU(2)} =  (0,0,0,1,0^{27}). 
\label{SU2no1T4Z4Z4}
\end{align}
The $-2$ curve corresponding to $C_{132}^{(E_7, SO(7))}$ is 
\begin{align}
b_{132}^{SU(2)} =  (0^{13}, 1, 0^{17}). 
\label{SU2no2T4Z4Z4}
\end{align}
Direct computation shows \eqref{SU2no1T4Z4Z4} and \eqref{SU2no2T4Z4Z4} with \eqref{redomegaT4Z4Z4} yield \eqref{anomalySU2T6Z4Z4}. Similarly the other $-2$ curves also satisfy \eqref{anomalySU2T6Z4Z4}. 

$-3$ curves with an $SO(7)$ gauge group appears either in $(E_7, E_7)$ conformal matter or correspond to the gluing curves $C^{(-3)}_3$ and $C'^{(-3)}_3$. For example, the $-3$ curve of $C_{113}^{(E_7, E_7)}$ is
\begin{align}
b^{SO(7)}_{113} =  (0,0,0,0,1,0^{26}). 
\label{SO7no1T4Z4Z4}
\end{align}
The $-3$ curve corresponding to $C^{(-3)}_{3}$ is 
\begin{align}
b^{SO(7)}_3 =  (1, 0^{25}, 
-1, -1, -1, -1, -1), \label{SO7no2T4Z4Z4}
\end{align}
and the $-3$ curve of $C'^{(-3)}_3$ is 
\begin{align}
b'^{SO(7)}_3= (0, 1, 0^{10}, 
-1, -1, 0^{10}, 
-1, -1, 0, 0, 0, 0, -1)\label{SO7no3T4Z4Z4}
\end{align}
Indeed \eqref{SO7no1T4Z4Z4}, \eqref{SO7no2T4Z4Z4} and \eqref{SO7no3T4Z4Z4} with \eqref{redomegaT4Z4Z4} reproduce \eqref{anomalySO7T6Z4Z4}. It is straightforward to check that the other $-3$ curves also satisfy \eqref{anomalySO7T6Z4Z4}. 

There are also hypermultiplets in the representation $\frac{1}{2}({\bf 2}, {\bf 8})$ of $SU(2) \times SO(7)$. For example the intersection between \eqref{SU2no1T4Z4Z4} and \eqref{SO7no1T4Z4Z4} gives
\begin{align}
b^{SU(2)}_{112} \cdot b^{SO(7)}_{113} = 1,
\end{align}
which is consistent with \eqref{anomalyBFT6Z4Z4}. Hypermultiplets also arise from the intersection between a $-2$ curve in the $(SO(7), E_7)$ conformal matter theory and a $-3$ curve for the $SO(7)$ gauging. For example the intersection between \eqref{SU2no2T4Z4Z4} and \eqref{SO7no3T4Z4Z4} is 
\begin{align}
b_{132}^{SU(2)} \cdot b'^{SO(7)}_3 = 1.
\end{align} 

Finally we look at the intersection numbers involving the $-8$ curves for the $E_7$ gaugings. A $-8$ curve corresponding to $C_1^{(-8)}$ is 
\begin{align}
b^{E_7}_1 = (1, 0, -3, -2, -1, -1, -1, -3, -2, -1, -1, -1, -2, -1, 0, \cdots, 0) \label{E7T4Z4Z4} 
\end{align}
The intersection number with the canonical class \eqref{aT4Z4Z4} and also the self-intersection number reproduces the condition \eqref{anomalyE7T6Z4Z4}. We can also check that the intersection numbers involving the other $-8$ curves satisfy \eqref{anomalyE7T6Z4Z4}.

\subsubsection{$T^6/\mathbb{Z}_6\times \mathbb{Z}_6$}
\begin{figure}
  \centering
  \includegraphics[width=.85\linewidth]{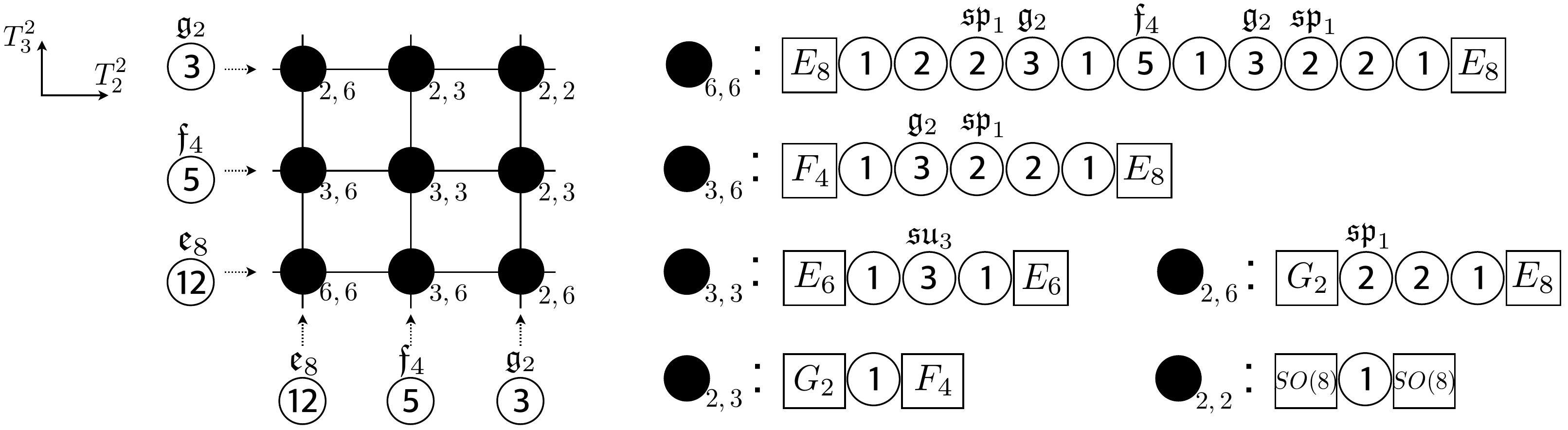}
  \caption{6d gravity model of $T^6/\mathbb{Z}_6\times \mathbb{Z}_6$ geometry. }
  \label{fig:T4-Z6Z6}
\end{figure}

The field theory sector of the 6d gravity theory realized by F-theory compactified on $T^6/\mathbb{Z}_6\times \mathbb{Z}_6$ is drawn in Figure \ref{fig:T4-Z6Z6}. 
The base $T^4/\mathbb{Z}_6\times \mathbb{Z}_6$ has nine fixed points of the orbifold action in (\ref{eq:orbifold-action}) with $\alpha^6=w^6=1$. There are two $\mathbb{Z}_2\times \mathbb{Z}_3$ fixed points, two $\mathbb{Z}_2\times \mathbb{Z}_6$ fixed points, two $\mathbb{Z}_3\times \mathbb{Z}_6$ fixed points, one $\mathbb{Z}_2\times \mathbb{Z}_2$ fixed point, one $\mathbb{Z}_3\times \mathbb{Z}_3$ fixed point, and one $\mathbb{Z}_6\times \mathbb{Z}_6$ fixed point. The 6d field theory localized at each fixed point is the $(G,G')$ conformal matter theory of the corresponding orbifold $\mathbb{Z}_m\times \mathbb{Z}_n$ given in Figure \ref{fig:orbifold}. These local 6d conformal matter theories are glued by two $G_2$ gauge symmetries on $-3$ curves and two $F_4$ gauge symmetries on $-5$ curves and two $E_8$ gauge symmetries on $-12$ curves as illustrated in Figure \ref{fig:T4-Z6Z6}. 
We conjecture that this 6d field theory in Figure \ref{fig:T4-Z6Z6} is the non-gravitational sector of the 6d supergravity associated to $T^6/\mathbb{Z}_6\times \mathbb{Z}_6$. 

\subsubsection*{Gravitational anomaly}
The intersection matrix $\Omega$ of the 6d field theory is a $39\times 39$ symmetric matrix. Its diagonal elements are
\begin{equation}
	\text{diag} \ \Omega = -\left(1,2,2,3,1,5,1,3,2,2,1,1,3,1,1,\{1,3,2,2,1\}^2,\{2,2,1\}^2,1^2,3^2,5^2,12^2\right) \ , \nonumber
\end{equation}
and off-diagonal elements are given by the rule in (\ref{eq:intersect-matrix-off}). This matrix has signature
\begin{equation}
	(-,+,0) = (34,1,4) \ .
\end{equation}
This tells us that there are four null tensors and the number of dynamical tensors is
\begin{equation}
	h^{1,1}(B)=35 \ , \quad T=34 \ .
\end{equation}
The rank of the gauge groups and number of vector multiplets are
\begin{equation}
	r(V) = 48 \ , \quad V = 762 \ .
\end{equation}
So the field theory counting $h^{1,1}(X)=h^{1,1}(B)+r(V)+1=94$ agrees with the topological numbers of the $T^6/\mathbb{Z}_6\times\mathbb{Z}_6$ geometry given in Table \ref{tb:orbifold2}:
\begin{equation}
	h^{1,1}(X) = 94 \ , \quad h^{2,1}(X) = 0 \ .
\end{equation}
The number of hypermultiplets is
\begin{equation}
	H = 48 + 1 =49 \ .
\end{equation}
Therefore, the field theory model for the gravity theory associated to $T^6/\mathbb{Z}_6\times\mathbb{Z}_6$ satisfies gravitational anomaly cancellation:
\begin{equation}
	H-V+29T-273= 49-762+29\times 34-273=0 \ .
\end{equation} 

\subsubsection*{Gauge/gravity mixed anomaly}

The anomaly cancellation conditions are given by
\begin{align}
		a \cdot a &= K_B^2 = 9 - T  = -25,\label{anomaly1T6Z6Z6}
\end{align}
and \eqref{anomaly0T6Z4Z4}, \eqref{GSanomaly4}, \eqref{GSanomaly5} and \eqref{bifundT6Z4Z4} where $I$ now labels the curves on which we have gauge groups for the $T^6/\mathbb{Z}_6\times \mathbb{Z}_6$ model. We also have matter in the representation of $\frac{1}{2}({\bf 2}, {\bf 7}) + \frac{1}{2}({\bf 2}, {\bf 1})$ under $SU(2) \times G_2$. The anomaly cancellation condition associated to curves with $SU(2)$ is given by \eqref{anomalySU2T6Z4Z4} and \eqref{anomaly0T6Z4Z4} is satisfied. As for the $G_2$ gauge groups, the adjoint representation of $G_2$ gives 
\begin{align}
A_{\text{adj}} = 4, \quad B_{\text{adj}} = 0, \quad C_{\text{adj}} = \frac{5}{2}, \quad \lambda_{G_2} = 2, 
\end{align}
and the fundamental representation of $G_2$ gives
\begin{align}
A_{\text{fund}} = 1, \quad B_{\text{fund}} = 0, \quad C_{\text{fund}} = \frac{1}{4}, 
\end{align}
Hence the anomaly cancellation condition for curves with $G_2$ is 
\begin{align}
a \cdot b^{G_2}_I &= 1, \quad b_I^{G_2} \cdot b_I^{G_2}  = -3, \label{anomalyG2T6Z6Z6}
\end{align}
and \eqref{anomaly0T6Z4Z4} is trivially satisfied. The anomaly cancellation condition for the bifundamental \eqref{bifundT6Z4Z4} becomes
\begin{align}
b_I^{SU(2)} \cdot b_J^{G_2}  = 1, \label{anomalyBFT6Z6Z6}
\end{align}
where the curves correspond to $b_I^{SU(2)}$ and $b_J^{G_2}$ are next to each other.
We also have $F_4$ gauge groups and the adjoint representation of $F_4$ yields 
\begin{align}
A_{\text{adj}} = 3, \quad B_{\text{adj}} = 0, \quad C_{\text{adj}} = \frac{5}{12}, \quad \lambda_{F_4} = 6. \label{numbersF4}
\end{align}
Hence the anomaly cancellation conditions become
\begin{align}
a \cdot b_I^{F_4} &= 3, \quad b_I^{F_4} \cdot b_I^{F_4} = -5, \label{anomalyF4T6Z6Z6}
\end{align}
and \eqref{anomaly0T6Z4Z4} is satisfied. Finally, the adjoint representation of $E_8$ gives
\begin{align}
A_{\text{adj}} = 1, \quad B_{\text{adj}} = 0, \quad C_{\text{adj}} = \frac{1}{100}, \quad \lambda_{F_4} = 60.
\end{align}
Hence \eqref{anomaly0T6Z4Z4} is satisfied and non-trivial anomaly cancellation conditions are
\begin{align}
a \cdot b_I^{E_8} &= 10, \quad b_I^{E_8} \cdot b_I^{E_8}  = -12. \label{anomalyE8T6Z6Z6}
\end{align}
 
Let us then see the anomaly cancellation conditions \eqref{anomaly1T6Z6Z6}. \eqref{anomalySU2T6Z4Z4}, \eqref{anomalyG2T6Z6Z6}, \eqref{anomalyBFT6Z6Z6} and \eqref{anomalyF4T6Z6Z6} are indeed satisified in the quiver model for the $T^6/\mathbb{Z}_6 \times \mathbb{Z}_6$. 
Again not all the curves are independent as they are subject to the following constraints: 
\begin{align}
C\equiv &C^{(-12)}_1 +\Big(\big(\sum_{k=1}^4(6-k)C^{(E_8, E_8)}_{11k}\big) + 3C^{(E_8, E_8)}_{1115} + C^{(E_8, E_8)}_{116} + 2C^{(E_8, E_8)}_{117} + \sum_{k=8}^{11}C^{(E_8, E_8)}_{11k}\Big) \cr
&+ \Big(\big(\sum_{k=1}^4(5-k)C^{(E_8, F_4)}_{12k}\big) + C^{(E_8, F_4)}_{125}\Big) + \sum_{k=1}^3(4-k)C^{(E_8, G_2)}_{13k}\nonumber\\
=&C^{(-5)}_2 + \Big(2C^{(F_4, E_8)}_{211} + \sum_{k=2}^5C^{(F_4, E_8)}_{21k}\Big) + \Big(2C^{(F_4, F_4)}_{221} + C^{(F_4, F_4)}_{222} + C^{(F_4, F_4)}_{223}\Big) +  C^{(F_4, G_2)}_{331} \nonumber\\
=&C^{(-3)}_3 + \Big(\sum_{k=1}^3 C^{(G_2, E_8)}_{31k}\Big) + C^{(G_2, F_4)}_{321} + C^{(G_2, G_2)}_{331},
\end{align}
and 
\begin{align}
C'\equiv &C'^{(-12)}_1+\Big(\big(\sum_{k=1}^4C^{(E_8, E_8)}_{11k}\big) + 2C^{(E_8, E_8)}_{115} + C^{(E_8, E_8)}_{116} + 3C^{(E_8, E_8)}_{117} + \sum_{k=8}^{11}(k-6)C^{(E_8, E_8)}_{11k}\Big) \cr
&+ \Big(C^{(F_4, E_8)}_{211} + \big(\sum_{k=2}^5(k-1)C^{(F_4, E_8)}_{21k}\big)\Big) + \sum_{k=1}^3kC^{(G_2, E_8)}_{31k}\nonumber\\
=&C'^{(-5)}_2 + \Big(\big(\sum_{k=1}^4C^{(E_8, F_4)}_{12k}\big) + 2C^{(E_8, F_4)}_{125}\Big) + \Big(C^{(F_4, F_4)}_{221} + C^{(F_4, F_4)}_{222} + 2C^{(F_4, F_4)}_{223}\Big) +  C^{(G_2, F_4)}_{321} \nonumber\\
=&C'^{(-3)}_3 + \Big(\sum_{k=1}^3 C^{(E_8, G_2)}_{13k}\Big) + C^{(F_4, G_2)}_{231} + C^{(G_2, G_2)}_{331}.
\end{align}
We can then choose the basis 
\begin{align}
\Big(C, C', \{C^{(E_8, E_8)}_{11k}\}, \{C^{(E_8, F_4)}_{12k}\}, C^{(E_8, G_2)}_{131}, \{C^{(F_4, E_8)}_{21k}\}, &\{C^{(F_4, F_4)}_{22k}\}, C^{(F_4, G_2)}_{231},\cr
& \{C^{(G_2, E_8)}_{31k}\}, C^{(G_2, F_4)}_{321}, C^{(G_2, G_2)}_{331}\Big),\cr
\end{align}
which leads to the following reduced intersection matrix:
\begin{eqnarray}
\tilde{\Omega} &=& \scalemath{0.75}{\left(
\begin{array}{cc}
0 & 1 \\
1 & 0
\end{array}
\right)\oplus 
\left(
\begin{array}{ccccccccccc}
-1 & 1 & 0 & 0 & 0 & 0 & 0 & 0 & 0 & 0 & 0\\
1 & -2 & 1 & 0 & 0 & 0 & 0 & 0 & 0 & 0 & 0\\
0 & 1 & -2 & 1 & 0 & 0 & 0 & 0 & 0 & 0 & 0\\
0 & 0 & 1 & -3 & 1 & 0 & 0 & 0 & 0 & 0 & 0\\
0 & 0 & 0 & 1 & -1 & 1 & 0 & 0 & 0 & 0 & 0\\
0 & 0 & 0 & 0 & 1 & -5 & 1 & 0 & 0 & 0 & 0\\
0 & 0 & 0 & 0 & 0 & 1 & -1 & 1 & 0 & 0 & 0\\
0 & 0 & 0 & 0 & 0 & 0 & 1 & -3 & 1 & 0 & 0\\
0 & 0 & 0 & 0 & 0 & 0 & 0 & 1 & -2 & 1 & 0\\
0 & 0 & 0 & 0 & 0 & 0 & 0 & 0 & 1 & -2 & 1\\
0 & 0 & 0 & 0 & 0 & 0 & 0 & 0 & 0 & 1 & -1
\end{array}
\right)
 \oplus \left(
\begin{array}{ccccc}
-1 & 1 & 0 & 0 & 0\\
1 & -2 & 1 & 0 & 0\\
0 & 1 & -2 & 1 & 0\\
0 & 0 & 1 & -3 & 1\\
0 & 0 & 0 & 1 & -1
\end{array}
\right)\oplus \left(
\begin{array}{ccc}
-1 & 1 & 0\\
1 & -2 & 1\\
0 & 1 & -2
\end{array}
\right)}\nonumber\\
&&\scalemath{0.75}{ \oplus  \left(
\begin{array}{ccccc}
-1 & 1 & 0 & 0 & 0\\
1 & -3 & 1 & 0 & 0\\
0 & 1 & -2 & 1 & 0\\
0 & 0 & 1 & -2 & 1\\
0 & 0 & 0 & 1 & -1
\end{array}
\right)\oplus \left(
\begin{array}{ccc}
-1 & 1 & 0\\
1 & -3 & 1\\
0 & 1 & -1
\end{array}
\right) \oplus (-1) \oplus \left(
\begin{array}{ccc}
-2 & 1 & 0\\
1 & -2 & 1\\
0 & 1 & -1
\end{array}
\right) \oplus (-1) \oplus (-1)}. \label{omegaT4Z6Z6}
\end{eqnarray}

One can confirm the anomaly cancellation conditions \eqref{anomaly1T6Z6Z6}, \eqref{anomalySU2T6Z4Z4}, \eqref{anomalyG2T6Z6Z6}, \eqref{anomalyBFT6Z6Z6} and \eqref{anomalyF4T6Z6Z6} are satisfied by using \eqref{omegaT4Z6Z6}. The canonical class $a$ can be computed from \eqref{afromselfint} and it is given by
\begin{align}
a = (-2, -2, 5, 4, 3, 2, 4, 1, 4, 2, 3, 4, 5, 4, 3, 2, 1, 2, 3, 2, 1, 2, 1, 2, 3, 4, 2, 1, 2, 1, 1, 2, 3, 1, 1), \label{aT4Z6Z6}
\end{align}
which satisfies \eqref{anomaly1T6Z6Z6}.
The $-2$ curve for $C^{(E_8, E_8)}_{113}$and the $-3$ curve of $C^{(E_8, E_8)}_{114}$, which are adjacent, are given by
\begin{align}
b^{SU(2)}_{113} &= (0,0,0,0,1,0,\cdots, 0),~~b^{G_2}_{114}  = (0,0,0,0,0,1,0,\cdots, 0). 
\end{align}
These curves satisfy the relevant anomaly cancellation conditions \eqref{anomalySU2T6Z4Z4}, \eqref{anomalyG2T6Z6Z6}, \eqref{anomalyBFT6Z6Z6}. The $-5$ curve corresponding to $C^{(E_8, E_8)}_{116}$ in the $(E_8, E_8)$ conformal matter theory is 
\begin{align}
b^{F_4}_{116} = (0^7, 1, 0^{27}), 
\end{align}
reproducing \eqref{anomalyF4T6Z6Z6}. 

We also have gauging curves $C^{(-12)}_1, C'^{(-12)}_1, C^{(-5)}_2, C'^{(-5)}_2, C^{(-3)}_3, C'^{(-3)}_3$, with classes given by
\begin{align}
b^{E_8}_1 =(&1, 0, -5, -4, -3, -2, -3, -1, -2, -1^4, -4, -3, -2, -1^2, -3, -2, -1,0^{14}),\\
b'^{E_8}_1 =(&0, 1, -1^4, -2, -1, -3, -2, -3, -4, -5, 0^8, -1^2, -2, -3, -4, 0^4, -1, -2, -3, 0^2),\\
b^{F_4}_2 =(&1, 0^{20}, -2, -1^4, -2, -1^3, 0^5),\\
b'^{F_4}_2 =(&0, 1, 0^{11}, -1^4, -2, 0^8, -1^2, -2, 0^4, -1, 0),\\
b^{G_2}_3 =(&1, 0^{29}, -1^5), \label{G2gaugingT4Z4Z4}\\
b'^{G_2}_3 =(&0, 1, 0^{16}, -1^3, 0^8, -1, 0^4 -1), \label{G2gauging2T4Z4Z4}
\end{align}
respectively. With the reduced intersection matrix \eqref{omegaT4Z6Z6}, it is possible to check that the above classes satisfy \eqref{anomalyE8T6Z6Z6}, \eqref{anomalyF4T6Z6Z6} and \eqref{anomalyG2T6Z6Z6}. The anomaly cancellation condition of \eqref{anomalyBFT6Z6Z6} is also satisfied by the curve \eqref{G2gaugingT4Z4Z4} or \eqref{G2gauging2T4Z4Z4} with the $-2$ curve connected to it.

\subsubsection{$T^6/\mathbb{Z}_2\times \mathbb{Z}_4$}
\begin{figure}
  \centering
  \includegraphics[width=.6\linewidth]{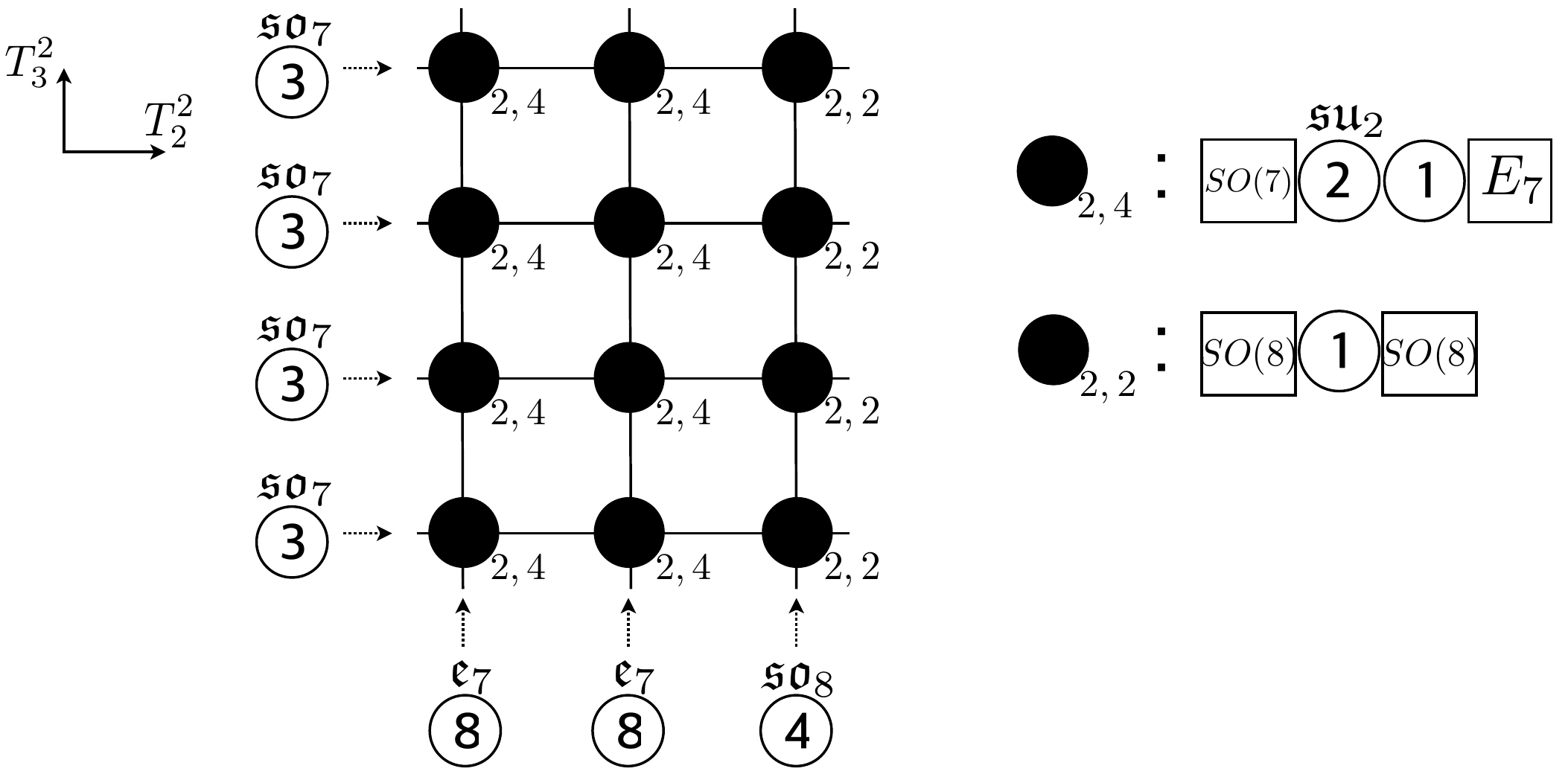}
  \caption{6d gravity model of $T^6/\mathbb{Z}_2\times \mathbb{Z}_4$ geometry. }
  \label{fig:T4-Z2Z4}
\end{figure}

The field theory sector for the 6d gravity theory realized by F-theory compactified on $T^6/\mathbb{Z}_2\times \mathbb{Z}_4$ is given in Figure \ref{fig:T4-Z2Z4}.
On the base $T^4/\mathbb{Z}_2\times \mathbb{Z}_4$, there are 12 fixed points of the orbifold action generated by $g$ and $h$ in (\ref{eq:orbifold-action}) with $\alpha^2=1$ and $w^4=1$. Among these, eight fixed points are local geometries $\mathbb{C}^2/\mathbb{Z}_2\times \mathbb{Z}_4$ described by the $(SO(7),E_7)$ conformal matter theory and four fixed points are local geometries $\mathbb{C}^2/\mathbb{Z}_2\times \mathbb{Z}_2$ described by $(SO(8),SO(8))$ conformal matter theories. These conformal matter theories are glued by four $SO(7)$ gauge symmetries on $-3$ curves, two $E_7$ gauge symmetries on $-8$ curves, and one $SO(8)$ gauge symmetry on a $-4$ curve as drawn in Figure \ref{fig:T4-Z2Z4}. We conjecture that the 6d field theory constructed in this way realizes the non-gravitational field theory sector of the 6d gravity theory coming from $T^6/\mathbb{Z}_2\times \mathbb{Z}_4$ in F-theory.

\subsubsection*{Gravitational anomaly}
The intersection matrix $\Omega$ of the 6d field theory is a $27\times 27$ symmetric matrix, with diagonal elements
\begin{equation}
	\text{diag} \ \Omega = -\left(\{2,1\}^8,1^4,3^4,8^2,4\right) \ , \nonumber
\end{equation}
and off-diagonal elements given by the rule in (\ref{eq:intersect-matrix-off}). The matrix $\Omega$ has signature
\begin{equation}
	(-,+,0) = (21,1,5) \ .
\end{equation}
This means that we have five null tensors and the number of dynamical tensor multiplets is 
\begin{equation}
	h^{1,1}(B)=22 \ , \quad T = 21 \ .
\end{equation}
This theory has the following gauge symmetries: eight $SU(2)$'s, four $SO(7)$'s, two $E_7$'s, one $SO(8)$.
Thus the rank of the gauge groups and number of the vector multiplets are, respectively, 
\begin{equation}
	r(V) = 38 \ , \quad V = 402 \ .
\end{equation}
The field theory counting $h^{1,1}(X)=h^{1,1}(B)+r(V)+1=61$ agrees with the geometric data of $T^6/\mathbb{Z}_2\times\mathbb{Z}_4$ in Table \ref{tb:orbifold2}, namely
\begin{equation}
	h^{1,1}(X) = 61 \ , \quad h^{2,1}(X) = 1 \ .
\end{equation}
There are 16 half hypermultiplets in each intersection between $SO(7)$ and $SU(2)$ gauge groups.
So the number of hypermultiplets is $H=8\times 8 +h^{2,1}(Y)+1 = 66$. Therefore our field theory model satisfies gravitational anomaly cancellation, as
\begin{equation}
	H-V+29T -273 = 66 -402 + 29\times 21 - 273=0 \ .
\end{equation} 

\subsubsection*{Gauge/gravity mixed anomaly}

As stated above, the gauge groups in this model are $E_7, SO(7), SU(2), SO(8)$ and the corresponding anomaly cancellation conditions are \eqref{anomalyE7T6Z4Z4}, \eqref{anomalySO7T6Z4Z4}, \eqref{anomalySU2T6Z4Z4}, and \eqref{anomaly2T6Z2Z2} with 
\begin{align}
a \cdot a = K_B^2 = 9 - T = -12. \label{anomaly1T6Z2Z4}
\end{align}
We also have hypermultiplets in the representation $\frac{1}{2}({\bf 2}, {\bf 8})$ of $SU(2) \times SO(7)$, which leads to an additional condition \eqref{anomalyBFT6Z4Z4}. As usual, to show the anomaly cancellation conditions are indeed satisfied in this model, we use a reduced intersection matrix. 
The constraints among the curves in Figure \ref{fig:T4-Z2Z4} are
\begin{align}
C\equiv & C^{(-3)}_i + \sum_{j=1,k=1}^2C_{ijk}^{(SO(7), E_7)} + C_{i31}^{(SO(8), SO(8))},~~ i = 1, \dots, 4
\end{align}
and 
\begin{align}
C' \equiv & C_1'^{(-8)} + \sum_{i=1}^3\left(C_{i11}^{(SO(7), E_7)} + 2C_{i12}^{(SO(7), E_7)}\right)\nonumber\\
=&C_2'^{(-8)} + \sum_{i=1}^3\left(C_{i21}^{(SO(7), E_7)} + 2C_{i22}^{(SO(7), E_7)}\right)\nonumber\\
=&C'^{(-4)}_3 + \sum_{i=1}^4 C_{i31}^{(SO(8), SO(8))}.
\end{align}
Thus we choose the basis 
\begin{align}
\big(C, C', &\{C_{11k}^{(SO(7), E_7)}\}, \{C_{12k}^{(SO(7), E_7)}\}, C_{131}^{(SO(8), SO(8))}, \{C_{21k}^{(SO(7), E_7)}\}, \{C_{22k}^{(SO(7), E_7)}\}, C_{231}^{(SO(8), SO(8))}, \cr
&\{C_{31k}^{(SO(7), E_7)}\}, \{C_{32k}^{(SO(7), E_7)}\}, C_{331}^{(SO(8), SO(8))}, \{C_{41k}^{(SO(7), E_7)}\}, \{C_{42k}^{(SO(7), E_7)}\}, C_{431}^{(SO(8), SO(8))}\big).
\end{align}
The reduced intersection matrix in this basis is
\begin{eqnarray}
\tilde{\Omega} = \left(
\begin{array}{cc}
0 & 1 \\
1 & 0
\end{array}
\right) \oplus  \left\{\left(
\begin{array}{cc}
-2 & 1 \\
1 & -1
\end{array}
\right) \oplus \left(
\begin{array}{cc}
-2 & 1 \\
1 & -1
\end{array}
\right)  \oplus (-1)\right\}^4. \label{redomegaT4Z2Z4}
\end{eqnarray}

We then check the anomaly cancellation conditions \eqref{anomaly1T6Z2Z4}, \eqref{anomalyE7T6Z4Z4}, \eqref{anomalySO7T6Z4Z4}, \eqref{anomalySU2T6Z4Z4}, \eqref{anomaly2T6Z2Z2} and \eqref{anomalyBFT6Z4Z4} by using the reduced intersection matrix \eqref{redomegaT4Z2Z4}. From \eqref{afromselfint}, the canonical class is given by
\begin{align}
a = (-2, -2, 1, 2, 1, 2, 1, 1, 2, 1, 2, 1, 1, 2, 1, 2, 1, 1, 2, 1, 2, 1). \label{aT6Z2Z4}
\end{align}
and indeed satisfies  \eqref{anomaly1T6Z2Z4}. The $-8$ curve of $C_1^{(-8)}$ is 
\begin{align}
b_1'^{E_7} = (0, 1, -1, -2, 0, 0, 0, -1, -2, 0, 0, 0, -1, -2, 0, 0, 0, -1, -2, 0, 0, 0). \label{E7T6Z2Z4}
\end{align}
Note that \eqref{E7T6Z2Z4} and \eqref{redomegaT4Z2Z4} leads to \eqref{anomalyE7T6Z4Z4}. The $-3$ curve of $C_1^{(-3)}$ is given by
\begin{align}
b^{SO(7)}_1 = (1, 0, -1,-1,-1,-1,-1,0^{15}), 
\label{SO7T6Z2Z4}
\end{align}
and reproduces \eqref{anomalySO7T6Z4Z4}. The $-2$ curve of $C^{(SO(7), E_7)}_{111}$ is 
\begin{align}
b^{SU(2)}_{111} = (0, 0, 1,0^{19}), 
\label{SU2T6Z2Z4}
\end{align}
and satisfies \eqref{anomalySU2T6Z4Z4}. Furthermore, the intersection between \eqref{SO7T6Z2Z4} and \eqref{SU2T6Z2Z4} yields \eqref{anomalyBFT6Z4Z4}. Finally, the $-4$ curve $C'^{(-4)}_3$ is 
\begin{align}
b^{SO(8)}_3=(0, 1, 0, 0, 0, 0, -1, 0, 0, 0, 0, -1, 0, 0, 0, 0, -1, 0, 0, 0, 0, -1),
\end{align}
reproducing \eqref{anomaly2T6Z2Z2}. Although we have only demonstrated anomaly cancellation using one curve of each type in the full set of $-8$, $-3$, and $-2$ curves, it is straightforward to check that the other curves also satisfy \eqref{anomalyE7T6Z4Z4}, \eqref{anomalySO7T6Z4Z4}, \eqref{anomalySU2T6Z4Z4}, \eqref{anomaly2T6Z2Z2}, and \eqref{anomalyBFT6Z4Z4}.

\subsubsection{$T^6/\mathbb{Z}_2\times \mathbb{Z}_6$}
\begin{figure}
  \centering
  \includegraphics[width=.6\linewidth]{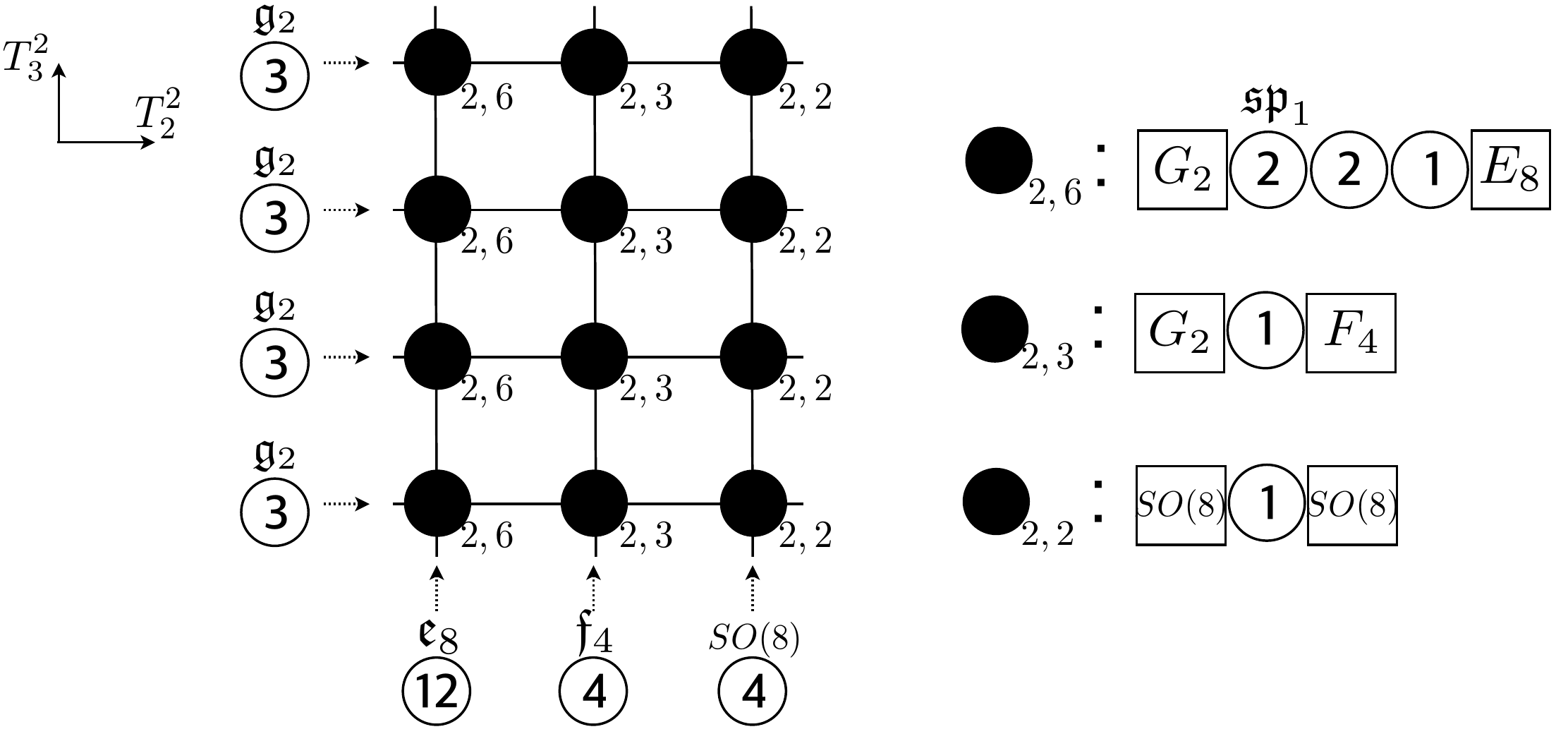}
  \caption{6d gravity model of $T^6/\mathbb{Z}_2\times \mathbb{Z}_6$ geometry. }
  \label{fig:T4-Z2Z6}
\end{figure}

The field theory sector of the 6d gravity theory realized by F-theory compactified on $T^6/\mathbb{Z}_2\times \mathbb{Z}_6$ is given in Figure \ref{fig:T4-Z2Z6}.
The base $B=T^4/\mathbb{Z}_2\times \mathbb{Z}_6$ of the elliptic fibration has 12 fixed points under the orbifold action $\mathbb{Z}_2\times \mathbb{Z}_6$ of type in (\ref{eq:orbifold-action}). Four of them are locally $\mathbb{C}^2/\mathbb{Z}_2\times \mathbb{Z}_6$ geometries described by the $(G_2,E_8)$ conformal matter theory, another four fixed points are locally $\mathbb{C}^2/\mathbb{Z}_2\times\mathbb{Z}_3$ geometries described by the $(G_2,F_4)$ conformal matter theory, and the remaining four fixed points are $\mathbb{C}^2/\mathbb{Z}_2\times \mathbb{Z}_2$ geometries associated to the $(SO(8),SO(8))$ conformal matter system. These local 6d theories are glued by seven tensor nodes with 7-branes carrying five $G_2$ gauge groups, one $E_8$ gauge group, and one $F_4$ gauge group as drawn in Figure \ref{fig:T4-Z2Z6}. We emphasize here that the 7-brane of the $F_4$ gauge group is wrapping a $-4$ curve, so this tensor node couples to $N_f=1$ fundamental hypermultiplet of the $F_4$ gauge group \cite{Heckman:2015bfa}. This is consistent with the LST formed by four $(G_2,F_4)$ conformal matter theories glued by the $F_4$ gauge group in the second vertical line in Figure \ref{fig:T4-Z2Z6}. Note that the LST involves only one null tensor multiplet, which is only possible when the $F_4$ gauge 7-brane wraps the $-4$ curve. We will also see that this is consistent with the anomaly cancellation of the full compact geometry.

\subsubsection*{Gravitational anomaly}
The intersection matrix $\Omega$ of the 6d field theory is a $27\times 27$ symmetric matrix. Its diagonal elements are
\begin{equation}
	\text{diag} \ \Omega = -\left(\{2,2,1\}^4,1^4,1^4,3^4,4,4,12\right) \ , \nonumber
\end{equation}
and off-diagonal elements are given by the rule in  (\ref{eq:intersect-matrix-off}). The matrix $\Omega$ has signature 
\begin{equation}
	(-,+,0) = (21,1,5) \ .
\end{equation}
This means that we have five null tensors and the number of dynamical tensor multiplets is 
\begin{equation}
	h^{1,1}(B)=22 \ , \quad T = 21 \ .
\end{equation}
This theory has the following gauge symmetries: four $SU(2)$'s, four $G_2$'s, one $E_8$, one $F_4$, and one $SO(8)$.
Thus the rank of the gauge groups and the number of the vector multiplets are, respectively,
\begin{equation}
	r(V) = 28 \ , \quad V = 396 \ .
\end{equation}
The field theory counting $h^{1,1}(X)=h^{1,1}(B)+r(V)+1=51$ agrees with the geometric data of $T^6/\mathbb{Z}_2\times\mathbb{Z}_6$ in Table \ref{tb:orbifold2}, namely
\begin{equation}
	h^{1,1}(X) = 51 \ , \quad h^{2,1}_{\text{untwisted}}(X) = 1 \ .
\end{equation}
There are 16 half hypermultiplets in each intersection between $G_2$ and $SU(2)$ gauge groups, and also we have one fundamental, and so 26 hypermultiplets of the $F_4$ gauge group.
This tells us that the number of hypermultiplets is $H=4\times 8 +26+h^{2,1}_{\text{untwisted}}(X)+1 = 60$. Therefore our field theory model has no gravitational anomaly
\begin{equation}
	H-V+29T -273 = 60 -396 + 29\times 21 - 273=0 \ .
\end{equation}

We remark that the $F_4$ gauge group should be put on $-4$ curve with one fundamental hypermultiplet, rather than a $-5$ curve, to be consistent with the gravity theory. With this choice, the field theory has the correct numbers of dynamical tensors and hypermultiplets. In addition, the two twisted massless hypermultiplets corresponding to $h^{2,1}_{\rm twisted}(X)=2$ in the geometry $T^6/\mathbb{Z}_2\times \mathbb{Z}_6$ are precisely the two singlets of $U(1)^4\subset F_4$ in the fundamental hypermultiplet of the $F_4$ gauge group. This provides a novel understanding of the geometric data $h^{2,1}_{\rm twisted}(X)$ as the modes arising from 6d local CFT degrees of freedom.

\subsubsection*{Gauge/gravity mixed anomaly}

In addition to 
\begin{align}
a \cdot a &=K_B^2 = 9 - T =  -12,\label{anomaly1T6Z2Z6}
\end{align}
the anomaly cancellation conditions for $SU(2), G_2, , SO(8), E_8$, namely \eqref{anomalySU2T6Z4Z4}, \eqref{anomalyG2T6Z6Z6}, \eqref{anomaly2T6Z2Z2}, and \eqref{anomalyE8T6Z6Z6} (respectively) along with \eqref{AC2} are satisfied. As for the $F_4$ gauge group, we have a hypermultiplet in the fundamental representation. Using
\begin{align}
A_{\text{fund}} = 1, \quad B_{\text{fund}} = 0, \quad C_{\text{fund}} = \frac{1}{12}, 
\end{align}
with \eqref{numbersF4} yields
\begin{align}
a \cdot b_I^{F_4} &=  2, \quad b_I^{F_4} \cdot b_I^{F_4} = -4, \label{anomalyF4T6Z2Z6}
\end{align}
and \eqref{AC2} is satisified. 

We will reproduce \eqref{anomaly1T6Z2Z6}, \eqref{anomalySU2T6Z4Z4}, \eqref{anomalyG2T6Z6Z6}, \eqref{anomaly2T6Z2Z2}, \eqref{anomalyE8T6Z6Z6}, and \eqref{anomalyF4T6Z2Z6} from the intersection numbers of the quiver model in Figure \ref{fig:T4-Z2Z6}. 
The constraints between the curves in Figure \ref{fig:T4-Z2Z6} are
\begin{align} 
C\equiv C_i^{(-3)} + \sum_{k=1}^3C^{(G_2, E_8)}_{i1k} + C^{(G_2, F_4)}_{i21} + C^{(G_2, SO(8))}_{i31},~~ i = 1, \dots, 4
\end{align}
and 
\begin{align}
C'\equiv & C'^{(-12)}_1 + \sum_{i=1}^4\big(C^{(G_2, E_8)}_{i11} + 2C^{(G_2, E_8)}_{i12} + 3C^{(G_2, E_8)}_{i13} + 4C^{(G_2, E_8)}_{i14}\big),\nonumber\\
=&C'^{(-4)}_{2} + \sum_{i=1}^4C^{(G_2, F_4)}_{i21},\nonumber\\
=&C'^{(-4)}_{3} + \sum_{i=1}^4C^{(G_2, SO(8))}_{i31}.
\end{align}
Then we can choose the basis
\begin{align}
(C, C', &\{C^{(G_2, E_8)}_{11k}\}, C^{(G_2, F_4)}_{121}, C^{(G_2, SO(8))}_{131}, \{C^{(G_2, E_8)}_{21k}\}, C^{(G_2, F_4)}_{221}, C^{(G_2, SO(8))}_{231},\cr
&\{C^{(G_2, E_8)}_{31k}\}, C^{(G_2, F_4)}_{321}, C^{(G_2, SO(8))}_{331},\{C^{(G_2, E_8)}_{41k}\}, C^{(G_2, F_4)}_{421}, C^{(G_2, SO(8))}_{431},), \label{baseT4Z2Z6}
\end{align}
and correspdonding reduced intersection matrix
\begin{eqnarray}\label{omegaT4Z2Z6}
\tilde{\Omega} = \left(
\begin{array}{cc}
0 & 1\\
1 & 0
\end{array}
\right) \oplus \left\{\left(
\begin{array}{ccc}
-2 & 1 & 0\\
1 & -2 & 1\\
0 & 1 & -1
\end{array}
\right) \oplus (-1) \oplus (-1)\right\}^4.
\end{eqnarray}

With the basis \eqref{baseT4Z2Z6}, the canonical class computed from \eqref{afromselfint} is
\begin{align}
a=(-2, -2, 1, 2, 3, 1, 1, 1, 2, 3, 1, 1, 1, 2, 3, 1, 1, 1, 2, 3, 1, 1),
\end{align}
and satisfies \eqref{anomaly1T6Z2Z6}. The $-2$ curve of $C_{111}^{(G_2, E_8)}$with $SU(2)$ is 
\begin{align}
b^{SU(2)}_{111} = (0, 0, 1, 0^{19}), 
\label{bSU2T6Z2Z6}
\end{align}
and gives \eqref{anomalySU2T6Z4Z4}. The class $b$ associated to the curve $C^{(-3)}_1$ is 
\begin{align}
b^{G_2}_1 = (1, 0, -1, -1, -1, -1, -1, 0^{15}), 
\label{bG2T6Z2Z6}
\end{align}
and satisfies \eqref{anomalyG2T6Z6Z6}. Furthermore, \eqref{bSU2T6Z2Z6} and \eqref{bG2T6Z2Z6} reproduce \eqref{anomalyBFT6Z6Z6}. One can check that the other curves with $SU(2)$ or $G_2$ also satisfy \eqref{anomalySU2T6Z4Z4}, \eqref{anomalyG2T6Z6Z6}, and \eqref{anomalyBFT6Z6Z6}. The curve $C'^{(-4)}_{2}$ is given by
\begin{align}
b'^{F_4}_2 = (0, 1, 0, 0, 0, -1, 0, 0, 0, 0, -1, 0, 0, 0, 0, -1, 0, 0, 0, 0, -1, 0).
\end{align}
and yields \eqref{anomalyF4T6Z2Z6}. Finally, the curve $C'^{(-4)}_{3}$ is 
\begin{align}
b'^{SO(8)}_3 = (0, 1, 0, 0, 0, 0, -1, 0, 0, 0, 0, -1, 0, 0, 0, 0, -1, 0, 0, 0, 0, -1),
\end{align}
and satisfies \eqref{anomaly2T6Z2Z2}.

\subsubsection{$T^6/\mathbb{Z}_3\times \mathbb{Z}_6$}
\begin{figure}
  \centering
  \includegraphics[width=.65\linewidth]{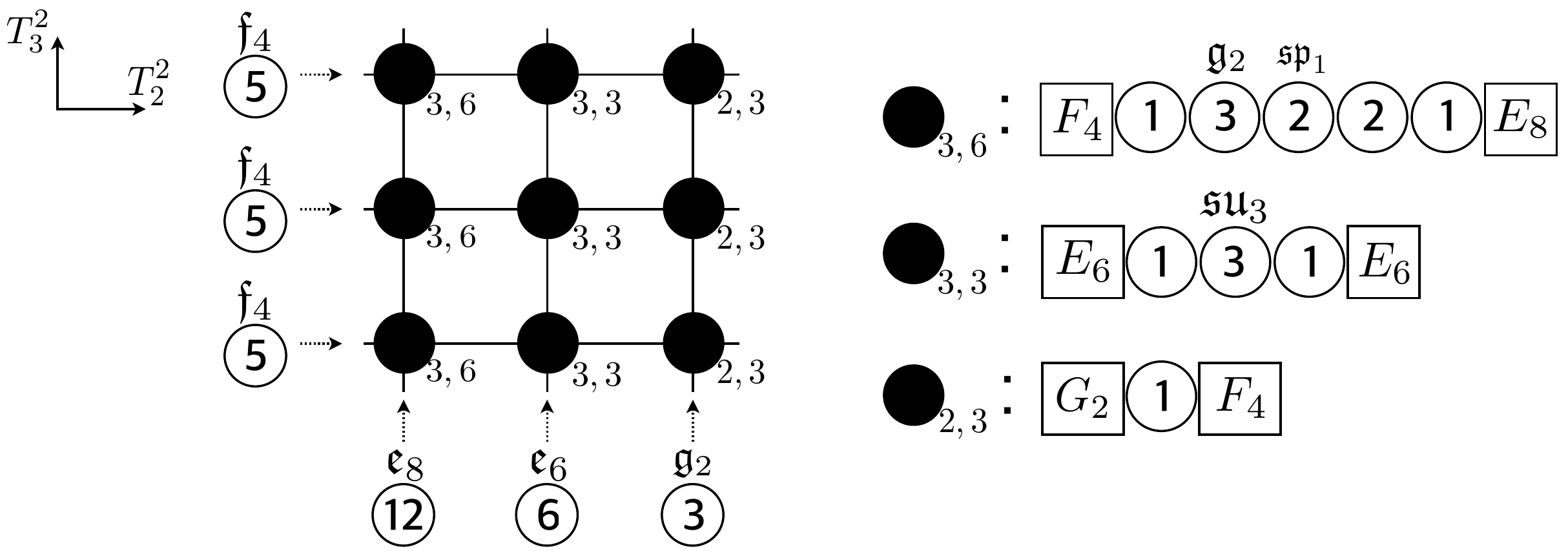}
  \caption{6d gravity model of $T^6/\mathbb{Z}_3\times \mathbb{Z}_6$ geometry. }
  \label{fig:T4-Z3Z6}
\end{figure}

Our last example is the 6d gravity theory obtained from F-theory on $T^6/\mathbb{Z}_3\times \mathbb{Z}_6$. The field theory sector for this gravity theory is drawn in Figure \ref{fig:T4-Z3Z6}. The base $B=T^4/\mathbb{Z}_3\times \mathbb{Z}_6$ has nine fixed points. Three of these fixed points are locally $\mathbb{C}^2/\mathbb{Z}_3\times \mathbb{Z}_6$ described by the $(F_4,E_8)$ conformal matter theory, another three fixed points are locally $\mathbb{C}^2/\mathbb{Z}_3\times \mathbb{Z}_3$ described by the $(E_6,E_6)$ conformal matter theory, and the remaining three fixed points are locally $\mathbb{C}^2/\mathbb{Z}_2\times \mathbb{Z}_6$ described by the $(G_2,F_4)$ conformal matter theory. These theories are glued by three $F_4$ gauge symmetry on $-5$ curves, one $E_8$ gauge symmetry on a $-12$ curve, one $E_6$ gauge symmetry on a $-6$ curve, one $G_2$ gauge symmetry on a $-3$ curve as drawn in Figure \ref{fig:T4-Z3Z6}. Note here that the $G_2$ gauge node gluing three $(G_2,F_4)$ theories contains one fundamental hypermultiplet.
We claim this theory realizes the non-gravitational field theory sector of 6d supergravity of $T^6/\mathbb{Z}_3\times \mathbb{Z}_6$.

\subsubsection*{Gravitational anomaly}
The intersection matrix $\Omega$ of the 6d field theory is a $33\times 33$ symmetric matrix, with diagonal elements 
\begin{equation}
	\text{diag} \ \Omega = -\left(\{1,3,2,2,1\}^3,\{1,3,1\}^3,1^3,5^3,3,6,12\right) \ , \nonumber
\end{equation}
and off-diagonal elements given in (\ref{eq:intersect-matrix-off}). The matrix $\Omega$ has signature
\begin{equation}
	(-,+,0) = (28,1,4) \ .
\end{equation}
This implies that there are four null tensor multiplets and the number of dynamical tensor multiplets is
\begin{equation}
	h^{1,1}(B)=29 \ , \quad T=28 \ .
\end{equation}
The full field theory has the following gauge symmetries: one $G_2$, three $Sp(1)$'s, three $SU(3)$'s, three $F_4$'s, one $E_6$, and one $E_8$. Therefore we have
\begin{equation}
	r(V) = 43 \ , \quad V=571 \ .
\end{equation}
The Hodge numbers of the geometry in Table \ref{tb:orbifold2} are
\begin{equation}
	h^{1,1}(X)=73 \ , \quad h^{2,1}_{\text{untwisted}}(X) = 0\ , \quad h^{2,1}_{\rm twisted}(X) = 1 \ .
\end{equation}
The number of K\"ahler parameters in the field theory, $h^{1,1}(X)=h^{1,1}(B)+r(V)+1 = 73$ therefore agrees with the above Hodge numbers.

There are 16 half hypermultiplets localized at each intersection between the $G_2$ and $Sp(1)$ gauge groups and in addition there is one fundamental hypermultiplet of the $G_2$ gauge group gluing three $(G_2,F_4)$ conformal matters. The total number of hypermultiplets is thus $H=3\times 8 + 7 + 1 = 32$. Again, our field theory model gives a natural interpretation for $h^{2,1}_{\rm twisted}(X)=1$ in this geometry---this corresponds to the singlet field of $U(1)^2\subset G_2$ in the fundamental representation of the $G_2$.

One can now show that the gravitational anomaly vanishes for our model:
\begin{equation}
	H-V+29T-273 = 32 -571+29\times 28-273=0 \ .
\end{equation}

\subsubsection*{Gauge/gravity mixed anomaly}

The anomaly cancellation of this model requires
\begin{align}
a \cdot a & = K_B^2 =9 - T  = -19,\label{anomaly1T6Z3Z6}
\end{align}
and also \eqref{anomalySU2T6Z4Z4}, \eqref{anomalyG2T6Z6Z6}, \eqref{anomaly2T6Z3Z3}, \eqref{anomaly3T6Z3Z3}. \eqref{anomalyE8T6Z6Z6} for the $SU(2), G_2, SU(3), E_6, E_8$ gauge groups. 
The curves in Figure \ref{fig:T4-Z3Z6} are subject to the conditions 
\begin{align}
C\equiv  C^{(-5)}_i + \big(2C^{(F_4, E_8)}_{i11} + \sum_{k=2}^5C^{(F_4, E_8)}_{i1k}\big) + \big(2C^{(F_4, E_6)}_{i21} + \sum_{k=2}^3C^{(F_4, E_8)}_{i2k}\big) + C^{(F_4, G_2)}_{i31},
\end{align}
for $i = 1, 2, 3$, and 
\begin{align}
C'\equiv &C'^{(-12)}_1 + \sum_{i=1}^3\big(C^{(F_4, E_8)}_{i11} + \sum_{K=2}^5(k-1)C^{(F_4, E_8)}_{i1k}\big)\nonumber\\
=&C'^{(-6)}_2 + \sum_{i=1}^3\big(C^{(F_4, E_6)}_{i21} + \sum_{k=2}^3(k-1)C^{(F_4, E_6)}_{i2k}\big)\nonumber\\
=&C'^{(-3)}_3 + \sum_{i=1}^3C^{(F_4, G_2)}_{i31}.
\end{align}
Thus we can choose a basis
\begin{align}
\big(C, C'. \{C^{(F_4, E_8)}_{11k}\}, \{C^{(F_4, E_6)}_{12k}\}, C^{(F_4, G_2)}_{131}, &\{C^{(F_4, E_8)}_{21k}\}, \{C^{(F_4, E_6)}_{22k}\}, C^{(F_4, G_2)}_{231},\cr &\{C^{(F_4, E_8)}_{31k}\}, \{C^{(F_4, E_6)}_{32k}\}, C^{(F_4, G_2)}_{331}\big), \label{baseT4Z3Z6}
\end{align}
and the corresponding reduced intersection matrix  
\begin{eqnarray}
\tilde{\Omega} &=& \left(
\begin{array}{cc}
0 & 1 \\
1 & 0
\end{array}
\right) \oplus \left\{\left(
\begin{array}{ccccc}
-1 & 1 & 0 & 0 & 0\\
1 & -3 & 1 & 0 & 0\\
0 & 1 & -2 & 1 & 0\\
0 & 0 & 1  & -2 & 1\\
0 & 0 & 0 & 1 & -1
\end{array}
\right) \oplus \left(
\begin{array}{ccc}
-1 & 1 & 0\\
1 & -3 & 1 \\
0 & 1 & -1
\end{array}
\right) \oplus (-1)\right\}^3. \label{omegaT4Z3Z6}
\end{eqnarray}

It is now possible to confirm the anomaly cancellations conditions using \eqref{baseT4Z3Z6} and \eqref{omegaT4Z3Z6}. The canonical class can be computed from \eqref{afromselfint}:
\begin{align}
a = (-2, -2, 2, 1, 2, 3, 4, 2, 1, 2, 1, 2, 1, 2, 3, 4, 2, 1, 2, 1, 2, 1, 2, 3, 4, 2, 1, 2, 1),
\end{align}
and the self-intesrection number of this class yields \eqref{anomaly1T6Z3Z6}. The curves $C_{113}^{(F_4, E_8)}$ and $C_{112}^{(F_4, E_8)}$ are
\begin{align}
b^{SU(2)}_{113} = (0, 0, 0, 0, 1, 0^{24}), \qquad
b^{G_2}_{112} = (0, 0, 0, 1,0^{25}),
\end{align} 
and they satisfy \eqref{anomalySU2T6Z4Z4}, \eqref{anomalyG2T6Z6Z6}, and \eqref{anomalyBFT6Z6Z6}. We can check that the other $G_2$ curves and the $SU(2)$ curves in the $(F_4, E_8)$ conformal matter theories also satisfy \eqref{anomalySU2T6Z4Z4}, \eqref{anomalyG2T6Z6Z6}, and \eqref{anomalyBFT6Z6Z6}. We have another $G_2$ curve given by
\begin{align}
b'^{G_2}_3 = (0, 1, 0^8, 
-1, 0^8, 
-1, 0^8, 
-1),
\end{align}
reproducing \eqref{anomalyG2T6Z6Z6}. The curve $C^{(-5)}_1$ corresponds to 
\begin{align}
b^{F_4}_1 = (1, 0, -2, -1, -1, -1, -1, -2, -1, -1, -1, 0^{18}),
\end{align}
and satisfies \eqref{anomalyF4T6Z6Z6}. It is possible to show that the other curves of $C^{(-5)}_i, i=2, 3$ also reproduce \eqref{anomalyF4T6Z6Z6}. The $E_8$ curve $C'^{(-12)}_1$ and the $E_6$ curve $C'^{(-6)}_2$ are
\begin{align}
b'^{E_8}_1 =& (0, 1, -1^2, -2, -3, -4, 0^4, -1^2, -2, -3, -4, 0^4, -1^2, -2, -3, -4, 0^4),\\
b'^{E_6}_2 =& (0, 1, 0^5, -1^2, -2, 0^6, -1^2, -2, 0^6, -1^2, -2, 0).
\end{align}
and they satisfy \eqref{anomalyE8T6Z6Z6} and \eqref{anomaly3T6Z3Z3}.

\subsection{6d strings and BPS black holes}
The self-dual strings in the 6d $(1,0)$ theories describe motions of D3-branes wrapping compact 2-cycles in the elliptic CY 3-fold.
The worldsheet theories on the strings are 2d SCFTs preserving $\mathcal{N}=(0,4)$ supersymmetry. We shall consider the self-dual string worldsheet theories in 6d supergravity engineered by F-theory on a compact elliptic CY 3-fold $X=T^6/\mathbb{Z}_m\times\mathbb{Z}_n$. In particular, we are interested in 5d supersymmetric black holes that descend from the self-dual string states in the 6d supergravity theory realized by F-theory on $X\times S^1$.

In the context of F-theory, spinning black holes are generated by string states arising from D3-branes wrapped on a genus $g$ curve $C$ in the elliptic 3-fold and carrying KK modes as studied in \cite{Haghighat:2015ega} (see also the older work \cite{Vafa:1997gr} and the more recent work \cite{Couzens:2017way}). Let us first
focus on the string itself.  As discussed in \cite{Haghighat:2015ega} (see also \cite{Couzens:2017way}) in this case we expect a holographic duality between the 2d worldsheet theory and $AdS_3\times S^3\times B$ where $B$ is the base of the 3-fold; in our specific case $B=T^4/\mathbb{Z}_m\times\mathbb{Z}_n$.
The 2d worldsheet theory on these strings has $SO(4)=SU(2)_L\times SU(2)_R$ symmetry which rotates the transverse $\mathbb{R}^4$ directions in the 6d theory. The $SU(2)_R$ symmetry becomes the R-symmetry of the $\mathcal{N}=(0,4)$ superconformal algebra in the IR SCFT. The $SU(2)_L$ symmetry realizes a left-moving current algebra with level $k_L=g(C)$ where $g(C)$ is the genus of the curve $C$. The left-moving central charge $c_L$ and right-moving central charge $c_R$ of the worldsheet theory are determined by the genus $g(C)$ as \cite{Haghighat:2015ega}
\begin{equation}
	c_L = 6g(C) + 12 c_1(B)\cdot C \ , \quad c_R=6g(C)+6c_1(B) \cdot C \ ,
\end{equation}
respectively. Here $c_1(B)$ is the first Chern class of the tangent bundle of the base $B$.

Using the adjunction formula
\begin{equation}
	g(C) = \frac{1}{2}(C\cdot C - c_1(B)\cdot C) + 1 \ ,
\end{equation}
the central charges can then be rewritten purely in terms of intersection numbers as
\begin{equation}\label{eq:cenral-charges}
	c_L = 3C\cdot C + 9c_1(B)\cdot C + 6 \ , \quad c_R = 3C\cdot C + 3c_1(B)\cdot C + 6 \ .
\end{equation}

Even though one expects this holographic duality to be true for general $B$ and arbitrary class $C$ defining the string charge, an explicit description of the $(0,4)$ supersymmetric worldsheet theory is not known in general.  As we will discuss later for the case of $B=T^4/\mathbb{Z}_2\times\mathbb{Z}_2$ we propose a candidate worldsheet quiver gauge theory description.

We shall also consider 5d spinning BPS black hole states arising from D3-branes wrapping $S^1$ of $X\times S^1$ as well as the curve class $C$, and carrying $n$ units of KK momentum along $S^1$.
 These states can now be viewed as 5d black holes.  From the perspective of M-theory compactified on the elliptic 3-fold such black hole states can be viewed as M2-branes wrapping
 the curve class $n[T^2]+[C]$ carrying angular momentum $J_L$ for $SU(2)_L$.
The microscopic entropy of these spinning black holes can be obtained from the central charges of the 2d CFT on the spinning strings from D3-branes. The Cardy formula tells us that the black hole entropy is given by \cite{Breckenridge:1996is,Haghighat:2015ega}
\begin{equation}\label{eq:entropy}
	S = 2\pi \sqrt{\frac{c_L}{6}\left(n-\frac{J_L^2}{4k_L}\right)} \ .
\end{equation}

In the following subsections, we argue that the 2d worldsheet CFTs on self-dual strings in our 6d field theory models for compact CY 3-folds $X=T^6/\mathbb{Z}_m\times \mathbb{Z}_n$ are holographically dual to black strings (or black holes when compactified on a circle) in type IIB string theory on the background $AdS_3\times S^3 \times B$, constructed by F-theory on an elliptic 3-fold with the base  $B=T^4/\mathbb{Z}_m\times \mathbb{Z}_n$, at least in certain limit of complex structure moduli of $B$. More precisely, the IR CFT on the Higgs branch of the worldsheet theory describes the black string states in the 6d supergravity (or the black hole states with $n$ units of charge in the 5d supergravity).
We will confirm this conjecture by comparing the central charges of the 2d worldsheet theories with the central charges (\ref{eq:cenral-charges}) of spining strings in the 6d supergravity theory.
We will also discuss some subtle issues about multi-string sectors in the 2d field theory appearing when the curve $C$ degenerates to multiple curves \cite{Haghighat:2015ega}. 

\subsubsection{$T^6/\mathbb{Z}_2\times \mathbb{Z}_2$ model}
For this model, we propose a concrete 2d quiver gauge theory which has many of the needed ingredients to be the CFT living on self-dual strings in the 6d gravity theory at low energy, though as we will discuss the proposed theory will lack some necessary features. The 6d field theory for this model consists of 16 $\mathcal{O}(-1)$ theories joined together by eight $\mathcal{O}(-4)$ theories. The 2d $\mathcal{N}=(0,4)$ gauge theory for self-dual strings in the $\mathcal{O}(-1)$ theory is proposed in \cite{Kim:2014dza,Kim:2015fxa} and also the 2d theory of the strings in the $\mathcal{O}(-4)$ theory is proposed in another reference \cite{Haghighat:2014vxa}. These gauge theories describe string worldsheet theories of the local 6d SCFTs embedded in our compact model. This suggests that the 2d quiver gauge theory on the self-dual strings in the full 6d compact model can be constructed by gluing these local worldsheet theories in an appropriate manner. Gluing local 2d gauge theories which live on self-dual strings in 6d SCFTs and LSTs being comprised only of $\mathcal{O}(-1)$ and $\mathcal{O}(-4)$ theories are studied in \cite{Haghighat:2014vxa} and \cite{Kim:2017xan} respectively. Here, two adjacent $\mathcal{O}(-1)$ and $\mathcal{O}(-4)$ string theories are connected by coupling to bifundamental matter fields charged under the gauge groups of both theories. The resulting quiver theory will contain a collection of bifundamental matter interacting with the matter fields which already exist in both $\mathcal{O}(-1)$ and $\mathcal{O}(-4)$ string models through the superpotentials described in \cite{Gadde:2015tra}. 

We expect that this gluing procedure works in each local region of the compact base $B$ where the  geometry can be approximated as the geometry for a CFT or an LST. This suggests that this gluing procedure involving bifundamental matter may be reliable at least in regions of the moduli space of 6d self-dual strings where the vevs of the operators coming from bifundamental matters are suitably small compared to the size of the curve class $C$. We later discuss some subtle issues that occur when the vevs of bifundamental fields become large and therefore additional ingredients are needed to realize the global structure of the compact $X$. Therefore, we propose that combining these 16 $+$ 8 worldsheet theories together with bifundamental matters and their interactions, we can construct a 2d quiver gauge theory capturing the dynamics of the 6d self-dual strings in our field theory model of $X=T^6/\mathbb{Z}_2\times \mathbb{Z}_2$, at least in certain regions of the moduli space.

Let us first briefly review the 2d gauge theory descriptions for self-dual strings in 6d theories on $-1$ and $-4$ curves. The strings in the 6d $\mathcal{O}(-1)$ minimal SCFT admit two different gauge theory descriptions, given separately in \cite{Kim:2014dza} and \cite{Kim:2015fxa}. We find it more suitable for our purposes to use the $O(k)$ gauge theory description given in \cite{Kim:2014dza}. The theory on $k$ self-dual strings is described by an $O(k)$ gauge theory preserving  2d $\mathcal{N}=(0,4)$ supersymmetry and consists of the following multiplets \cite{Kim:2014dza}:
\begin{eqnarray}\label{eq:E-string-matters}
	{\rm vector} \ &:& \ O(k) \ {\rm antisymmetric} \ (A_\mu, \lambda_+^{\dot{\alpha}A}) \nonumber \\
	{\rm hyper} \ &:& \ O(k) \ {\rm symmetric} \ (\varphi_{\alpha\dot{\beta}}, \lambda_-^{\alpha A})  \nonumber \\
	1/2 \ {\rm Fermi} \ &:& \ O(k)\times SO(16) \ {\rm bifundamental} \ \Psi_{+l} \ .
\end{eqnarray}
Note that the hypermultiplet is a real hypermultiplet (satisfying a reality condition) and here, $+$ (or $-$) in $\lambda$ denotes left (or right) worldsheet chirality.

This theory has $SU(2)_R\times SU(2)_I$ R-symmetry whose doublet indices are $\dot{\alpha}$ and $A$ respectively. The global symmetry is $SU(2)_L\times SO(16)$ and $\alpha$ is the doublet index of $SU(2)_L$. The $SO(4)=SU(2)_L\times SU(2)_R$ is the 6d Lorentz symmetry transverse to the 2d strings and $SU(2)_I$ corresponds to the R-symmetry of the 6d SCFT. The global symmetry $SO(16)$ acting only on the Fermi multiplet $\Psi_l$ is the maximal subgroup of $E_8$ global symmetry of the 6d E-string theory. We expect that this symmetry enhances to $E_8$ at the low energy CFT limit by quantum effects.

The $n$ self-dual strings in the 6d $\mathcal{O}(-4)$ minimal SCFT are described by a 2d $\mathcal{N}=(0,4)$ $Sp(n)$ gauge theory with the following matter content \cite{Haghighat:2014vxa}:
\begin{eqnarray}\label{eq:so8-string-matters}
	{\rm vector} \ &:& \ Sp(n) \ {\rm symmetric} \ (\tilde{A}_\mu, \tilde{\lambda}_+^{\dot{\alpha}A}) \nonumber \\
	{\rm hyper} \ &:& \ Sp(n) \ {\rm antisymmetric} \ (\tilde{\varphi}_{\alpha\dot{\beta}}, \tilde\lambda_-^{\alpha A})  \nonumber \\
	{\rm hyper} \ &:& \ Sp(n)\times SO(8) \ {\rm bifundamental} \ (q_{\dot{\alpha}}, \psi_-^A) \ .
\end{eqnarray}
These two types of hypermultiplets are both half-hypermultiplets subject to a reality condition.
This theory has $SO(8)$ global symmetry as well as $SU(2)_L\times SU(2)_R\times SU(2)_I$ symmetry. This $SO(8)$ symmetry is the gauge symmetry of the 6d $\mathcal{O}(-4)$ theory.

Lastly, at each intersection between a $-1$ curve and a $-4$ curve, there exists additional bifundamental matter given by \cite{Gadde:2015tra}
\begin{eqnarray}\label{eq:bifundamental}
	1/2 \ {\rm twisted \ hyper} \ &:& \ O(k)\times Sp(n) \ {\rm bifundamental} \ (\Phi_{A}, \eta_-^{\dot{\alpha}}) \nonumber \\
	1/2 \ {\rm Fermi} \ &:& \ O(k)\times Sp(n) \ {\rm bifundamental} \ (\chi_{+\alpha}) \ .
\end{eqnarray}
These bifundamental fields couple to the fields in the $O(k)$ and $Sp(n)$ gauge nodes through $\mathcal{N}=(0,4)$ superpotentials as described in \cite{Gadde:2015tra} that identify the $SO(8)$ global symmetry of the $Sp(n)$ gauge theory with one of $SO(8)\times SO(8)\subset SO(16)$ global symmetry of the $O(k)$ gauge theory.

\begin{figure}
  \centering
  \includegraphics[width=.55\linewidth]{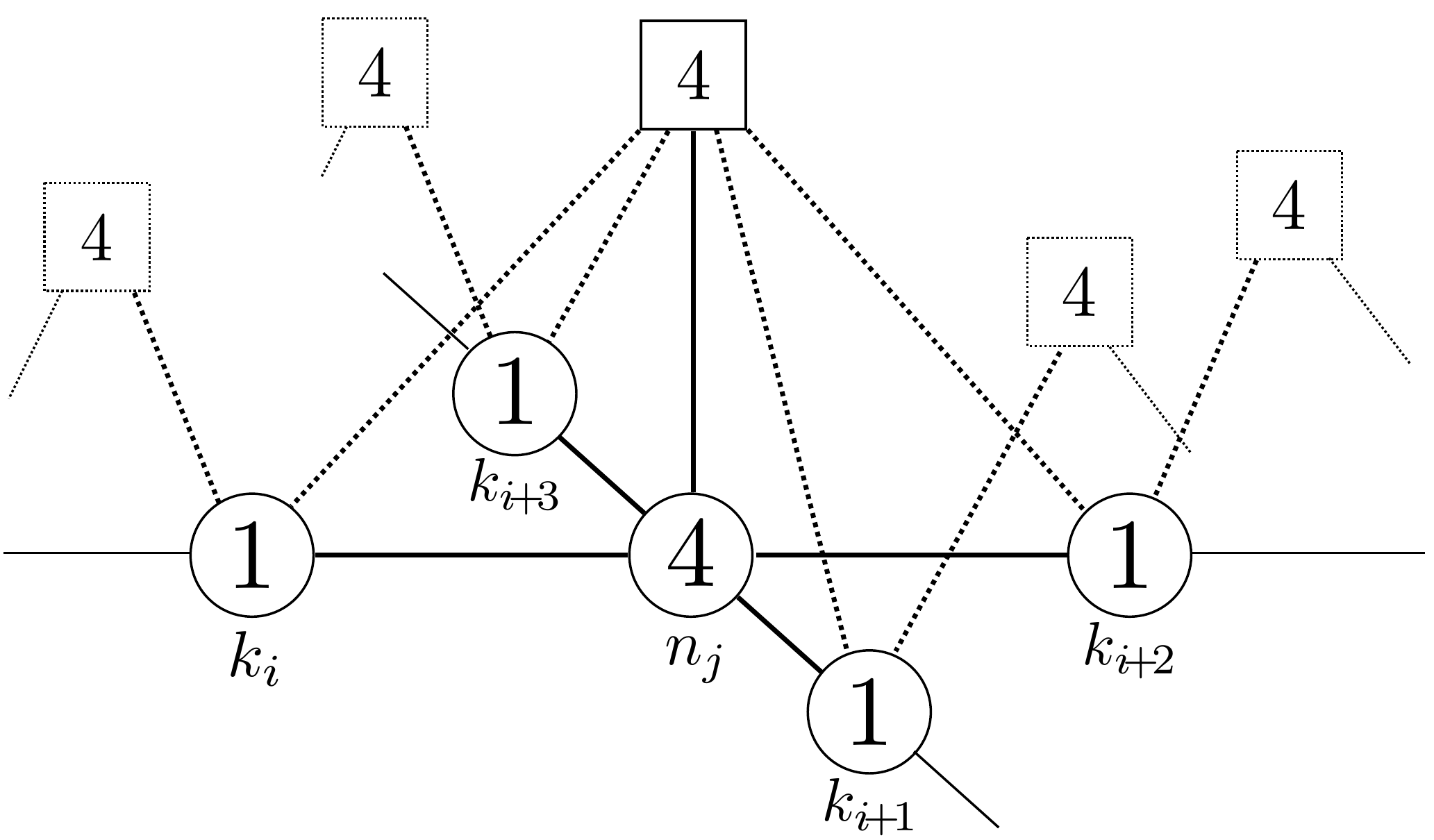}
  \caption{Quiver diagram near a $-4$ curve which intersects four $-1$ curves. Note that the  diagram represents a subset of the full quiver diagram for 6d strings. Self-dual strings over a $-4$ curve (denoted by a circle with $4$) are described by an $Sp(n)$ gauge theory, while strings over a $-1$ curve (denoted by  circles with $1$) have an $O(k)$ gauge theory description. An $Sp(n)$ gauge node includes a real antisymmetric hypermultiplet and four fundamental hypermultiplets denoted by a vertical solid line. An $O(k)$ gauge node includes a real symmetric hypermultiplet. A solid line between two gauge node denotes a twisted half-hypermultiplet and Fermi multiplet in the bifundamental representation. The dotted lines stand for Fermi multiplets. When all other $\mathcal{O}(-4)$ tensor multiplets are decoupled (i.e. when the external lines in the diagram are removed), the diagram describes the LST realized by F-theory on $T^4\times \mathbb{C}/\mathbb{Z}_2\times \mathbb{Z}_2$.}
  \label{fig:quivers}
\end{figure}

As we have seen, the 6d theory we are interested in involves gluing of $\mathcal{O}(-1)$ and $\mathcal{O}(-4)$ theories. 
Thus the 2d gauge theory for the strings in this 6d theory will be realized by a quiver gauge theory connected by bifundamental matters given in (\ref{eq:bifundamental}).
The emergence of bifundamental fields is geometrically natural given that the D3 branes wrapping the corresponding $\mathbb P^1$ see the neighboring $\mathbb P^1$ if and only if they intersect.   We are interested in the description of the strings wrapping the corresponding $\mathbb P^1$'s given by classes $(k_i,n_j)$ with $i=1,\cdots, 16$ and $j=1,\cdots,8$ for the full gravity model, namely for D3 branes wrapping an arbitrary configuration of $\mathbb P^1$'s in $X$.  The matter content of the corresponding 2d field theory is easy to argue:  Given what we have discussed it is natural to consider the gauge theory to be the product of gauge theories for each of the individual $\mathbb P^1$'s.
In other words, the 2d gauge theory with gauge group $G=\prod_{i=1}^{16}O(k_i)\times  \prod_{i=1}^8 Sp(n_i)$ realizes the worldsheet theory on self-dual strings in the 6d field theory model of $X$.  The structure of the bifundamental matters should also follow from the local description given above. Our 2d quiver gauge theory contains 32 copies of the bifundamental matter fields together with their superpotentials. A neighborhood of a $-4$ curve is described by the quiver diagram in Figure \ref{fig:quivers}.

More precisely, the 2d CFT on the Higgs branch of this quiver gauge theory implements the moduli space of self-dual strings coming from D3-branes wrapped on a genus $g$ curve class $C$ labelled by $(k_i,n_j)$ in the compact 3-fold with base $ B=T^4/\mathbb{Z}_2\times \mathbb{Z}_2$. For a given $C=(k_i,n_j)$, we can construct the corresponding 2d quiver gauge theory using the above construction. Since this description covers all possible curve classes $C$, we propose that the 2d quiver gauge theory described above engineers the worldsheet theory on self-dual strings in the 6d supergravity theory of an elliptic 3-fold with base $B$. This leads us to ask if our 2d quiver theory is the holographic dual of type IIB strings on $AdS_3\times S^3 \times B$.


We remark that we used 2d gauge theories for strings on individual $\mathbb{P}^1$'s and glued them with bifundamental matters in a manner consistent with the construction of CFTs and LSTs. This local gluing prescription may not be enough to realize the full moduli space of strings in the compact 3-fold. There is the possibility that additional interactions emerge when all of the nodes are coupled together. In fact, we show below that we need additional interactions in order to precisely realize the full moduli space of strings in the compact 3-fold.
However, we also show that even without these additional ingredients, the central charges lead to the expected results for self-dual strings in this choice of compact 3-fold \cite{Haghighat:2015ega}.

We find that the 2d worldsheet quiver theory exhibits some surprising structure in the IR.
The 2d $\mathcal{N}=(0,4)$ gauge theories in general flow to different interacting fixed points in the IR.
In this paper, we consider the CFTs on the Higgs branch that we expect to describe the moduli space of D3-branes bound to curve classes $C$ in the base $B$.
The Higgs branch of vacua in our 2d quiver gauge theory is rather complicated. It is parametrized by the vacuum expectation values of the scalar fields in the hypermultiplets. We have two types of hypermultiplets in 2d $\mathcal{N}=(0,4)$ theory: one is the usual hypermultiplet whose scalar field $q_{\dot\alpha}$ is charged under the $SU(2)_R$ symmetry and another one is the twisted hypermultiplet whose scalar field $\Phi_A$ carries charges of $SU(2)_I$ symmetry. Roughly, these two different types of hypermultiplets lead to two distinct Higgs branches at infinite distance which only meet at the origin.

The reason for this structure is as follows. The UV $\mathcal{N}=(0,4)$ quiver gauge theory has two non-abelian right-moving R-symmetries, $SU(2)_R\times SU(2)_I$. However, we expect that the IR CFTs have a small $\mathcal{N}=(0,4)$ superconformal algebra containing only one right-moving $SU(2)$ R-symmetry current. Therefore, only one combination of the right-moving $SU(2)_R\times SU(2)_I$ symmetries participates in the IR superconformal algebra. As discussed in \cite{Witten:1997yu}, the right-moving R-symmetry in the IR CFT cannot act on the scalar fields in the moduli space of vacua. This means that the 2d quiver theory can flow to two quantum mechanically disinct IR CFTs on two different Higgs branches. One Higgs branch is parametrized by the hypermultiplet scalar fields. In this branch, the right-moving R-symmetry in the IR superconformal algebra is determined to be the $SU(2)_I$ symmetry since it acts trivially on these scalars. There is also another Higgs branch parametrized by the twisted hypermultiplet scalars. In this second branch, $SU(2)_R$ becomes the right-moving R-symmetry of the superconformal algebra as the scalars are neutral under this symmetry. Thus, two theories on these two branches must be different CFTs as the two branches have different R-symmetries. Indeed, as we will see soon, these two CFTs have different central charges meaning that they have different numbers of interacting degrees of freedom.

In the 2d quiver theory, one can easily see the decoupling of the two Higgs branches: The hypermultiplets in this theory come from the instanton strings in each tensor node, while the twisted hypermultiplets arise from the interactions between two tensors.
There are quartic superpotential couplings between the hypermultiplets and the twisted hypermultiplets \cite{Gadde:2015tra}. 
When scalar fields of either the hypermultiplets or twisted hypermultiplets take nonzero vacuum expectation values, the other scalar fields acquire masses from their quartic interactions. Therefore, giving vevs to scalars in both hypermultiplets and twisted hypermultiplets is prohibited due to the quartic couplings.  This tells us that two Higgs branches are decoupled.

Of course, we can consider mixed branches where disjoint components of both hypermultiplet scalars and twisted hypermultiplet scalars get non-trivial vevs. The 2d theory contains these mixed branches as well as the two particular Higgs branches of interest discussed above. However, the resulting IR theory cannot be a single CFT in these mixed branches. It is because IR superconformal algebra must include a right-moving $SU(2)$ R-current, but both $SU(2)_R\times SU(2)_I$ symmetries cannot be the R-symmetry in the conformal algebra as they both act on the scalar fields of the moduli space. This implies that when scalar fields in both hypermultiplets and twisted hypermultiplets acquire nonzero vevs, which is possible only when they are disjoint components so that they do not meet through the quartic couplings, the low energy theory should be a collection of disconnected CFTs having two different $(0,4)$ algebras. Since we are interested in a single interacting CFT in the IR, we will not consider these mixed branches.

The 2d CFT corresponding to the black strings in the 6d supergravity lives in the second Higgs branch parametrized by bifundamental scalar fields in the twisted hypermultiplets. In string theory, we expect that these strings are from D3-branes wrapping a genus $g>0$ curve $C$ in the base. This curve $C$ is an irreducible curve formed by a collection of intersecting curve classes. As noted in \cite{Haghighat:2015ega}, when $C$ is non-degenerate, the worldsheet degrees of freedom localized at the intersection of the curve with D7-branes are all in the left-moving sector. This means that the $SO(8)$ instanton moduli space in each $-4$ curve, which represents string states smearing deep in the $SO(8)$ gauge orbit over D7-branes, cannot participate in the string states of the non-degenerate curve $C$. Similarly, we expect that the instanton string states localized in a $\mathcal{O}(-1)$ theory cannot contribute to the string states in the gravity theory. We thus conclude that the Higgs branch of the bifundamental scalar fields realizes the strings from D3-branes on a curve $C$ in the 6d supergravity. This is consistent with the fact that the vevs of bifundamental scalars lift such instanton Higgs branches of local 6d CFTs.
This is also natural since D3-branes can wrap a non-degenerate curve $C$ only when two theories on two tensor nodes are tightly bound by interactions between degrees of freedom on two tensor nodes and these interactions are mainly realized by bifundamental fields and their interactions. In addition, the superconformal R-symmetry in this branch is chosen to be the $SU(2)_R$ symmetry that is a part of the 6d Lorentz symmetry. This agrees with the right-moving R-symmetry of spinning strings in the 6d gravity theory. We will now show that the central charges of the Higgs branch of the bifundamentals indeed agree with the central charges (\ref{eq:cenral-charges}) of the 2d CFTs living on the strings of the 6d supergravity theory.

Let us compute the central charges of the 2d theory on $(k_i,n_j)$ strings. The left and right central charges of a 2d CFT are defined by
\begin{equation}
	c_R = 3{\rm Tr}(\gamma^3R_{\rm cft}^2) \ , \quad c_R-c_L = {\rm Tr}(\gamma^3) \ ,
\end{equation}
where $R_{\rm cft}$ is the right-moving R-charge in the IR superconformal algebra and $\gamma^3$ is the 2d chirality projection operator acting on chiral fermions $\psi_\pm$ as $\gamma^3\psi_\pm = \mp \psi_\pm$.

The $SU(2)_R\times SU(2)_I$ anomalies can be computed using our UV gauge theory description. The worldsheet theory on  $k_i$ strings in $i$-th $\mathcal{O}(-1)$ theory has the anomalies
\begin{eqnarray}
	&&c_I= 3{\rm Tr}(\gamma^3SU(2)_I^2) =  6k_i \ , \quad 
	c_R = 3{\rm Tr}(\gamma^3SU(2)_R^2) =  -3k_i(k_i-1) \ , \nonumber \\
	&& c_{R_{\rm cft}} -c_L = {\rm Tr}\gamma^3 = -6k_i \ .
\end{eqnarray}
Similarly, the worldsheet theory on $n_j$ strings in $j$th $\mathcal{O}(-4)$ theory has the anomalies
\begin{eqnarray}
	&&c_I=   36n_i \ , \quad 
	c_R =  -6n_i(2n_i+1) \ , \quad  c_{R_{\rm cft}} -c_L = 12n_i \ .
\end{eqnarray}

At each intersection between a $-1$ curve and a $-4$ curve, we have additional contributions to the anomlies coming from bifundamental fields. For the bifundamental fields of $O(k_i)\times Sp(n_j)$ gauge groups, we find
\begin{eqnarray}
	c_I=0 \ , \quad c_R=6k_in_j \ , \quad c_{R_{\rm cft}}-c_L = 0 \ .
\end{eqnarray}

We remark here that our 2d theory includes an extra real hypermultiplet $(\varphi_{\alpha \dot{\alpha}}, \lambda_-^{\alpha A})_{\text{com}}$ corresponding to center-of-mass degrees. The center-of-mass fields decoupe from the IR CFT. Thus, in order to compute the correct central charges of interactinge CFT degrees of freedom, we need to elliminate the center-of-mass contributions:
\begin{equation}
	c_I^{\text{com}} = 6 \ , \quad c_R^{\text{com}} = 0 \ , \quad c_{R_{\rm cft}}^{\text{com}} - c_L^{\text{com}} =  2 \ ,
\end{equation}
which come from the free hypermultiplet  $(\varphi_{\alpha \dot{\alpha}}, \lambda_-^{\alpha A})_{\text{com}}$.

Therefore, by summing over all these anomaly contributions, the total anomalies are
\begin{equation}\label{eq:centrals-Z2Z2}
	c_I = 6\sum_{i=1}^{16}k_i + 36\sum_{i=1}^8n_i  - 6\ ,  \quad c_R = 3\sum_{i,j=1}^{24}\Omega^{ij}C_iC_j + 3 \sum_{i=1}^{24}a^i C_i \ , \quad
	c_{R_{\rm cft}}-c_L = -6 \sum_{i=1}^{24}a^i C_i - 2\ ,
\end{equation}
where $C=(\vec{k},\vec{n})$ is the vector of string numbers, $\Omega^{ij}$ is the intersection matrix for 24 tensor nodes, and $a^i=2+\Omega^{ii}$. 
When an elliptic 3-fold is compact, $a^i$ can be identified with $i$-th component of the first Chern class $c_1(B)$ of the base $B$. So we can rewrite the $SU(2)_R$ anomaly in terms of geometric quantities as
\begin{equation}
	c_R = 3C\cdot C + 3c_1(B)\cdot C = 6g + 6c_1(B)\cdot C-6 \  ,
\end{equation}
where $C\cdot C=\Omega^{ij}C_i C_j$ and $g$ is the genus of $C$, and $C$ is the curve class wrapped by D3-branes of the self-dual strings.

Now it is obvious that we have two distinct CFTs having different central charges on the Higgs branches of the 2d gauge theory. In the first Higgs branch leading to a 2d CFT of instanton strings in the local 6d CFTs,
the right-moving central charge of the IR CFT is given by $c_{R_{\rm cft}}=c_I$. On the other hand, the 2d IR CFT in the second branch, which we expect to be dual to the 6d black strings, has the central charges
\begin{equation}
	c_{R_{\rm cft}}=c_R = 3C\cdot C + 3c_1(B)\cdot C  \  , \quad
	c_L = 3C\cdot C + 9c_1(B)\cdot C  +2 \ .
\end{equation}
This result shows the perfect agreement with the central charges of the spinning strings in (\ref{eq:cenral-charges}) obtained from the gravity computation up to constant factors $+6$ in $c_R$ and $+4$ in $c_L$. These extra constant factors come from the center-of-mass degrees of freedom. The central charges in (\ref{eq:cenral-charges}) involves the center-of-mass contributions from 4 left-moving bosons and $4+4$ right-moving  boson and fermion pairs. These are exactly the extra factors $+4$ in $c_L$ and $+6$ in $c_R$.  When compactified on a circle (and after removing the center-of-mass contributions), these central charges give rise to the expected microscopic entropy (\ref{eq:entropy}) of the 5d spinning BPS black holes with charge $n$ along the KK circle.
We find that the agreement of central charges shows that our 2d worldsheet quiver theory captures an important part of the physics of the dual CFT to type IIB on an $AdS_3\times S^3\times B$ background.

One can also compute the $SU(2)_L$ anomaly from the matter contents in (\ref{eq:E-string-matters}), (\ref{eq:so8-string-matters}), and (\ref{eq:bifundamental}). The result is
\begin{equation}
	k_L = \frac{1}{2}(C\cdot C - c_1(B)\cdot C) = g-1 \ .
\end{equation}
This also agrees with the expected level $k_L=g$ of the $SU(2)_L$ current algebra of 6d black hole strings when we take into account the center-of-mass contribution $k^{CM}_L=-1$ as noted in \cite{Haghighat:2015ega}.

Let us discuss the second Higgs branch in some detail. We focus on the  worldsheet theory of the string wrapping a $\mathbb{P}^1$ inside $T^4/\mathbb{Z}_2\times \mathbb{Z}_2$ with class $C_{\text{LST}} = C_i^{(-4)} + \sum_{j=1}^4 C_{ij}^{(SO(8),SO(8))}$.
The corresponding quiver theory is given by the $O(1)^4\times Sp(1)$ gauge theory with matter. The second Higgs branch is parametrized by the bifundamental scalar fields $\Phi^{I} \equiv \Phi^I_{A=1}$ (with $I=1,2,3,4$) between the $O(1)_I$ and $Sp(1)$ gauge nodes satisfying the F-term conditions
\begin{equation}
	\Phi^I_{(\alpha} \Phi^I_{\beta)} = 0 \ ,
\end{equation}
where $\alpha,\beta$ are the $Sp(1)$ gauge indices and also the D-term conditions.
In this branch we have $O(4)$ global symmetry exchanging these four scalar fields. The $O(1)_I=\mathbb{Z}_2$ invariance requires the $I$-th scalar field to satisfy
\begin{equation}
	\Phi_\alpha^I = -\Phi_\alpha^I \ .
\end{equation}
One can thus expect that this moduli space is identical to the moduli space of a single instanton in the $SO(4)$ gauge theory with additional $\mathbb{Z}_2$ gauge symmetries. In terms of the $Sp(1)$ invariant operator $M_{ab}$ defined as
\begin{equation}
	M_{ab} \equiv \Phi_{\alpha a\dot{a}}\Phi_{\beta b\dot{b}}\epsilon^{\alpha\beta}\epsilon^{\dot{a}\dot{b}} \ ,
\end{equation}
where $(a,\dot{a})$ are doublet indices of $SU(2)\times SU(2)=SO(4)$ symmetry, the moduli space is therefore given by the space of $M_{ab}$ subject to the constraint
\begin{equation}\label{eq:A1}
	M_{11}M_{22} = M_{12}^2 \ ,
\end{equation} 
together with $\mathbb{Z}_2^1\times \mathbb{Z}_2^2$ orbifolds acting on $M_{ab}$ as
\begin{eqnarray}\label{eq:orbifoldonM}
	&&\mathbb{Z}_2^1 \ : \ (M_{11},M_{22},M_{12}) \ \rightarrow (M_{11},M_{22},M_{12}) \ , \nonumber \\ 
	&& \mathbb{Z}_2^1 \ : \ (M_{11},M_{22},M_{12}) \ \rightarrow (M_{22},M_{11},-M_{12}) \ .
\end{eqnarray}
Combining (\ref{eq:A1}) and (\ref{eq:orbifoldonM}), we find that the second Higgs branch of the $O(1)^4\times Sp(1)$ quiver theory is given by
\begin{equation}\label{eq:moduli-Higgs}
	\mathbb{C}^2/\Gamma_{D_4} \ ,
\end{equation}
where $D_4$ is the dihedral group of order 8.



The second (i.e. twisted) Higgs branch of the 2d theory describing strings wrapping the class $C_{\text{LST}}$, which is simply the normal geometry of  $C_{\text{LST}}$ inside the compact 3-fold, is obviously compact. However, the second Higgs branch of the 2d quiver gauge theory, namely the surface singularity (\ref{eq:moduli-Higgs}), is non-compact---in other words, our quiver description fails to realize the correct moduli space of the corresponding self-dual string. This tells us that our quiver gauge theory description is incomplete and it cannot capture the full moduli space of self-dual strings in the compact 3-fold $X$, even though it contains the right quivers describing the physics of strings in any local CFTs and LSTs embedded in $X$. This failure may be related to our gluing prescription for each pair of two adjacent quiver nodes. As discussed, we may need to introduce additional interactions when all of local $\mathbb{P}^1$'s contained in the compact base $B$ are glued together. 
Our 2d quiver gauge theory may describe only particular corners of the moduli space of strings in the 6d supergravity.

Regarding this, we propose the following two possibilities. The first possibility is that the 2d quiver gauge theory can capture only subregions of the moduli space of strings in the gravity theory where the vevs of the bifundamental scalar fields are much smaller than the size of the curve class $C$. In these subregions we can trust our gluing prescription because small bifundamental vevs cannot see the global strucutre of the compact base $B$ and thus the intersections between two $\mathbb{P}^1$ cycles can be well-approximated as those in the local CFTs or LSTs. This can explain the non-compactness of the second Higgs branch of the quiver theory in (\ref{eq:moduli-Higgs}) since, in this limit, the local moduli space formed by the bifundamental scalars reduces to that of the strings wrapped on the $\mathbb{P}^1$ associated to the elliptic class $T^2$ in the LST arising from the non-compact base $T^2\times \mathbb{C}/\mathbb{Z}_2\times\mathbb{Z}_2$ at low energies. The $SU(2)_I$ internal symmetry which is absent in the supergravity theory also appears to be restored in this limit. So the 2d quiver gauge theory has this $SU(2)_I$ symmetry.

Some protected quantities of self-dual strings in the supergravity theory can be computed in these subregions. Indeed, we have already checked that the central charges of our quiver gauge theory agrees with those of self-dual strings in the supergravity. Moreover, we expect that the elliptic genus of the string worldsheet CFT in the compact base $T^4/\mathbb{Z}_2\times\mathbb{Z}_2$ can be exactly computed using our 2d quiver theory. The reason is as follows: Suppose we compute the elliptic genus of the 2d CFT using localization. For the localization, we will turn on equivariant parameters including the holonomies for the corresponding D3-branes. These parameters lift all charged matter and thus the bifundamental fields are localized around the origin of the moduli space where all vevs of charged scalar fields vanish. The localization result is determined by small fluctuations of the fields near the origin. Our 2d quiver theory covers this region of the moduli space. Therefore, we expect that the elliptic genus of our quiver gauge theory agrees with the localization result of the 2d string CFT in the supergravity. This can be supported from the fact that the elliptic genus of the 2d quiver gauge theory, which we will compute soon , has the correct modular anomaly of the self-dual strings
since the modular anomaly is fixed by the anomaly polynomial of the string worldsheet theory \cite{Bobev:2015kza,DelZotto:2016pvm,Braun:2018fdp} and the anomaly polynomial of this 2d quiver theory is equal to that of the strings in the supergravity theory.

The second possibility is that the above 2d quiver theory describes the 6d strings only in a particular limit of the complex structure parameters of the compact base $B$, which we expect to be $\tau\rightarrow \infty$. In this limit, the additional interactions we need to complete the 2d quiver theory may become irrelevant. So the moduli space of 6d strings in this limit may be well-described by our quiver gauge theory. In addition, since the BPS quantities are independent of the complex structure deformation, we can again claim that the elliptic genus of this quiver theory will give the correct elliptic genus of the self-dual strings in the supergravity.

Before we move on, let us discuss other possible branches localized at the origin of the Higgs branches. It is conjectured in \cite{Haghighat:2015ega} that the Higgs branch of the 2d theory on 6d strings can meet other phases called `multi-string branches' at the origin. In this context, the CFT of the second Higgs branch discussed above is `single string branch'. In F-theory, multi-string branches correspond to the situation where the curve $C$ wrapped by D3-branes degenerates to a sum of lower genus curves with multiple degeneracies. This branch exists only when the D3-branes carry non-zero KK momenta after we wrap the 6d theory on a circle. So, this situation describes an interesting phase structure for 5d black hole states---see \cite{Haghighat:2015ega} for more details.


\subsubsection*{Elliptic genus}

In this section we compute the elliptic genus of our 2d quiver gauge theory. The elliptic genus of the $\mathcal{N}=(0,4)$ worldsheet theory for $(k_i,n_j)$ strings is defined as 
\begin{equation}
	Z_{k_i,n_j}(\tau,\epsilon_+,\epsilon_-,m) = {\rm Tr}\left[(-1)^F q^{H_+}\bar{q}^{H_-} e^{2\pi i \epsilon_+ (J_R+J_I)} e^{2\pi i \epsilon_- J_L} \prod_{a=1}^{32} e^{2\pi i m_a F_a}\right] \ ,
\end{equation}
where $q=e^{2\pi i\tau}$ and $H_\pm = \frac{H\pm P}{2}$ with Hamiltonian $H$ and momentum $P$, and $J_L,J_R,J_I$ are Cartan generators of the $SU(2)_L\times SU(2)_R\times SU(2)_I$ symmetry. Note also that $F_a$, $a=1, \dots, 32$, are the Cartan generators of the $SO(8)^8$ symmetries. The elliptic genus is the partition function of BPS states saturating the BPS bound as $H_-=0$, and is thus independent of $\bar{q}$.

The elliptic genus for the quiver theory acquires contributions from each $\mathcal{O}(-1)$ and $\mathcal{O}(-4)$ tensor node, as well as from bifundamental fields. Combining these, we manage to write the full elliptic genus is given by the form of a contour integral expression
\begin{eqnarray}\label{eq:elliptic-genus}
	Z_{\vec{k},\vec{n}^H,\vec{n}^V} &=& \oint \prod_{i,j=1}^{4}Z^{\mathcal{O}(-1)}_{k_{ij}}(\varphi_{ij}, m_i^H,m_j^V) \times \prod_{i=1}^4 Z^{\mathcal{O}(-4)}_{n^H_i}(\tilde{\varphi}_{i}^H,m_{i}^H)\times \prod_{j=1}^4 Z^{\mathcal{O}(-4)}_{n^V_j}(\tilde{\varphi}^V_i,m^V_i) \nonumber \\
	&& \quad \times \prod_{i,j=1}^4 Z^{\mathcal{O}(-1)\times \mathcal{O}(-4)}_{k_{ij},n^H_i}(\varphi_{ij},\tilde\varphi_i^H) \times Z_{k_{ij},n^V_j}^{\mathcal{O}(-1)\times \mathcal{O}(-4)}(\varphi_{ij},\tilde\varphi_j^V) \ ,
\end{eqnarray}
where $\varphi_{ij}$ denotes the $O(k)$ gauge holonomies of $(i,j)$-th $\mathcal{O}(-1)$ string and $\varphi^H_i$ and $\varphi^V_j$ denote the $Sp(n)$ gauge holonomies of the $i$-th horizontal $\mathcal{O}(-4)$ string and of the $j$-th vertical $\mathcal{O}(-4)$ string respectively. Note that $m^H_i$ and $m^V_j$ are the $i$-th horizontal and $j$-th vertical $SO(8)$ holonomies, respectively.
The elliptic genus of each $\mathcal{O}(-1)$ string theory is given by
\begin{eqnarray}
	Z^{\mathcal{O}(-1)}_{k}(\varphi, m) &=& \frac{1}{|W_k|} \prod_{I=1}^r \left(\frac{d\varphi_I}{2\pi i} \cdot \frac{\theta_1(2\epsilon_+)}{i\eta}\right)\prod_{e\in {\rm root}}\frac{\theta_1(e(\varphi))\theta_1(2\epsilon_++e(\varphi))}{i^2\eta^2} \nonumber \\
	&& \times \prod_{\rho\in {\rm sym}}\frac{i^2\eta^2}{\theta_1(\epsilon_{1,2}+\rho(\varphi))} \prod_{\rho \in {\rm fund}} \prod_{a=1}^8\frac{\theta_1(m_a+\rho(\varphi))}{i\eta} \ ,
\end{eqnarray}
where $r$ is the rank and $W_k$ is the Weyl group of the $O(k)$ gauge group.
The elliptic genus of an $\mathcal{O}(-4)$ string theory is 
\begin{eqnarray}
	Z^{\mathcal{O}(-4)}_{n}(\tilde\varphi, m) &=& \frac{1}{|W_n|} \prod_{I=1}^n \left(\frac{d\tilde\varphi_I}{2\pi i} \cdot \frac{\theta_1(2\epsilon_+)}{i\eta}\right)\prod_{e\in {\rm root}}\frac{\theta_1(e(\tilde\varphi))\theta_1(2\epsilon_++e(\tilde\varphi))}{i^2\eta^2} \nonumber \\
	&& \times \prod_{\rho\in {\rm anti}}\frac{i^2\eta^2}{\theta_1(\epsilon_{1,2}+\rho(\tilde\varphi))} \prod_{\rho \in {\rm fund}} \prod_{p=1}^4\frac{i\eta}{\theta_1(\epsilon_++\rho(\tilde\varphi)\pm m_p)} \ ,
\end{eqnarray}
where $W_n$ is the Weyl group of the $Sp(n)$ gauge group. The contributions from the bifundamental multiplets are given by
\begin{eqnarray}
	Z_{k,n}^{\mathcal{O}(-1)\times \mathcal{O}(-4)}(\phi,\varphi) = \prod_{\substack{\rho\in O(k) \, {\rm fund}, \\ w\in Sp(n) \, {\rm fund}}} \frac{\theta_1(-\epsilon_- + \rho(\varphi) - w(\tilde\varphi))}{\theta_1(-\epsilon_+ + \rho(\varphi) - w(\tilde\varphi))} \ .
\end{eqnarray}
The contour integral (\ref{eq:elliptic-genus}) can be evaluated by using the Jeffrey-Kirwan residue prescription \cite{Benini:2013nda,Benini:2013xpa}.

This elliptic genus contains contributions from all different branches to which the UV gauge theory can flow.  The different branches correspond to different residue contributions to the above contour integral.   This includes the single string branch forming a macroscopic 5d black hole with KK charge $H_+=n$ and also all the other mixed branches. The totality of BPS states in 5d, including their spins, is captured by the topological string partition function  \cite{Gopakumar:1998ii,Gopakumar:1998jq,Katz:1999xq}.  As discussed in \cite{Haghighat:2015ega} this implies that
\begin{equation}
	Z_{top}(t,t',m_a; g_s=\epsilon_-)=\sum_{k_i,n_j} \left.{\rm exp}(-k_it^i-n_j {t'}^j) Z^{ell}_{k_i,n_j} (\tau,m_a,\epsilon_+,\epsilon_-)\right|_{\epsilon_+=0}\ .
\end{equation}
Here $g_s$ denotes the topological string coupling constant.  Also the refinement parameter $\epsilon_+$ has been set to zero to obtain the unrefined topological string.  
Note that there are $18$ inequivalent $k,n$'s, $32$ mass parameters $m_\alpha$ and one $\tau$ giving a total of $51$ K\"ahler parameters of the CY 3-fold 
$T^6/\IZ_2\times \IZ_2$.

\subsubsection{Other models}
Unfortunately, we could not find 2d gauge theory descriptions for self-dual strings in the other gravity models discussed in this paper since they involve local 6d CFTs such as minimal $\mathcal{O}(-n)$ CFTs with $n>4$ whose self-dual strings currently have no known gauge theory realization. However, we can extract some useful information of the self-dual string states without knowing the explicit worldsheet gauge theory realization, for example the central charges of self-dual strings in our gravity models. In this subsection, we will compute the central charges of string worldsheet theories from the local field theory data in the gravity models and show that they agree with the central charges in (\ref{eq:cenral-charges}) from the gravity computation.

The central charges of the string states are encoded in the anomaly polynomial of the 2d worldsheet theory. In \cite{Kim:2016foj,Shimizu:2016lbw}, anomaly polynomials of 2d self-dual string theories in general 6d CFTs are obtained by the anomaly inflow mechanism. In the presence of self-dual strings, the Green-Schwarz term in the 6d CFT induces anomaly inflows toward the 2d string worldsheet theory. The anomaly polynomial of the 2d worldsheet theory is fixed by requiring it to cancel this anomaly inflow contribution from the 6d bulk theory. The 4-form anomaly polynomial $I_4$ of the 2d worldsheet theory in a 6d CFT is \cite{Kim:2016foj,Shimizu:2016lbw}~\footnote{The intersection matrix $\Omega^{ij}$ in this paper differs from that in \cite{Kim:2016foj} by overall minus sign.}
\begin{equation}\label{eq:2d-anomaly-pol}
	I_4 = -\sum_\alpha \left(\frac{F^2}{2}+\frac{p_1(T_2)}{24}\right)= -k_i I^i_{6d}+\frac{1}{2}\Omega^{ij}k_ik_j\, \chi(T_4) \ ,
\end{equation}
where $\alpha$ runs over all chiral fermions in the 2d theory. $F$ denotes the background curvature of the symmetry groups and $p_1(T_2)$ is the first Pontryagin class of the 2d worldsheet.
$I^i_{6d}$ is the 4-form appearing in the 6d Green-Schwarz coupling $\int B_i\wedge I^i_{6d}$ and $\chi(T_4) = c_2(L)-c_2(R)$ is the Euler class of the tangent bundle transverse to the strings in the 6d theory. Here $c_2(\mathrm{r})$ denote the 2nd Chern class of a $SU(2)_{\mathrm{r}}$ bundle.

This 2d anomaly polynomial was derived for 6d SCFTs based on the 6d anomaly computation in \cite{Ohmori:2014kda}, so it may be necessary to modify the polynomial to apply to the field theory models of the 6d supergravity theories discussed in this paper.
In particular, the inverse 4-form $I_{6d,j}=(\Omega^{-1})_{ij} I_{6d}^j$, which appears in the original 2d anomaly formula in \cite{Kim:2016foj}, is not well-defined in the gravity models since $\Omega$ is not invertible. However, we conjecture that this formula (\ref{eq:2d-anomaly-pol}) works for any 6d field theory models embedded in gravity constructed by local 6d SCFTs, such as our field theory models. Instead, we will use only the 4-form $I_{6d}^i$ with upper index, not the inverse 4-form $I_{6d,i}$ as in the case of the original formula in \cite{Kim:2016foj}. Here, $I_{6d}^i$ is defined as the 4-form for $i$-th tensor node when all other tensors are taken to be non-dynamical. More precisely, we conjecture that the anomaly polynomial of self-dual string theories in our gravity models can be obtained by summing over anomaly contributions from all local 2d string theories (in local 6d SCFTs), given by $-k_iI^i_{6d}$, and from the degrees of freedom localized at the intersection between two adjacent 2d theories, given by $ \frac{1}{2}\Omega^{ij}k_ik_j\chi(T_4)$.\footnote{It may be possible to compute anomaly polynomial of 2d strings by studying anomaly inflows computed with respect to the reduced tensor bases and corresponding Dirac pairings which we defined above to verify gauge/gravity mixed anomaly cancellation. But this would require an independent anomaly inflow analysis for each gravity model, and hence we do not pursue this approach.}

This conjecture is sensible because any local 6d SCFT contained in a gravity model must have strings with the anomaly polynomial (\ref{eq:2d-anomaly-pol}) in the field theory limit in which we zoom in near the CFT locus and decouple other tensors from the 6d SCFT. Also, this is consistent with the locality of 2d worldsheet theories, which we expect to be constructed by gluing 2d string theories in local 6d CFTs embedded in the gravity theory.
We will provide more evidence for this conjecture below by comparing the the anomaly polynomial (\ref{eq:2d-anomaly-pol}) with the expected central charges of the strings in the gravity theories.

The central charges $c_I,c_R,c_L$ of a 2d theory are encoded in the coefficients of the following terms in the anomaly polynomial :
\begin{equation}
	I_4 = c_I\left(-\frac{c_2(I)}{6}\right)+c_R\left(-\frac{c_2(R)}{6}\right)+(c_R-c_L)\left(-\frac{p_1(T_2)}{24}\right) + \cdots \ .
\end{equation}
Using this formula, we can compute the central charges of self-dual strings in the 6d gravity models.

For $T^6/\mathbb{Z}_n\times \mathbb{Z}_m$ with $T$ tensor fields, we compute
\begin{equation}\label{eq:centrals-ZnZm}
	c_I = 6\sum_{i=1}^T H^i C_i -6  \ , \quad
	c_R = 3C\cdot C + 3 c_1(B)\cdot C \ , \quad
	c_{R_{\rm cft}}-c_L = -6 c_1(B)\cdot C -2 \ ,
\end{equation}
where $H=h^\vee_G$ is the dual Coxeter number of group $G$ for the tensor nodes with non-trivial gauge groups $G$, and $H=1$ otherwise. In these formulae, we have already subtracted the center-of-mass contributions.
One can easily check that the above result applied to the specific case of $T^6/\mathbb{Z}_2\times \mathbb{Z}_2$ matches the central charges in (\ref{eq:centrals-Z2Z2}) obtained from the UV 2d gauge theory realization of this model.

As discussed, the worldsheet theories are expected to contain various CFT branches and accordingly these CFTs have different central charges. The single string branch, which corresponds to black strings coming from D3-branes wrapping a non-degenerate curve $C$, gives rise to a $(0,4)$ CFT in IR and the right-moving R-symmetry in its superconformal algebra is the $SU(2)_R$ symmetry. In this branch, the central charges $c_R,c_L$ in (\ref{eq:centrals-ZnZm}) of our gravity models perfectly agree with those of single string states computed in \cite{Haghighat:2015ega}, up to constant factors $+6$ in $c_R$ and $+4$ in $c_L$ coming from center-of-mass modes as discussed above. This provides more strong evidence in support of our field theory models of 6d supergravity theories realized by F-theory on $T^6/\mathbb{Z}_n\times \mathbb{Z}_m$.

\section{5d Perspective on Compact Calabi-Yau 3-folds}
\label{sec:5d}
In the previous section we argued that topological string amplitudes may be captured by the elliptic genus of certain 2d quiver theories.  The basic idea was to decompose the theory into local contributions associated to 6d strings, and then to glue these contributions.  However this only works for elliptic CY 3-folds, which have a 6d F-theory realization.  It is natural to ask if similar ideas could work for other compact CY 3-folds which are not elliptic, such as the quintic 3-fold.
In this section we address this question and argue that a similar decomposition into 5d SCFTs exists.  We first review some known examples of circle compactifications of 6d SCFTs which can be viewed as gauging multiple copies of 5d SCFTs and then proceed to study several non-elliptic CY 3-folds using this 5d perspective.  

\subsection{Review: Circle compactification of 6d minimal SCFTs}
\label{sec:5drev}

The possibility of a 6d system decomposition into 5d SCFTs is strongly motivated by the results of \cite{Hayashi:2017jze}.  There it is quantitatively shown that 6d $(1,0)$ $\mathcal O(-n)$ theories compactified on a circle lead to 5d systems which can be constructed by gluing (gauging) local 5d SCFTs associated to toric CY 3-folds (for more geometric perspective, see also \cite{DelZotto:2015rca,DelZotto:2017pti}).  The geometries associated to these theories are of the form
$T^2\times \mathbb{C}^2/\Gamma$.  This is very much in the spirit of what we have already done in this paper using 6d SCFTs.

\begin{figure}
\begin{center}
\begin{tikzpicture}
\node(L) at (0,0) {$
\begin{tikzpicture}[scale=1.1]
	\draw[ultra thick] (0,0) -- (1,0);
	\draw[ultra thick] (1,0) -- (1.707107,0.707107);
	\draw[ultra thick] (1,0) -- (1.707107,-0.707107);
	\draw[ultra thick] (1.707107,-0.707107) -- (1.707107,0.707107);
	\draw[ultra thick] (1.707107,0.707107) --++ (.707107,1+0.707107);
	\draw[ultra thick] (1.707107,-0.707107) --++ (.707107,-1-0.707107);
	\draw[ultra thick] (0,0) -- (-0.5, 0.866025 );
	\draw[ultra thick] (-0.5,0.866025) -- (-0.5+0.258819, 0.866025+0.965926);
	\draw[ultra thick] (-0.5,0.866025)-- (-0.5-0.965926, 0.866025+ 0.258819);
	\draw[ultra thick] (-0.5+0.258819, 0.866025+0.965926) -- (-0.5-0.965926, 0.866025+ 0.258819);
	\draw[ultra thick] (-0.5+0.258819, 0.866025+0.965926) --++ (1.48356, 1.67303);
	\draw[ultra thick] (-0.5-0.965926, 0.866025+ 0.258819) --++ (-2.19067, -0.448288);
	\draw[ultra thick] (0,0) -- (-0.5, - 0.866025);
	\draw[ultra thick] (-0.5, - 0.866025) -- (-0.5+0.258819 , - 0.866025-0.965926);
	\draw[ultra thick] (-0.5, - 0.866025) -- (-0.5-0.965926, - 0.866025-0.258819 );
	\draw[ultra thick]  (-0.5+0.258819 , - 0.866025-0.965926) -- (-0.5-0.965926, - 0.866025-0.258819 );
	\draw[ultra thick]  (-0.5+0.258819 , - 0.866025-0.965926) --++ (1.48356, -1.67303);
	\draw[ultra thick]  (-0.5-0.965926, - 0.866025-0.258819 ) --++ (-2.19067, 0.448288);
	\draw [blue] plot [smooth cycle,tension=1] coordinates {(1,0) (0.166667, 0.288675) (-0.5, 0.866025 )(-0.333333, 0) (-0.5, - 0.866025) (0.166667, -0.288675)};
	\draw[blue] (1,0) to [bend left] (1.707107,0.707107) to [bend left] (1,0);
	\draw[blue] (1,0) to [bend left] (1.707107,-0.707107) to [bend left] (1,0);
	\draw[blue](1.707107,0.707107) to [bend left=20] (1.707107,-0.707107) to [bend left=20] (1.707107,0.707107);
	\draw[blue] (1.707107,0.707107)  to [bend right=10] (2.61421, 2.33137);
	\draw[blue] (1.707107,0.707107) to [bend left = 10] (2.21421, 2.49706);
	\draw[blue] (1.707107,-0.707107) to [bend left = 10] (2.61421, -2.33137);
	\draw[blue] (1.707107,-0.707107) to [bend right = 10] (2.21421, -2.49706);
	\draw[blue] (-0.5,0.866025) to [bend left] (-0.5+0.258819, 0.866025+0.965926) to [bend left](-0.5,0.866025);
	\draw[blue] (-0.5,0.866025) to [bend left]  (-0.5-0.965926, 0.866025+ 0.258819) to [bend left](-0.5,0.866025);
	\draw[blue] (-0.5+0.258819, 0.866025+0.965926) to [bend left=20]  (-0.5-0.965926, 0.866025+ 0.258819) to [bend left=20] (-0.5+0.258819, 0.866025+0.965926);
	\draw[blue] (-0.5+0.258819 , - 0.866025-0.965926) to [bend left=10] (1.44238, -3.32763);
	\draw[blue] (-0.5+0.258819 , - 0.866025-0.965926) to [bend right=10] (1.04238, -3.68233);
	\draw[blue] (-0.5-0.965926, 0.866025+ 0.258819) to [bend left=10] (-3.6066, 0.432219);
	\draw[blue] (-0.5-0.965926, 0.866025+ 0.258819) to [bend right=10] (-3.7066, 0.920893);
	\draw[blue] (-0.5, - 0.866025) to [bend left] (-0.5+0.258819 , - 0.866025-0.965926) to [bend left] (-0.5, - 0.866025);
	\draw[blue] (-0.5, - 0.866025) to [bend left] 	(-0.5-0.965926, - 0.866025-0.258819 ) to [bend left] (-0.5, - 0.866025);
	\draw[blue] (-0.5+0.258819 , - 0.866025-0.965926)to [bend left=20] (-0.5-0.965926, - 0.866025-0.258819 ) to [bend left=20] (-0.5+0.258819 , - 0.866025-0.965926);
	\draw[blue] (-0.5-0.965926, - 0.866025-0.258819 ) to [bend right=10] (-3.6066, -0.432219);
	\draw[blue] (-0.5-0.965926, - 0.866025-0.258819 ) to [bend left=10] (-3.7066, -0.920893);
	\draw[blue] (-0.5+0.258819, 0.866025+0.965926) to [bend right=10] (1.39238, 3.37197);
		\draw[blue] (-0.5+0.258819, 0.866025+0.965926) to [bend left=10] (1.09238, 3.63799);
	\node at (.35,.5) {$1^-$};
	\node at (2.1,.7) {$1^+$};
\end{tikzpicture}
$};
\node(R) at (8,0) 
{
\begin{tikzpicture}
	\node[](a) at (0,0) {$SU(1)$};
	\node(b) at (2,0) {$\mathbb P^2$};
	\node(c) at (120:2) {$\mathbb P^2$};
	\node(d) at (240:2) {$\mathbb P^2$};
	\draw (a) -- (b);
	\draw (a) -- (c);
	\draw (a) -- (d);
\end{tikzpicture}
};
\end{tikzpicture}
\end{center}
\caption{5d perspective on the 6d minimal $\mathcal O(-3)$ theory compactified on $S^1$. From a 5d field theoretic perspective, this theory can be viewed as three local $\mathbb P^2$ theories glued together by a three-punctured sphere which we view as $SU(1)$ gauging. From the standpoint of the topological vertex, this three-punctured sphere is an example of the $1^-$ vertex, which is the mirror of the typical $1^+$ vertex. In the left diagram above, two examples of the mirror $1^{\pm{}}$ vertices are labeled for contrast. The diagram on the right is a schematic graph of the 5d decomposition, where the nodes are 5d SCFTs and the trivalent vertex is the $SU(1)$ gauging.}
\label{fig:O(-3)}
\end{figure}
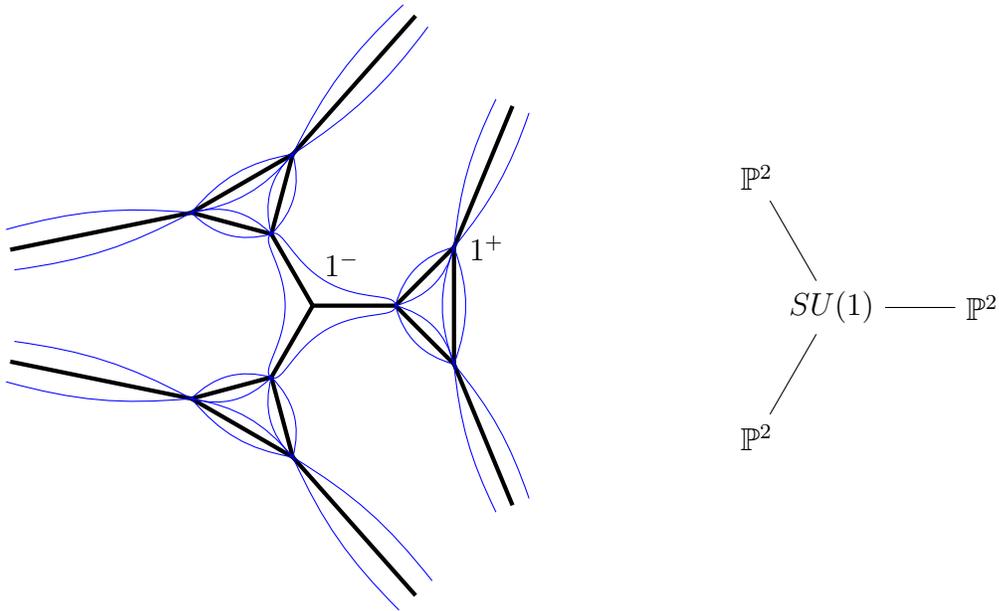

From the point of view of the topological vertex, the conceptually simplest example is the $\mathcal O(-3)$ theory, for which the orbifold action of $\Gamma = \mathbb Z_3$ is generated by $g=(\omega^2 ; \omega^{-1}, \omega^{-1})$ with $\omega^3=1$. The group $\Gamma$ has three fixed points on $T^2$, and the local geometry in the vicinity of each fixed point is $\mathbb C^3/\mathbb Z_3$. The full geometry can therefore be viewed as three local $\mathbb P^2$'s joined by a three-punctured sphere, namely the $\mathbb P^1$ containing the three fixed points of the $\mathbb Z_3$ action. A web picture is depicted in Figure \ref{fig:O(-3)}. An alternative method to obtain this geometry is to resolve the elliptic 3-fold engineering the circle compactification of the $\mathcal O(-3)$ theory, which consists of three Hirzebruch surfaces $\mathbb F_1$ glued along a common $\mathbb P^1$ with normal bundle $\mathcal O(-1) \oplus \mathcal O(-1)$. One then performs a flop transition by blowing down the $\mathbb P^1$ and subsequently blowing up a curve inside each of the three resulting $\mathbb P^2$'s---see for example \cite{DelZotto:2017pti}. Notice that each local $\mathbb P^2$ geometry engineers a 5d $E_0$ theory with a single Coulomb branch modulus, and hence the resulting 5d field theory consists of three $E_0$ theories ``coupled'' to one another by a common state in the massive BPS spectrum with mass depending on all three moduli. This is an example of the ``$SU(1)$'' gauging described in \cite{Hayashi:2017jze}. 

The remaining circle compactified $\mathcal O(-n)$ theories with $n= 4,6,8,12$ can also be assembled from toric 3-folds, in an analogous manner to $\mathcal O(-3)$, by using the non-abelian generalization of the $SU(1)$ gauging to $SU(2)$ gauging. As described in \cite{DelZotto:2015rca} these theories can be realized geometrically as $T^2 \times \mathbb C^2/\Gamma_{\text{orb}}$ where the orbifold group $\Gamma_{\text{orb}} = \langle \mathbb Z_{2p} , \Gamma_{\text{ADE}} \rangle$ is generated by $(\alpha^2, \alpha^{-1} \Gamma_{\text{ADE}})$, with $\alpha^{2p}=1$ and $\Gamma_{\text{ADE}}$ a finite subgroup of $SU(2)$. Note that $p$ must be restricted to the values $p = 2,3,4,6$ to ensure that the action of $\Gamma_{\text{orb}}$ is compatible with the isometries of $T^2$, given a choice of complex structure. As before, the local geometry in the vicinity of each $T^2/G$ fixed point is of the form $\mathbb C^3/\Gamma_{\text{orb}}$, and engineers the 5d theory $\hat{D}_p(G)$.\footnote{The theories $\hat D_p(G)$ are the 5d lifts of generalized 4d $\mathcal N =2$ D-type Argyres-Douglas theories $D_p(G)$.} Since there is a non-compact locus of singularities arising as the fixed point set of $\Gamma_{\text{ADE}}$, the global symmetry group of the 5d theory is the simply laced group $G$ corresponding to the ADE subgroup $\Gamma_{\text{ADE}}$. 

In the full geometry $T^2 \times \mathbb C^2/ \Gamma_{\text{orb}}$, the locus of ADE singularities is now compact and the global symmetry $G$ associated to each local toric 3-fold is consequently gauged. Furthermore, each $T^2$ may contain several fixed points with local geometry $\mathbb C^3/\Gamma_{\text{orb}}$ engineering the 5d theory $\hat{D}_{m_i}(G)$, where $m_i$ is the order of the stabilizer group of the $i$-th fixed point. The compact locus $T^2/\mathbb Z_{2p}$ containing the fixed points is a geometric realization of a ``$G$'' gauging, where $G$ is the diagonal subgroup of product of global symmetry groups associated to each $T^2$ fixed point.

\begin{figure}
\begin{center}
$
\begin{tikzpicture}
	\node(a) at (0,0) {$\begin{tikzpicture}[scale=1.1]
	\draw[ultra thick,double] (0,0) -- (1,0);
	\draw[ultra thick] (1,0) -- (1.707107,0.707107);
	\draw[ultra thick] (1,0) -- (1.707107,-0.707107);
	\draw[ultra thick] (1.707107,-0.707107) -- (1.707107,0.707107);
	\draw[ultra thick] (1.707107,0.707107) --++ (.707107,1+0.707107);
	\draw[ultra thick] (1.707107,-0.707107) --++ (.707107,-1-0.707107);
	\draw[ultra thick] (2.41421, 2.41421) --++ (0.212132, 0.812132);
	\draw[ultra thick]  (2.41421, -2.41421) --++ (0.212132, -0.812132);
	\draw[ultra thick] (2.41421, 2.41421) -- (2.41421, -2.41421);
	\draw[blue] (2.41421, 2.41421)  to  [bend right=10] (2.82634, 3.1741);
	\draw[blue] (2.41421, 2.41421)  to  [bend left=10] (2.42634, 3.27858);
	\draw[blue] (2.41421, -2.41421) to [bend left=10] (2.82634, -3.1741);
	\draw[blue] (2.41421, -2.41421)	to [bend right =10] (2.42634, -3.27858);
	\draw[blue] (1.707107,0.707107) to [bend left=11] (2.41421, 2.41421) to [bend left=11] (1.707107,0.707107);
	\draw[blue] (1.707107,-0.707107) to [bend left=11]  (2.41421, -2.41421) to [bend left=11]  (1.707107,-0.707107) ;
	\draw[blue] (2.41421, 2.41421) to [bend left=6]  (2.41421, -2.41421) to [bend left=6] (2.41421, 2.41421);
	\draw[ultra thick,double] (0,0) -- (-0.5, 0.866025 );
	\draw[ultra thick] (-0.5,0.866025) -- (-0.5+0.258819, 0.866025+0.965926);
	\draw[ultra thick] (-0.5,0.866025)-- (-0.5-0.965926, 0.866025+ 0.258819);
	\draw[ultra thick] (-0.5+0.258819, 0.866025+0.965926) -- (-0.5-0.965926, 0.866025+ 0.258819);
	\draw[ultra thick] (-0.5+0.258819, 0.866025+0.965926) --++ (1.48356, 1.67303);
	\draw[ultra thick] (-0.5-0.965926, 0.866025+ 0.258819) --++ (-2.19067, -0.448288);
	\draw[ultra thick] (-3.6566, 0.676556) --  (1.24238, 3.50498);
	\draw[ultra thick] (-0.5+0.258819, 0.866025+0.965926) --++ (1.48356, 1.67303);
	\draw[ultra thick] (-3.6566, 0.676556)  --++ (-1.02463, -0.346618);
	\draw[blue](-3.6566, 0.676556) to [bend right=10]  (-4.73123, 0.477742);
	\draw[blue](-3.6566, 0.676556) to [bend left=10]  (-4.63123, 0.182134);
	\draw[ultra thick]  (1.24238, 3.50498)--++ (0.812493, 0.714041);
	\draw[blue]  (1.24238, 3.50498) to [bend right=10] (2.15487, 4.10523);
	\draw[blue]  (1.24238, 3.50498) to [bend left=10] (1.95487, 4.33281);
	\draw[ultra thick,double] (0,0) -- (-0.5, - 0.866025);
	\draw[ultra thick] (-0.5, - 0.866025) -- (-0.5+0.258819 , - 0.866025-0.965926);
	\draw[ultra thick] (-0.5, - 0.866025) -- (-0.5-0.965926, - 0.866025-0.258819 );
	\draw[ultra thick]  (-0.5+0.258819 , - 0.866025-0.965926) -- (-0.5-0.965926, - 0.866025-0.258819 );
	\draw[ultra thick]  (-0.5+0.258819 , - 0.866025-0.965926) --++ (1.48356, -1.67303);
	\draw[ultra thick]  (-0.5-0.965926, - 0.866025-0.258819 ) --++ (-2.19067, 0.448288);
	\draw[ultra thick]  (-0.5+0.258819 +1.48356 , - 0.866025-0.965926 -1.67303) -- (-0.5-0.965926-2.19067, - 0.866025-0.258819 +0.448288);
	\draw[ultra thick] (-0.5+0.258819 +1.48356 , - 0.866025-0.965926 -1.67303) --++ (.3*2.70831, -.3*2.38014);
	\draw[blue] (-0.5+0.258819 +1.48356 , - 0.866025-0.965926 -1.67303) to [bend left=15] (2.15487, -4.10524);
	\draw[blue] (-0.5+0.258819 +1.48356 , - 0.866025-0.965926 -1.67303) to [bend right=15] (1.95487, -4.33281);
	\draw[ultra thick] (-0.5-0.965926-2.19067, - 0.866025-0.258819 +0.448288) --++ (-.3*3.41542, .3*1.1554);
	\draw[blue] (-0.5-0.965926-2.19067, - 0.866025-0.258819 +0.448288)  to [bend right=15] (-4.63122, -0.182134);
	\draw[blue] (-0.5-0.965926-2.19067, - 0.866025-0.258819 +0.448288)  to [bend left=15] (-4.73122, -0.477738);
	\draw [blue] plot [smooth cycle,tension=1] coordinates {(1,0) (0.166667, 0.288675) (-0.5, 0.866025 )(-0.333333, 0) (-0.5, - 0.866025) (0.166667, -0.288675)};
	\draw[blue] (1,0) to [bend left] (1.707107,0.707107) to [bend left] (1,0);
	\draw[blue] (1,0) to [bend left] (1.707107,-0.707107) to [bend left] (1,0);
	\draw[blue](1.707107,0.707107) to [bend left=20] (1.707107,-0.707107) to [bend left=20] (1.707107,0.707107);
	\draw[blue] (-0.5,0.866025) to [bend left] (-0.5+0.258819, 0.866025+0.965926) to [bend left](-0.5,0.866025);
	\draw[blue] (-0.5,0.866025) to [bend left]  (-0.5-0.965926, 0.866025+ 0.258819) to [bend left](-0.5,0.866025);
	\draw[blue] (-0.5+0.258819, 0.866025+0.965926) to [bend left=20]  (-0.5-0.965926, 0.866025+ 0.258819) to [bend left=20] (-0.5+0.258819, 0.866025+0.965926);
	\draw[blue] (-0.5+0.258819 , - 0.866025-0.965926) to [bend right=10] (-0.5+0.258819 +1.48356 , - 0.866025-0.965926 -1.67303) to [bend right=10] (-0.5+0.258819 , - 0.866025-0.965926);
	\draw[blue] (-0.5+0.258819 +1.48356 , - 0.866025-0.965926 -1.67303) to [bend right=5]  (-0.5-0.965926-2.19067, - 0.866025-0.258819 +0.448288) to [bend right=5] (-0.5+0.258819 +1.48356 , - 0.866025-0.965926 -1.67303);
	\draw[blue] (-0.5-0.965926, 0.866025+ 0.258819) to [bend left=10](-3.6566, 0.676556) to [bend left=10] (-0.5-0.965926, 0.866025+ 0.258819);
	\draw[blue] (-0.5, - 0.866025) to [bend left] (-0.5+0.258819 , - 0.866025-0.965926) to [bend left] (-0.5, - 0.866025);
	\draw[blue] (-0.5, - 0.866025) to [bend left] 	(-0.5-0.965926, - 0.866025-0.258819 ) to [bend left] (-0.5, - 0.866025);
	\draw[blue] (-0.5+0.258819 , - 0.866025-0.965926)to [bend left=20] (-0.5-0.965926, - 0.866025-0.258819 ) to [bend left=20] (-0.5+0.258819 , - 0.866025-0.965926);
	\draw[blue] (-0.5-0.965926, - 0.866025-0.258819 ) to [bend right=10] (-0.5-0.965926-2.19067, - 0.866025-0.258819 +0.448288) to [bend right=10] (-0.5-0.965926, - 0.866025-0.258819 );
	\draw[blue] (-0.5+0.258819, 0.866025+0.965926) to [bend right=10]  (1.24238, 3.50498)to [bend right=10](-0.5+0.258819, 0.866025+0.965926);
	\draw[blue] (-3.6566, 0.676556) to [bend left=5]  (1.24238, 3.50498) to [bend left =5] (-3.6566, 0.676556);
	\node at (.43,.43) {$2^-$};
	\node at (1.5,.8) {$1^+$};
\end{tikzpicture}
$};
\node(b) at (8,0) {$
	\begin{tikzpicture}
			\node[](a) at (0,0) {$SU(2)$};
	\node(b) at (1.5,0) {$\mathbb P^2_{(1,1,2)}$};
	\node(c) at (120:1.5) {$\mathbb P^2_{(1,1,2)}$};
	\node(d) at (240:1.5) {$\mathbb P^2_{(1,1,2)}$};
		\node(b1) at (3,0) {$\mathbb F_4$};
	\node(c1) at (120:3) {$\mathbb F_4$};
	\node(d1) at (240:3) {$\mathbb F_4$};
	\draw (a) -- (b);
	\draw (a) -- (c);
	\draw (a) -- (d);
	\draw (b1) -- (b);
	\draw (c1) -- (c);
	\draw (d1) -- (d);
	\end{tikzpicture}
$};
\end{tikzpicture}
$
\end{center}
\caption{5d perspective on the 6d minimal $\mathcal O(-6)$ theory compactified on $S^1$. This theory can be viewed as three local $\hat D_{3}(SU(2))$ theories glued together by a three-punctured sphere (i.e. a $\mathbb P^1$ with three marked points) which is a geometric realization of $SU(2)$ gauging. In contrast to $\mathcal O(-3)$ case, the punctures on the central $\mathbb P^1$ correspond to three distinct $A_1$ singularities. In the left figure above, the central trivalent vertex is an example of the $2^-$ vertex. We have also labeled an example of the usual $1^+$ vertex for contrast. The figure on the right is a schematic diagram of the 5d decomposition where the central $SU(2)$ represents the three-punctured sphere.}
\label{fig:O(-6)}
\end{figure}
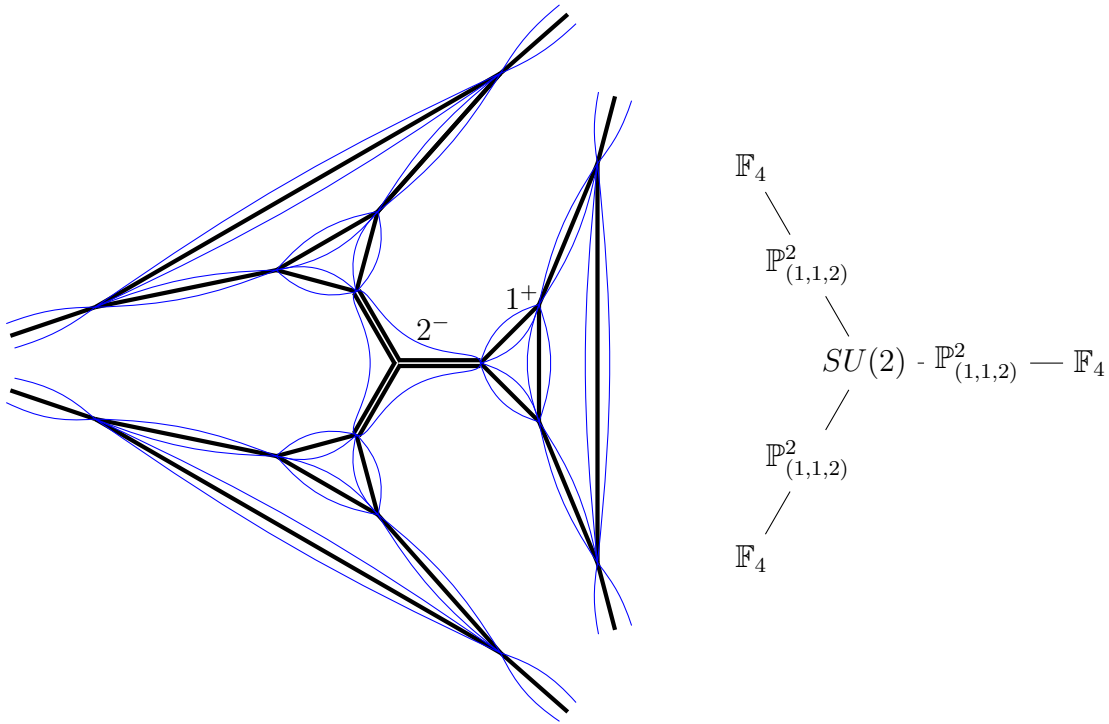

When $G = SU(2)$ the construction described above realizes $\mathcal O(-2p)$ theories compactified on a circle. The situation is similar to the case of $\mathcal O(-3)$ with the important difference being that the 5d theories acting as the building blocks of these theories are now coupled by trivalent\footnote{Or tetravalent, in the case of $\mathcal O(-4)$.} $SU(2)$ gauging, which is in a sense the non-abelian generalization of the $SU(1)$ gauging necessary to describe the $\mathcal O(-3)$ case. As an example, consider the specific case of $\mathcal O(-6)$ compactified on a circle. In the geometry associated to this theory, the $T^2$ factor has three $\mathbb Z_3$ fixed points, and hence the global description consists of three copies of the non-compact 3-fold $\mathbb C^3 / \Gamma_{\text{orb}}$ coupled by trivalent $SU(2)$ gauging, where the orbifold action of $\Gamma_{\text{orb}} = \mathbb Z_6$ is generated by $(\alpha^2; \alpha^{-1}, \alpha^{-1})$ with $\alpha^6 = 1$ (note that the group $\Gamma_{\text{ADE}} = \mathbb Z_2$ generated by $(1;-1,-1)$ is a subgroup of $\mathbb Z_6$.) The local 3-fold $\mathbb C^3/ \mathbb Z_6$ with this orbifold action can be viewed as a neighborhood of the singular K\"ahler surface $\mathbb F_4 \cup \mathbb F_2 \cup \mathbb F_0$ and the full geometry is three copies of $\mathbb F_4 \cup \mathbb F_2$ glued along three distinct 
fibers in a common $\mathbb F_0$. To bring this geometry into a form similar to the geometric description of the $\mathcal O(-3)$, we can blow down the $\mathbb F_0$ along its ruling to obtain a $\mathbb P^1$ meeting three weighted projective planes $\mathbb P^2_{(1,1,2)}$ at their respective $A_1$ singularities; see \cite{DelZotto:2017pti}. The vertex decomposition of this geometry is depicted in Figure \ref{fig:O(-6)}.

In the cases of $\mathcal O(-3)$ and $\mathcal O(-4)$, a number of consistency checks of the above decomposition into 5d SCFTs coupled by $SU(2)$ gauging were performed in \cite{Hayashi:2017jze}, most notably comparisons of the Nekrasov partition function with the elliptic genus. Moreover, expressions for the Nekrasov partition functions of the $\mathcal O(-6)$, $\mathcal O(-8)$, and $\mathcal O(-12)$ theories compactified on a circle were also proposed in \cite{Hayashi:2017jze}. The existence of trivalent gaugings compatible with the Nekrasov partition function offers the tantalizing possibility that the usual topological vertex can be extended to non-toric 3-folds by including an additional vertex which accounts for trivalent gluing of multiple numbers of external legs in the dual $(p,q)$ 5-brane web diagrams. In the case of $SU(N)$ gauging we refer to a trivalent vertex consisting of $N$ external legs as an ``$N^-$ vertex''. We explore this point in more detail in Section \ref{sec:gentopver}, after we describe several examples of compact 3-folds that also admit a decomposition into toric 3-folds glued by various types of $SU(N)$ gaugings.

\subsection{Mirror Fermat Calabi-Yau 3-folds} 
\label{sec:singularloci}

Here we attempt to address cases for CY 3-folds without an elliptic fibration such as the mirror quintic and ask whether or not they admit a complete 5d gluing prescription.   We find that there is evidence that the 5d local theories glued together capture the key ingredients of the compact model, and we conjecture that they contain many (but not all) of the needed ingredients to capture the all-genus topological string amplitudes for certain compact CY 3-folds, including the mirror quintic. 
We illustrate this point by considering orbifolds of Fermat hypersurfaces in weighted projective spaces.

Consider a Calabi-Yau 3-fold $M$ defined as a degree $d$ Fermat hypersurface of the weighted projective space  $\mathbb P^4_{\vec w}$. We denote such hypersurfaces by $M = \mathbb P^4_{\vec w}[d]$, where $\vec w$ has components $w_i$ satisfying $\sum w_i = d$:
	\begin{align}
		M = \mathbb P^4_{\vec w}[d]= \left\{ \sum_{i=1}^5 z_{i}^{p_i}   = 0 \right\} \subset \mathbb P^4,~~ p_i = \frac{d}{w_i}. 
	\end{align}
	Let us count the number of K\"ahler parameters associated to the mirror $W$ defined by 
	\begin{align}
		W= M \slash G,~~ G =\prod_{k} \mathbb Z_{q_k}.
		\label{eq:CYdef}
	\end{align}	
The group $G$ is the maximal group of abelian symmetries whose matrix representatives have unit determinant:
	\begin{align}
		G ~:~ \vec z \rightarrow g \vec z,~~ \text{det}(g) = 1.  
	\end{align}
  The CY 3-fold $W$ is the mirror dual to $M$, which from the above equations we see is defined as the hypersurface in the non-orbifolded version of the same weighted projective space \cite{Greene:1990ud}.  The reason we have modded out by the maximal allowed abelian symmetry is that this creates the most singular 3-fold possible and thus has the best chance of being described as a collection of local singularities.  

Similar to what we discussed in the 6d case, in order to capture the local physics, we need to concentrate on the local singularities which lead to 5d SCFTs and gauge symmetries. We shall see exactly these two ingredients appear for us in this case as well.
Thus, we consider M-theory on the above CY 3-fold and consider its singular loci.  The point singularities give rise to 5d SCFTs, while curve singularities correspond to gauge symmetries.
 
\begin{figure}
	\begin{center}
	\scalebox{.75}{$
	\begin{tikzpicture}[scale=.25]
	\node[color=white](1) at (0,0) {1};
	\node[color=white](2) at (1,1) {2};
	\node[color=white](3) at (3,3) {3};
	\node[color=white](4) at (4,4) {4};
	\node[color=white](5) at (6,6) {5};
	\node[color=white](1+) at (1,2) {1+};
	\node[color=white](2+) at (2,3) {2+};
	\node[color=white](1+) at (1,2) {1+};
	\node[color=white](3+) at (4,5) {3+};
	\node[color=white](4+) at (5,6) {4+};
	\node[color=white](1++) at (2,4) {1++};
	\node[color=white](2++) at (3,5) {2++};
	\node[color=white](3++) at (5,7) {3++};
	\node[color=white](1+++) at (3,6) {1+++};
	\node[color=white](2+++) at (4,7) {2+++};
	\node[color=white] (1++++) at (4,8) {1++++};
	\node[color=white](1-) at (2,1) {1-};
	\node[color=white](2-) at (3,2) {2-};
	\node[color=white](1-) at (2,1) {1-};
	\node[color=white](3-) at (5,4) {3-};
	\node[color=white](4-) at (6,5) {4-};
	\node[color=white](1--) at (4,2) {1--};
	\node[color=white](2--) at (5,3) {2--};
	\node[color=white](3--) at (7,5) {3--};
	\node[color=white](1---) at (6,3) {1---};
	\node[color=white](2---) at (7,4) {2---};
	\node[color=white] (1----) at (8,4) {1----};
	\draw[ultra thick] (1+.center) -- (2+.center);
	\draw[ultra thick] (1.center) -- (2.center);
	\draw[ultra thick] (3.center) -- (4.center);
	\draw[ultra thick] (2.center) -- (1+.center);
	\draw[ultra thick] (2+.center) -- (1++.center);
	\draw[ultra thick] (2++.center) -- (1+++.center);
	\draw[ultra thick] (4.center) -- (3+.center);
	\draw[ultra thick] (4+.center) -- (3++.center);
	\draw[ultra thick] (2+.center) -- (3.center);
	\draw[ultra thick] (2++.center) -- (3+.center);
	\draw[ultra thick] (2+++.center) -- (3++.center);
	\draw[ultra thick] (4+.center) -- (5.center);
	\draw[ultra thick] (2+++.center) -- (1++++.center);
	\draw[ultra thick] (1++.center) -- (2++.center);
	\draw[ultra thick] (3+.center) -- (4+.center);
	\draw[ultra thick](1+++.center) -- (2+++.center);
	\draw[ultra thick] (1-.center) -- (2-.center);
	\draw[ultra thick] (1.center) -- (2.center);
	\draw[ultra thick] (3.center) -- (4.center);
	\draw[ultra thick] (2.center) -- (1-.center);
	\draw[ultra thick] (2-.center) -- (1--.center);
	\draw[ultra thick] (2--.center) -- (1---.center);
	\draw[ultra thick] (4.center) -- (3-.center);
	\draw[ultra thick] (4-.center) -- (3--.center);
	\draw[ultra thick] (2-.center) -- (3.center);
	\draw[ultra thick] (2--.center) -- (3-.center);
	\draw[ultra thick] (2---.center) -- (3--.center);
	\draw[ultra thick] (4-.center) -- (5.center);
	\draw[ultra thick] (2---.center) -- (1----.center);
	\draw[ultra thick] (1--.center) -- (2--.center);
	\draw[ultra thick] (3-.center) -- (4-.center);
	\draw[ultra thick](1---.center) -- (2---.center);
	
	\node[color=white](1U) at (0-14,0+16) {1};
	\node[color=white](2U) at (1-14,1+16) {2};
	\node[color=white](3U) at (3-14,3+16) {3};
	\node[color=white](4U) at (4-14,4+16) {4};
	\node[color=white](5U) at (6-14,6+16) {5};
	\node[color=white](1+U) at (1-14,2+16) {1+};
	\node[color=white](2+U) at (2-14,3+16) {2+};
	\node[color=white](1+U) at (1-14,2+16) {1+};
	\node[color=white](3+U) at (4-14,5+16) {3+};
	\node[color=white](4+U) at (5-14,6+16) {4+};
	\node[color=white](1++U) at (2-14,4+16) {1++};
	\node[color=white](2++U) at (3-14,5+16) {2++};
	\node[color=white](3++U) at (5-14,7+16) {3++};
	\node[color=white](1+++U) at (3-14,6+16) {1+++};
	\node[color=white](2+++U) at (4-14,7+16) {2+++};
	\node[color=white] (1++++U) at (4-14,8+16) {1++++};
	\node[color=white](1-U) at (2-14,1+16) {1-};
	\node[color=white](2-U) at (3-14,2+16) {2-};
	\node[color=white](1-U) at (2-14,1+16) {1-};
	\node[color=white](3-U) at (5-14,4+16) {3-};
	\node[color=white](4-U) at (6-14,5+16) {4-};
	\node[color=white](1--U) at (4-14,2+16) {1--};
	\node[color=white](2--U) at (5-14,3+16) {2--};
	\node[color=white](3--U) at (7-14,5+16) {3--};
	\node[color=white](1---U) at (6-14,3+16) {1---};
	\node[color=white](2---U) at (7-14,4+16) {2---};
	\node[color=white] (1----U) at (8-14,4+16) {1----};
	\draw[ultra thick] (1+U.center) -- (2+U.center);
	\draw[ultra thick] (1U.center) -- (2U.center);
	\draw[ultra thick] (3U.center) -- (4U.center);
	\draw[ultra thick] (2U.center) -- (1+U.center);
	\draw[ultra thick] (2+U.center) -- (1++U.center);
	\draw[ultra thick] (2++U.center) -- (1+++U.center);
	\draw[ultra thick] (4U.center) -- (3+U.center);
	\draw[ultra thick] (4+U.center) -- (3++U.center);
	\draw[ultra thick] (2+U.center) -- (3U.center);
	\draw[ultra thick] (2++U.center) -- (3+U.center);
	\draw[ultra thick] (2+++U.center) -- (3++U.center);
	\draw[ultra thick] (4+U.center) -- (5U.center);
	\draw[ultra thick] (2+++U.center) -- (1++++U.center);
	\draw[ultra thick] (1++U.center) -- (2++U.center);
	\draw[ultra thick] (3+U.center) -- (4+U.center);
	\draw[ultra thick](1+++U.center) -- (2+++U.center);
	\draw[ultra thick] (1-U.center) -- (2-U.center);
	\draw[ultra thick] (1U.center) -- (2U.center);
	\draw[ultra thick] (3U.center) -- (4U.center);
	\draw[ultra thick] (2U.center) -- (1-U.center);
	\draw[ultra thick] (2-U.center) -- (1--U.center);
	\draw[ultra thick] (2--U.center) -- (1---U.center);
	\draw[ultra thick] (4U.center) -- (3-U.center);
	\draw[ultra thick] (4-U.center) -- (3--U.center);
	\draw[ultra thick] (2-U.center) -- (3U.center);
	\draw[ultra thick] (2--U.center) -- (3-U.center);
	\draw[ultra thick] (2---U.center) -- (3--U.center);
	\draw[ultra thick] (4-U.center) -- (5U.center);
	\draw[ultra thick] (2---U.center) -- (1----U.center);
	\draw[ultra thick] (1--U.center) -- (2--U.center);
	\draw[ultra thick] (3-U.center) -- (4-U.center);
	\draw[ultra thick](1---U.center) -- (2---U.center);

	\node[color=white](1L) at (-0-20,0) {1};
	\node[color=white](2L) at (-1-20,1) {2};
	\node[color=white](3L) at (-3-20,3) {3};
	\node[color=white](4L) at (-4-20,4) {4};
	\node[color=white](5L) at (-6-20,6) {5};
	\node[color=white](1+L) at (-1-20,2) {1+};
	\node[color=white](2+L) at (-2-20,3) {2+};
	\node[color=white](1+L) at (-1-20,2) {1+};
	\node[color=white](3+L) at (-4-20,5) {3+};
	\node[color=white](4+L) at (-5-20,6) {4+};
	\node[color=white](1++L) at (-2-20,4) {1++};
	\node[color=white](2++L) at (-3-20,5) {2++};
	\node[color=white](3++L) at (-5-20,7) {3++};
	\node[color=white](1+++L) at (-3-20,6) {1+++};
	\node[color=white](2+++L) at (-4-20,7) {2+++};
	\node[color=white] (1++++L) at (-4-20,8) {1++++};
	\node[color=white](1-L) at (-2-20,1) {1-};
	\node[color=white](2-L) at (-3-20,2) {2-};
	\node[color=white](1-L) at (-2-20,1) {1-};
	\node[color=white](3-L) at (-5-20,4) {3-};
	\node[color=white](4-L) at (-6-20,5) {4-};
	\node[color=white](1--L) at (-4-20,2) {1--};
	\node[color=white](2--L) at (-5-20,3) {2--};
	\node[color=white](3--L) at (-7-20,5) {3--};
	\node[color=white](1---L) at (-6-20,3) {1---};
	\node[color=white](2---L) at (-7-20,4) {2---};
	\node[color=white] (1----L) at (-8-20,4) {1----};
	\draw[ultra thick] (1+L.center) -- (2+L.center);
	\draw[ultra thick] (1L.center) -- (2L.center);
	\draw[ultra thick] (3L.center) -- (4L.center);
	\draw[ultra thick] (2L.center) -- (1+L.center);
	\draw[ultra thick] (2+L.center) -- (1++L.center);
	\draw[ultra thick] (2++L.center) -- (1+++L.center);
	\draw[ultra thick] (4L.center) -- (3+L.center);
	\draw[ultra thick] (4+L.center) -- (3++L.center);
	\draw[ultra thick] (2+L.center) -- (3L.center);
	\draw[ultra thick] (2++L.center) -- (3+L.center);
	\draw[ultra thick] (2+++L.center) -- (3++L.center);
	\draw[ultra thick] (4+L.center) -- (5L.center);
	\draw[ultra thick] (2+++L.center) -- (1++++L.center);
	\draw[ultra thick] (1++L.center) -- (2++L.center);
	\draw[ultra thick] (3+L.center) -- (4+L.center);
	\draw[ultra thick](1+++L.center) -- (2+++L.center);
	\draw[ultra thick] (1-L.center) -- (2-L.center);
	\draw[ultra thick] (1L.center) -- (2L.center);
	\draw[ultra thick] (3L.center) -- (4L.center);
	\draw[ultra thick] (2L.center) -- (1-L.center);
	\draw[ultra thick] (2-L.center) -- (1--L.center);
	\draw[ultra thick] (2--L.center) -- (1---L.center);
	\draw[ultra thick] (4L.center) -- (3-L.center);
	\draw[ultra thick] (4-L.center) -- (3--L.center);
	\draw[ultra thick] (2-L.center) -- (3L.center);
	\draw[ultra thick] (2--L.center) -- (3-L.center);
	\draw[ultra thick] (2---L.center) -- (3--L.center);
	\draw[ultra thick] (4-L.center) -- (5L.center);
	\draw[ultra thick] (2---L.center) -- (1----L.center);
	\draw[ultra thick] (1--L.center) -- (2--L.center);
	\draw[ultra thick] (3-L.center) -- (4-L.center);
	\draw[ultra thick](1---L.center) -- (2---L.center);
	
	\node[color=white](1D) at (0,-0-20) {1};
	\node[color=white](2D) at (1,-1-20) {2};
	\node[color=white](3D) at (3,-3-20) {3};
	\node[color=white](4D) at (4,-4-20) {4};
	\node[color=white](5D) at (6,-6-20) {5};
	\node[color=white](1+D) at (1,-2-20) {1+};
	\node[color=white](2+D) at (2,-3-20) {2+};
	\node[color=white](1+D) at (1,-2-20) {1+};
	\node[color=white](3+D) at (4,-5-20) {3+};
	\node[color=white](4+D) at (5,-6-20) {4+};
	\node[color=white](1++D) at (2,-4-20) {1++};
	\node[color=white](2++D) at (3,-5-20) {2++};
	\node[color=white](3++D) at (5,-7-20) {3++};
	\node[color=white](1+++D) at (3,-6-20) {1+++};
	\node[color=white](2+++D) at (4,-7-20) {2+++};
	\node[color=white] (1++++D) at (4,-8-20) {1++++};
	\node[color=white](1-D) at (2,-1-20) {1-};
	\node[color=white](2-D) at (3,-2-20) {2-};
	\node[color=white](1-D) at (2,-1-20) {1-};
	\node[color=white](3-D) at (5,-4-20) {3-};
	\node[color=white](4-D) at (6,-5-20) {4-};
	\node[color=white](1--D) at (4,-2-20) {1--};
	\node[color=white](2--D) at (5,-3-20) {2--};
	\node[color=white](3--D) at (7,-5-20) {3--};
	\node[color=white](1---D) at (6,-3-20) {1---};
	\node[color=white](2---D) at (7,-4-20) {2---};
	\node[color=white] (1----D) at (8,-4-20) {1----};
	\draw[ultra thick] (1+D.center) -- (2+D.center);
	\draw[ultra thick] (1D.center) -- (2D.center);
	\draw[ultra thick] (3D.center) -- (4D.center);
	\draw[ultra thick] (2D.center) -- (1+D.center);
	\draw[ultra thick] (2+D.center) -- (1++D.center);
	\draw[ultra thick] (2++D.center) -- (1+++D.center);
	\draw[ultra thick] (4D.center) -- (3+D.center);
	\draw[ultra thick] (4+D.center) -- (3++D.center);
	\draw[ultra thick] (2+D.center) -- (3D.center);
	\draw[ultra thick] (2++D.center) -- (3+D.center);
	\draw[ultra thick] (2+++D.center) -- (3++D.center);
	\draw[ultra thick] (4+D.center) -- (5D.center);
	\draw[ultra thick] (2+++D.center) -- (1++++D.center);
	\draw[ultra thick] (1++D.center) -- (2++D.center);
	\draw[ultra thick] (3+D.center) -- (4+D.center);
	\draw[ultra thick](1+++D.center) -- (2+++D.center);
	\draw[ultra thick] (1-D.center) -- (2-D.center);
	\draw[ultra thick] (1D.center) -- (2D.center);
	\draw[ultra thick] (3D.center) -- (4D.center);
	\draw[ultra thick] (2D.center) -- (1-D.center);
	\draw[ultra thick] (2-D.center) -- (1--D.center);
	\draw[ultra thick] (2--D.center) -- (1---D.center);
	\draw[ultra thick] (4D.center) -- (3-D.center);
	\draw[ultra thick] (4-D.center) -- (3--D.center);
	\draw[ultra thick] (2-D.center) -- (3D.center);
	\draw[ultra thick] (2--D.center) -- (3-D.center);
	\draw[ultra thick] (2---D.center) -- (3--D.center);
	\draw[ultra thick] (4-D.center) -- (5D.center);
	\draw[ultra thick] (2---D.center) -- (1----D.center);
	\draw[ultra thick] (1--D.center) -- (2--D.center);
	\draw[ultra thick] (3-D.center) -- (4-D.center);
	\draw[ultra thick](1---D.center) -- (2---D.center);

	\draw[ultra thick] (1L.center) -- (1.center);
	\draw[ultra thick] (1+L.center) -- (1+.center);
	\draw[ultra thick] (1++L.center) -- (1++.center);
	\draw[ultra thick] (1+++L.center) -- (1+++.center);
	\draw[ultra thick] (1++++L.center) -- (1++++.center);
	\draw[ultra thick] (1D.center) -- (1.center);
	\draw[ultra thick] (1-D.center) -- (1-.center);
	\draw[ultra thick] (1--D.center) -- (1--.center);
	\draw[ultra thick] (1---D.center) -- (1---.center);
	\draw[ultra thick] (1----D.center) -- (1----.center);
	
	\draw[ultra thick] (1U.center) --++ (0,-8);
	\draw[ultra thick] (1-U.center) --++ (0,-9);
	\draw[ultra thick] (1--U.center) --++ (0,-10);
	\draw[ultra thick] (1---U.center) --++ (0,-11);
	\draw[ultra thick] (1----U.center) --++ (0,-12);
	
	\draw[ultra thick] (1++++U.center) --++ (4.5,4.5);
	\draw[ultra thick,dashed] ([shift=({4.5,4.5})]1++++U.center) --++ (2.5,2.5);
	\draw[ultra thick] (3++U.center) --++ (4.5,4.5);
	\draw[ultra thick,dashed] ([shift=({4.5,4.5})]3++U.center) --++ (2.5,2.5);
	\draw[ultra thick] (5U.center) --++ (4.5,4.5);
	\draw[ultra thick,dashed] ([shift=({4.5,4.5})]5U.center) --++ (2.5,2.5);
	\draw[ultra thick] (3--U.center) --++ (4.5,4.5);
	\draw[ultra thick,dashed] ([shift=({4.5,4.5})]3--U.center) --++ (2.5,2.5);
	\draw[ultra thick] (1----U.center) --++ (4.5,4.5);
	\draw[ultra thick,dashed] ([shift=({4.5,4.5})]1----U.center) --++ (2.5,2.5);

	\draw[ultra thick] (1U.center) --++ (-4,0);
	\draw[ultra thick,dashed] ([shift=({-4,0})]1U.center) --++ (-4,0);
	\draw[ultra thick] (1+U.center) --++ (-5,0);
	\draw[ultra thick,dashed] ([shift=({-5,0})]1+U.center) --++ (-4,0);
	\draw[ultra thick] (1++U.center) --++ (-6,0);
	\draw[ultra thick,dashed] ([shift=({-6,0})]1++U.center) --++ (-4,0);
	\draw[ultra thick] (1+++U.center) --++ (-7,0);
	\draw[ultra thick,dashed] ([shift=({-7,0})]1+++U.center) --++ (-4,0);
	\draw[ultra thick] (1++++U.center) --++ (-8,0);
	\draw[ultra thick,dashed] ([shift=({-8,0})]1++++U.center) --++ (-4,0);
	
	\draw[ultra thick] (1L.center) --++ (0,-4);
	\draw[ultra thick,dashed] ([shift=({0,-4})]1L.center) --++ (0,-4);
	\draw[ultra thick] (1-L.center) --++ (0,-5);
	\draw[ultra thick,dashed] ([shift=({0,-5})]1-L.center) --++ (0,-4);
	\draw[ultra thick] (1--L.center) --++ (0,-6);
	\draw[ultra thick,dashed] ([shift=({0,-6})]1--L.center) --++ (0,-4);
	\draw[ultra thick] (1---L.center) --++ (0,-7);
	\draw[ultra thick,dashed] ([shift=({0,-7})]1---L.center) --++ (0,-4);
	\draw[ultra thick] (1----L.center) --++ (0,-8);
	\draw[ultra thick,dashed] ([shift=({0,-8})]1----L.center) --++ (0,-4);
	
	\draw[ultra thick] (1++++L.center) --++ (-4.5,4.5);
	\draw[ultra thick,dashed] ([shift=({-4.5,4.5})]1++++L.center) --++ (-2.5,2.5);
	\draw[ultra thick] (3++L.center) --++ (-4.5,4.5);
	\draw[ultra thick,dashed] ([shift=({-4.5,4.5})]3++L.center) --++ (-2.5,2.5);
	\draw[ultra thick] (5L.center) --++ (-4.5,4.5);
	\draw[ultra thick,dashed] ([shift=({-4.5,4.5})]5L.center) --++ (-2.5,2.5);
	\draw[ultra thick] (3--L.center) --++ (-4.5,4.5);
	\draw[ultra thick,dashed] ([shift=({-4.5,4.5})]3--L.center) --++ (-2.5,2.5);
	\draw[ultra thick] (1----L.center) --++ (-4.5,4.5);
	\draw[ultra thick,dashed] ([shift=({-4.5,4.5})]1----L.center) --++ (-2.5,2.5);
	
	\draw[ultra thick] (1++++.center) --++ (4.5,4.5);
	\draw[ultra thick,dashed] ([shift=({4.5,4.5})]1++++.center) --++ (2.5,2.5);
	\draw[ultra thick] (3++.center) --++ (4.5,4.5);
	\draw[ultra thick,dashed] ([shift=({4.5,4.5})]3++.center) --++ (2.5,2.5);
	\draw[ultra thick] (5.center) --++ (4.5,4.5);
	\draw[ultra thick,dashed] ([shift=({4.5,4.5})]5.center) --++ (2.5,2.5);
	\draw[ultra thick] (3--.center) --++ (4.5,4.5);
	\draw[ultra thick,dashed] ([shift=({4.5,4.5})]3--.center) --++ (2.5,2.5);
	\draw[ultra thick] (1----.center) --++ (4.5,4.5);
	\draw[ultra thick,dashed] ([shift=({4.5,4.5})]1----.center) --++ (2.5,2.5);
	
	\draw[ultra thick] (1++++D.center) --++ (4.5,-4.5);
	\draw[ultra thick,dashed] ([shift=({4.5,-4.5})]1++++D.center) --++ (2.5,-2.5);
	\draw[ultra thick] (3++D.center) --++ (4.5,-4.5);
	\draw[ultra thick,dashed] ([shift=({4.5,-4.5})]3++D.center) --++ (2.5,-2.5);
	\draw[ultra thick] (5D.center) --++ (4.5,-4.5);
	\draw[ultra thick,dashed] ([shift=({4.5,-4.5})]5D.center) --++ (2.5,-2.5);
	\draw[ultra thick] (3--D.center) --++ (4.5,-4.5);
	\draw[ultra thick,dashed] ([shift=({4.5,-4.5})]3--D.center) --++ (2.5,-2.5);
	\draw[ultra thick] (1----D.center) --++ (4.5,-4.5);
	\draw[ultra thick,dashed] ([shift=({4.5,-4.5})]1----D.center) --++ (2.5,-2.5);
	
	\draw[ultra thick] (1D.center) --++ (-4,0);
	\draw[ultra thick,dashed] ([shift=({-4,0})]1D.center) --++ (-4,0);
	\draw[ultra thick] (1+D.center) --++ (-5,0);
	\draw[ultra thick,dashed] ([shift=({-5,0})]1+D.center) --++ (-4,0);
	\draw[ultra thick] (1++D.center) --++ (-6,0);
	\draw[ultra thick,dashed] ([shift=({-6,0})]1++D.center) --++ (-4,0);
	\draw[ultra thick] (1+++D.center) --++ (-7,0);
	\draw[ultra thick,dashed] ([shift=({-7,0})]1+++D.center) --++ (-4,0);
	\draw[ultra thick] (1++++D.center) --++ (-8,0);
	\draw[ultra thick,dashed] ([shift=({-8,0})]1++++D.center) --++ (-4,0);
	\node[draw,circle,ultra thick,fill=white,scale=2.2] at (-10,4) {$SU(5)$};
	\draw[ultra thick] (4,-14) --++ (9,0);
	\draw[ultra thick,dashed] (13,-14) --++ (4,0);
	\draw[ultra thick] (4,-6) --++ (9,0);
	\draw[ultra thick,dashed] (13,-6) --++ (4,0);
	\draw[ultra thick] (4,-10) --++ (9,0);
	\draw[ultra thick,dashed] (13,-10) --++ (4,0);
	\draw[ultra thick] (4,-8) --++ (9,0);
	\draw[ultra thick,dashed] (13,-8) --++ (4,0);
	\draw[ultra thick] (4,-12) --++ (9,0);
	\draw[ultra thick,dashed] (13,-12) --++ (4,0);
	\node[draw,circle,ultra thick,fill=white,scale=2.2] at (4,-10) {$SU(5)$};
	
	\draw[ultra thick] (13,13) --++ (0,8);
	\draw[ultra thick] (11,13) --++ (0,8);
	\draw[ultra thick] (15,13) --++ (0,8);
	\draw[ultra thick] (9,13) --++ (0,8);
	\draw[ultra thick] (17,13) --++ (0,8);
	\draw[ultra thick,dashed] (13,21) --++ (0,3);
	\draw[ultra thick,dashed] (11,21) --++ (0,3);
	\draw[ultra thick,dashed] (15,21) --++ (0,3);
	\draw[ultra thick,dashed] (9,21) --++ (0,3);
	\draw[ultra thick,dashed] (17,21) --++ (0,3);
	\draw[ultra thick] (13,15) --++ (8,0);
	\draw[ultra thick,dashed] (21,15) --++ (3,0);
	\draw[ultra thick] (13,13) --++ (8,0);
	\draw[ultra thick,dashed] (21,13) --++ (3,0);
	\draw[ultra thick] (13,11) --++ (8,0);
	\draw[ultra thick,dashed] (21,11) --++ (3,0);
	\draw[ultra thick] (13,17) --++ (8,0);
	\draw[ultra thick,dashed] (21,17) --++ (3,0);
	\draw[ultra thick] (13,9) --++ (8,0);
	\draw[ultra thick,dashed] (21,9) --++ (3,0);
	\node[draw,circle,ultra thick,fill=white,scale=2.2] at (13,13) {$SU(5)$};
\end{tikzpicture}$}
	\end{center}
	\caption{Schematic example of a collection of 5d $T_{ijk}$ theories with trivalent $SU(n_{ij})$ gaugings of the curves $C_{ij}$. Depicted above are three $SU(5)$ gaugings of four local $T_5$ theories, as described in the 5d model of the mirror quintic. Each $T_5$ theory has (not accounting for possible symmetry enchancements) $SU(5) \times SU(5) \times SU(5)$ global symmetry, which can be viewed as arising from three separate stacks of parallel non-compact branes. The dotted lines connect to either additional $SU(5)$ gaugings or local SCFTs $T_{i'j'k'}$. The global structure of the system is described in Figure \ref{fig:pentagon}.}
	\label{eq:toricdiagram}
	\end{figure}
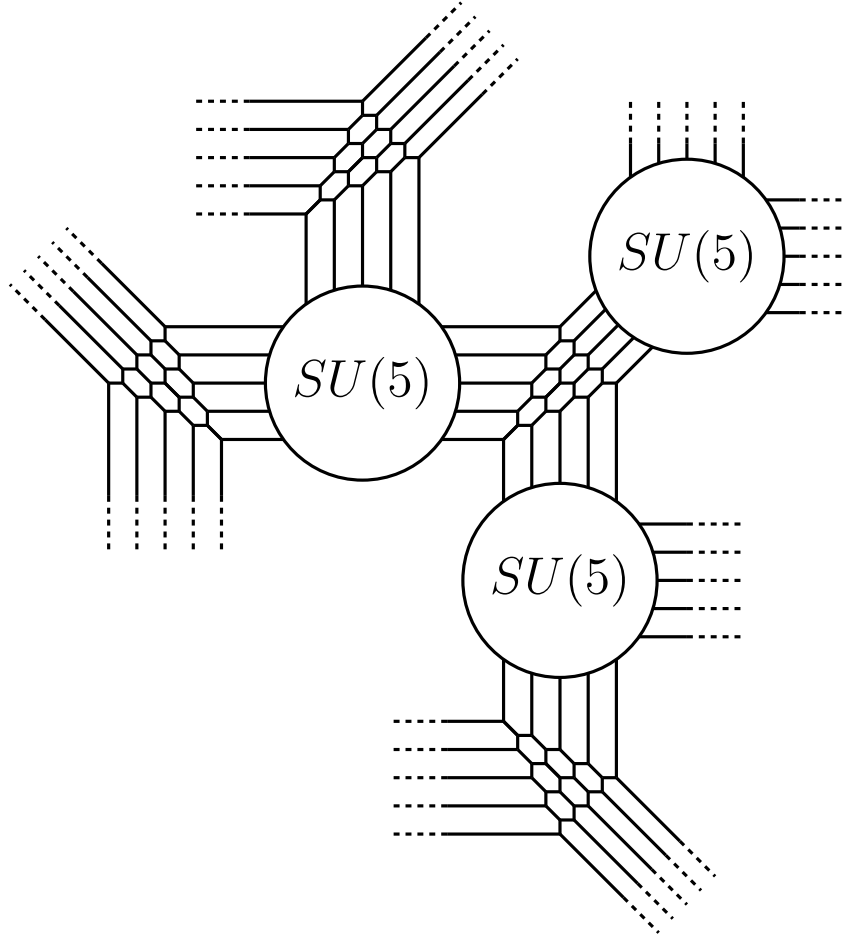

The 3-folds we consider have singular loci arising as the fixed point sets of $G$. The fixed points due to the abelian symmetries of $G$ which do not involve the intrinsic symmetry of the underlying weighted projective space are generally curves or points. Moreover, weighted projective spaces have intrinsic singularities of the form:
	\begin{align}
		\text{Sing}_S = \{ z_i = 0, \forall   i \in \{ 1 , \dots, n +1 \} \backslash S \} 
	\end{align}
provided the subset of weights $\{ w_i \}_{i \in S}$ has nontrivial gcd $n_S$.  The examples we consider below are well-formed hypersurfaces. A complete intersection $M$ given by the zero locus of $m$ homogeneous polynomials is well formed if $M$ does not contain any codimension $m +1$ singular loci of the weighted projective space and if furthermore the weights of any set of $n$ projective coordinates are coprime. In our case, $n=4$ and $m= 1$, and thus it follows that all singularities of our 3-fold hypersurfaces are curves or points, for which the normal geometry is (respectively) $\mathbb C^2/\mathbb Z_{n_0}$ or $\mathbb C^3 / \mathbb Z_{n_1}\times \mathbb Z_{n_2}$, with some $n_0,n_1,n_2$.   The former produces a gauge symmetry $SU(n_0)$, while the latter produces a 5d CFT labeled by $T(n_1,n_2)$. Curve singularities appear over loci $z_i=z_j=0$ and are denoted $C_{ij}$. The normal geometry of $C_{ij}$ is $\mathbb C^2/\mathbb Z_{n_{ij}}$ (with $n_{ij} \equiv \text{gcd}(p_i,p_j)$)
and is responsible for $SU(n_{ij})$ gauge symmetry localized over $C_{ij}$ with gauge parameter $1/g^2$ proportional to the area of $C_{ij}$. We can understand this in more precise terms as follows: the orbifold action of $G$ on the tangent directions $\sum_{k \ne i,j} z_k^{p_k} = 0$ produces a genus 0 curve $C_{ij}$ while the action of $G$ on the normal directions $z_i, z_j$ is given by $\mathbb Z_{n_{ij}}$ and hence generates $n_{ij}$ images leading to $SU(n_{ij})$ gauging, associated to the following number of K\"ahler parameters:
	\begin{align}
		N(C_{ij} ) = n_{ij}-1.	
	\end{align}
On the other hand, singularities of the type $\mathbb C^3 / \mathbb Z_{n_1}\times \mathbb Z_{n_2}$ arise at points given by
$z_i=z_j=z_k=0$; we sometimes label the corresponding 5d SCFT by $T_{ijk}$.
Since each of these points is the intersection of three curves $C_{ij} \cap C_{jk} \cap C_{ki}$ described above, we should think of each point as a local $T(n_1,n_2)$ theory defined by a toric diagram with $n_{ij}$ branes crossing one edge, $n_{jk}$ branes crossing the second edge, and $n_{ki}$ branes crossing the third edge; see Figure \ref{eq:toricdiagram}.
 The $T_{ijk}$ theory exhibits $SU(n_{ij})\times SU(n_{jk})\times
SU(n_{ik})$ global symmetries which are gauged by the curves $C_{ij},C_{jk},C_{ik}$ respectively; one can then select the two largest numbers $n_1, n_2$ from the set $n_{ij}, n_{jk}, n_{ki}$ to determine the orbifold singularity $\mathbb C^3/\mathbb Z_{n_1} \times \mathbb Z_{n_2}$ engineering the local SCFT $T(n_1,n_2)$. One can use Pick's theorem count the number of internal points in the above toric diagram to determine the number of K\"ahler parameters (i.e. internal points of the toric diagram) associated to each $T_{ijk}$:\footnote{Notice that Pick's formula reduces to the usual degree-genus formula for a curve of genus $g$ and degree $d$ in $\mathbb P^2$ when $n_{ij} = n_{jk} = n_{ki} = n $. Specifically, we obtain
	\begin{align}
		N(T_{ijk}) = \frac{1}{2}(n-1)(n-2),
	\end{align}
where in the above formula $N = g$ and $n = d$.
}
	\begin{align}
	\begin{split}
		N(T_{ijk}) = N(T(n_1,n_2)) &=1+\frac{1}{2} \left( n_1 n_2   - n_{ij} - n_{jk} - n_{kl} \right)\\
		&=1+\frac{1}{2} \left( \frac{n_{ij} n_{jk} n_{kl}}{\text{min}(n_{ij}, n_{jk} , n_{kl} ) }  - n_{ij} - n_{jk} - n_{kl} \right).
	\end{split}
	\end{align}
The resulting 5d system can be depicted by a pentagon with diagonals drawn as in Figure \ref{fig:pentagon}.
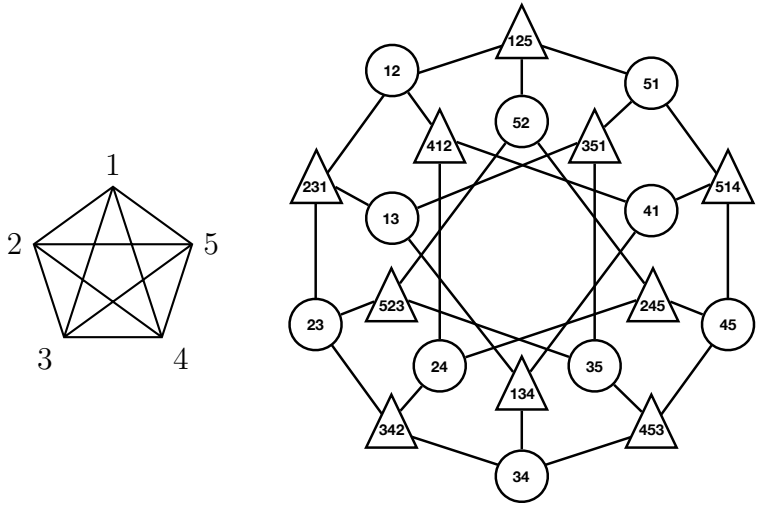
\begin{figure}[t]
	\centering
	\begin{tikzpicture}[thick, scale =1.3, baseline = -120]
		\draw (0,0) coordinate (1) -- ++(216:1) coordinate (2) -- ++(288:1) coordinate (3) -- ++(0:1) coordinate (4) -- ++(72:1) coordinate (5) --cycle;
		\draw[rounded corners=.7] (1)-- (3) --(5)--(2)--(4)--cycle;
		\node[anchor = south]  at (1) {$1$};
		\node[anchor = east]  at (2) {$2$};
		\node[anchor = north east]  at (3) {$3$};
		\node[anchor = north west]  at (4) {$4$};
		\node[anchor = west]  at (5) {$5$};
	\end{tikzpicture}
	\hspace{.5cm}
	\includegraphics[width=.4\linewidth]{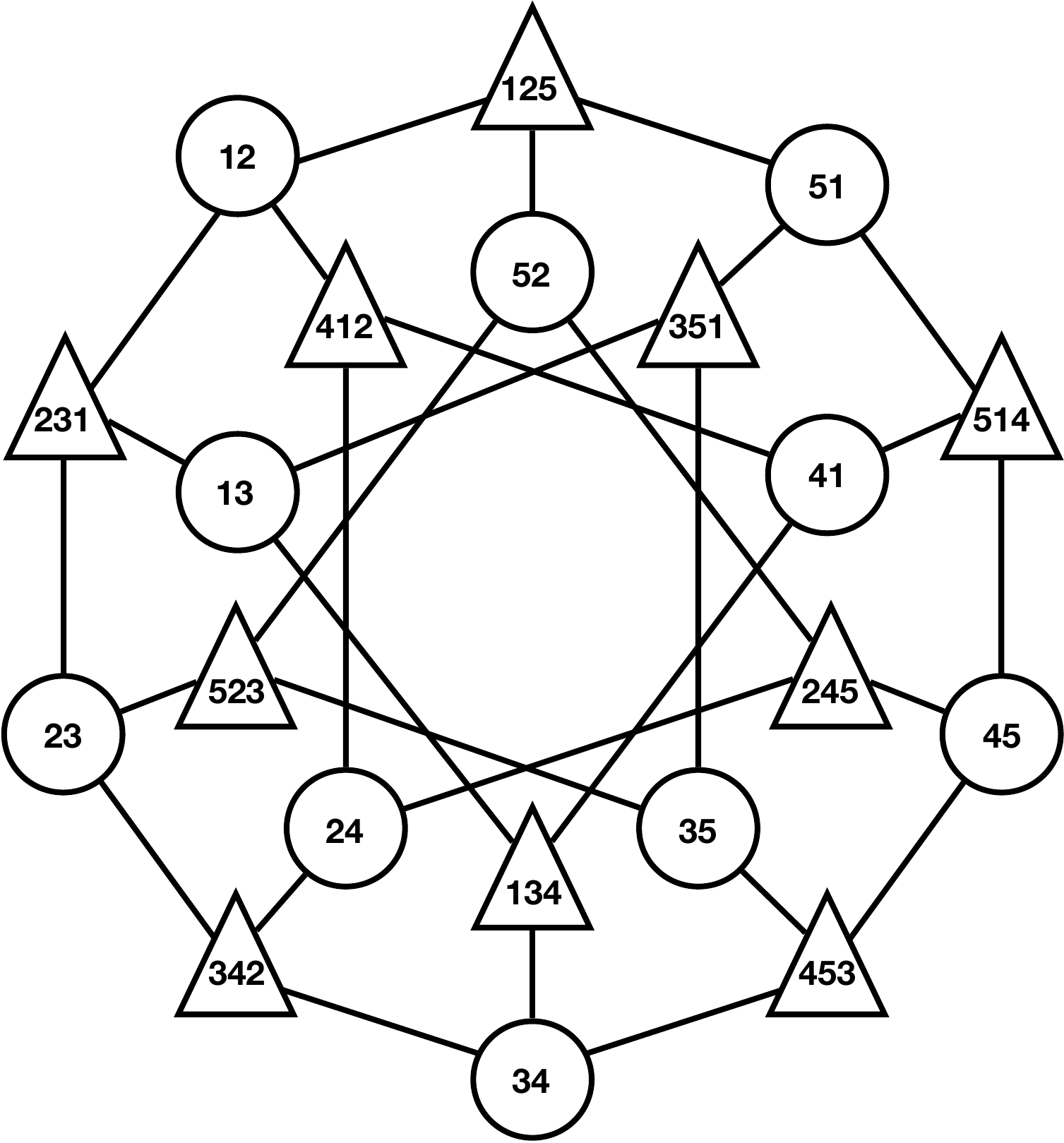}
	\caption{ A graphical representation of the putative 5d system describing the geometry $W$ defined by  \eqref{eq:CYdef}.
		Each triangle $(ijk)$ in the pentagon corresponds to a conformal system $T_{ijk}$ and each edge $(ij
		)$ correspond to $SU(n_{ij})$ gauging all the $T_{ijk}$ which share that edge.
		In the right graph each triangle corresponds to a conformal system $T_{ijk}$ and each circle corresponds to $SU(n_{ij})$ gauging.
	}
	\label{fig:pentagon}
\end{figure}

To check if this model captures all, we have to see if we can recover the counting of the massless degrees of freedom of M-theory on such manifolds correctly.
For all these examples the moduli space of complex structures is one dimensional (given by the deformation of the defining polynomial by the term $\prod_i z_i$).   Just as in the 6d case we expect this to deform the gluing of the 5d theories. However the BPS degeneracies of the 5d theories (which are captured by topological string amplitudes) do not depend on this. 

The other things to count would involve the K\"ahler moduli of the 3-fold and to see if we can recover this counting from the 5d model.  This would involve counting the $N(T_{ijk})$ Coulomb branch parameters of the 5d SCFTs $T_{ijk}$, the $n_{ij}-1$ gauge parameters of the curves $C_{ij}$, and the overall K\"ahler moduli of the weighted projective space which fixes the gauge coupling associated to the curves $C_{ij}$ of gauge symmetries (this K\"ahler modulus corresponds to the hyperplane class of the weighted projective space).
Note that although naively we have ten couplings corresponding to each $S(n_{ij})$, but geometrically it seems natural to impose relations between them.
More detailed analysis is performed in Appendix~\ref{sec:prepotential} for the mirror quintic.
If we admit the existence of the relation among the couplings, the total count is
\[N_{\text{K\"ahler}}=\sum_{ijk}N(T_{ijk}) +\sum_{ij} N(C_{ij})+1 \ .\]
Below we shall check that this indeed agrees with the count of the K\"ahler parameters of several examples of one parameter compact CY 3-folds \cite{Klemm:1992tx}.  For example as we shall discuss, for the mirror quintic we have $10$ theories all of $ T_5 = T(5,5)$ type (with $\mathbb{C}^3/\IZ_5\times \IZ_5$ local geometry) and they are glued via $10$ $SU(5)$ gauge symmetries. Since $N(C_{ij}) = 4$ and $N(T_5)=6$, we obtain
\[N_{\text{K\"ahler}}=101\]
which agrees with the number of K\"ahler parameters of the mirror quintic.
%
%

A curious property of the 5d model is its Higgs branch.
Consider a special point of the K\"ahler moduli where all exceptional divisors have been shrunk to points and the geometry is literally described by \eqref{eq:CYdef}.
At that special point of the K\"ahler moduli space, we expect an enlarged set of complex structure moduli, and hence a geometric transition, because we can perform complex deformations of singular locus $C_{ij}$ whose normal geometry is the ALE space $\mathbb{C}/\IZ_{n_{ij}}$.

After the geometric transition, the number of K\"ahler moduli should be $\tilde{N}_\text{K\"ahler} =1$, because all the K\"ahler parameters other than the total volume of the weighted projective space are frozen after the transition. Let us count the number of complex structure moduli $\tilde{N}_\text{complex}$ after the transition using the 5d model we have described.
This mode should correspond to the Higgs branch of the corresponding 5d theory, which opens when all the parameters associated to each conformal system $T_{ijk}$ are turned off.
The quartanionic dimension of the Higgs branch of this system, which should be equal to the number of complex structure deformations of the geometry after the geometric transition, is calculated by
\[ \tilde{N}_\text{complex} = \sum_{ijk} \mathrm{dim}_{\mathbb{H}} \text{Higgs}(T_{ijk}) - \sum_{ij} \mathrm{dim} SU(n_{ij})+N_\text{complex},  \]
where $N_\text{complex}=1$ is the number of the complex structure deformations which already existed before the geometric transition.
In the case of mirror quintic, the 5d SCFT $T_{ijk}$ is the $T_5$ system, for which the dimension of Higgs branch is shown in \cite{Gaiotto:2009gz,Benini:2009gi} to be
\[ \mathrm{dim}_\mathbb{H} \text{Higgs}(T_5) = 34.\]
Therefore, for the mirror quintic model, the dimension of complex structure moduli after the geometric transition is
\begin{equation}
	\tilde{N}_\text{complex} = 10\times 34 -10 \times 24 +1 =101,
\end{equation}
which suggests that the geometry after the geometric transition is the original quintic.

In \cite{Chiang:1995hi}, it was shown that the 7555 CY 3-folds realizable as hypersurfaces in projective space are related by geometric transitions. This result essentially relies on the fact that hypersurfaces in projective space are a special case of complete intersections in toric varieties. Given a pair of 3-folds admitting toric descriptions, one can use appropriate representations of the toric data to construct a third toric 3-fold interpolating between the two, thus demonstrating the existence of a geometric transition. Since both the quintic and the mirror quintic can be realized as such hypersurfaces, it follows immediately from this result that there exists a transition relating the two. What is described above is a field theoretic perspective on this type of transition which leans heavily on the fact that much of the local physics of our 5d model is encoded in the singularities of the mirror quintic. 

It is also possible to explicitly see the above geometric transition from the mirror quintic to the quintic in terms of algebraic equations. Let $y_i=z_i^5$ where $z_i, i=1, \cdots, 4$ are homogeneous coordinates of $\mathbb{P}^4$.  Then the $y_i$'s are invariant under the $\mathbb{Z}_5 \times \mathbb{Z}_5$ orbifold action of the mirror quintic. We also introduce a complex coordinate $\rho = \prod_i z_i$. Then the defining equation of the mirror quintic can then be described as 
\begin{align}
\sum_i y_i = 5\psi \rho, \label{mirrorquintic.eq1}
\end{align}
where $\psi$ is the complex structure modulus of the mirror quintic. Note that $y_i, i=1, \cdots, 5$ and $\rho$ are not independent of each other but are constrained by
\begin{align}
\prod_{i} y_i = \rho^5. \label{mirrorquintic.eq2}
\end{align} 
Since $z_i, i=1, \cdots, 4$ are homogeneous coordinates of $\mathbb{P}^4$, $(y_i, \rho)$ may be thought of homogeneous coordinates of $\mathbb{P}^5$ and hence the mirrror quintic can be also realized by the complete intersection given by \eqref{mirrorquintic.eq1} and \eqref{mirrorquintic.eq2} inside $\mathbb{P}^5$. 
%
%
From this viewpoint, the $C_{ij}$ correspond to $y_i=y_j=\rho=0$ and $C_{ijk}$ correspond to $y_i=y_j=y_k=\rho=0$. The geometry corresponds to a special region of the K$\ddot{\text{a}}$hler moduli space where all the exceptional divisors are collapsed to points. We can desingularize the geometry by adding degree five monomials of the coordinates $(y_i, \rho)$ to the equation \eqref{mirrorquintic.eq2}. Furthermore, since \eqref{mirrorquintic.eq1} is a linear equation we may erase one coordinate, for example $y_1$, and then the equation \eqref{mirrorquintic.eq2} after turning on deformations of degree five monomials becomes the defining equation for the quintic. Hence adding general degree five monomials to \eqref{mirrorquintic.eq2} corresponds to the Higgsing transition.

In each of the following examples, which are the mirrors of all hypersurface 3-folds with a single K\"ahler modulus, there is a single gauge coupling which fixes the (equal) volumes of the curves $C_{ij}$. Consequently, when counting $h^{1,1}$ we must in each case add to the K\"ahler moduli associated to the singularities of the corresponding 3-fold an additional K\"ahler modulus controlling the volume of $C_{ij}$. 

\subsubsection{$\mathbb P^4_{(1,1,1,1,1)}[5]/G$}
We have already described the quintic above but we repeat the results here for completeness. The mirror $W=\mathbb P^4_{(1,1,1,1,1)}[5]/G$ of the quintic has the following Hodge numbers:
	\begin{align}
		h^{1,1}(W) = 101,~~ h^{2,1}(W) = 1.
	\end{align}
The fixed curves $C_{ij}$ all have $n_{ij} = 5$ and thus carry $SU(5)$ symmetries. Hence
	\begin{align}
		\sum_{ij} N(C_{ij}) = 10 \cdot 4 = 40.
	\end{align}
The fixed points $T_{ijk}$ are all $T(5,5)$ theories, with $N(T(5,5)) = 6$, and therefore
	\begin{align}
		\sum_{ijk} N(T_{ijk})=  10 \cdot 6 = 60. 
	\end{align}
The total number of K\"ahler parameters is therefore 
	\begin{align}
		N_{\text{K\"ahler}}   = 101. 
	\end{align}
\subsubsection{$\mathbb P^4_{(2,1,1,1,1)}[6]/G$}
The 3-fold $M =\mathbb P^4_{(2,1,1,1,1)}[6]$ is defined by 
	\begin{align}
		x_1^3 + x_2^6 + x_3^6 + x_4^6 + x_5^6 = 0
	\end{align}
in $ \mathbb P^4_{(2,1,1,1,1)}$. The mirror $W$ has the following Hodge numbers:
	\begin{align}
		h^{1,1}(W) = 102,~~~ h^{2,1}(W) = 1.
	\end{align}
In this case, there are two types of gauge symmetries: four $SU(3)$ symmetries and six $SU(6)$ symmetries. For example, the curve $C_{12}$ has $n_{12} = 3$ and hence carries $SU(3)$ gauge symmetry, while $C_{23}$ carries $SU(6)$ gauge symmetry. Accounting for all singular curves, we have
	\begin{align}
		\sum_{ij} N(C_{ij}) = 4 \cdot 2 + 6 \cdot 5 = 38.
	\end{align}
There are also two types of fixed points $T_{ijk}$, namely six fixed points with associated conformal system $T(3,6)$, with $N(T(3,6)) = 4$ and four fixed points associated to $T(6,6)$ with $N(T(6,6)) = 10$. The total contribution of these fixed points is 
	\begin{align}
		\sum_{ijk} N(T_{ijk}) = 6 \cdot 4 + 4 \cdot 10 = 64,
	\end{align}
and thus the total number of K\"ahler parameters is 
	\begin{align}
		N_{\text{K\"ahler}}= 102.
	\end{align}

\subsubsection{$\mathbb P^4_{(4,1,1,1,1)}[8]/G$}
The 3-fold $W = \mathbb P^4_{(4,1,1,1,1)}[8]/G$ has the following Hodge numbers:
	\begin{align}
		h^{1,1}(W) = 149,~~~ h^{2,1}(W) = 1. 
	\end{align}
The fixed curves contribute
	\begin{align}
		\sum_{ij} N(C_{ij}) = 4 \cdot 1 + 6 \cdot 7=  46.
	\end{align}
There are two types of fixed points: six fixed points $T(2,8)$ and four fixed points $T(8,8)$. These fixed points contribute the following numbers of K\"ahler moduli:
	\begin{align}
		\sum_{ijk} N(T_{ijk}) = 	6 \cdot 3 + 4 \cdot 21 =  102.
	\end{align}
The total number of K\"ahler parameters is thus
	\begin{align}
		N_{\text{K\"ahler}} = 149. 
	\end{align}

\subsubsection{$\mathbb P^4_{(5,2,1,1,1)}[10]/G$}
	The 3-fold $W = \mathbb P^4_{(5,2,1,1,1)}[10]$ has Hodge numbers
		\begin{align}
			 h^{1,1}(W) = 145,~~~ h^{2,1}(W) = 1. 
		\end{align}
	The fixed curves contribute
		\begin{align}
			\sum_{ij} N(C_{ij}) = 3 \cdot 1 + 3 \cdot 4 + 3 \cdot 9 = 42.
		\end{align}
There are four types of fixed points, namely $T(2,5), T(2,10), T(5,10), T(10,10)$. These fixed points contribute the following numbers of K\"ahler moduli:
		\begin{align}
			\sum_{ijk} N(T_{ijk}) = 3 \cdot 2 + 3 \cdot 4 + 3 \cdot 16 + 1 \cdot 36 = 102.
		\end{align}
	The total number of K\"ahler parameters is thus
		\begin{align}
			N_{\text{K\"ahler}}  = 145. 
		\end{align}

\subsection{A generalization of the topological vertex}
\label{sec:gentopver}

We have seen in this section that to discuss the compact CY 3-folds from the perspective of local singularities, we need in addition to the usual topological vertex a way to describe gauging global symmetries of local models.  In particular, we need to gauge a diagonal subgroup $H = (H \times H \times H)_{\text{diag}}$ of the global symmetries $H \subset G_{i=1,2,3}$ of three local 5d SCFTs. So in this section we present a generalization of the topological vertex \cite{Aganagic:2003db} which will be able to 
describe a trivalent $SU(N)$ gauging for three toric geometries, based on \cite{Hayashi:2017jze}. This method also gives more geometric insights into the 5d models described above.

\begin{figure}
\centering
\includegraphics[width=7cm]{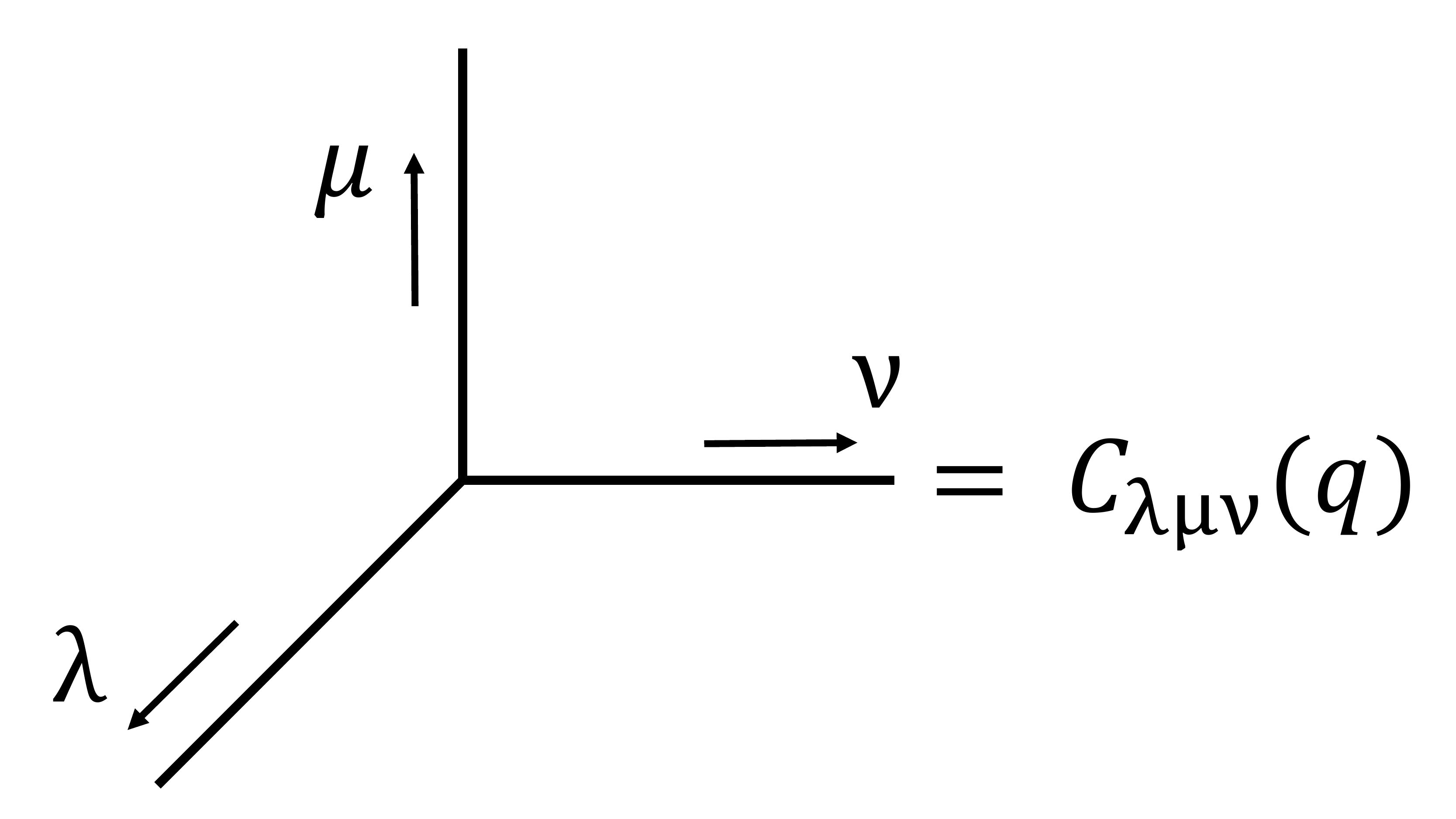}
\caption{An assignment of Young diagrams and arrows for a vertex.}
\label{fig:topvertex}
\end{figure}

\begin{figure}
\centering
\includegraphics[width=7cm]{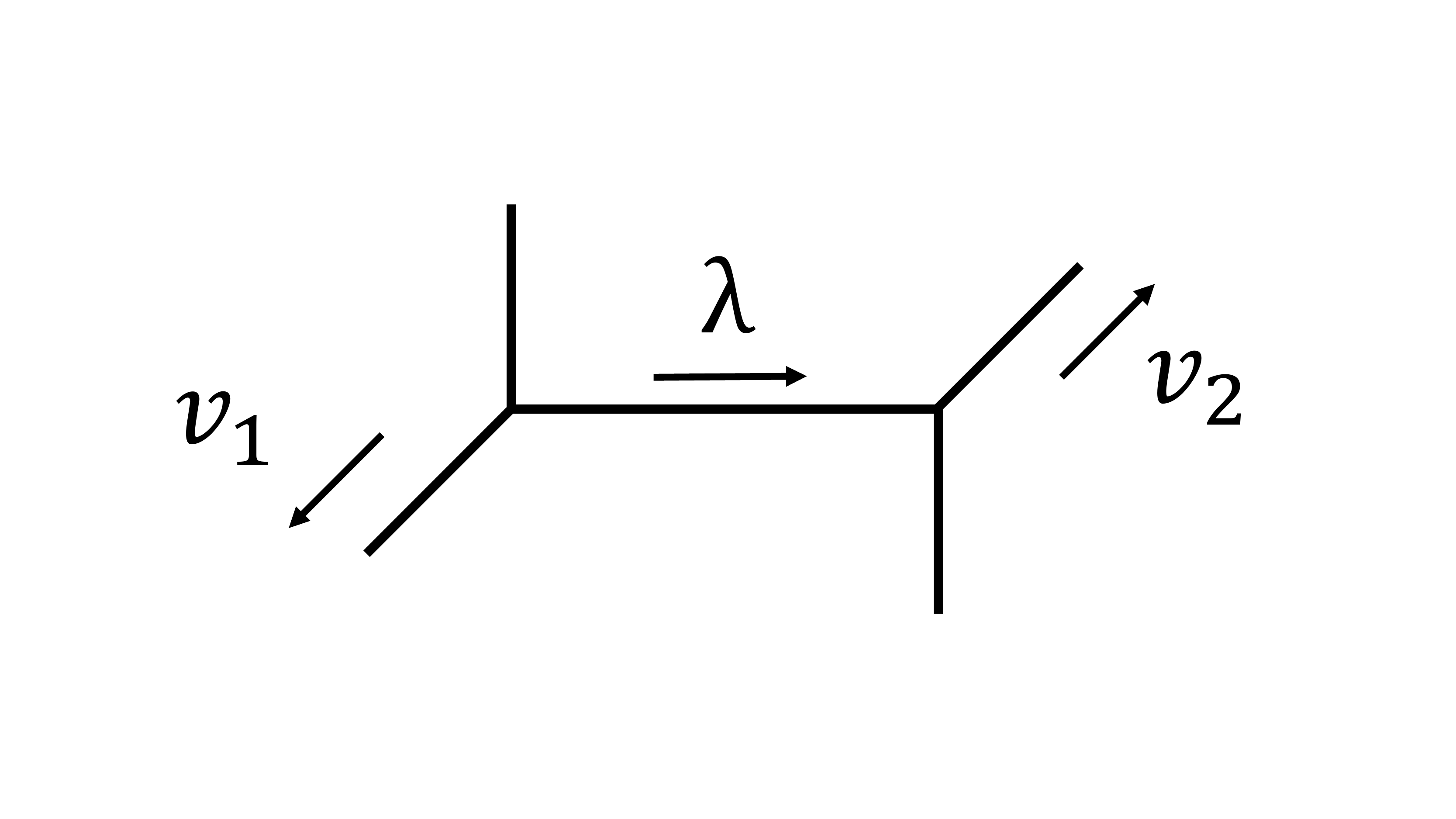}
\caption{An example of gluing along a leg with a Young diagram $\lambda$.}
\label{fig:framing}
\end{figure}

First note that since the local geometry of $T_{ijk}$ is toric 
it is possible to compute its all-genus topological string partition function using the standard topological vertex, which we call ``+'' vertices. We decompose the toric diagram into vertices with three legs where Young diagrams are assigned with arrows to each leg. The (standard) topological vertex is given by
\begin{align}\label{topvertex}
C_{\lambda\mu\nu}(y) = y^{-\frac{||\mu^t||^2}{2} + \frac{||\mu||^2}{2} + \frac{||\nu||^2}{2}}\tilde{Z}_{\nu}(y)\sum s_{\lambda^t/\eta}(y^{-\rho - \nu})s_{\mu/\eta}(y^{-\rho - \nu^t}),
\end{align}
for a vertex with Young diagrams $\lambda, \mu, \nu$ assigned in a clockwise manner and arrows oriented outward as in Figure \ref{fig:topvertex}. 
Here, we define $y = e^{-s}$ where $s$ is the topological string coupling and $||\mu||^2 = \sum_i\mu_i^2$ for a Young diagram $\mu$. $\tilde{Z}_{\nu}(y)$ is given by 
\begin{align}
\tilde{Z}_{\nu}(y) = \prod_{(i, j)\in \nu}\left(1 - y^{\nu_i -j + \nu_j^t - i +1}\right)^{-1}.
\end{align}
$s_{\mu/\nu}(x)$ is the skew Schur function and $-\rho =  i - \frac{1}{2}, i=1, 2, \cdots$. When we glue vertices along a leg with a Young diagram $\lambda$, we include a factor
\begin{align}
(-Q)^{|\lambda|}f_{\lambda}(y)^n,
\label{eq:framingfactor}
\end{align}
where $Q$ is a K\"ahler parameter for the gluing leg and $f_{\lambda}(q) $ is framing factor
\begin{align}
f_{\lambda}(y) = (-1)^{|\lambda|}y^{\frac{||\lambda^t||^2 - ||\lambda||^2}{2}}.
\end{align}
We also define $|\lambda|  = \sum_i\lambda_i$ and $n = \det(v_1, v_2)$ where $v_1, v_2$ are vectors as in Figure \ref{fig:framing}.
Note that we need to assign non-trivial Young diagrams for external legs of the $T_{ijk}$ diagrams since we further need to glue the external legs of $T_{ijk}$ with the external legs of other $T_{i'j'k'}$ diagrams. 

Although the partition function for each $T_{ijk}$ can be computed using the standard topological vertex, we also need a different type of vertex which glues three toric diagrams by connecting their external legs. 
Gluing together three collections of $N$ parallel external legs such that the external legs are compactified physically corresponds to an $SU(N)$ gauging, and it is because of this $SU(N)$ gauging that the geometry differs from the geometry of toric 3-folds. The procedure of a single $SU(N)$ gauging of three or four toric geometries is developed in \cite{Hayashi:2017jze} and we summarize the rule below.

\begin{figure}[t]
	\centering
	\includegraphics[width=.4\linewidth]{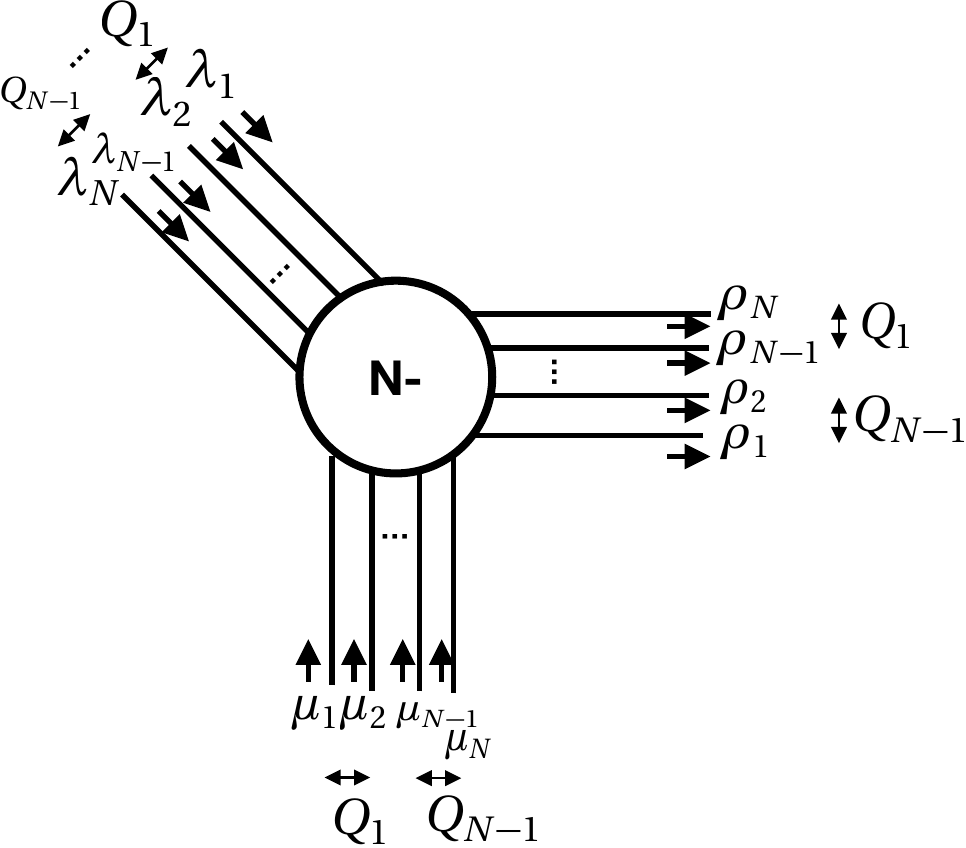}
	\caption{The $N^-$ vertex. The vertex have $3N$ legs, with a Young diagram, $\lambda_a$, $\mu_a$, or $\rho_a$ in the picture, assigned to each leg. The Young diagrams $\lambda_a$ and $\mu_a$ are assinged inwords, but $\rho_a$ are assinged outwards. This is just to reproduce q-YM result when $N=1$ as in \eqref{1minus}. The contribution from the vertex is \eqref{eq:Nminus} and it is proportional to a product of delta functions which imposes $\lambda_a, \mu_a, \rho_a$ to be the same for each $a$. 
	The angles of the legs have no meaning other than that there are three sets of legs. The K\"ahler parameters parametrizing the distances between the parallel legs in one direction should be identified the distances between the parallel legs in other directions as indicated in the figure.}
	\label{fig:Nminus}
\end{figure}

We consider a rule for a local CY 3-fold which is constructed by a single trivalent $SU(N)$ gauging. In order to formulate the prescription, it may be useful to think of a trivalent $SU(N)$ gauging with a Chern-Simons level $k$ as a new type of vertex, which we call an ``$N^-_k$'' vertex\footnotemark\footnotetext{This gauging is more like a $U(N)$ gauging with the Coulomb branch parameter for the overall $U(1)$ turned off. In this sense, we can also consider a ``$1^-_k$'' vertex, i.e. an $N^-_k$ vertex with $N=1$ and $k=0, 1$.}.
The $N^-_k$ vertex has three sets of $N$ tuples of legs as depicted in Figure~\ref{fig:Nminus}.
We assign a Young diagram for each leg and also arrows where two of them are in an inward direction and one is in an outward direction.
The contribution to the $N^-_k$ vertex is 
\begin{equation}
	C^{(N^-_k)}_{\vec\lambda,\vec\mu,\vec\rho}(\vec{Q}; y)= \frac{\prod_{a=1}^Nf_{\lambda_a}(y)^{-l(N,k,a)} \delta_{\lambda_a,\mu_a, \rho_a}} {Z^\text{half vector}_{SU(N),\vec{\lambda}}(\vec{Q}; y)},
	\label{eq:Nminus}
\end{equation}
where $\vec\lambda,\vec\mu,\vec\rho$ mean the sets of Young diagrams $\{\lambda_a\}, \{\mu_a\}, \{\nu_a\}$ where $a=1, \cdots, N$. $\delta_{\lambda,\mu,\rho}$ is $1$ when all the three Young diagrams are the same and $0$ otherwise.
The power $l(N,k,a)\in \mathbb{Z}$ of the framing factor $f_{\lambda_a}$ is the effective level of the corresponding effective $U(1)$ gauge field coming from the $SU(N)_k$ gauging. This number depends on the chamber of the enlarged K\"ahler moduli space, and therefore it jumps when a flop transition occurs.
Although the explicit form of $l(N,k,a)$ will not be important in this paper, it can be determined by writing down a diagram describing the geometry around the $N_k^-$ vertex. In Appendix~\ref{sec:SU5gauging}, the numbers $l(N,k,a)$ are determined for the case of the mirror quintic in a chamber. 
The denominator factor $Z^\text{half vector}_{SU(N), \vec\lambda}$ is roughly a ``half'' of the contribution of the $SU(N)$ vector multiplet to the Nekrasov partition function, which can be computed using the toric diagram in Figure \ref{fig:Zhalf}. The explicit expression is given by
\begin{figure}[t]
	\centering
	\begin{tikzpicture}[scale=.6, thick]
		\draw (0,0) coordinate (0) -- ++(0,-1) coordinate (1) -- ++(-1,-1) coordinate (2) -- ++ (-1,-.5) coordinate (3);
		\draw[dotted] (3) ++ (-.5,-.25) -- ++ (-1,-1) ++ (-.5,-.25) coordinate (4);
		\draw (4) --++ (-1.5,-.5) coordinate (5) --++(-4,-1) coordinate (6) -- ++(-1.25,-.25) coordinate(7);
		\draw (1) --++(1,0) coordinate (1a);
		\draw (2) --++(2,0) coordinate (2a);
		\draw (5) --++(6.5,0) coordinate (5a);
		\draw (6) --++(10.5,0) coordinate (6a);
		\draw[dotted] (2a) ++ (.5,-.7) --++(0,-1);
		\node[anchor=south] at (0) {$\varnothing$};
		\node[anchor=west] at (1a) {$\lambda_1$};
		\node[anchor=west] at (2a) {$\lambda_2$};
		\node[anchor=west] at (5a) {$\lambda_{N-1}$};
		\node[anchor=west] at (6a) {$\lambda_{N}$};
		\node[anchor=east] at (7) {$\varnothing$};
		\draw[<->] (1a) ++ (1,0) coordinate (1b) -- ++(0,-1) coordinate (2b);
		\draw[<->] (5a) ++ (2,0) coordinate (5b) -- ++(0,-1) coordinate (6b);
		\node[anchor=west] at ($(1b)!.5!(2b)$) {$Q_{1}$};
		\node[anchor=west] at ($(5b)!.5!(6b)$) {$Q_{N-1}$};
		\draw[<-] ($(1a)+(0,.25)$) -- ++(-.5,0); 
		\draw[<-] ($(2a)+(0,.25)$) -- ++(-.5,0); 
		\draw[<-] ($(5a)+(0,.25)$) -- ++(-.5,0); 
		\draw[<-] ($(6a)+(0,.25)$) -- ++(-.5,0); 
	\end{tikzpicture}
	\caption{The diagram that defines $Z^\text{half vector}_{SU(N),\vec{\lambda}}(\vec{Q}; y)$. Each vertex represents the usual unrefined topological vertex, which is also called the ``+'' vertex here.}
	\label{fig:Zhalf}
\end{figure}
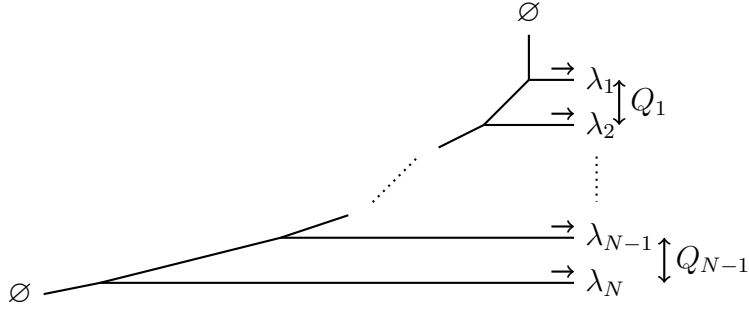
\begin{align}\label{eq:Zhalf}
	Z^\text{half vector}_{SU(N), \vec\lambda}(\vec{Q}; y) &= y^{\frac{1}{2}\sum_{a=1}^N||\lambda_a||^2}\prod_{a=1}^N\tilde{Z}_{\lambda_a}(y)\cr
	&\prod_{1 \leq a < b \leq N-1}\prod_{i,j=1}^{\infty}\left(1 - Q_aQ_{a+1}\cdots Q_{b}y^{i+j-\lambda_{a,i} - \lambda_{b+1,j}^t-1}\right)^{-1}. 
\end{align}
When the arrows for the all the three Young diagrams are in an inward direction, the $N^-_k$ vertex becomes 
\begin{equation}
	\mathcal{C}^{(N^-_k)}_{\vec\lambda,\vec\mu,\vec\rho}(\vec{Q};y)= \frac{\prod_{a=1}^N(-1)^{|\lambda_a|}f_{\lambda_a}(y)^{-l(N,k,a)+1} \delta_{\lambda_a,\mu_a, \rho_a}} {Z^\text{half vector}_{SU(N),\vec{\lambda}}(\vec{Q}; y)},
	\label{eq:Nminus1}
\end{equation}
due to the identity 
\begin{align}
C_{\lambda\mu\nu}(y) = (-1)^{|\lambda| + |\mu| + |\nu|}f_{\lambda}^{-1}(y)f_{\mu}^{-1}(y)f_{\nu}^{-1}(y)C_{\mu^t\lambda^t\nu^t}(y).
\end{align}

For the $N^-_k$ vertex, we also assign the following weight of K\"ahler parameters
\begin{align}
\prod_{a=1}^N(-Q_{B_a})^{|\lambda_a|},
\label{eq:weight}
\end{align}
where 
$Q_{B_a}$ are the effective couplings of $U(1)$ gauge fields which arise on the Coulomb branch of $SU(N)$.
Note that these effective couplings are not independent---only one among $Q_{B_a}, a=1, \cdots, N-1$ is an independent parameter and it is related to the gluing parameter (i.e. the gauge coupling) of the $SU(N)$.
The precise relation among the $Q_{B_a}$ can also be read off from the geometry around the $N^-_k$ vertex and the explicit relation can be found in Appendix~\ref{sec:SU5gauging} for the $SU(5)$ gauging used in the computation for the mirror quintic.
Note that the Nekrasov partition function for the pure $SU(N)$ gauge theory with the $k$ Chern-Simons level is 
\begin{align}
	Z_{SU(N)_k}(u_\text{instanton}, \vec{Q}; y) = \sum_{\vec\lambda}\left[\prod_{a=1}^N(-Q_{B_a})^{|\lambda_a|}(-1)^{|\lambda_a|}f_{\lambda_a}(y)^{N-k+2-2a}\right]Z^\text{half vector}_{SU(N), \vec\lambda}(\vec{Q}; y) ^2,
\end{align}
when we identify $s = \epsilon_1 = -\epsilon_2$. In the above expression, the instanton fugacity is given by 
\begin{align}
	u_\text{instanton} = Q_{B_1}\prod_{a=1}^{N-1}Q_a^{-2N+2k+2a\frac{N-k}{N}}.
\end{align}
Therefore, if we glue three copies of the diagram of Figure \ref{fig:Zhalf} by the $N_k^-$ vertex, we obtain the Nekrasov partition function $Z_{SU(N)_k}(u,\vec{Q};y)$ of the pure $SU(N)$ gauge theory with Chern-Simons level $k$. This is the motivation for the definition \eqref{eq:Nminus}.

\begin{figure}
	\begin{center}
		$
		\begin{tikzpicture}
		\node[label={above:$1^+$}](A) at (0,0) {$
			\begin{tikzpicture}
				\draw[ultra thick] (0,0) -- (1,0);
				\draw[ultra thick] (0,0) -- (-0.5, 0.866025 );
				\draw[blue] (-0.240192, 1.01603) to [bend left=10] (0,0);
				\draw[blue] (1,.3) to [bend right=10] (0,0);
				\draw[blue] (-0.759808, 0.716025) to [bend right=10] (0,0);
				\draw[blue] (-0.759808, -0.716025) to [bend left=10] (0,0);
				\draw[ultra thick] (0,0) -- (-0.5,-0.866025 );
				\draw[blue] (-0.240192, -1.01603) to [bend right =10] (0,0);
				\draw[blue] (1,-.3) to [bend left =10] (0,0);
				\node(au) at (1.1,.4) {};
				\node(ad) at (1.1,-.4) {};
				\draw[rotate=0,blue] (1,0) ellipse (.15*1 and .15*2);
				\draw[rotate=120,blue] (1,0) ellipse (.15*1 and .15*2);
				\draw[rotate=240,blue] (1,0) ellipse (.15*1 and .15*2);
			\end{tikzpicture}
			$};
		\node[label={above:$1^-$}](B) at (5,0){$
		\begin{tikzpicture}
				\draw[ultra thick] (0,0) -- (1,0);
				\draw[ultra thick] (0,0) -- (-0.5, 0.866025 );
				\draw[blue] (-0.240192, 1.01603) to [bend right] (1,.3);
				\draw[blue] (-0.759808, 0.716025) to [bend left] (-0.759808, -0.716025);
				\draw[ultra thick] (0,0) -- (-0.5,-0.866025 );
				\draw[blue] (-0.240192, -1.01603) to [bend left] (1,-.3);
				\node(au) at (1.1,.4) {};
				\node(ad) at (1.1,-.4) {};
				\draw[rotate=0,blue] (1,0) ellipse (.15*1 and .15*2);
				\draw[rotate=120,blue] (1,0) ellipse (.15*1 and .15*2);
				\draw[rotate=240,blue] (1,0) ellipse (.15*1 and .15*2);
			\end{tikzpicture}
			$};
		\draw[big arrow] (A) -- (B);
		\draw[big arrow] (B) -- node[above,midway]{mirror} (A);
		\end{tikzpicture}
		$	
	\end{center}
	\caption{Diagrams of $1^{\pm{}}$ topological vertices. The $1^+$ vertex is the usual (unrefined) topological vertex, while the $1^-$ vertex, which is responsible for $SU(1)$ gauging, is a three-punctured sphere. These two vertices are exchanged by mirror symmetry.}
	\label{fig:mirver}
\end{figure}
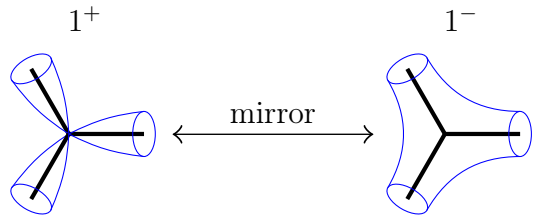

In the simplest case, $N=1$, the vertex \eqref{eq:Nminus} reduces to\footnote{Since the meaning of $l(N,k,a)$ in \eqref{eq:Nminus} is the effective level of an effective $U(1)$ coming from $SU(N)$ gauging, when $N=1$, $l(1,k,1)$ should be set to be $k$.}
\begin{equation}
	C^{(1^-_k)}_{\lambda,\mu,\rho}(y)= \frac{f_{\lambda}(y)^{-k} \delta_{\lambda,\mu,\rho}} {y^{\frac{1}{2}||\lambda||^2}\tilde{Z}_{\lambda}(y)}.\label{1minus}
\end{equation}
The above expression is the $1^-_k$ vertex, and has already been discussed in \cite{Aganagic:2004js} as a building block for computing the topological string partition function for a local 3-fold which can be described as the total space of the rank two bundle 
\begin{align}
\mathcal{L}_1 \oplus \mathcal{L}_2 \to \Sigma_g, \label{L1L2Sigmag}
\end{align}
where $\mathcal{L}_{1, 2}$ are line bundles with degrees $2g-2+p$ and $-p$ respectively and $\Sigma_g$ is a genus $g$ Riemann surface (see also \cite{Bryan:2004iq}).  The topological string partition function for the 3-fold given by \eqref{L1L2Sigmag} can be computed by decomposing the genus $g$ Riemann surface $\Sigma_g$ into caps, annuli and pants. We can define two pants contributions denoted by $P^{(0, 1)}$ and $P^{(1, 0)}$ where the superscript denotes the degrees $(d_1,d_2)$ of the two line bundles. Then the contribution for the two types of pants gives rise to the factor \eqref{1minus} for $k=0$ or $1$ corresponding to the $1^-_k$ vertex.   As was noted in \cite{Aganagic:2004js} this vertex may be thought of as a mirror to the standard topological vertex, because a pair of pants in $\Sigma_g$ is the mirror of a vertex in a toric diagram for a toric 3-fold.   In other words, the $1^-$ vertex may be thought of as the mirror of $1^+$ vertex---see Figure \ref{fig:mirver}.

At this stage, we can make a connection with the Strominger-Yau-Zaslow (SYZ) picture of a CY 3-fold as a $T^3$-fibration \cite{Strominger:1996it, MR1876066, MR1876067, MR1821145, MR1876075, MR1959057, MR2015547, MR2681793}. 
In the references, it is conjectured that there is a topological $T^3$-fibration structure of a 3-fold, and the discriminant locus of the $T^3$ fibration of the 3-fold forms a trivalent graph $\Gamma$, equipped with two types of vertices: one is called the ``positive vertex'' and the other is called the ``negative vertex''. Under the mirror symmetry, those two types of vertices are exchanged. 
For example, the graph $\Gamma$ for the quintic is as follows:
$\Gamma$ contains 10 copies of $T_5$ graphs but with only negative vertices. These 10 copies $T_5$ are glued by positive vertices in the manner depicted in Figure \ref{fig:pentagon}.
Applying mirror symmetry, we conclude that we can regard Figure \ref{fig:pentagon} as the discriminant locus graph $\Gamma$ of the mirror quintic, interpreting each triangle as a $T_5$ graph with positive vertices, and each circle as a set of five negative vertices.

Therefore, the vertex structure of the SYZ picture is compatible with what we have found here. Namely, we would like to relate the ``positive'' vertex to the usual topological vertex, or ``$1^+$'' vertex, and the ``negative'' vertex to the $1^-$ vertex. The graph $\Gamma$ only knows topological information, but as we have seen the $1^\pm$ vertices do not seem to give the full answer for the topological string partition function for a local 3-fold constructed involving  trivalent gaugings, and thus the structure must be enlarged by including the additional $N^-_k$ vertex. 
However, we will see in the next subsection that the $N^-_k$ is not still enough to capture the full topological string partition function for the compact 3-folds constructed in Section \ref{sec:singularloci}. 

As an additional remark, note that it is natural to expect that there should also exist an ``$N^+_k$'' vertex which is exchanged with the $N^-_k$ vertex by mirror symmetry:
\begin{equation}
	N^+_k \stackrel{\text{mirror}}{\Longleftrightarrow} N^-_k.
\end{equation}
The $N^+_k$ vertices should in principle be related to the $T_N$ models in the most singular limit of K\"ahler moduli, which would be interesting to develop.

\subsection{Towards the topological string partition function for the mirror quintic}
\label{sec:topmirrorquintic}

In the previous subsection, we defined the $N^-_k$ vertex \eqref{eq:Nminus} or \eqref{eq:Nminus1} in addition to the standard $1^+$ vertex \eqref{topvertex} for the computation of the topological string partition function for a local CY 3-fold given by a trivalent gauging of SCFTs. In Section \ref{sec:singularloci}, on the other hand, we saw that the mirror quintic can be thought as a collection of a trivalent $SU(5)_0$ gaugings, each involving three $T_5$ theories.  Therefore, one would expect that a straightforward application of the $5^-_0$ vertex to each $SU(5)_0$ gauging would be able to reproduce the full topological string partition function for the mirror quintic. 
However, it turns out that naive application of the $5^-_0$ vertices is not enough to describe the mirror quintic and that an additional modification is required.
Although we could not compute the full partition function for the mirror quintic, we nevertheless propose a concrete formalism which can compute a large part of the full partition function but which is missing some information localized to the five corners of thse pentagon in Figure \ref{fig:pentagon}. We carry out this proposal by modifying the computation illustrated in the previous subsection using insights from our geometric picture of the resolved mirror quintic. We propose some higher genus GV invariants using this formalism below.

\begin{figure}
	\centering
	\begin{tikzpicture}[scale=1, ultra thick, baseline = -65]
		\draw (0,0) coordinate (0) -- ++(0,-1/2) coordinate (1) -- ++(-1/2,-1/2) coordinate (2) -- ++(0,-1/2) coordinate (3) -- ++(-1/3,-1/3) coordinate (4) -- ++(0,-1/3) coordinate (5) -- ++(1/3,-1/3) coordinate (6) -- ++(0,-1/2) coordinate (7) -- ++(1/2,-1/2) coordinate (8) -- ++(0,-1/3) coordinate(9) -- ++(1/3,-1/3) coordinate (10)--++(2/3,-1/3) coordinate(11) --++(1/2,-1/2) coordinate(12);
		\draw (0) ++ (2/3,0) coordinate (a) -- ++(1/2,-1/2) coordinate (b) --++(2/3,-1/3) coordinate (c) --++(1/3,-1/3) coordinate(d) --++(0,-1/3) coordinate(e) --++(1/2,-1/2) coordinate(f) -- ++ (0,-1/2) coordinate(g) --++(1/3,-1/3) coordinate(h) --++(0,-1/3) coordinate(i)--++(-1/3,-1/3) coordinate(j)--++(0,-1/2) coordinate(k)--++(-1/2,-1/2) coordinate(l)--++(0,-1/2) coordinate(m);
		\foreach \x/\y in {1/b,3/e,6/g,8/j,11/l}
		{\draw (\x) --(\y);}
		\foreach \x in {2,4,5,7,9,10}
		{\draw (\x) -- ++(-1,0) coordinate (\x l);}
		\foreach \x in {c,d,f,h,i,k}
		{\draw (\x) -- ++(1,0) coordinate (\x r);}
		\foreach \x/\y/\z in {2/d/1,5/f/2,7/i/3,10/k/4}
		{\node at ($(\x)!.5!(\y)$) {$S_\z$};}
		\node[anchor=south] at ($(1)!.4!(b)$) {$S_0$};
		\node[anchor=north] at ($(11)!.7!(l)$) {$S_5$};
	\end{tikzpicture}
	\caption{The diagram that describes the geometry around a triavalent $SU(5)$ gauging of three $T_5$ theories. Each of the four surfaces $S_1$, $S_2$, $S_3$ and $S_4$ 
	 are glued together along a $\mathbb{P}^1$. The K$\ddot{\text{a}}$hler parameter for the surfaces corresponds to a Coulomb branch parameter of the $SU(5)$. Furthermore, $S_1$ and $S_4$ are glued to additional surfaces $S_0$ and $S_5$ respectively. Note that the external edges of this diagram cannot be extended to be infinitely long, which indicates that the geometry makes sense only as a part of a larger geomery.}
	\label{fig:SU5geom}
\end{figure}
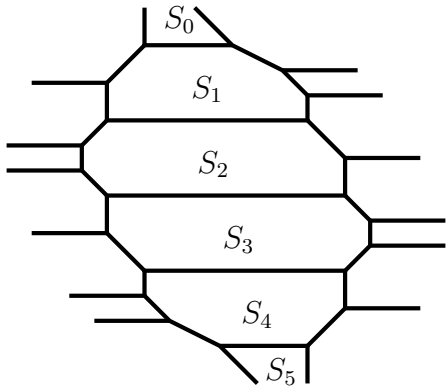

To see the problem we need to understand the geometric structure of the mirror quintic in more detail. We first explain more clearly the geometry around a trivalent $SU(5)$ gauging in Figure~\ref{eq:toricdiagram}, which corresponds to each edge of the pentagon in Figure \ref{fig:pentagon}; more details about the local geometry of the $SU(5)$ gauging can be found in Appendix \ref{sec:mirrorquintic}. 
While we cannot write a single toric diagram (i.e. a planar diagram consisting only of $1^+$ vertices) including a trivalent $SU(5)$ gauging and three $T_5$'s, a complex structure deformation of the local geometry of the $SU(5)$ gauging can be represented by the diagram in Figure \ref{fig:SU5geom}. We explain below how to apply a modification of the topological vertex formalism to this geometry, which is a proxy for the actual local geometry of the $SU(5)$ gauging.

Let us describe the proxy diagram in Figure \ref{fig:SU5geom} in more detail, and in particular why it makes sense to regard this diagram as a (modified) $SU(5)$ gauging. Since the $T_5$ theory may be described in an appropriate limit as a linear quiver gauge theory $[SU(5)] - SU(4) - SU(3) - SU(2) - [2]$ \cite{Bergman:2014kza, Hayashi:2014hfa}, the local theory around one trivalent gauging of three $T_5$ theories is an $SU(5)$ gauge theory with $4 \times 3 = 12$ flavors, for which the twelve external horizontal lines in Figure \ref{fig:SU5geom} imply the twelve flavors and the five internal horizontal lines yield the $SU(5)$ vector multiplets. 
The five internal horizontal lines are identified with the five external legs of a pair of $T_5$'s glued together, and the twelve external horizontal lines are identified with the $4\times 3$ internal lines of each $T_5$, parallel and next to the glued external lines.
Geometrically, the compact four faces bounded by the internal lines in Figure \ref{fig:SU5geom} correspond to 
compact complex surfaces. More concretely, $S_1, S_2, S_3, S_4$ yield four pseudo-del Pezzo surfaces PdP$_4$ which are related to four pseudo-del Pezzo surfaces PdP$_4$' of other types, which we expect to be in the mirror quintic, by certain complex structure deformations. 
Note that the diagram in Figure \ref{fig:SU5geom} is incomplete in the sense that the two lines in the upper part or similarly in the lower part will meet each other. Therefore, the diagram should be thought of as a local piece of the full mirror quintic.

\begin{figure}[t]
	\centering
	\includegraphics[width=.6\linewidth]{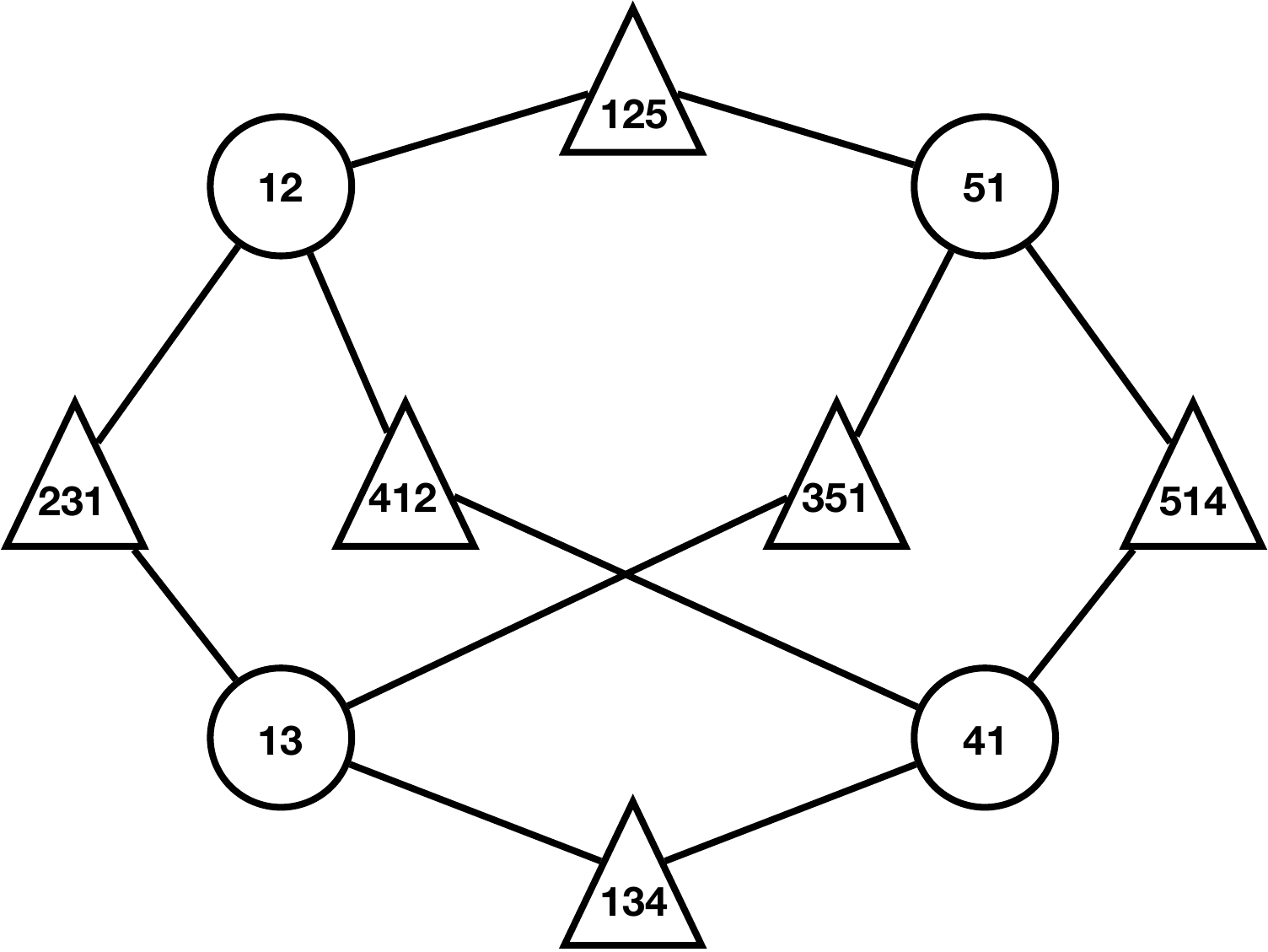}
	\caption{A graphical expression of the local structure around the corner $1$ in Figure \ref{fig:pentagon}. The triangles are $1^+$ vertices and the circles represent the $1^-_{-2}$ vertices. 
	While we believe that this Figure describes the geometry of the $T^3$ fibration around the corner, it will turn out that naively applying the $1^\pm$ vertices does not give a correct answer.}
	\label{fig:corner}
\end{figure}

There is another local geometry yet to be identified which affects this analysis. First note that the external lines going in the upper directions or in the lower directions should also be identified with a part of external lines of the $T_5$ theories. 
It follows that those lines are a part of other $SU(5)$ gauging lines. Let us then concentrate on local structure around $S_0$ in Figure \ref{fig:SU5geom}, which corresponds to focusing on one corner of the pentagon diagram in Figure \ref{fig:pentagon}. Suppose we focus on the corner $1$ in Figure \ref{fig:pentagon}. 
From the $SU(5)$ gauging procedure, the geometry around the corner would be described by the combinations of $1^+$ and $1^-_{-2}$ vertices given in Figure \ref{fig:corner}. The effective level $k=-2$ can be read off from the framing factor for the top horizontal internal line in Figure \ref{fig:SU5geom}.
Since we are looking at one gauging line out of five gauge lines of an $SU(5)$ gauging, the local picture in Figure \ref{fig:corner} is given by $U(1)$ gaugings where the gauging lines are represented by the circles. 
In fact, a combination of the $U(1)$ gauging lines makes the region $S_0$ bounded and hence yields another surface. For example, starting from the gauging line denoted by the circle with $12$, one can go back to the original gauging line by a combination of other gauging lines. Therefore, the region $S_0$ in Figure \ref{fig:SU5geom} should be a compact face, implying the existence of a compact complex surface which is schematically depicted in Figure \ref{fig:corner}. This explains a compactification at the corner of the pentagon by the four gauging lines originated from the four trivalent $SU(5)$ gaugings at one corner.

In order to understand the geometry of the mirror quintic, we need to identify the local geometry $S_0$ represented in Figure \ref{fig:corner}. We argue that the $S_0$ (and similarly $S_5$) in Figure \ref{fig:SU5geom} is given by a $\mathbb{P}^2$ which essentially comes from the $\mathbb{P}^2$ given by $y_1 = \rho = 0$ along with \eqref{mirrorquintic.eq1}, namely $y_2 + y_3 + y_4 + y_5 = 0$, inside $\mathbb{P}^5$. The four $U(1)$ gauging lines originate from the four lines in $\mathbb{P}^2$ on which there are $A_4$ singularities before the resolution. Furthermore, the fact that the four gauging lines are inside a $\mathbb{P}^2$ may constrain the length of the gauging lines. 
A careful identification of the complex surfaces comprising the geometry of the compact ``faces'' of a trivalent $SU(5)$ gauging is presented in Appendix \ref{sec:mirrorquintic}. 
The identification of the geometry around each corner of the pentagon with a $\mathbb{P}^2$ completes a geometric picture of the mirror quintic. That is to say, the mirror quintic may be decomposed into a gluing of a collection of local 3-folds, which includes ten $T_5$ geometries, 
ten $SU(5)$ gaugings each of which is given by the diagram in Figure \ref{fig:SU5geom} 
and five local $\mathbb{P}^2$'s which correspond to each corner of the pentagon in Figure \ref{fig:pentagon}. 

A complication affecting this relatively simple picture is that straightforward application of $1^+$ vertices and the $1^-_{-2}$ vertices to the diagram in Figure \ref{fig:corner} does not yield the topological string partition function for a local $\mathbb{P}^2$. Since the power of the framing factor associated to each $U(1)$ gauging is $-2$, a loop with three $U(1)$ gaugings and three $1^+$ vertices in Figure \ref{fig:corner} indeed gives a local $\mathbb{P}^2$. This also means that the compact curve coming from the $U(1)$ gauging is a rational curve with self-intersection $-3$ in $\mathbb{P}^2$. So the class of this curve can naturally be identified with the hyperplane (curve) class in the $\mathbb{P}^2$ of $S_0$. However, the diagram in Figure \ref{fig:corner} is not given by a single loop but by several loops. In particular the number of the gauging lines is four and hence the genus zero Gopakumar-Vafa (GV) invariant of the lowest order becomes $4$, instead of $3$ which is the correct GV invariant for a local $\mathbb{P}^2$. Therefore the method in Section \ref{sec:gentopver} applied to the gluing picture in Figure \ref{fig:corner} does not give the topological string partition function for a local $\mathbb{P}^2$. 
The naive GV invariants computed by applying the methods in Section \ref{sec:gentopver} to five copies\footnote{Five copies are necessary because there are five copies of the geometry in Figure \ref{fig:corner}, each associated to a different corner of the pentagon in Figure \ref{fig:pentagon}.} of the picture in Figure \ref{fig:corner} are summarized in Table \ref{tb:putativeGV}. Although one-fifth of the numbers in Table \ref{tb:putativeGV} do not agree with the GV invariants of local $\mathbb{P}^2$, the vanishing structure is the same as that of local $\mathbb{P}^2$.
\begin{table}[t]
\centering
\begin{tabular}{c|c|c|c|c|c|c|c|c}
	\diagbox[height=.8cm,width=1.3cm]{$d$~}{$g$~} & $0$ & $1$ & $2$ & $3$ & $4$ & $5$ & $6$ & $7$\\
\hline
$1$ & $20$ & $0$ & $0$ & $0$ & $0$ & $0$ & $0$ & $0$\\
\hline 
$2$ & $-50$ & $0$ & $0$ & $0$ & $0$ & $0$ & $0$ & $0$\\
\hline
$3$ & $280$ & $-100$ & $0$ & $0$ & $0$ & $0$ & $0$ & $0$\\
\hline
$4$ & $-2410$ & $2785$ & $-1200$ & $175$ & $0$ & $0$ & $0$ & $0$\\
\hline
$5$ & $25540$ & $-63780$ & $75160$ & $-49740$ & $18560$ & $-3600$ & $280$ & $0$\\
\hline
\end{tabular}
\caption{Naive GV invariants obtained by applying the method in Section \ref{sec:gentopver} to five copies of the diagram in Figure \ref{fig:corner}. $g$ is a genus and $d$ is a degree of a curve class of the gauging line. Since the K$\ddot{\text{a}}$hler parameters of the gauging lines are all the same due the geometric constraint explained in Appendix \ref{sec:prepotential}, the GV invariants are labeled by the single degree $d$. Although $\frac{1}{5}$ of the numbers do not agree with the GV invariants of a local $\mathbb{P}^2$, the vanishing structure agree with that of a local $\mathbb{P}^2$. }
\label{tb:putativeGV}
\end{table}

Note that a similar problem had also occurred in the trivalent $SU(N)$ gauging in Figure \ref{fig:Nminus} when we simply summed over the Young diagrams without using the $N^-_k$ vertex.
For example, if we had not been careful to include the denominator in \eqref{eq:Nminus},
we would have obtained $-3$ for the genus zero contribution associated to the degree one curve with K$\ddot{\text{a}}$hler class $Q_1$, and not the correct GV invariant $-2$.  The denominator in \eqref{eq:Nminus} modifies the naive answer to give the correct answer.
In the case of the diagram in Figure \ref{fig:corner}, we still need a mechanism which can reduce the GV invariants computed using the $1^+$ and $1^-_{-2}$ vertices to the GV invariants for local $\mathbb{P}^2$.
Namely, since a combination of the four trivalent $SU(5)$ gaugings automatically yields a gluing along $\mathbb{P}^2$ and the gluing is not merely a collection of indepedent $SU(5)$ gaugings but rather 10 $SU(5)$ gaugings connected by five $\mathbb{P}^2$ gluings, we need to supplement the $5_0^-$ vertex in a manner which accounts for the presence of the $\mathbb{P}^2$ gluings.

Thus far we have not been able to find a modification which extends the four $5_0^-$ vertices by including the $\mathbb{P}^2$ gluings. Nevertheless, we constructed another vertex formalism which can capture the GV invariants of the mirror quintic modulo some missing curve classes. This formalism, which uses the details of the resolution of the mirror quintic described in this paper, is presented below. 

\begin{figure}
\centering
\includegraphics[width=140mm]{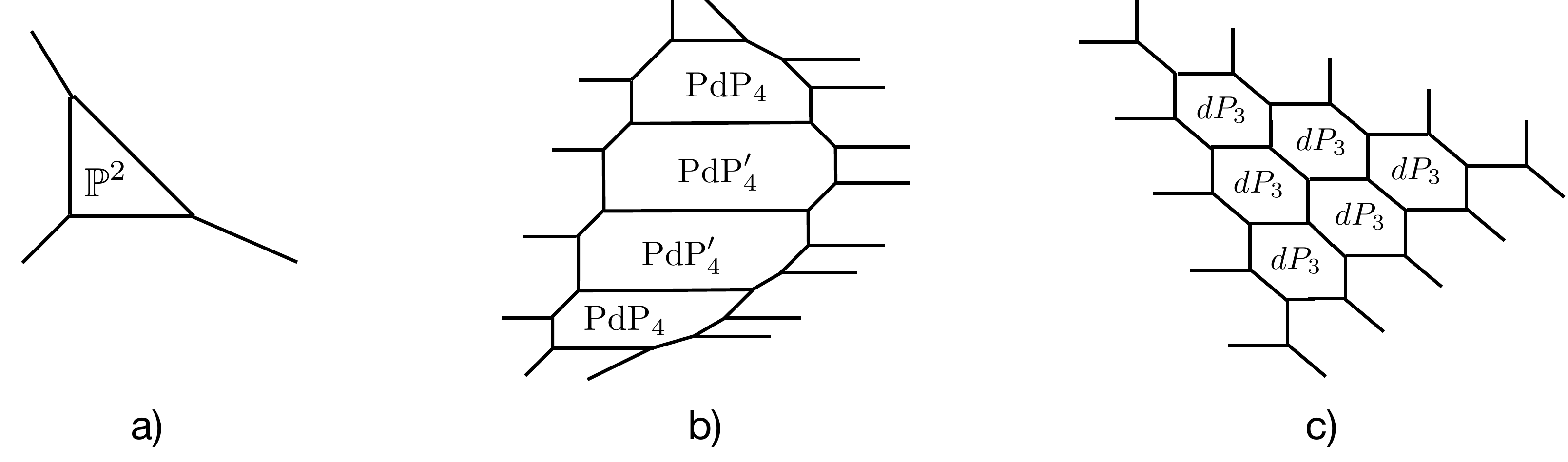} 
\caption{a) The vertex diagram for a local $\mathbb{P}^2$. b) The vertex diagram for a $SU(5)$ gauging. c) The vertex diagram for a $T_5$ geometry.}
\label{fig:P2-SU5-T5}
\end{figure}

First, we use the fact that the mirror quintic is comprised of five $\mathbb{P}^2$'s and ten $T_5$'s glued by ten $SU(5)$ gaugings. Each component can be described as one of the vertex diagrams in Figure \ref{fig:P2-SU5-T5}. So the vertex formalism for the mirror quintic should involve all these ingredients properly glued to each other. Let us explain how to glue them together by using local properties of the mirror quintic.

\begin{figure}
\centering
\includegraphics[width=130mm]{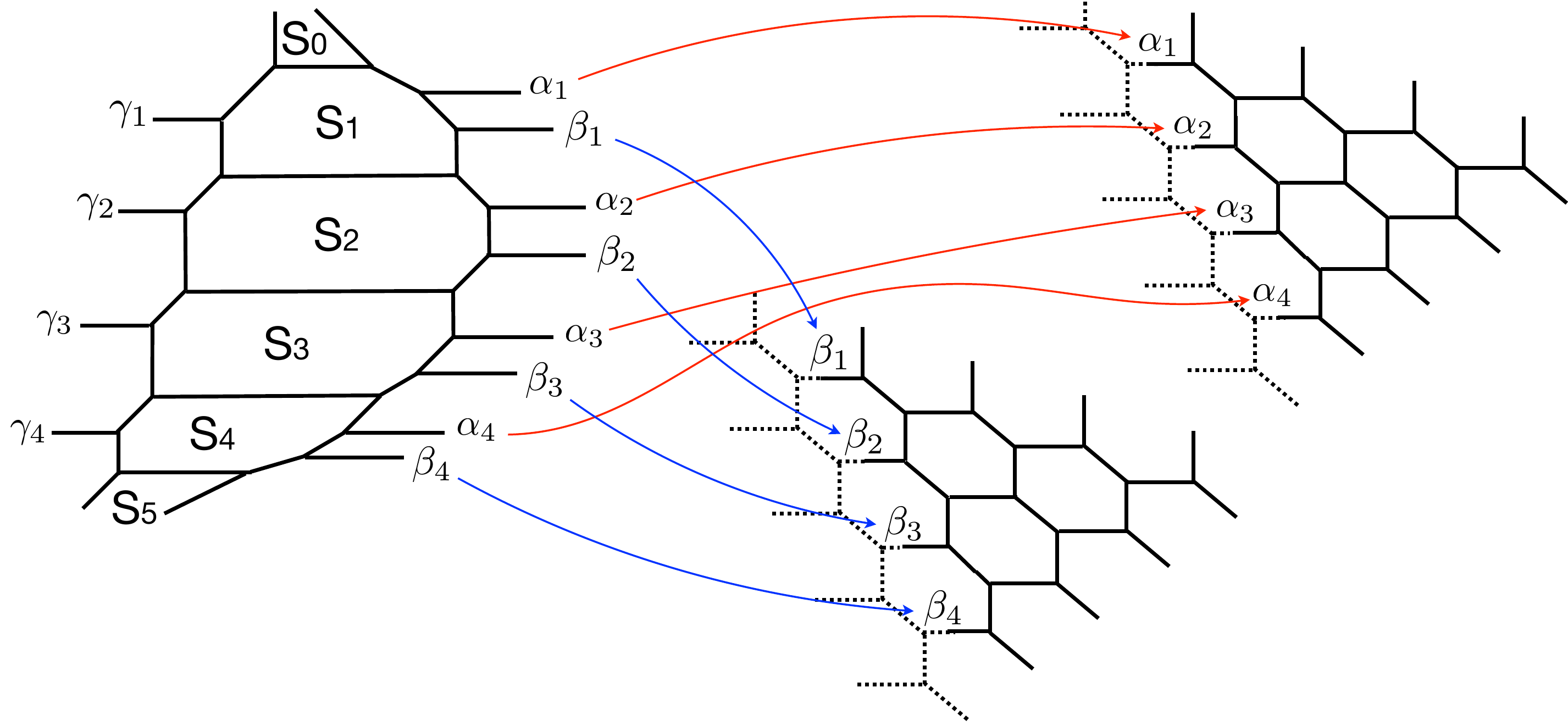} 
\caption{A plane $SU(5)$ vertex connected to two $T_5$ diagrams. The edges with $\gamma_i$ are also connected to a 3rd $T_5$ diagram.}
\label{fig:SU5-deformed}
\end{figure}

In the vertex formalism, the $5^-_0$ vertices for $SU(5)$ gaugings are replaced by the planar $SU(5)$ vertex diagrams given in diagram b) of Figure \ref{fig:P2-SU5-T5}. This diagram is related to the $SU(5)$ diagram in Figure \ref{fig:SU5geom} by Hanany-Witten transitions (or equivalently complex structure deformations) which move two external legs on the left side to the right side of the diagram. Diagram b) of Figure \ref{fig:P2-SU5-T5} is useful for simultaneously incorporating the geometries in diagrams a) and c). Then, the remaining external legs in the diagram are connected to other parts of the mirror quintic geometry. 
A vertex diagram for a single $SU(5)$ gauging should be glued to two $\mathbb{P}^2$'s and three $T_5$ vertex diagrams.
For example, a planar $SU(5)$ vertex diagram glued to two $T_5$ vertex diagrams is illustrated in Figure \ref{fig:SU5-deformed}. Note here that the dotted lines in the $T_5$ diagrams are also a part of the planar $SU(5)$ vertex diagram after the gluing.
The legs labelled by $\gamma_i$ will also be connected to the third $T_5$ vertex diagram in the same manner as the edges labeled by $\alpha_i$. The remaining external edges on the top and the bottom of the diagram are then glued to two local $\mathbb{P}^2$'s, respectively.

We remark however that the planar $SU(5)$ diagram is not equivalent to the $5^-_0$ vertex. As noted above, it describes a complex structure deformation of the geometry for an $SU(5)$ gauging. This deformation introduces some additional $-2$ curves to the planar $SU(5)$ diagram which are not a part of the $SU(5)$ gaugings in  the mirror quintic. They provide extra contributions to the GV invariants in our computation. To compute the actual mirror quintic GV invariants, one must therefore subtract the spurious contributions of the extra $-2$ curves.

We can identify the extra $-2$ curve contributions in the planar $SU(5)$ diagram as follows.
In Figure \ref{fig:SU5-deformed}, the external edges with Young diagrams $\alpha_i$ are connected to a $T_5$ diagram and the edges with Young diagrams $\beta_i$ are connected to another $T_5$ diagram.  In this gluing the $-2$ curve classes between $\alpha_i$ and $\alpha_j$ are identified with the self-intersection 0, genus zero curve classes of the dP$_3$'s in the first $T_5$ geometry, and similarly the $-2$ curves between $\beta_i,\beta_j$ are identified with the self-intersection 0, genus zero curve classes in the second $T_5$ geometry. Note that the framing factors for the sums over Young diagrams $\alpha_i,\beta_j$ are all trivial because the corresponding curves are all $-1$ curves. On the other hand, the $-2$ curve classes between $\alpha_i$ and $\beta_j$ are in fact absent in the mirror quintic. These $-2$ curves are the extra curves introduced by complex structure deformations. We thus need to remove their contributions from the GV invariants.

An extra $-2$ curve with K\"ahler parameter $Q$ yields roughly a ``half'' of the contribution of the $SU(2)$ vector multiplet to the result. More concretely, we define the extra factor as
\begin{equation}
	Z^{(-2)}_{(\mu,\nu)}(Q) \equiv \prod_{i,j=1}^{\infty}\left(1 - Q
y^{i+j-\mu_{i} - \nu_{j}^t-1}\right)^{-1},
\end{equation}
where $\mu,\nu$ are Young diagrams of the external edges connecting this $-2$ curve to the other part. The directions of the arrows for the Young diagrams $\mu$ and $\nu$ are chosen in the same directions. 
Compared to \eqref{eq:Zhalf}, the factor $y^{\frac{1}{2}\left(||\mu||^2 + ||\nu||^2\right)}\tilde{Z}_{\mu}(y)\tilde{Z}_{\nu}(y)$ is removed since this factor is necessary as the Cartan parts of the two $SU(4)$ gaugings for the edges with $\alpha_i$ and also for the edges with $\beta_j$ in Figure \ref{fig:SU5-deformed}. Therefore the extra contribution we need to remove from a single $SU(5)$ diagram depicted in Figure \ref{fig:SU5-deformed} is
\begin{equation}\label{eq:extra-2}
	Z^{\rm extra}_{SU(5)}(\vec{\alpha},\vec{\beta}) = \left(\prod_{1 \leq i \leq j \leq 4} Z^{(-2)}_{(\alpha_i,\beta_j)}(Q_{\alpha_i\beta_j})\right)\left(\prod_{1 \leq i < j \leq 4} Z^{(-2)}_{(\beta_i, \alpha_j)}(Q_{\alpha_j\beta_i})\right),
\end{equation}
where $Q_{\alpha_i\beta_j}$ denotes the K\"ahler parameter for the curve class between $\alpha_i$ and $\beta_j$. In the full mirror quintic geometry, there exist in total 10 $SU(5)$ gaugings. Each $SU(5)$ gauging will be implemented by the planar $SU(5)$ vertex diagram we are discussing here, namely diagram b) of Figure \ref{fig:P2-SU5-T5}. We need to subtract the extra contribution given in equation (\ref{eq:extra-2}) for each $SU(5)$ diagram. This means we need to divide the full vertex formula for the mirror quintic by this extra factor before we perform Young diagram summations in the vertex computation.

One may wonder if this prescription can subtract all the extra contributions from the $SU(5)$ gaugings in the vertex formalism discussed in this section. We do not have a concrete proof for this. However, 
we can provide some non-trivial evidence that this prescription correctly subtracts all extra contributions. 

First, when a planar $SU(5)$ vertex diagram is not connected to a local $\mathbb{P}^2$ while all other external edges are non-trivial (and hence connected to $T_5$ diagrams), the vertex formula factorizes into the contribution to the mirror quintic and the extra contributions given in equation (\ref{eq:extra-2}). Specifically, when we assign Young diagrams $\nu_i \; (i=1, \cdots, 5)$ and the K\"ahler parameter $Q_{B_i}\; (i=1, \cdots, 5)$ to the horizontal internal lines from the top to the bottom of diagram a) in Figure \ref{fig:P2-SU5-T5}, applying the topological vertex to the diagram gives
\begin{align}
& \sum_{\nu_i, \alpha_j, \beta_k, \gamma_l}q^{\frac{1}{2}\left(\sum_{i=1}^4\left(||\alpha_i||^2 + ||\beta_i||^2 + ||\gamma^t_i||^2\right)+ \sum_{i=1}^5\left(||\nu_i||^2 + ||\nu_i^t||^2\right)\right)}\prod_{i=1}^5\left(f_{\nu_i}(y)^{-3+i}\left(-Q_{B_i}\right)^{|\nu_i|}\right)\nonumber\\
&\left(\prod_{1 \leq a \leq b \leq 4}\prod_{i,j=1}^{\infty}\left(1 - Q_{\nu_a\alpha_b}q^{i + j - \nu_{a, i} - \alpha^t_{b, j} + 1}\right)\right)\left(\prod_{1 \leq a < b \leq 5}\prod_{i,j=1}^{\infty}\left(1 - Q_{\nu_b\alpha_a}q^{i + j - \alpha_{a, i} - \nu^t_{b, j} + 1}\right)\right)\nonumber\\
&\left(\prod_{1 \leq a \leq b \leq 4}\prod_{i,j=1}^{\infty}\left(1 - Q_{\nu_a\beta_b}q^{i + j - \nu_{a, i} - \beta^t_{b, j} + 1}\right)\right)\left(\prod_{1 \leq a < b \leq 5}\prod_{i,j=1}^{\infty}\left(1 - Q_{\nu_b\beta_a}q^{i + j - \beta_{a, i} - \nu^t_{b, j} + 1}\right)\right)\nonumber\\
&\left(\prod_{1 \leq a \leq b \leq 4}\prod_{i,j=1}^{\infty}\left(1 - Q_{\nu_a\gamma_b}q^{i + j - \nu_{a, i} - \gamma^t_{b, j} + 1}\right)\right)\left(\prod_{1 \leq a < b \leq 5}\prod_{i,j=1}^{\infty}\left(1 - Q_{\nu_b\gamma_a}q^{i + j - \gamma_{a, i} - \nu^t_{b, j} + 1}\right)\right)\nonumber\\
&\left(\prod_{1 \leq a < b \leq 5}\prod_{i,j=1}^{\infty}\left(1 - Q_{\nu_a\nu_b}q^{i + j - \nu_{a, i} - \nu^t_{b, j} + 1}\right)^{-2}\right)\left(\prod_{1 \leq a < b \leq 4}\prod_{i,j=1}^{\infty}\left(1 - Q_{\alpha_a\alpha_b}q^{i + j - \alpha_{a, i} - \alpha^t_{b, j} + 1}\right)^{-1}\right)\nonumber\\
&\left(\prod_{1 \leq a < b \leq 4}\prod_{i,j=1}^{\infty}\left(1 - Q_{\beta_a\beta_b}q^{i + j - \beta_{a, i} - \beta^t_{b, j} + 1}\right)^{-1}\right)\left(\prod_{1 \leq a < b \leq 4}\prod_{i,j=1}^{\infty}\left(1 - Q_{\gamma_a\gamma_b}q^{i + j - \gamma_{a, i} - \gamma^t_{b, j} + 1}\right)^{-1}\right)\nonumber\\
&\left(\prod_{1 \leq a \leq b \leq 4}\prod_{i,j=1}^{\infty}\left(1 - Q_{\alpha_a\beta_b}q^{i + j - \alpha_{a, i} - \beta^t_{b, j} + 1}\right)^{-1}\right)\left(\prod_{1 \leq a < b \leq 4}\prod_{i,j=1}^{\infty}\left(1 - Q_{\alpha_b\beta_a}q^{i + j - \beta_{a, i} - \alpha^t_{b, j} + 1}\right)^{-1}\right), \label{top.SU5deformed}
\end{align}
where $Q_{\lambda\mu}$ denotes the K\"ahler parameter for the curve class between $\lambda$ and $\mu$. Note that not all the K\"ahler parameters appearing in \eqref{top.SU5deformed} are independent from one another. A more detailed explanation of a parameterization for an $SU(5)$ gauging diagram is given in Appendix \ref{sec:SU5gauging}. The direction of the arrows for the Young diagrams are chosen in the right direction in the diagram in Figure \ref{fig:P2-SU5-T5} b). 
The last line in \eqref{top.SU5deformed} is precisely the factor given in \eqref{eq:extra-2} and dividing the partition function computed from the vertex formalism by the factor \eqref{eq:extra-2} simply removes the last line in \eqref{top.SU5deformed}. Furthermore, when we remove the factors in the last line of \eqref{top.SU5deformed} the result in fact agrees with the topological string partiton function of a part of gluing three $T_5$ diagrams using the $5^-_0$ vertex. 

Second, we compute below GV invariants using our vertex prescription and compare the result against those from a geometric counting up to finite order in a two-parameter expansion. The comparison shows a perfect agreement between the two results. With this supporting evidence, we conjecture that our prescription removes all extra contributions to the $SU(5)$ gaugings and leaves only the correct GV invariants for the mirror quintic.
From now on we shall assume that the contributions (\ref{eq:extra-2}) from the extra $-2$ curves in the planar $SU(5)$ vertex diagrams have  been subtracted using the above prescription.

\begin{figure}
\centering
\includegraphics[width=155mm]{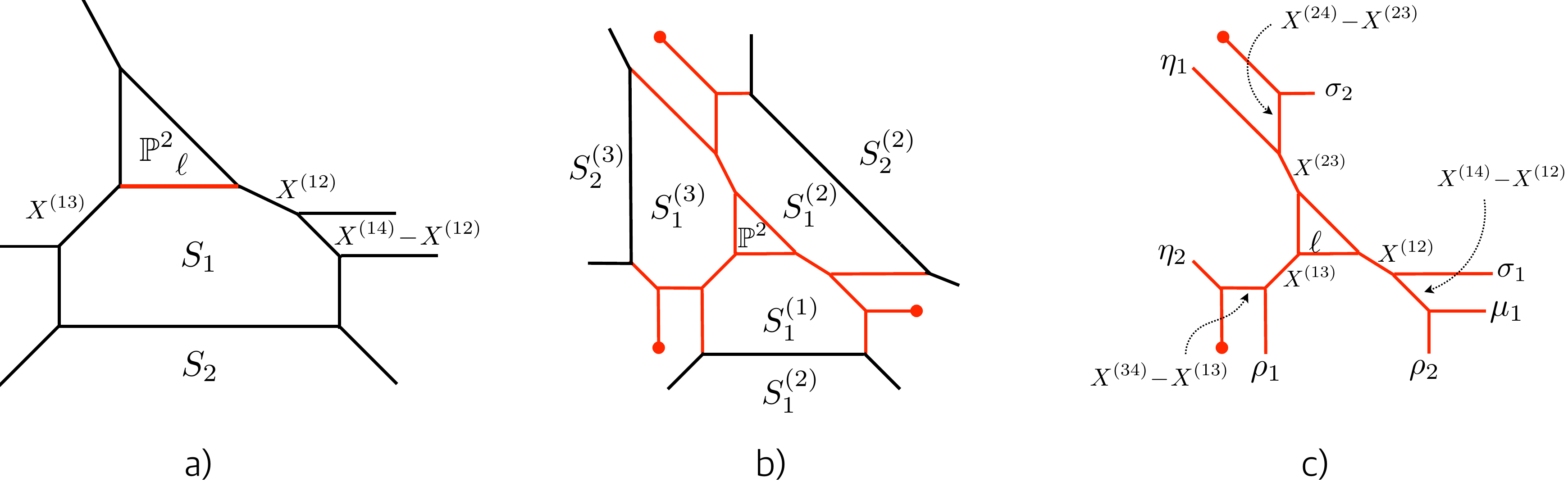} 
\caption{a) Local geometry near a $\mathbb{P}^2$ intersecting the first surface $S_1$ in the $SU(5)$ vertex diagram in Figure \ref{fig:SU5geom}. b) A vertex diagram where a $\mathbb{P}^2$ intersects with three $SU(5)$ vertex diagrams. c) A $T_3$ diagram which is a complex structure deformation of dP$_6$.}
\label{fig:localP2dP4}
\end{figure}

We now consider a local $\mathbb{P}^2$ which meets four copies of the $SU(5)$ diagram in Figure \ref{fig:SU5-deformed}. The local $\mathbb{P}^2$ is glued to the first surface $S_1$ in each planar $SU(5)$ vertex diagram. This gluing is performed by identifying the hyperplane (curve) class $\ell$ (with $\ell^2=1$) in $\mathbb{P}^2$ with the class $e$ (with $e^2=-3$) in $S_1$. The surface $S_1$ is a complex structure deformation of dP$_4$, and contains three additional exceptional curves $X^{(1i)}$ with $i=2,3,4$, each of which intersects $\mathbb{P}^2$ at a point. The local geometry for this gluing is given by the web diagram a) in Figure \ref{fig:localP2dP4}. In diagram a), the gluing curve, which is $\ell$ in $\mathbb{P}^2$ and $e$ in $S_1$, is indicated by the red line between the $\mathbb{P}^2$ and $S_1$.
This web diagram shows how $\mathbb{P}^2$ intersects a planar $SU(5)$ vertex diagram.

Attaching two more planar $SU(5)$ vertex diagrams to this web diagram is straightforward. Using the description of $\mathbb{P}^2\cup S_1$ in diagram a) of Figure \ref{fig:localP2dP4}, we find that the vertex diagram for the local geometry of three planar $SU(5)$ vertices attached to local $\mathbb{P}^2$ can be represented by diagram b) in Figure \ref{fig:localP2dP4}. Here, three $-3$ curves $e$ in the surfaces $S_1^{(i=1,2,3)}$ of the $SU(5)$ gaugings are glued to the curve $\ell$ in $\mathbb{P}^2$.

There are six $-1$ curves, which we denote by $X^{(ij)}$, each of which intersects the $\mathbb{P}^2$ at a point. Moreover, the curve $X^{(ij)}$ glues together two surfaces $S_1^{(i)}$ and $S_1^{(j)}$ in the four $SU(5)$ gaugings. All six of these $-1$ curves are involved in diagram b) of Figure \ref{fig:localP2dP4}. However, the surface $S_1^{(4)}$ in the the fourth $SU(5)$ gauging is not visible here. Note that the geometry of a $\mathbb{P}^2$ intersecting with 6 exceptional curves is the del Pezzo surface dP$_6$. Indeed a dP$_6$ is embedded in diagram b), which is the sub-diagram indicated by the red lines. More precisely, this sub-diagram is a complex structure deformation of dP$_6$, namely the so-called $T_3$ geometry depicted in diagram c) of Figure \ref{fig:localP2dP4}. In the diagram c), we assign non-trivial Young diagrams on the seven external legs, as they are connected to the other parts of the mirror quintic. The remaining two external legs have empty Young diagrams. The complex structure deformation adds 9 extra $-2$ curves  corresponding to the curve classes between the external edges in diagram c); they are
\begin{equation}\label{extram2}
\begin{aligned}
	&\ell + X^{(12)}+ X^{(13)}+ X^{(14)} \ , \quad \ell + X^{(12)}+X^{(23)}+X^{(24)} \ , \quad \ell + X^{(13)}+X^{(23)}+X^{(34)} \ , \\
	&\ell+X^{(12)}+X^{(14)}+X^{(34)} \ , \quad \ell + X^{(23)}+X^{(24)}+X^{(14)} \ , \quad \ell+X^{(13)}+X^{(34)}+X^{(24)} \ , \\
	& X^{(14)}-X^{(12)} \ , \quad X^{(24)}-X^{(23)} \ , \quad X^{(34)}-X^{(13)} \ .
\end{aligned}
\end{equation}
These curve classes are not holomorphic classes in dP$_6$, but rather are holomorphic in a complex structure deformation of dP$_6$.

Among these $-2$ curves, the first three curves in \eqref{extram2} are necessary in the mirror quintic geometry to form the three surfaces $S_1^{(i=1,2,3)}$ depicted in diagram b) of Figure \ref{fig:localP2dP4}.
The other six curves are not components in the full geometry and thus their contributions to the GV invariants should be subtracted. The extra contributions from the last three curves are involved in the extra factors in equation (\ref{eq:extra-2}) that we already subtracted from the result.  The $-2$ curves in the second line in \eqref{extram2} also produce extra contributions in diagram c) in Figure \ref{fig:localP2dP4} that we need to subtract.
These extra contributions are the same type as those in the $SU(5)$ gaugings given in equation (\ref{eq:extra-2}).
So we find that the extra $-2$ curve contribution in diagram c) of Figure \ref{fig:localP2dP4} is
\begin{equation}\label{eq:extra-T3}
	Z^{\rm extra}_{T_3}(\vec{\rho},\vec{\sigma},\vec{\eta},\mu_1) =  Z^{(-2)}_{(\phi,\rho_2)}(uQ_{12}Q_{14}Q_{34})\cdot Z^{(-2)}_{(\mu_1,\sigma_2)}(uQ_{23}Q_{24}Q_{14}) \cdot Z^{(-2)}_{(\phi,\eta_2)}(uQ_{13}Q_{34}Q_{24}) \ ,
\end{equation}
where $Q_{ij}$ are K\"ahler parameters for the $X^{(ij)}$ curve and $\phi$ stands for an empty Young diamgram. The extra factors of this type in the local $T_3$ diagrams around the $\mathbb{P}^2$'s as well as the extra factors in (\ref{eq:extra-2}) should be removed from the final vertex result. We will again assume they have all been subtracted in the discussion below.

In the vertex diagram b) in Figure \ref{fig:localP2dP4}, the three solid dots correspond to 7-branes when we regard the diagram as a $(p,q)$ 5-brane web. This means that, when we pull these three dots out to infinity with the branch cuts of the 7-branes taken into account, we can use this diagram to compute the topological string partition function for the geometry of a local $\mathbb{P}^2$ meeting three PdP$_4$'s, though we will keep the dots at finite distance for later use.

Note that we assign a Young diagram $\mu_1$ to one of the 7-branes in  diagram c) of Figure \ref{fig:localP2dP4}, while the other two come with empty Young diagrams. This 7-brane leg with $\mu_1$ will be used to connect some part of the fourth $SU(5)$ gauging which will be discussed in detail soon. We notice that for the fourth $SU(5)$ gauging we need to introduce one more $-2$ curve class, which is $\ell+X^{(14)}+X^{(24)}+X^{(34)}$, to diagram b) or diagram c) by means of a complex structure deformation. This class is necessary to form the surface $S^{(4)}_1$, but it is absent in the $T_3$ diagram. We claim that this curve class can be introduced by multiplying the vertex formula for the geometry around each $\mathbb{P}^2$ depicted in diagram b) by the factor
\begin{equation}
	Z^{(-2)}_{\mu_1,\phi}(uQ_{14}Q_{24}Q_{34}) \ .
\end{equation}

\begin{figure}
\centering
\includegraphics[width=120mm]{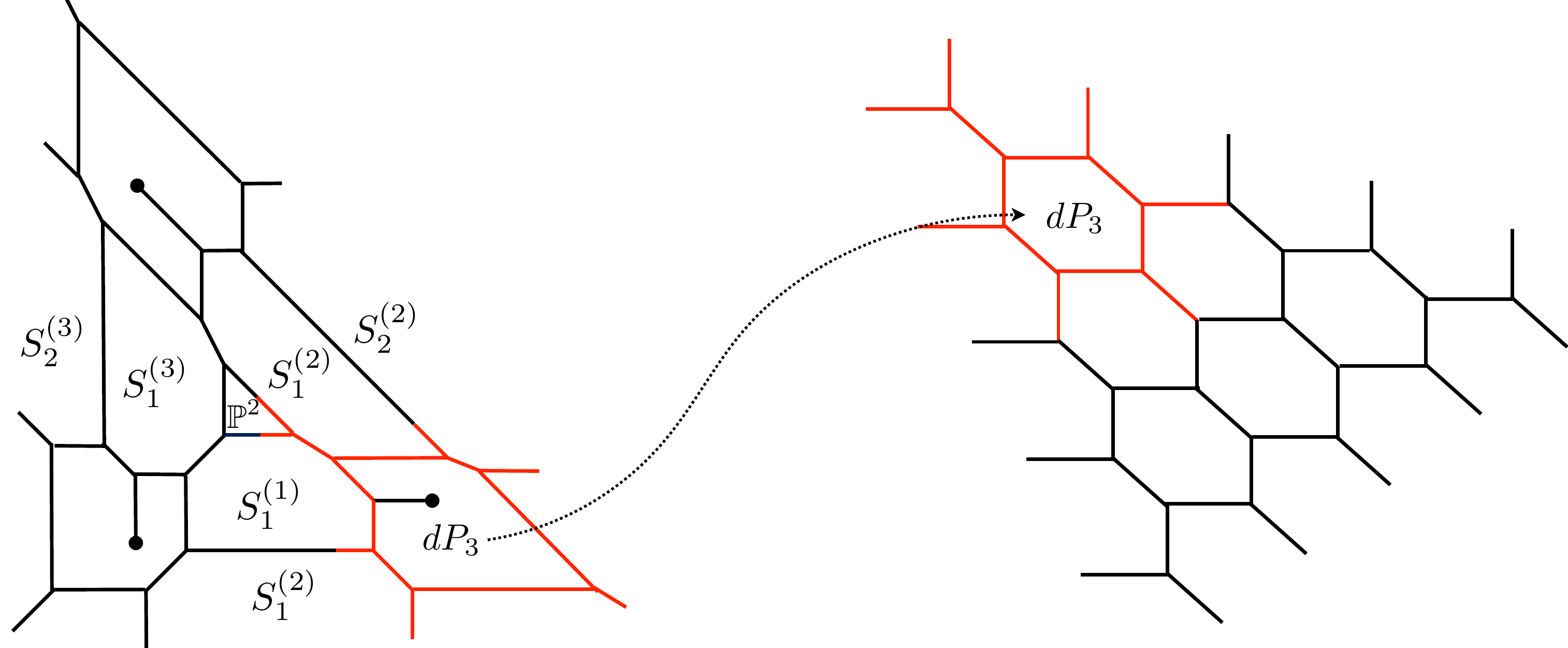} 
\caption{A local $\mathbb{P}^2$ intersecting with three plane $SU(5)$ diagrams and three $T_5$ diagrams. The surface dP$_3$ formed by the red lines in the left diagram is mapped to the first dP$_3$ in the $T_5^{(12)}$ vertex diagram between the 1st and the 2nd $SU(5)$ gaugings.}
\label{fig:P2-3dP4}
\end{figure}

Other external edges in diagram b) of Figure \ref{fig:localP2dP4} are connected to other parts of the planar $SU(5)$ vertex diagrams and also three $T_5$ diagrams like those in Figure \ref{fig:P2-3dP4}. In this figure, the diagrams denoted by red edges are mapped to the first dP$_3$ diagram in the $T_5^{(12)}$ between the first and the second $SU(5)$ gaugings. We denote the $T_5$ vertex diagram between the $i$-th and $j$-th SU(5) gaugings by $T_5^{(ij)}$. 
The gluing rule between an $SU(5)$ gauging and the adjacent $T_5$ vertex diagrams is already given in Figure \ref{fig:SU5-deformed}. We can use this rule to extend the vertex diagram in Figure \ref{fig:P2-3dP4} to the remaining surfaces in the planar $SU(5)$ diagrams, as well as the $T_5$ diagrams.
This explains the vertex configuration describing the local geometry of a $\mathbb{P}^2$ intersecting with three $SU(5)$ gaugings and three $T_5$'s embedded in the mirror quintic.

We still need to incorporate the fourth $SU(5)$ gauging and its three associated $T_5$ vertex diagrams (i.e. $T_5^{(14)}, T_5^{(24)}, T_5^{(34)}$) into our vertex construction. Unfortunately, we could not find a consistent vertex configuration simultaneously incorporating all these ingredients. Instead, we
 propose the following vertex configuration which misses two curve classes in the vicinity of each local $\mathbb{P}^2$ in the full mirror quintic. 

\begin{figure}
\centering
\includegraphics[width=140mm]{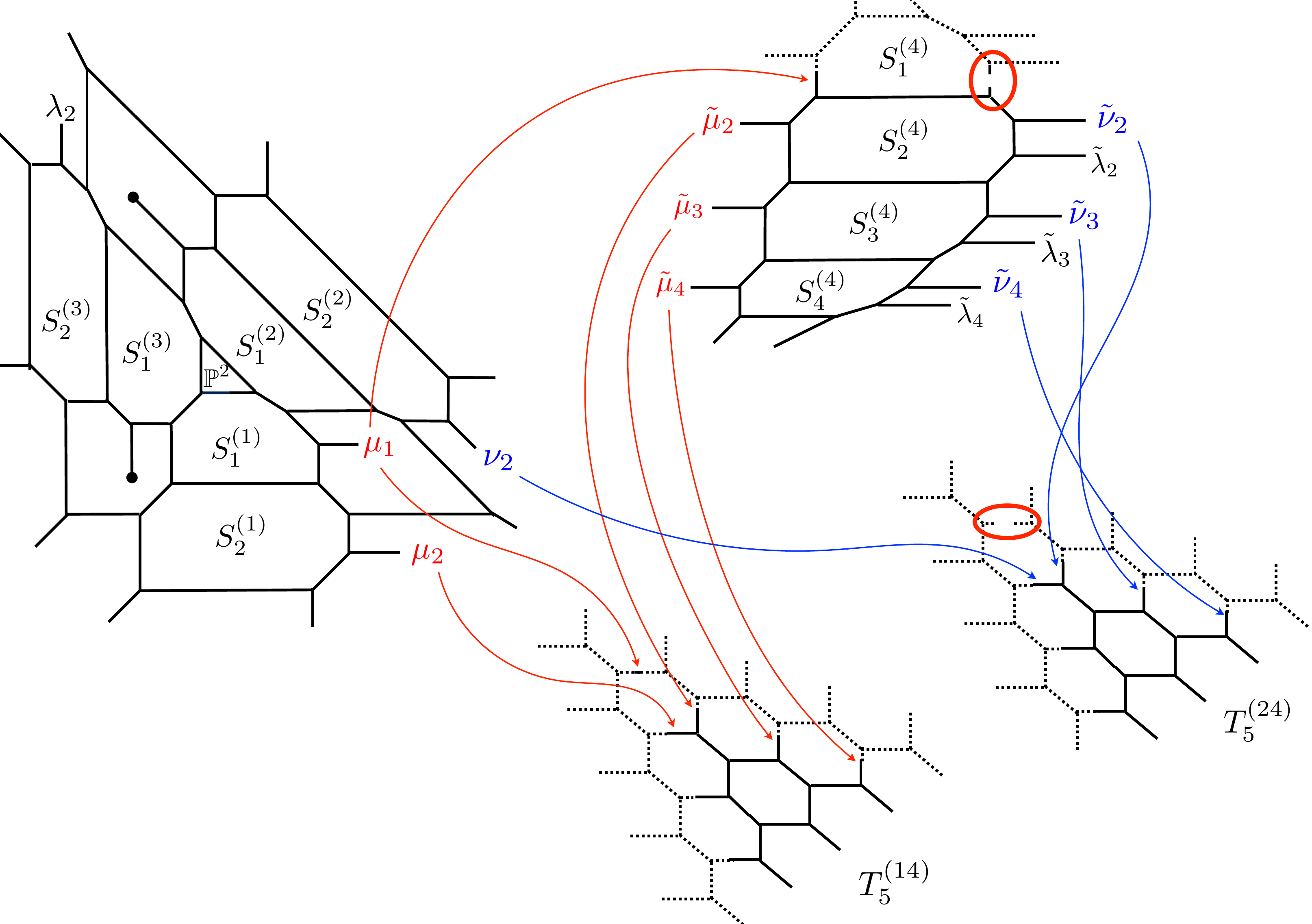} 
\caption{The vertex diagram for the 4-th $SU(5)$ gauging. The curve class denoted by a red circle is the missing curve class.}
\label{fig:P2-4SU5}
\end{figure}

We first glue one of the 7-brane legs, labeled by $\mu_1$, to the first surface $S^{(4)}_1$ in the fourth $SU(5)$ gauging as drawn in Figure \ref{fig:P2-4SU5}. After the gluing, the edge of the Young diagram $\mu_1$ becomes the proper transform $f^{(4)}_1-X^{(14)}$ of the self–intersection 0, genus zero curve class in the $S^{(4)}_1$ intersecting with $X^{(14)}$ at a point, denoted $f^{(4)}_1$.
Here one notices that the dotted lines in the $S_1^{(4)}$ are already included in the left vertex diagram of a $\mathbb{P}^2$ and $S_i^{(1,2,3)}$. The left vertex diagram is an extension of diagram b) in Figure \ref{fig:localP2dP4} and, as discussed above, the three exceptional curves $X^{(14)},X^{(24)},X^{(34)}$ in the $S_1^{(4)}$ are all contained in this diagram.
Other external legs with Young diagrams $\tilde{\mu}_i,\tilde{\nu}_i,\tilde{\lambda}_i$ in the fourth $SU(5)$ diagram are then connected to the $T_5^{(14)},T_5^{(24)},T_5^{(34)}$ diagrams, respectively, as illustrated in Figure \ref{fig:P2-4SU5}.

Note however that the line denoted by a red circle in the $S^{(4)}_1$ is disconnected. The Young diagram on this leg is empty. The geometry implies that this leg must be connected to some part of the left diagram, but we could not determine how to implement this in our formalism. This means that the curve classes assocated to this line are missing in our vertex formalism.
We note that there are precisely two missing curves around each local $\mathbb{P}^2$, which we denote $\mathcal{C}_1$ and $\mathcal{C}_2$. They are primitive\footnote{A primitive curve class is a class which cannot be expressed a positive linear combination of any other curve classes.} curve classes at the intersections of the surface $S_1^{(4)}$ with $T_5^{(24)}$ and $T_5^{(34)}$, i.e. $\mathcal{C}_1\subset S_1^{(4)}\cap T_5^{(24)}$ and $\mathcal{C}_2\subset S_1^{(4)} \cap T_5^{(34)}$. Their restrictions to $S_1^{(4)}$ are
$\mathcal{C}_1|_{S_1^{(4)}}=f^{(4)}_1-X^{(24)}$ and $\mathcal{C}_2|_{S_1^{(4)}}=f^{(4)}_1-X^{(34)}$ respectively, where $C|_S$ denotes the class of the curve $C$ inside the surface $S$. 
In the $T^{(24)}_5$, the missing curve $\mathcal{C}_1$ is the exceptional curve inside the red circle. Similarly, $\mathcal{C}_2$ is the missing exceptional curve in the $T^{(34)}$ at the same location.

This means that the GV invariants associated to curves whose classes can be expressed as non-negative linear combinations of primitive curve classes such that the coefficient of the primitive class $\mathcal{C}_1$ (and likewise $\mathcal{C}_2$) is positive cannot be computed using our formalism. For example, the GV invariants associated to a curve $C \subset S_1^{(4)}$ which can be expressed as 
	\begin{align}
		C =a \,\mathcal C_1 + \cdots,~~  a>0
	\end{align}
cannot be computed. A notable example of a curve in the surface $S_1^{(4)}$ whose associated GV invariants are not computable in this formalism is the class $f = (f_1^{(4)}- X^{(14)}) + (X^{(14)})$.

So now, apart from these two missing classes, all the components near a $\mathbb{P}^2$ are properly glued together following the intersection structure of the local geometry. Extending this procedure to the rest of the 3-fold, we can complete the vertex construction for the mirror quintic that unites all $\mathbb{P}^2$'s and $T_5$'s, and $SU(5)$ gaugings. As discussed above however this construction cannot realize two curve classes $\mathcal{C}_1,\mathcal{C}_2$ in each local geometry around a $\mathbb{P}^2$. The following is a summary of our vertex construction:

\begin{framed}
\begin{enumerate}
	\item The resolution of the  mirror quintic given in Figure \ref{fig:pentagon} consists of 5 $\mathbb{P}^2$'s and 10 $T_5$'s and 10 $SU(5)$ gaugings that are described by the vertex diagrams in Figure \ref{fig:P2-SU5-T5}.
	\item An $SU(5)$ vertex diagram is connected to three $T_5$ diagrams as described in Figure \ref{fig:SU5-deformed}.
	\item A $\mathbb{P}^2$ vertex diagram is glued to four $SU(5)$ vertex diagrams as described in Figure \ref{fig:localP2dP4} and Figure \ref{fig:P2-4SU5}.
	\item This vertex construction cannot realize two curve classes $\mathcal{C}_1|_{S_1^{(4)}}=f^{(4)}_1-X^{(24)}$ and $\mathcal{C}_2|_{S_1^{(4)}}=f^{(4)}_1-X^{(34)}$  around each $\mathbb{P}^2$.
\end{enumerate}
\end{framed}
We conjecture that the vertex formalism from this construction captures the GV invariants of the mirror quintic which are not positive combinations of the ten missing curve classes in total around five local $\mathbb{P}^2$'s. 
Note that the moduli of these curve classes may be related to the moduli of other curve classes in the mirror quintic by constraints which are implicit in the geometry. Because of this, we do not know the full set of curve classes which can be safely computed by our formalism. Roughly speaking, the full set of curves whose GV invariants can be computed are all curves whose classes cannot be expressed as positive linear combinations of primitive classes involving the ten excluded classes described above.

We present some leading orders of the topological string partition function computed using the vertex formalism described above. Since the mirror quintic has many K$\ddot{\text{a}}$hler parameters, we constrain these parameters in such a way that only two parameters remain, for simplicity of the computation. First we set all the blow up parameters to be equal to each other. Namely we restrict the length of each line of the $T_5$ diagram in the symmetric phase in Figure \ref{eq:toricdiagram} to be the same and denote it by $Q_{T_5} = e^{-\gamma}$ where $\gamma$ is the length of each line. We choose the other parameter as $u = e^{-\ell}$ where $\ell$ is the length of the top or bottom $U(1)$ gauging line of the $SU(5)$ gaugings. The results up to $\ell=3$ and $\gamma=4$ are summarized in Table \ref{tb:GV-mirror-quintic} for the genus zero GV invariants and also in Table \ref{tb:g1GV-mirror-quintic} for the genus one GV invariants. For computing the GV invariants in Table \ref{tb:GV-mirror-quintic} and Table \ref{tb:g1GV-mirror-quintic} we used a symmetry under exchanging the four PdP$_4$'s around each corner. 

\begin{table}[t]
\centering
\begin{tabular}{c|c|c|c|c|c}
	\diagbox[height=.8cm,width=1.3cm]{$\ell$~}{$\gamma$~} & $0$ & $1$ & $2$ & $3$ & $4$ \\
\hline
$0$ & $*$ & $300$ & $-440$ & $\textcolor{red}{850}$ & $\textcolor{red}{-2040}$ \\
\hline 
$1$ & $15$ & $-60$ & $155$ & $\textcolor{red}{-460}$ & $\textcolor{red}{1350}$ \\
\hline
$2$ & $\textcolor{blue}{-30}$ & $\textcolor{blue}{150}$ & $\textcolor{blue}{-500}$ & $\textcolor{red}{1710}$ & $\textcolor{red}{-4730}$ \\
\hline
$3$ & $\textcolor{red}{135}$ & $\textcolor{red}{-960}$ & $\textcolor{red}{4115}$ & $\textcolor{red}{-15780}$ & $\textcolor{red}{45685}$ 
\end{tabular}
\caption{Genus zero GV invariants of the mirror quintic obtained using our vertex formalism. The invariants in black, which had already been computed by Katz and Morrison, were confirmed by our vertex formalism. The invariants in blue were predicted by our vertex formalism and subsequently checked by Katz and Morrison. The invariants in red are predictions of our formalism which have yet to be reproduced by other means.}
\label{tb:GV-mirror-quintic}
\end{table}
\begin{table}[t]
  \centering
\begin{tabular}{c|c|c|c|c|c}
	\diagbox[height=.8cm,width=1.3cm]{$\ell$~}{$\gamma$~} & $0$ & $1$ & $2$ & $3$ & $4$ \\
\hline
$3$ & $\textcolor{red}{-50}$ & $\textcolor{red}{270}$ & $\textcolor{red}{-960}$ & $\textcolor{red}{3430}$ & $\textcolor{red}{-9750}$ 
\end{tabular}
\caption{Genus one GV invariants of the mirror quintic obtained using our vertex formalism. These invariants are predictions of our formalism which have yet to be reproduced by other means.}
\label{tb:g1GV-mirror-quintic}
\end{table}

It is instructive to look at the genus zero GV invariants of the class $a \ell + b\gamma$ where $0 \leq a \leq 1, 0 \leq b \leq 2$ in the vertex formalism in more detail. In this case the computation of the GV invariants reduces to counting lines in the diagrams. 
Since the K$\ddot{\text{a}}$hler parameter for the second line from the top or the bottom of an $SU(5)$ gauging is given by $uQ^3$ as shown in Appendix \ref{sec:SU5gauging}, we focus on the top or the bottom gauging line for the class  $\ell + b\gamma$ where $0 \leq b \leq 2$. 

\paragraph{$\ell$.}
The genus zero GV invariant for the class $\ell$ is the genus zero GV invariant of the minimum degree in a local $\mathbb{P}^2$, which is $3$. Since each corner is described by a $\mathbb{P}^2$ and hence the genus zero GV invariant for the class $\ell$ of the mirror quintic is $3 \times 5 = 15$. 

\paragraph{$\gamma$.}
The genus zero GV invariant for the class $\gamma$ can be computed by focusing on the $T_5$ diagrams. From one $T_5$ geometry, the genus zero GV invariant of the class $\gamma$ is $30$, which may be also obtained by counting lines in the $T_5$ diagram. Since we have ten $T_5$ diagrams, the genus zero GV invariant for the class $\gamma$ of the mirror quintic is $30 \times 10 = 300$.  

\begin{figure}
\centering
\includegraphics[width=80mm]{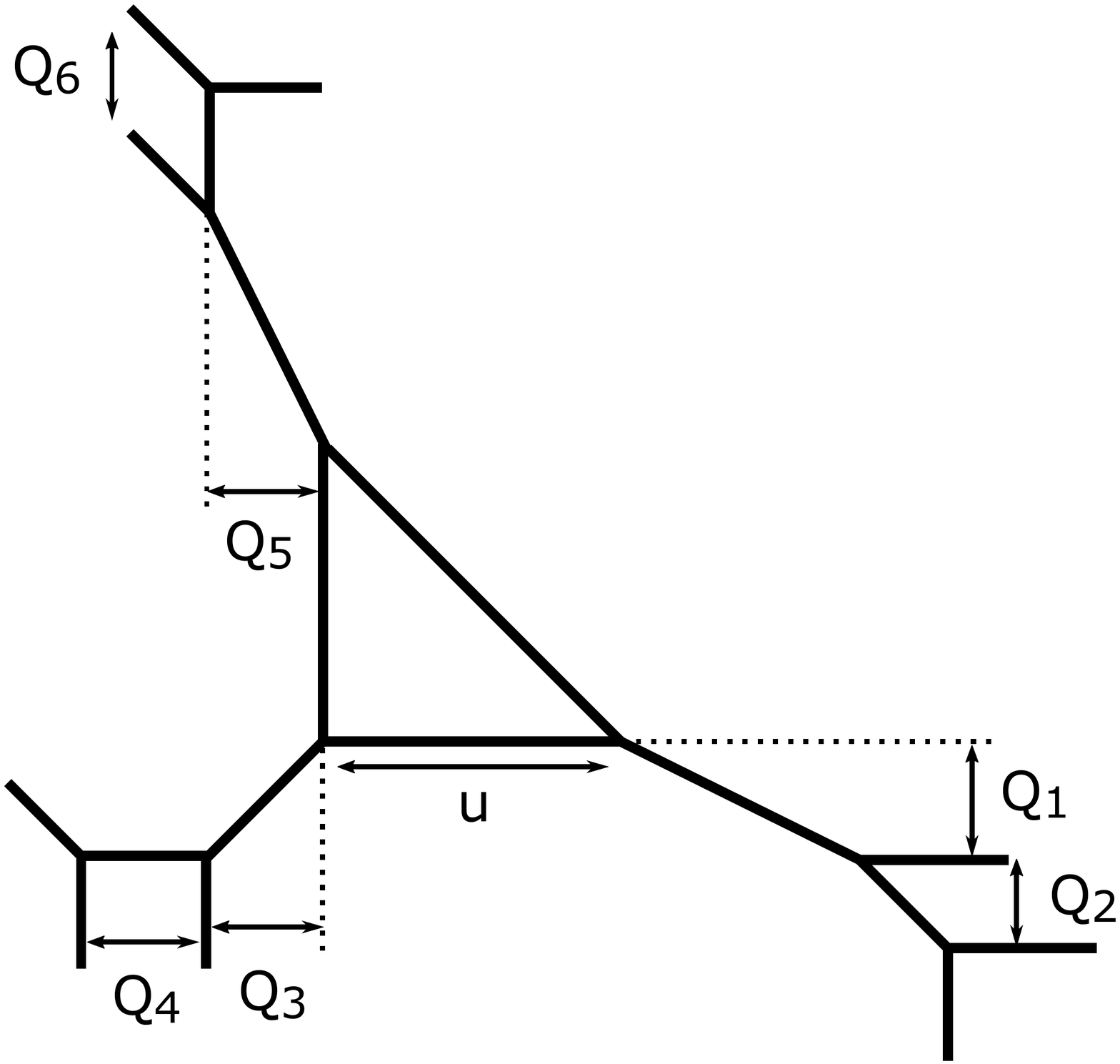} 
\caption{A $T_3$ diagram in a phase where a $\mathbb{P}^2$ can be explicitly seen. $u, Q_1, Q_2, Q_3, Q_4, Q_5, Q_6$ are the K$\ddot{\text{a}}$hler parameters and the restriction we are considering corresponds to $Q_1 = Q_3 = Q_5 = Q$ and $Q_2 = Q_4 = Q_6 = 1$.}
\label{fig:dP6v1}
\end{figure}

\begin{figure}
\centering
\includegraphics[width=80mm]{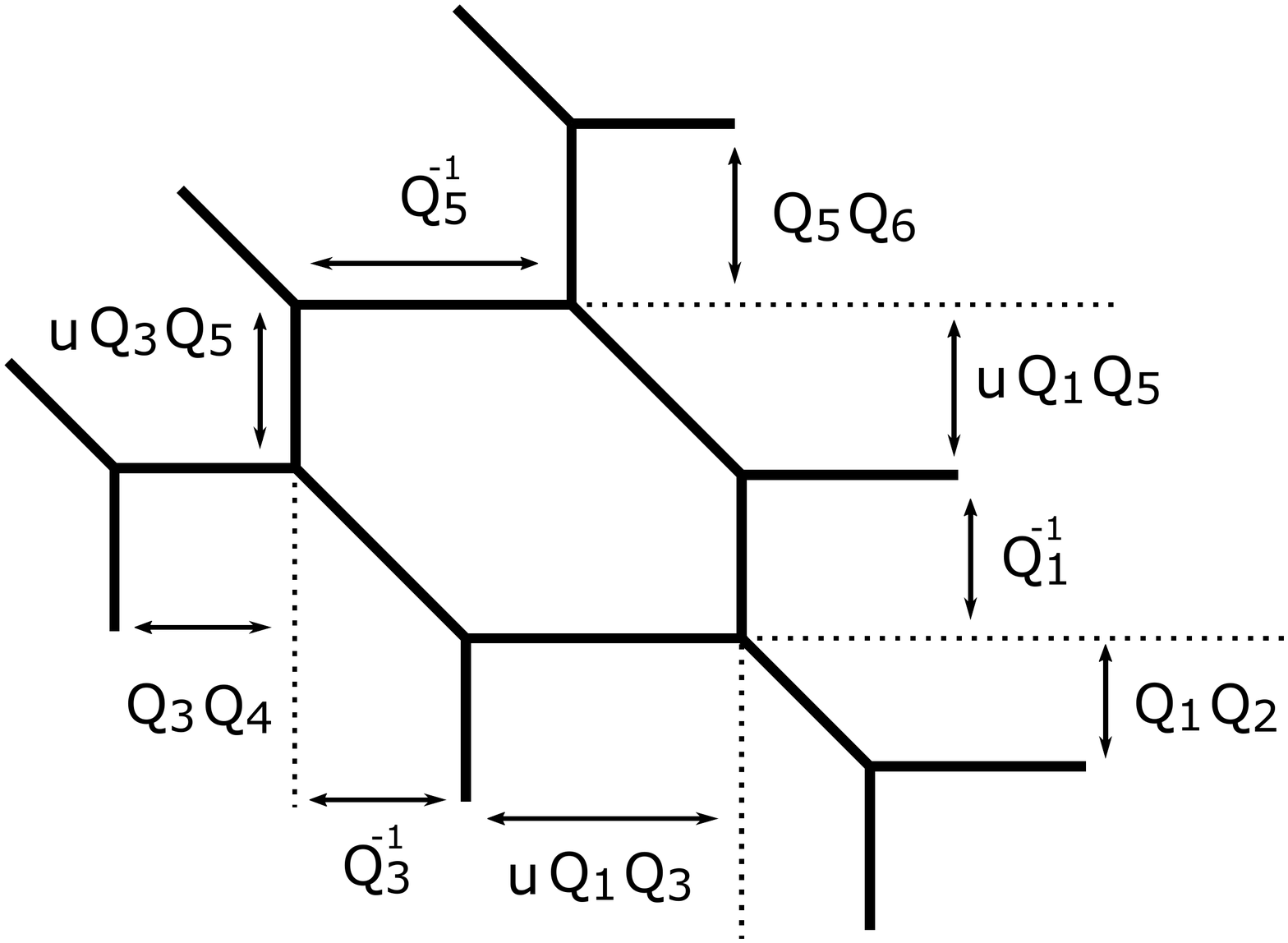} 
\caption{Another $T_3$ diagram in a phase where the counting of the genus zero GV invariant for the class $\ell + \gamma$, $\ell + 2\gamma$ can be carried out more explicitly. }
\label{fig:dP6v2}
\end{figure}

\paragraph{$\ell + \gamma$.}
We then consider the genus zero GV invariant for the class $\ell + \gamma$, which is a combination of a curve inside a $\mathbb{P}^2$ and a curve inside a $T_5$ geometry. 
The class $\ell + \gamma$ is included in a dP$_6$ around each corner of the pengaton.  A complex structure deformation of a dP$_6$ leads to a $T_3$ diagram depicted in Figure  \ref{fig:dP6v1}.
$u, Q_1, Q_2, Q_3, Q_4, Q_5, Q_6$ are the K$\ddot{\text{a}}$hler parameters for the $T_3$ and the restriction we are considering corresponds to $Q_1 = Q_3 = Q_5 = Q$ and $Q_2 = Q_4 = Q_6 = 1$. 
Instead of using the diagram in Figure \ref{fig:dP6v1}, we make use of a diagram in Figure \ref{fig:dP6v2} which is obtained by performing three flop transitions with respect to the curves with the K$\ddot{\text{a}}$hler parameters denoted by $Q_1, Q_3, Q_5$ in Figure \ref{fig:dP6v1}.
From the diagram in Figure \ref{fig:dP6v2}, the genus zero GV invariant for the class $\ell + \gamma$  can be computed by counting the lines with the K$\ddot{\text{a}}$hler parameter $uQ$ and it becomes $-12$ for one del Pezzo six surface. Hence the genus zero GV invariant for the class $\ell + \gamma$ for the mirror quintic becomes $-12 \times 5 = - 60$. 

\paragraph{$2\gamma$.}
The genus zero curve of the class $2\gamma$ is included in the $T_5$ geometry. However, we need to be careful of the $2\gamma$ curve which is also included in the $SU(5)$ gauging. First the genus zero contribution for the curve class $2\gamma$ which is included only in one $T_5$ geometry gives $-36$. Hence, ten $T_5$ diagrams yield $-36 \times 10 = -360$ in total. On the other hand, the genus zero GV invariant for the class $2\gamma$ which is inlcuded in one $SU(5)$ gauging gives $-8$. Since we have ten $SU(5)$ gaugings, the contribution becomes $-8 \times 10 = -80$. Note that this number is different from a naive counting from the viewpoint of the $T_5 $ diagram. From each $T_5$ diagram the contribution for the class $2\gamma$ which is also included in the $SU(5)$ gauging is $-4 \times 3 = -12$. Therefore, the total number might look like $-12 \times 10 = -120$. However, this is overcounting since from the $5^-_0$ vertex language, we need to remove the factor in the denominator in \eqref{eq:Nminus}, which amounts to reducing $-120$ to $-80$. This removal can be automatically taken into account when one uses the planar $SU(5)$ vertex diagram b) in Figure \ref{fig:P2-SU5-T5}, which was an advantage of making use of the $SU(5)$ vertex diagram in our vertex formalism. Therefore, the final result for the genus zero GV invariant for the class $2\gamma$ is $(-360) + (-80) = -440$.

\begin{figure}
\centering
\includegraphics[width=70mm]{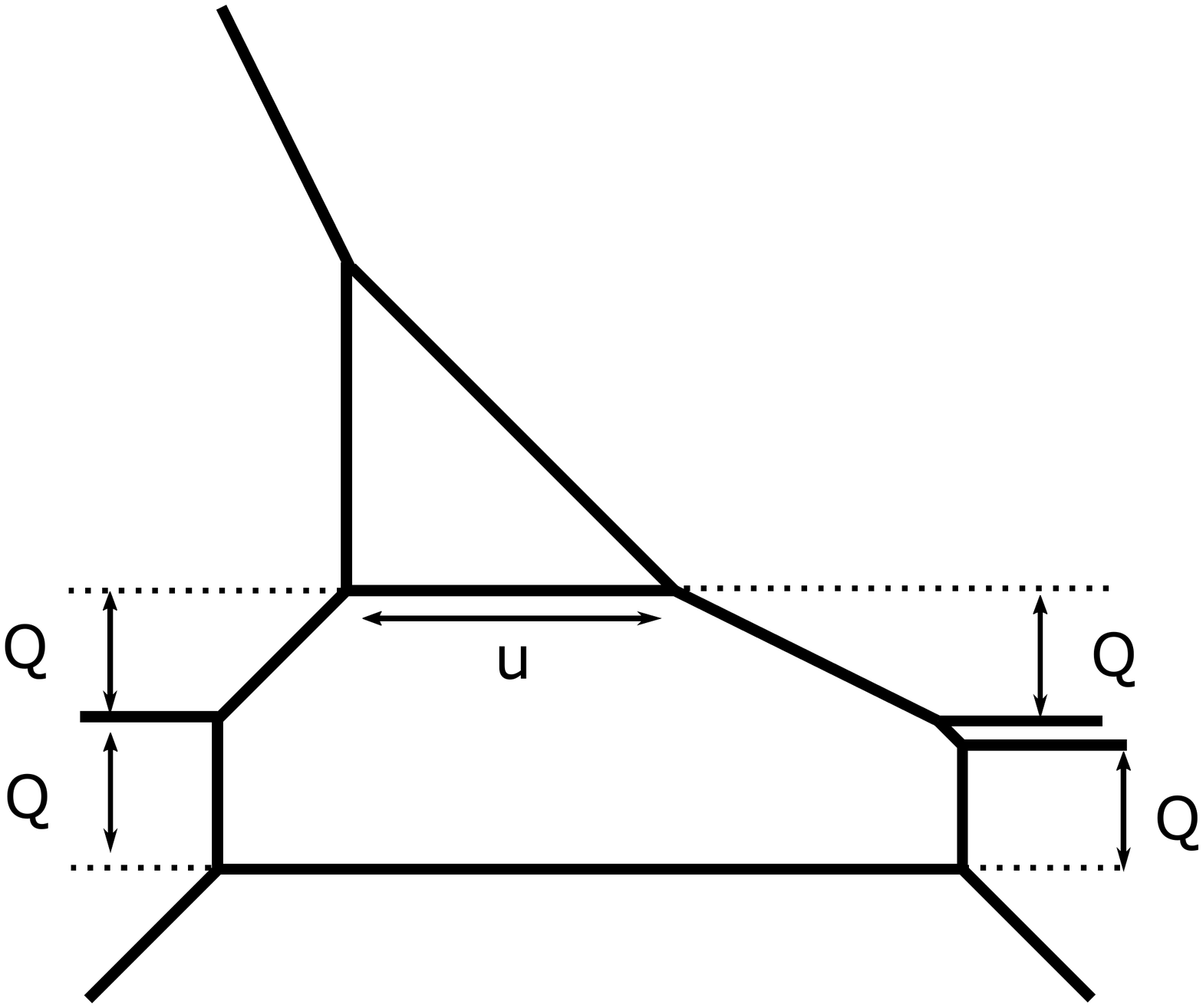} 
\caption{A toric diagram for a local 3-fold which is obtained by gluing a $\mathbb{P}^2$ with a PdP$_4$ along a $\mathbb{P}^1$. }
\label{fig:P2dP4}
\end{figure}

\paragraph{$\ell + 2\gamma$.}
The last example is the genus zero GV invariant for the class $\ell + 2\gamma$. The curve in this class is included in a $T_3$ in Figure \ref{fig:dP6v2} or a $\mathbb{P}^2$ glued with a PdP$_4$ which is one of the four faces in the diagram of an $SU(5)$ gauging in Figure \ref{fig:SU5geom} or the diagram b) in Figure \ref{fig:P2-SU5-T5}. A diagram for the latter geometry is depicted in Figure \ref{fig:P2dP4}. 
The genus zero GV invariant of the class $\ell + 2\gamma$ can be computed from the diagram in Figure \ref{fig:dP6v2} and the one in Figure \ref{fig:P2dP4}. The two types of the curves are independent from each other and we can compute the GV invariant separately. By counting the lines with the K$\ddot{\text{a}}$hler class $\ell + 2\gamma$ in Figure \ref{fig:dP6v2}, we obtain the genus zero GV invariant $15$ from one $T_3$. On the other hand, the number of the lines in the class $\ell + 2\gamma$ in Figure \ref{fig:P2dP4} is four. Since four PdP$_4$ are glued with a $\mathbb{P}^2$ at one corner, the genus zero GV invariant in the class $\ell + 2\gamma$ of this type from one corner becomes $4 \times 4 = 16$. Hence, the GV invariant from one corner is $15 + 16 = 31$. Since the contribution from each corner of the pentagon is the same, the genus zero GV invariant for the class $\ell + 2\gamma$ for the mirror quintic is $31 \times 5 = 155$. 

The results above explain the terms at some lowest orders in Table \ref{tb:GV-mirror-quintic}. The genus zero GV invariants for the mirror quintic for some low degrees were first obtained by S. Katz and D. Morrison in \cite{MorrisonKatz:2018} which are summarized in Table \ref{tb:GV2} .
Indeed, the disagreement between their result and the naive version of the topological vertex leading to  results in Table \ref{tb:putativeGV} was our motivation for extending the topological vertex formalism.  The comparison of our results in Table \ref{tb:GV-mirror-quintic}  based on the vertex formalism against the geometric results in Table \ref{tb:GV2} demonstrates that our formalism leads to the correct results at least up to the order $2\ell+2\gamma$. The results beyond this order are our prediction using the vertex formalism. We do not have strong evidences for the numbers and hence it would be interesting to check them.
\begin{table}[t]
\centering
\begin{tabular}{c|c|c|c}
	\diagbox[height=.8cm,width=1.3cm]{$\ell$~}{$\gamma$~} & $0$ & $1$ & $2$\\
\hline
$0$ & $\ast$ & $\textcolor{black}{300}$ & $\textcolor{black}{-440}$\\
\hline
$1$ & $\textcolor{black}{15}$ & $\textcolor{black}{-60}$ & $\textcolor{black}{155}$ \\
\hline
$2$ & $\textcolor{blue}{-30}$ & $\textcolor{blue}{150}$ & $\textcolor{blue}{-500}$ \\
\end{tabular}
\caption{Genus zero GV invariants of the mirror quintic computed by Katz and Morrison in \cite{MorrisonKatz:2018}. The invariants in black, which had already been computed by Katz and Morrison, were confirmed by our vertex formalism. The invariants in blue were predicted by our vertex formalism and subsequently checked by Katz and Morrison.}
\label{tb:GV2}
\end{table}

It would be especially interesting to see if we can modify the topological vertex rules along the lines above to obtain the full result including all the curve classes in the mirror quintic.

\section{Conclusions}
\label{sec:concl}
In this paper we propose a systematic construction of (non-gravitational) field theory sectors in 6d (1,0) supergravity theories from F-theory on compact elliptic 3-folds of type $T^6/\Gamma$ with $\Gamma=\mathbb{Z}_m\times \mathbb{Z}_n$. The orbifold action of $\Gamma$ leads to several fixed points on the base $B=T^4/\Gamma$ and each fixed point hosts a particular class of local 6d (1,0) SCFTs, called $(G,G')$ conformal matter theories in \cite{DelZotto:2014hpa,Heckman:2013pva,Heckman:2015bfa}.  These local SCFTs are glued together by introducing an additional local SCFT which gauges the global symmetries $H\subset G$ or $H'\subset G'$. We have shown that these 6d field theories have no gravitational and also gauge/gravity mixed anomalies, and therefore the 6d supergravity systems in which  field theories are embedded are consistent.

Two dimensional $\mathcal{N}=(0,4)$ SCFTs describing self-dual strings in 6d supergravity theories have proven to be of particular interest. We claim that these 2d self-dual string worldsheet theories flow to such 2d (0,4) SCFTs in the IR. When the corresponding D3-branes wrap higher genus 2-cycles in the base $B$ of the elliptic fibration, we conjecture the Higgs branch of the self-dual string theory to be holographically dual to type IIB theory on $AdS_3\times S^3 \times B$.  We have partially checked this conjecture by showing that the central charges of 2d theories in all our field theory models perfectly match those of the supergravity computations in \cite{Haghighat:2015ega}.   Moreover we argue the elliptic genera of these strings provide a means to compute the all-genus topological strings partition function for certain compact Calabi-Yau 3-folds.  

In this work, 
we have focused on a particular class of 3-folds described by orbifolds $T^6/\mathbb{Z}_m\times \mathbb{Z}_n$ at a special point in the moduli space. However, one can also consider 6d gravity theories associated to other types of compact orbifolds. It may be possible to construct their field theory sectors by suitably gluing local 6d SCFTs in a similar manner to that described in this paper.  For instance, the field theory sector in 6d supergravity from an elliptic 3-fold with base $\mathbb{P}^2/\mathbb{Z}_3$ was described in \cite{DelZotto:2014fia} in similar terms. 
The UV 2d quiver gauge theory we constructed for the base $B=T^4/\mathbb{Z}_2\times \mathbb{Z}_2$ allows us to have more control over this compact Calabi-Yau geometry or the corresponding 5d black hole states even though, as we argued, there are some features which seem to be missing from this quiver description of the corresponding strings. For example, we managed to write down a contour integral formula of the topological string partition function (equivalently, the elliptic genus) for this model and we expect that this partition function computes the degeneracies of 5d BPS spinning black hole states. However, we have not evaluated the contour integral and carried out any solid test for it apart from central charge matching. This is mainly because we are not aware of any known computation for this model with which to compare;
it would be interesting to compare our elliptic genus formula to an independent computation of the corresponding geometry's topological strings partition function. 

The considerations above rely on the 3-fold being elliptically fibered, and a direct generalization of the methods used in the above cases would not permit a computation of the topological string partition function for non-elliptic 3-folds.    Instead we find another way to study the topological string amplitudes for some certain non-elliptic 3-folds, including the mirror quintic.  The idea is to use 5d physics, where an analogous role was played by a coupled collection of 5d SCFTs in describing the non-gravitational field theory sector of a 5d supergravity theory. Furthermore, we describe a generalized topological vertex formalism which in principle has a potential of extending the conventional topological vertex to cover all of the compact 3-folds we described in this paper by gluing the local geometries using compact surfaces associated to trivalent gaugings.  This new gauging vertex is similar in spirit to what was done in \cite{Hayashi:2017jze} as a non-abelian generalization of the abelian gauging in \cite{Aganagic:2004js} and is  closely related to the ``negative vertex'' introduced in the reference \cite{MR2681793} in the context of the SYZ program.  However, we found that a precise definition of this `gauging vertex' requires some additional ingredients.

In particular, for the mirror quintic, we found that the proposed generalization was still not enough to compute the all-genus topological string amplitudes of the mirror quintic---even with the appropriate combination of trivalent gaugings, the vertex method needs further modification. Nevertheless we presented a modified vertex formalism and computed some low degree GV invariants for the mirror quintic inspired by the local geometry of the gaugings, which agrees with independent mathematical computations of the same GV invariants. Moreover, we conjecture that this modified topological vertex formalism correctly computes all GV invariants 
except for a limited number of curve classes in the mirror quintic. Further investigation in this direction  and trying to complete our gauging vertex would be very interesting.   It would also be quite interesting to explore the possibility of using the results of this work to obtain topological string amplitudes for other CY 3-folds not considered in this paper, for example the ordinary quintic.

\section*{Acknowledgements}

We are greatly indebted to Sheldon Katz and David Morrison for their computation of topological string invariants for some low degrees for the mirror quintic at our request.
We would also like to thank  Michele Del Zotto, Thomas Dumitrescu, and Seok Kim for useful comments and discussions. P.J., H.K., and C.V. thank the 2017 and 2018 SCGP Summer Workshop for hospitality during a part of this work. H.H. and K.O. would like to thank Harvard University and 2018 SCGP Summer Workshop for hospitality during a part of this work. The work of H.H. is supported in part by JSPS KAKENHI Grant Number JP18K13543. The research of H.K. and C.V. is supported in part by the NSF grant PHY-1067976. H.K. is supported by the POSCO Science Fellowship of POSCO TJ Park Foundation and the National Research Foundation of Korea (NRF) Grant 2018R1D1A1B07042934.
KO gratefully acknowledges support from IAS and NSF Grant PHY-1606531. P.J. is supported by the Harvard University Graduate Prize Fellowship.

\appendix

\section{Geometry of the mirror quintic in the symmetric chamber}  
\label{sec:mirrorquintic}
We begin by describing some features of the geometry of a particular resolution of the mirror quintic which is characterized by a large degree of symmetry. Recall that the mirror quintic can be described as the complete intersection
	\begin{align}
	\label{eqn:cicy}
		\sum_{i=0}^4 y_i - 5 \psi \rho = \prod_{i=0}^4 y_i - \rho^5 =0,
	\end{align} 
where $(y_i,\rho)$ are homogeneous coordinates parametrizing $\mathbb P^5$. The singular locus of the mirror quintic consists of 10 rational curves $C_{ij}$ defined by $y_i = y_j = 0$ with normal geometry $\mathbb C^2/\mathbb Z_5$, intersecting in 10 singular points $T_{ijk}= C_{ij} \cap C_{jk} \cap C_{ki} $ with normal geometry $\mathbb C^3/\mathbb Z_5 \times \mathbb Z_5$. The $C_{ij}$ act as the ``glue'' binding the 10 $T_{ijk}$ to one another (or, from a 5d field theory perspective, the $C_{ij}$ play the role of the $SU(5)$ gauging depicted in Figure \ref{eq:toricdiagram}.) In the following discussion we will abuse notation and use the same symbols $T_{ijk}, C_{ij}$, etc., to denote both the singularities and their total transforms in this particular resolution. 

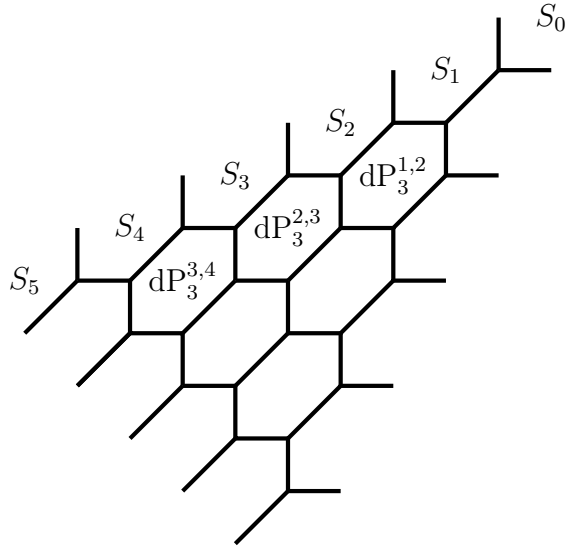
\begin{figure}
	\begin{center}
	$
	\begin{array}{c}
		\begin{tikzpicture}[scale=.7]
			\draw[ultra thick] (0,0) -- (0,1) -- (-1,1) -- (-1,2) -- (0,3) -- (1,3) -- (2,4) -- (3,4) -- (4,5) -- (5,5);
			\draw[ultra thick] (0,0) -- (1,0) -- (1,-1) -- (2,-1) -- (3,0) -- (3,1) -- (4,2) -- (4,3) -- (5,4) -- (5,5);
			\draw[ultra thick] (1,2) --++ (1,0);
			\draw[ultra thick] (1,2) --++ (0,1);
			\draw[ultra thick] (1,2) --++ (-1,-1);
			\draw[ultra thick] (2,1) --++ (1,0);
			\draw[ultra thick] (2,1) --++ (0,1);
			\draw[ultra thick] (2,1) --++ (-1,-1);
			\draw[ultra thick] (3,3) --++ (1,0);
			\draw[ultra thick] (3,3) --++ (0,1);
			\draw[ultra thick] (3,3) --++ (-1,-1);
			\draw[ultra thick] (-1,1) -- (-1-1,1-1);
			\draw[ultra thick] (0,0) -- (0-1,-1);
			\draw[ultra thick] (1,-1) -- (1-1,-1-1);
			\draw[ultra thick] (0,3) -- (0,3+1);
			\draw[ultra thick] (2,4) -- (2,4+1);
			\draw[ultra thick] (4,5) -- (4,5+1);
			\draw[ultra thick] (-1,1) -- (-1-1,1-1);
			\draw[ultra thick] (3,0) -- (3+1,0);
			\draw[ultra thick] (4,2) -- (4+1,2);
			\draw[ultra thick] (5,4) -- (5+1,4);
			\draw[ultra thick] (-2,2) --++ (1,0);
			\draw[ultra thick] (-2,2) --++ (-1,-1);
			\draw[ultra thick] (-2,2) --++ (0,1);
			\draw[ultra thick] (2,-2) --++ (1,0);
			\draw[ultra thick] (2,-2) --++ (-1,-1);
			\draw[ultra thick] (2,-2) --++ (0,1);
			\draw[ultra thick] (6,6) --++ (1,0);
			\draw[ultra thick] (6,6) --++ (-1,-1);
			\draw[ultra thick] (6,6) --++ (0,1);
			\node[] at (0,2) {$\text{dP}_3^{3,4}$};
			\node[] at (2,3) {$\text{dP}_3^{2,3}$};
			\node[] at (4,4) {$\text{dP}_3^{1,2}$};
			\node[] at (-3,2) {$S_5$};
			\node[] at (-1,3) {$S_4$};
			\node[] at (1,4) {$S_3$};
			\node[] at (3,5) {$S_2$};
			\node[] at (5,6) {$S_1$};
			\node[] at (7,7) {$S_0$};
		\end{tikzpicture}
			\end{array}
			$
	\end{center}
	\caption{Diagram which describes each resolved orbifold singuarlity $T_{ijk}$ in the mirror quintic.}
	\label{fig:symweb}
\end{figure}

The defining feature of this particular resolution of the mirror quintic is that each resolved orbifold singularity $T_{ijk}$ is described by the toric diagram in Figure \ref{fig:symweb}.  Taking note of the fact that this highly symmetric phase of $T_{ijk}$ can be viewed as six del Pezzo surfaces $\text{dP}_3$ glued together, one can use the diagram in Figure \ref{fig:symweb} in combination with the geometry of del Pezzo surfaces to determine the intersection structure between each $\text{dP}_3$ and the semi-compact exceptional divisors at the border of the diagram---for example, the surfaces $S_a$ arranged along the upper border, which belong to the total transform of the singular curve $C_{ij}$. Then, keeping in mind that these surfaces $S_a$ are in fact compact K\"ahler surfaces in the full mirror quintic, one can use the intersection structure with the $T_{ijk}$ to fully determine the geometry of these surfaces. 

We now go about the task of determining the $S_a$ belonging to $C_{ij}$. Once we have determined these surfaces, by symmetry we will have automatically determined the surfaces belonging to the remaining $C_{i'j'}$. Since the normal geometry of the singularity $C_{ij}$ is $\mathbb C^2/\mathbb Z_5$, we can expect that the total transform will consist of a collection of ruled surfaces $\cup_a S_{a}$ (possibly blown up at a finite number of points) joined together in a chain, which can be contracted along the base $\mathbb P^1$ of the ruling to obtain an $A_4$ singularity. 

Focusing on the interface between a single $\text{dP}_3^{a,a+1}$ and two of the component surfaces $S_a,S_{a+1}$, we can see that the curves along which these two surfaces are joined together must each have self-intersection $-1$ inside $\text{dP}_3$ and hence must also have self-intersection $-1$ inside $S_a$ and $S_{a+1}$ in order to satisfy the Calabi-Yau condition. Since each $C_{ij}$ meets three $T_{ijk'}$ in a symmetric fashion, each $S_a$ must therefore contain three pairs of irreducible $-1$ curves which can be permuted freely without modifying the remaining intersection structure. If we blow down these three $-1$ curves in a given surface $S_a$, we can expect to recover one of the ruled surfaces belonging to the total transform of the singularity $C_{ij}$.

We can say more about these $-1$ curves: In the above discussion, we considered a single $\text{dP}_3^{a,a+1}$ meeting two irreducible surfaces $S_a, S_{a+1}$. Suppose instead we consider a single $S_a$ meeting two del Pezzo surfaces $\text{dP}_3$. The situation in this case is completely analogous: there are two irreducible $-1$ curves inside $S_a$, each of which is identified with a distinct $-1$ curve on the two $\text{dP}_3$'s. Since these two $-1$ curves (call their numerical equivalence classes $e, e'$) intersect in a point inside $S_a$, the class associated to their sum must have self intersection $0$: 
	\begin{align}
		(e + e')^2 = e^2 + e'^2 + 2 e \cdot e' = -1 -1 + 2 =0. 
	\end{align}
 This implies that there exists an irreducible rational curve $f =e +e'$ in the blowdown of the surface $S_a$ with self-intersection $f^2= 0$. According to the cone theorem for surfaces (a well-known result of Mori theory), the curve $f$ is the fiber of the ruling of $S_a$. 
 
\begin{figure}
	\begin{center}
			\begin{tikzpicture}[scale=1.5]
		\draw[ultra thick] (0,0) -- node[above,midway]{$s + n_a f-\sum e_i$} (4,0);
		\draw[ultra thick] (1,0) -- node[left,midway] {$e_1$} (0,-1);
		\draw[ultra thick] (0,-1) -- node[left,midway] {$f-e_1$}(1,-2);
		\draw[ultra thick] (2,0) -- node[left,midway] {$e_2$}(1,-1);
		\draw[ultra thick] (1,-1)-- node[left,midway] {$f-e_2$}(2,-2);
		\draw[ultra thick] (3,0) -- node[left,midway] {$e_3$} (2,-1);
		\draw[ultra thick] (2,-1) -- node[left,midway] {$f-e_3$}(3,-2);
		\draw[ultra thick] (0,-2) -- node[below,midway] {$s$} (4,-2);
	\end{tikzpicture}
	\end{center}
	\caption{Geometry of the surfaces $S_a = \text{Bl}^*_3 \mathbb F_{n_a}$ for $a = 1, \dots, 4$. Each surface is the blowup of a Hirzebruch surface (i.e. a geometrically ruled surface over a rational curve) of degree $n_a$ at three points lying on a common line with numerical equivalence class $b = s+n_a f$. The curve class $b$ is the base rational curve of the ruling, while the class $f$ is the fiber.}
	\label{fig:ruled} 
\end{figure}
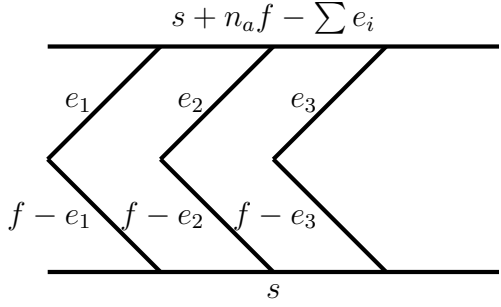

We therefore assume $S_a  = \text{Bl}_3^* \mathbb F_{n_a}$ for $a=1,\dots, 4$, namely the blowup of the Hirzebruch surface $\mathbb F_{n_a}$ at three points, where the superscript ${}^*$ is there to remind us that these three points lie along a common line\footnote{This special configuration is necessary to ensure that the three exceptional divisors $e_1,e_2,e_3$ can be permuted without affecting the intersection structure of the surface $S_a$.} and are hence in a special configuration---see Figure \ref{fig:ruled}. 

Using $(s,f,e_1,e_2,e_3)$ as a basis of classes of curves, the surfaces $S_a$ are characterized by the following intersection products:
	\begin{align}
		s^2 =- n_a,~~ s\cdot f = 1, ~~f^2 =0, ~~e_i \cdot e_j = - \delta_{ij}
	\end{align}  
with all other intersection products vanishing. We now specify the classes of the curves along which these surfaces $S_a$ are glued to each other in terms of the above basis. Clearly, we have two possible gluing configurations between a given pair of surfaces $S_a, S_{a+1}$ glued along a curve $c_{a,a+1} \equiv S_a \cap S_{a+1}$. One possibility is:
		\begin{align}
			\left. S_a \cap S_{a+1}  \right|_{S_a} \equiv \left. c_{a ,a+1} \right|_{S_a} = s + n_a f - \sum e_i,~~ \left. S_a \cap S_{a+1}  \right|_{S_{a+1}} \equiv \left. c_{a ,a+1} \right|_{S_{a+1}} = s
		\end{align}
where $s +n_a f$ is the class of the line containing the three points to which the exceptional divisors $e_i$ contract. The other possibility is 
	\begin{align}
		\left. S_a \cap S_{a+1}  \right|_{S_a} \equiv \left. c_{a ,a+1} \right|_{S_a} = s ,~~ \left. S_a \cap S_{a+1}  \right|_{S_{a+1}} \equiv \left. c_{a ,a+1} \right|_{S_{a+1}} = s.
	\end{align} 
To satisfy the Calabi-Yau condition, the curve representing the interface of $S_a$ and $S_{a+1}$ must satisfy
	\begin{align}
		(c_{a,a+1}^2)_{S_a} + (c_{a, a+1}^2)_{S_{a+1}} = 2g(c_{a,a+1}) -2 = -2,
	\end{align}
where in the above expression we have used the fact that $c_{a,a+1}$ is a rational curve. Hence, the non-negative integers $n_a, n_{a+1}$ characterizing the Hirzebruch surfaces $\mathbb F_{n_a}, \mathbb F_{n_{a+1}}$ must be chosen such that 
	\begin{align}
		n_a - n_{a+1} = 1~~\text{or} ~~ n_a + n_{a+1} = 2.
	\end{align}
	
To fully fix the geometry of the chain of surfaces $\cup S_a$ comprising $C_{ij}$, we still need to specify the geometry at the ``bottom'' and the ``top'' of these chains. Said differently, if we denote the chain of surfaces by $S_1 \cup S_2 \cup S_3 \cup S_4$, in the local picture of $T_{ijk}$ one can see there is an additional pair of identical K\"ahler surfaces $S_0 \cong S_5$ which are glued to $S_1$ and $S_4$ respectively; see Figure \ref{fig:symweb}. For concreteness, let us focus on the surface $S_0$. From Figure \ref{fig:symweb}, we can see that the two curves $C_{ij}$ and $C_{ik}$ both meet $S_0$. Comparing Figure \ref{fig:symweb} with Figure \ref{eq:toricdiagram}, it becomes clear that if we analyze the geometry of another singularity $T_{ij'k}$, the same conclusion must be true of $C_{ik}$ and $C_{ij'}$, namely that both $C_{ik}$ and $C_{ij'}$ intersect the compact surface $S_0$. 

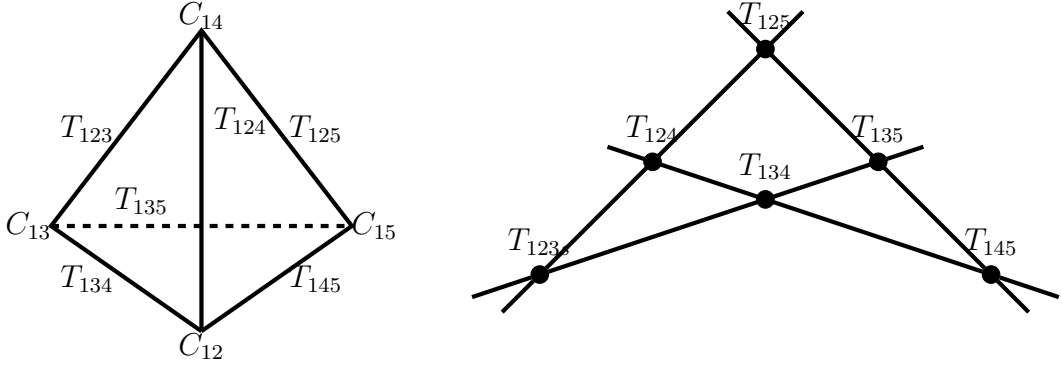
\begin{figure}
	\begin{center}
		\begin{tikzpicture}[scale=2.0]	
			\draw[ultra thick] (0,.3) -- node[left,midway]{$T_{134}$} (-1,1) --  node[left,midway]{$T_{123}$} (0,2.3) -- node[right,midway]{$T_{125}$} (1,1) -- node[right,midway]{$T_{145}$} (0,.3);
			\draw[ultra thick][dashed] (-1,1) -- node[above,pos=.3]{$T_{135}$} (1,1);
			\draw[ultra thick] (0,.3) -- node[right,pos=.7]{$T_{124}$} (0,2.3);
			\node[] at (0,.2) {$C_{12}$};
			\node[] at (-1.15,1) {$C_{13}$};
			\node[] at (1.15,1) {$C_{15}$};
			\node[] at (0,2.4) {$C_{14}$};
		\end{tikzpicture}	
~~~~
			\begin{tikzpicture}[]
				\node[draw,circle,fill=black,scale=.6,label={above:$T_{123s}$}](A) at (-3,0) {};
				\node[draw,circle,fill=black,scale=.6,label={above:$T_{145}$}](B) at (3,0) {};
				\node[draw,circle,fill=black,scale=.6,label={above:$T_{125}$}](C) at (0,3) {};
				\node[draw,circle,fill=black,scale=.6,label={above:$T_{134}$}](D) at (0,1) {};
				\draw[ultra thick] (A) --++ (3.5,3.5);
				\draw[ultra thick] (A) --++ (-.5,-.5);
				\draw[ultra thick] (B) --++ (-3.5,3.5);
				\draw[ultra thick] (B) --++ (.5,-.5);
				\draw[ultra thick] (A) --++ (1.7*3,1.7*1);
				\draw[ultra thick] (A) --++ (-.3*3,-.3*1);
				\draw[ultra thick] (B) --++ (-1.7*3,1.7*1);
				\draw[ultra thick] (B) --++ (.3*3,-.3*1);
				\node[draw,circle,fill=black,scale=.6,label={above:$T_{135}$}](E) at (-3 + 9/2,3/2) {};
				\node[draw,circle,fill=black,scale=.6,label={above:$T_{124}$}](F) at (3 - 9/2,3/2) {};
				\node at (0,-1) {\textcolor{white}{\text{$t$}}};
			\end{tikzpicture}
	\end{center}
	\caption{Left: A complete quadrangle, which represents incidence structure of configuration of singular curves $C_{1j=2,\dots,4}$ in the surface $S_0 \cong \mathbb P^2$ defined by $y_1 =0$ in (\ref{eqn:cicy}). Right: A complete quadrilateral, which is an equivalent depiction of the incidence structure of the singular curves $C_{1j} \subset S_0$. Note that the complete quadrangle and complete quadrilateral are projectively dual.}
	\label{fig:tetrahedron}
\end{figure}

For example, suppose we select $T_{123}$ as our starting point and focus on the surface $S_0$ which meets both $C_{12}$ and $C_{13}$. We would conclude that in fact $C_{12},C_{13},C_{14},C_{15}$ must all meet in the compact surface $S_0$. Clearly, these four singular curves all lie in the intersection of the hyperplane $y_1 = 0$ with the mirror quintic (\ref{eqn:cicy}), which is a $\mathbb P^2$. The four singular curves $C_{12},C_{13},C_{14},C_{15}$ intersect pairwise in the six points $T_{123},T_{124},T_{125},T_{134},T_{135},T_{145}$. The incidence geometry of these four curves can be represented as a tetrahedron where the curves $C_{ij}$ are the faces of the tetrahedron and the edges $C_{ij} \cap C_{ik}$ are the points $T_{ijk}$ as in the left figure Figure \ref{fig:tetrahedron}. This geometric configuration is known as a \emph{complete quadrangle}. An equivalent and perhaps more useful representation would be the configuration of four lines meeting 
in six points in a plane, known as a \emph{complete quadrilateral} which is depicted in the right figure in Figure \ref{fig:tetrahedron}. Note that the structure in Figure \ref{fig:tetrahedron} is essentailly the same as that in Figure \ref{fig:corner}, which we obtained from the corner of the pentagon picture in Figure \ref{fig:pentagon}. The circles and triangles in Figure \ref{fig:corner} are vertices and faces in Figure \ref{fig:tetrahedron} respectively. 

The upshot of this discussion is that the surface $S_0$ in the resolved geometry is a projective plane $\mathbb P^2$ blown up at the six points of intersection of the complete quadrilateral, and glued to the top component $S_1$ of each chain of surfaces $C_{ij}$ along an irreducible rational curve. Since the exceptional curves of this blowup are not involved in the gluing, the only curve class available for the gluing inside $\mathbb P^2$ is the class $\ell$ satisfying $\ell^2 =1$. The Calabi-Yau condition then requires
	\begin{align}
		(c_{0,1}^2)_{S_0} + (c_{0,1}^2)_{S_1} = 1 + (c_{0,1}^2)_{S_1} = -2
	\end{align}
from which we learn
	\begin{align}
		(c_{0,1}^2)_{S_1} =-3 = \begin{cases} &-n_1 \\ &~~n_1 - 3. \end{cases}
	\end{align}
Either of the above choices for $n_1$, both of which are equivalent, fixes the remaining integers $n_a$ for the entire configuration. For convenience, we choose $n_1 = 3$ so that the chain of surfaces is the configuration 
	\begin{align}
	\begin{split}
	\label{eqn:chain}
		C_{ij} ~:~ \cup_{a=0}^{5} S_a &= \text{Bl}_6^* \mathbb P^2 \overset{c_{01}}{\cup} \text{Bl}_3^* \mathbb F_3 \overset{c_{12}}{\cup} \text{Bl}_{3}^* \mathbb F_1 \overset{c_{23}}{\cup} \text{Bl}_{3}^* \mathbb F_1 \overset{c_{34}}{\cup} \text{Bl}_3^* \mathbb F_3 \overset{c_{45}}{\cup} \text{Bl}_6^* \mathbb P^2\\
		\left. c_{01} \right|_{\text{Bl}_6^* \mathbb P^2} & = \ell, ~~ \left. c_{01} \right|_{\text{Bl}_3^* \mathbb F_3 } = s\\
		\left. c_{12} \right|_{\text{Bl}_3^* \mathbb F_3} &= s + 3 f -\sum e_i ,~~\left. c_{12} \right|_{\text{Bl}_3^* \mathbb F_1} = s + f - \sum e_i\\
		\left. c_{23} \right|_{\text{Bl}_3^* \mathbb F_1} &= s
	\end{split}
	\end{align}
where in the above equation we omit the gluing curve classes in $S_3, S_4, S_5$ because they are identical to those of $S_0,S_1,S_2$.

We now have a complete understanding of the geometry of this choice of resolution of the mirror quintic. To summarize, each of the 10 surfaces $T_{ijk}$ can be viewed as six $\text{dP}_3$'s glued to each other, while each of the 10 surfaces $C_{ij}$ can be viewed as a chain of surfaces (\ref{eqn:chain}). In this setup, a given $T_{ijk}$ is glued to each of the three surfaces $C_{ij}, C_{jk}, C_{ki}$ along three pairs of $-1$ curves, where each pair of curves intersects in a single point.

Note that in practice rather than working directly with the complex surfaces $\text{Bl}^*_3 \mathbb F_{1,3}$, it is more convenient to compute the topological vertex using the pseudo del Pezzo surfaces PdP$_4$ depicted in Figure \ref{fig:SU5geom2}, which are related to $\text{Bl}_3^* \mathbb F_{1,3}$ by complex structure deformations. 

\section{$5^-_0$ vertex in the symmetric chamber}
\label{sec:SU5gauging}

\begin{figure}[t]
	\centering
	\begin{tikzpicture}[scale=1.0, ultra thick, baseline = -65]
		\draw (0,0) coordinate (0) -- ++(0,-1/2) coordinate (1) -- ++(-1/2,-1/2) coordinate (2) -- ++(0,-1/2) coordinate (3) -- ++(-1/3,-1/3) coordinate (4) -- ++(0,-1/3) coordinate (5) -- ++(1/3,-1/3) coordinate (6) -- ++(0,-1/2) coordinate (7) -- ++(1/2,-1/2) coordinate (8) -- ++(0,-1/3) coordinate(9) -- ++(1/3,-1/3) coordinate (10)--++(2/3,-1/3) coordinate(11) --++(1/2,-1/2) coordinate(12);
		\draw (0) ++ (2/3,0) coordinate (a) -- ++(1/2,-1/2) coordinate (b) --++(2/3,-1/3) coordinate (c) --++(1/3,-1/3) coordinate(d) --++(0,-1/3) coordinate(e) --++(1/2,-1/2) coordinate(f) -- ++ (0,-1/2) coordinate(g) --++(1/3,-1/3) coordinate(h) --++(0,-1/3) coordinate(i)--++(-1/3,-1/3) coordinate(j)--++(0,-1/2) coordinate(k)--++(-1/2,-1/2) coordinate(l)--++(0,-1/2) coordinate(m);
		\foreach \x/\y in {1/b,3/e,6/g,8/j,11/l}
		{\draw (\x) --(\y);}
		\foreach \x in {2,4,5,7,9,10}
		{\draw (\x) -- ++(-1,0) coordinate (\x l);}
		\foreach \x in {c,d,f,h,i,k}
		{\draw (\x) -- ++(1,0) coordinate (\x r);}
		\foreach \x/\y/\z in {3/e/2,6/g/3,8/j/4,11/l/5}
		{\draw[<->] ($(\x)+(0,0.15)$) -- node[midway,anchor=south] {$Q_{B_\z}$} ($(\y)+(0,0.15)$); }
		\draw[<->] ($(1)+(0,.6)$) -- node[midway,anchor=south] {$Q_{B_1}$} ($(b)+(0,.6)$);
		\foreach \x/\y/\z in {2l/1/1,4l/3/2,7l/6/3,9l/8/4}
		{\draw[<->] (\x |- \y) -- node[midway,anchor=east] {$Q_{(4,\z)}$} (\x);}
		\foreach \x/\y/\z in {cr/b/1,fr/e/2,hr/g/3,kr/j/4}
		{\draw[<->] (\x |- \y) -- node[midway,anchor=west] {$Q'_{(4,\z)}$} (\x);}
		\foreach \x/\y/\z in {cr/b/1,fr/e/2,hr/g/3,kr/j/4}
		{\draw[<->] (\x |- \y) -- node[midway,anchor=west] {$Q'_{(4,\z)}$} (\x);}
		\draw[<->] (5l) ++ (-1.3,0) coordinate (5ll) -- node[midway,anchor=east] {$Q''_{(4,2)}$} (5ll |- 3);
		\draw[<->] (10l) ++ (-1.6,0) coordinate (10ll) -- node[midway,anchor=east] {$Q''_{(4,4)}$} (10ll |- 8);
		\draw[<->] (dr) ++ (1,0) coordinate (drr) -- node[midway,anchor=west] {$Q''_{(4,1)}$} (drr |- b);
		\draw[<->] (ir) ++ (1.2,0) coordinate (irr) -- node[midway,anchor=west] {$Q''_{(4,3)}$} (irr |- g);
		\draw[<->] (1)-- node[midway,anchor=west] {$A_1$} (1 |- 3);
		\draw[<->] (3)-- node[midway,anchor=west] {$A_2$} (3 |- 6);
		\draw[<->] (8)-- node[midway,anchor=west] {$A_3$} (8 |- 6);
		\draw[<->] (11)-- node[midway,anchor=east] {$A_4$} (11 |- 8);
	\end{tikzpicture}
	\caption{The K\"ahler parameters around the geometry of Figure~\ref{fig:SU5geom}.
		These parameters are constrained by the relations \eqref{eq:AQPconstraint} and \eqref{eq:QBconstraint}.
	Here, $Q_{(4,a)}, a=1,2,3,4$ are the parameters specified in Figure~\ref{fig:T5} of one of the $T_5$ diagrams participating this $SU(5)$ gauging. $Q'_{(4,a)}$ and $Q''_{(4,a)}$ come from the other $T_5$'s. }
	\label{fig:SU5geom2}
\end{figure}
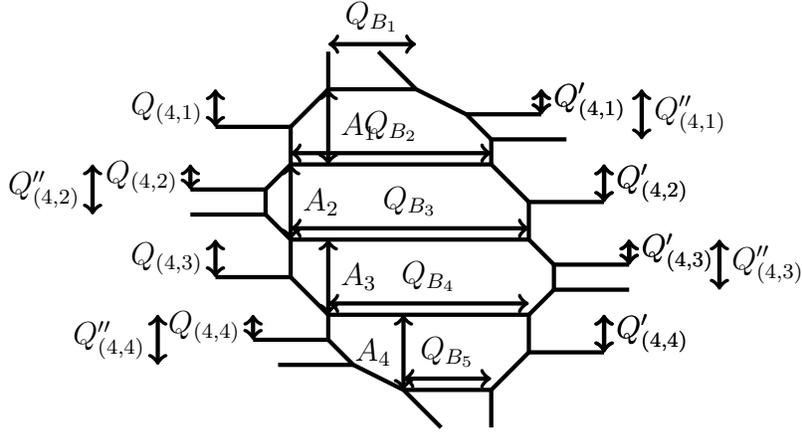

\begin{figure}[h]
	\centering
	\includegraphics[width=.75\linewidth]{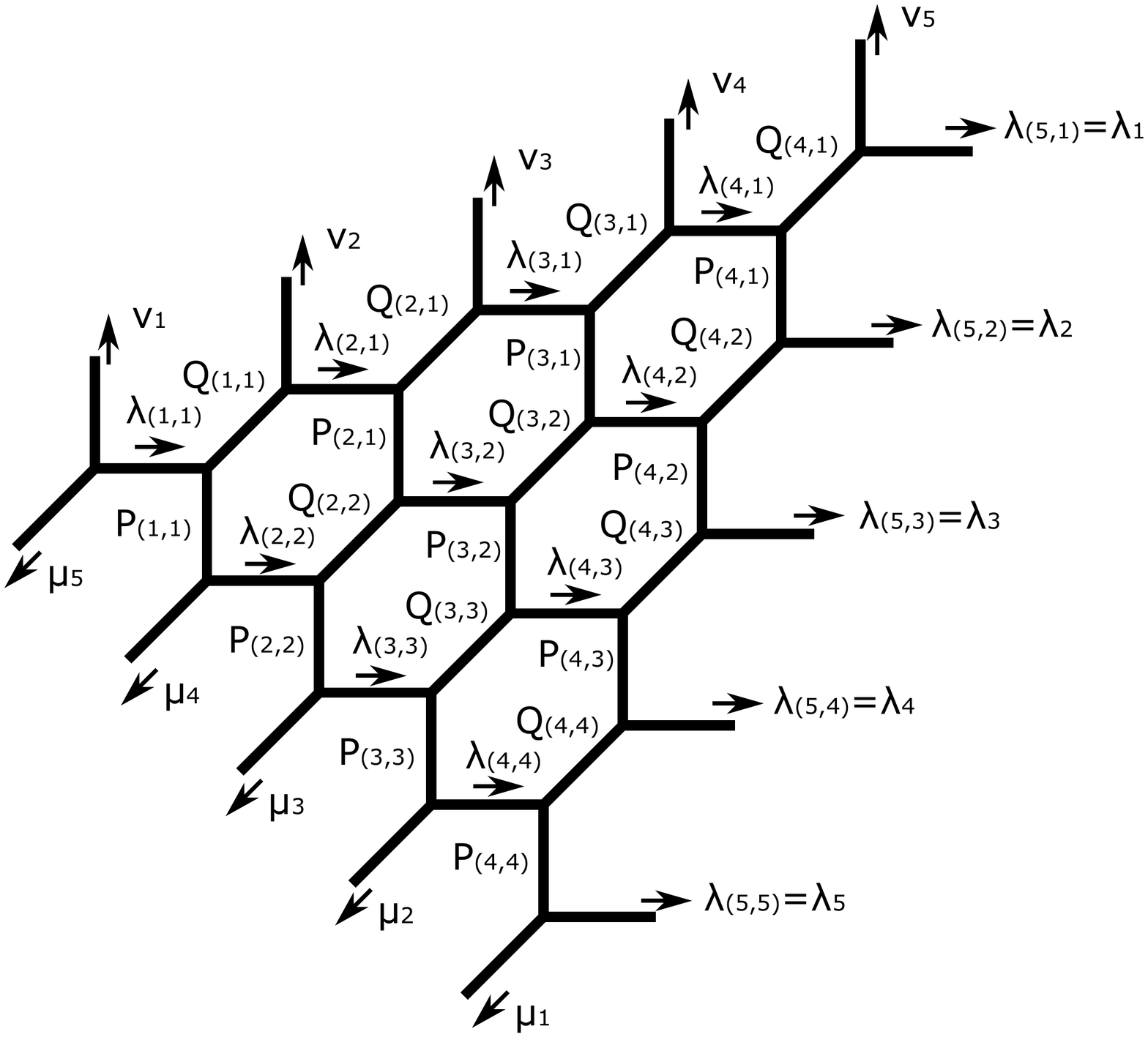}
	\caption{The $T_5$ diagram with an assignment of Young diagrams $\lambda_{k}, (1 \leq k \leq 4),$ and K\"ahler parameters $P_{(k, a)}, Q_{(k, a)}, R_{(k, a)} (1 \leq k \leq 4, 1 \leq a \leq k)$. The K\"ahler parameters are subject to the conditions $P_{(k+1, a)}Q_{(k+1, a+1)} = Q_{(k,a)}P_{(k,a)}$ and $Q_{(k, a)}R_{(k+1,a)} = R_{(k+1, a+1)}Q_{(k+1, a+1)}$ for $1 \leq k \leq 3$.}
	\label{fig:T5}
\end{figure}

In this Appendix we explain how to detemine the K\"ahler parameter weights and framing factors associated to the $5^-_0$ vertex in the mirror quintic geometry, which will differ depending on which chamber of the enlarged K\"ahler cone is considered for the glued $T_5$'s. Here we assume all of the $T_5$'s are in the chamber in Figure~\ref{fig:symweb}, which we call the symmetric chamber.
As explained in Subsection~\ref{sec:topmirrorquintic}, the geometry around a $5^-_0$ vertex can be described by the diagram depicted in Figure~\ref{fig:SU5geom2} after a complex structure deformation.
In the figure various K\"ahler parameters are named. The K\"ahler parameters $Q_{(4,a)}$ with $a=1,2,3,4$ should be identified with the K\"ahler parametrs of one of $T_5$'s connected to the $5^-_0$ vertex in Figure~\ref{fig:T5}.
$Q'_{(4,a)}$ and $Q''_{(4,a)}$ refer to the parameters of other $T_5$'s. Here we assume that the $5^-_0$ vertex is connected to the external legs extending toward the left in Figure~\ref{fig:T5}, and when another set of legs are glued, these parameters should be interchanged with parameters in other $T_5$'s accordingly.

The parameters $A_i$ with $i=1,2,3,4$ are the Coulomb branch parameters of the $SU(5)$ gauging, satisfying
\begin{equation}
	A_i=Q_{(4,i)}P_{(4,i)}=Q'_{(4,i)}P'_{(4,i)}=Q''_{(4,i)}P''_{(4,i)},
	\label{eq:AQPconstraint}
\end{equation}
where $P_{(4,a)},P'_{(4,a)}$ and $P''_{(4,a)}$ are again the parameters in each $T_5$'s specified by Figure~\ref{fig:T5}.
The parameters $Q_{B_i}$ are the effective couplings of effective $U(1)$ gauge fields coming from the $SU(5)$. Figure~\ref{fig:SU5geom2} indicates that there is a relation among them which is
\begin{equation}
	Q_{B_{i+1}} = A_{i}^{1-i}Q_{(4,i)}Q'_{(4,i)}Q''_{(4,i)}Q_{B,i}.
	\label{eq:QBconstraint}
\end{equation}
Therefore, among the four effective couplings, only one is independent as is the usual case in gauge theory.

The effective levels of effective $U(1)$ fields can also be read off from Figure~\ref{fig:SU5geom2}. For example, from the formula \eqref{eq:framingfactor}, the top horizontal internal edge in Figure \ref{fig:SU5geom2} gives a factor $f_{\mu}^{-2}$ where $\mu$ is the Young diagram assigned to the leg, which means the effective level is $-2$. 
By repeating the same procedure for other horizontal internal edges, we get the numbers $l(5,0,a)$ in \eqref{eq:Nminus} in this case to be
\begin{equation}
	l(5,0,a) = -3+a.
	\label{eq:efflevels}
\end{equation}

In summary, the contribution from the $5^-_0$ vertex in the symmetric chamber is
\begin{equation}
	\left(\prod_{a=1}^5 Q_{B_a} \right){C}^{(5_0^-)}_{\vec\lambda\vec\lambda'\vec\lambda''}(\vec A;y)
\end{equation}
where the K\"ahler parameters $Q_{B_i}$ are subject to the constraint \eqref{eq:QBconstraint}, and $C_{\vec{\lambda}\vec{\lambda}'\vec{\lambda}''}^{(5_0^-)}$ is given by \eqref{eq:Nminus} and \eqref{eq:efflevels}:
\begin{equation}
	C^{(5^-_0)}_{\vec\lambda,\vec\lambda',\vec\lambda''}(\vec{A};y)= \frac{\prod_{a=1}^5f_{\lambda_a}(y)^{-3+a} \delta_{\lambda_a,\lambda'_a, \lambda''_a}} {Z^\text{half vector}_{SU(N),\vec{\lambda}}(\vec{A}; y)}.
	\label{eq:5minustilde}
\end{equation}
Here $\lambda_a$ and $\lambda'_a$ are inward Young diagrams and $\lambda''_a$ are outwards.
If we flip the directions of $\lambda''_a$, then \eqref{eq:5minustilde} should be replaced by
\begin{equation}
	\mathcal{C}^{(5^-_0)}_{\vec\lambda,\vec\lambda',\vec\lambda''}(\vec{A};y)= \frac{\prod_{a=1}^5(-1)^{|\lambda_a|}f_{\lambda_a}(y)^{-2+a} \delta_{\lambda_a,\lambda'_a, \lambda''_a}} {Z^\text{half vector}_{SU(N),\vec{\lambda}}(\vec{A}; y)}.
	\label{eq:5minustilde2}
\end{equation}

Finally, let us investigate the K\"ahler weight $Q_{B_i}$ in the parameter specialization taken in Subsection~\ref{sec:topmirrorquintic}. There, all the parameters in $T_5$ are taken to be the same, called $\mathrm{e}^{-\gamma}=:Q_{T_5}$, and $Q_{B_1}$ is called $u=\mathrm{e}^{-\ell}$. In the specialization, $A_i=Q_{T_5}^2$ and thus $Q_{B_i}$ are
\begin{equation}
	Q_{B_1}= Q_{B_5}=u,\quad Q_{B_2}= Q_{B_4} =Q_{T_5}^3 u, \quad Q_{B_3}= Q_{T_5}^4 u.
\end{equation}
Hence, we can safely take the $Q_{T_5}\to0$ limit fixing $u$ without causing any flop transition from the symmetric chamber, obtaining 5 copies of $\mathbb{P}^2$ from the corner of the pentagon as assumed in Subsection~\ref{sec:topmirrorquintic}. Furthermore, up to the order $Q_{T_5}^2$, only the top and bottom lines among the horizontal lines in Figure~\ref{fig:SU5geom2} contribute.

\section{Triple intersections of mirror quintic}
\label{sec:prepotential}

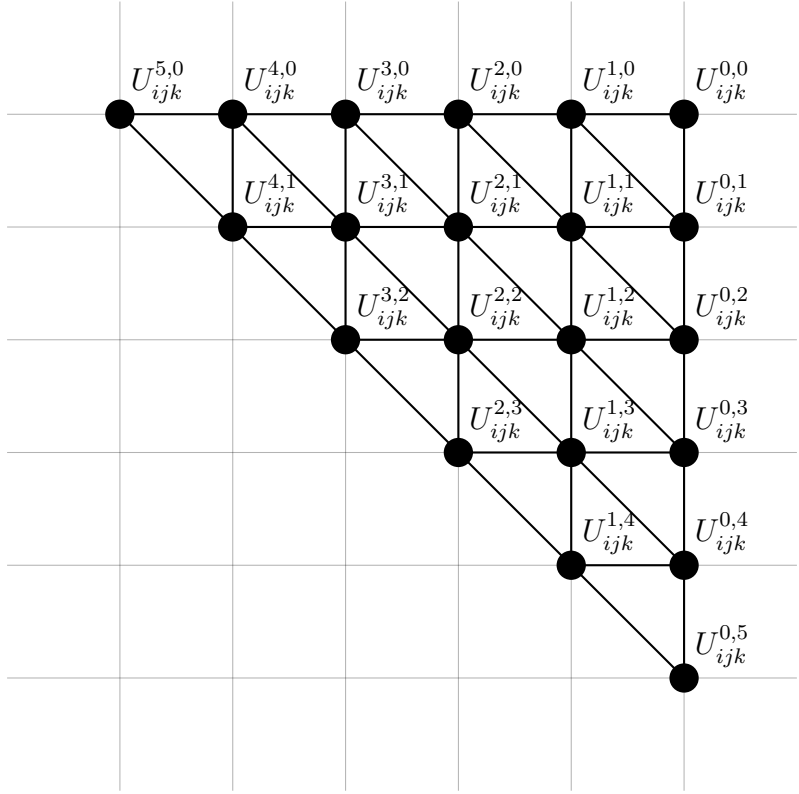
\begin{figure}
	\centering
	\begin{tikzpicture}[scale = 1.5]
		\foreach \x in {0,-1,-2,-3,-4,-5}
		{\draw[opacity =.3] (\x,-2) -- (\x,5);}
		\foreach \y in {-1,0,1,2,3,4}
		{\draw[opacity =.3] (1,\y) -- (-6,\y);}
		\foreach \fourpx in {0,1,2,3,4,5}{
			\foreach \y in {0,...,\fourpx}{
				\pgfmathtruncatemacro{\x}{5-\fourpx}
				\coordinate (\x\y) at (-\x,4-\y);
				\node[circle,fill=black] at (\x\y){};
				\node[anchor=south west] at (\x\y){$U_{ijk}^{\x,\y}$};
			}
		}
		\foreach \x in {0,...,4}{
			\pgfmathtruncatemacro{\xpp}{4-\x}
			\foreach \y in {0,...,\xpp}{
				\pgfmathtruncatemacro{\xp}{\x+1}
				\pgfmathtruncatemacro{\yp}{\y+1}
				\draw[thick] (\x\y) -- (\xp\y);
				\draw[thick] (\x\y) -- (\x\yp);
				\draw[thick] (\xp\y) -- (\x\yp);
			}
		}
	\end{tikzpicture}
	\caption{Dot diagram for $T_5$. We label the dots and corresponding divisors in $T_{ijk}$ as indicated in the above figure. Although in case of local $T_5$ geometry the divisors on the edge of the diagrams are non-compact, all of the dots are compact in the full mirror quintic geometry. }
	\label{fig:T5dot}
\end{figure}

In this Appendix we enumerate the divisors in the mirror quintic using the gauged $T_5$ perspective, and compute the triple intersections between them. 
For each $T_5$ system $T_{ijk}$ ($1\le i<j<k\le 5$) in Figure~\ref{fig:pentagon}, we label its divisors as in Figure~\ref{fig:T5dot}.
For notational convenience, we also introduce $U_{ijk}^{a,b}$ for distinct $i,j,k$ not satisfying $i<j<k$ and impose the condition
\begin{equation}
	U_{jik}^{a,b}=U_{ijk}^{a,5-a-b},\quad
	U_{kji}^{a,b}=U_{ijk}^{5-a-b,b}.
	\label{eq:rotateU}
\end{equation}
Here, the first relation represents flip of the vertical axis and the latter represents flip of the horizontal axis. From these it follows that
\begin{equation}
	U_{ikj}^{a,b}=U_{ijk}^{b,a}.
\end{equation}
Further, because those 10 $T_5$ are glued together, we have identifications
\begin{equation}
	S_{ij}^b\equiv U_{ijk}^{0,b}= U_{ijl}^{0,b}=S_{ji}^{5-b}.
\end{equation}
for all distinct $i,j,k,l=1,2,3,4,5$, and $b=0,1,2,3,4,5$.
Those $S_{ij}^b$, $b\ge 1$ can be interpreted as Coulomb branch parameters of $SU(5)_{ij}$,
and be identified to $S_b$ in Figure~\ref{fig:SU5geom} and Figure~\ref{fig:symweb}.

Note that we automatically have the relations among $S^0$'s:
\begin{equation}
	G_i\equiv  S_{ij}^0  =  S_{ik}^0.
\end{equation}
Thus, we have in total 105 divisors which are
\begin{equation}
	\begin{split}
	U_{ijk}^{a,b} \quad &(1\le i<j<k \le 5, 1\le a,b \le 3, 2\le a+b\le 4),\\
	S_{ij}^b \quad &(1\le i < j \le 5, 1\le b\le 4), \text{and}\\
	G_i \quad &(i=1,2,3,4,5).
	\end{split}
	\label{eq:T5divs}
\end{equation}
The K\"ahler parameters dual to the divisors $G_i$ can be thought as couplings of $SU(5)$'s.
At this stage it seems that 5 among 10 couplings might be independent, but we expect only 101 independent divisors, and therefore there should be 4 linear equivalences among the above divisors which relate $G_i$'s.

To find the linear equivalence, we need to investigate triple intersections among the divisors we have.
From the toric diagram Figure~\ref{fig:T5dot}, we can read off the triple-intersection numbers
\begin{align}
	U_{ijk}^{a,b}\cdot U_{ijk}^{c,d} \cdot U_{ijk}^{e,f}=1
	\label{eq:UUU}
\end{align}
when $(a,b),(c,d),(e,f)$ forms a triangle in the dot diagram, and
\begin{align}
	U_{ijk}^{a,b}\cdot (U_{ijk}^{c,d})^2 = -1
	\label{eq:U2U}
\end{align}
when $(a,b)$ and $(c,d)$ are connected by a edge and at least one of them is not on the edge of the diagram.
Furthermore, we can read triple-intersections involving only $S_{ij}^b$ can be determined by the web diagram Figure~\ref{fig:SU5geom}, or its dual dot diagram, which are
\begin{align}
	(S_{ij}^b)^2 \cdot S_{ij}^{b+1} &= b-3 \quad (b=0,1,2,3,4),\label{eq:SS1}\\
	(S_{ij}^b)^2 \cdot S_{ij}^{b-1} &= 2-b \quad (b=1,2,3,4,5).\label{eq:SS2}
\end{align}
Therefore, the non-vanishing curves have the form of :
\begin{equation}
	C_{ijk}^{a,b} \equiv  U_{i,j,k}^{a,b}\cdot U_{i,j,k}^{a+1,b} \quad (0\le a \le 4, 0\le b \le 5-a).
	\label{eq:T5curves}
\end{equation}
The curves $U_{i,j,k}^{a,b}\cdot U_{i,j,k}^{a,b+1}$ and $U_{i,j,k}^{a+1,b}\cdot U_{i,j,k}^{a,b+1}$ can also be written in the form of \eqref{eq:T5curves} using \eqref{eq:rotateU}.
Taking the relations among $U_{i,j,k}^{a,b}$ into account, there are 350 of those curves.
Denote $M$ a $105\times 350$ matrix whose components are intersection numbers between divisors \eqref{eq:T5divs} and \eqref{eq:T5curves}. Then we find 
\begin{equation}
	\mathrm{Rank}(M)=101,
\end{equation}
which implies there are 4 linear equivalences among the divisors \eqref{eq:T5divs} as expected.

The explicit linear equivalences among the divisors are
\begin{equation}
	X_i \sim X_j \quad (i,j=1,2,3,4,5),
	\label{eq:Xrel}
\end{equation}
where $X_i$ is defined by
\begin{equation}
	X_i = \sum_{\substack{j<k \\ j\neq i ,k\neq i}} \sum_{\substack{0\le a \le 5\\ 0\le b \le 5-a}} (5-a-b)U_{ijk}^{a,b}.
\end{equation}
One can compute the intersection of this divisor $X_i$ and $C_{klm}^{a,b}$ and can obtain
\begin{align}
	X_i \cdot C_{klm}^{a,b} &= 0,\qquad  (a\neq0),\\
	X_i \cdot C_{klm}^{0,b} &= 1,
\end{align}
which is independent of $i$.
Note that $X_i$ is $5G_i$ added by additional terms. 
In this sense, the K\"ahler parameter dual to $X_i$ can be thought as a dressed couplings of $SU(5)_{ij}$, and \eqref{eq:Xrel} is the relation among them.

Although we cannot directly read off the self-triple-intersections of the divisors from Figure~\ref{fig:SU5geom} and Figure~\ref{fig:T5dot}, we can obtain them by substituting the linear equivalence \eqref{eq:Xrel} into \eqref{eq:UUU}, \eqref{eq:U2U}, \eqref{eq:SS1}, and \eqref{eq:SS2}. The computation gives
\begin{equation}
	\begin{split}
		(U_{ijk}^{a,b})^3 &= 6 \quad (1\le i<j<k \le 5, 1\le a,b \le 3, 2\le a+b\le 4),\\
		(S_{ij}^b)^3 &= 5 \quad (1\le i < j \le 5, 1\le b\le 4), \text{and}\\
		(G_i)^3 &= 9 \quad (i=1,2,3,4,5).
	\end{split}
\end{equation}
Those intersections are consistent with the local geometries around each divisor, found in Appendix~\ref{sec:mirrorquintic}.
In particular, we reproduced the intersection number $(G_i)^3=9$ which is the same as the self-triple-intersection of local $\mathbb{P}^2$.

Finally, we can easily compute the prepotential given the triple-intersections.
We choose the independent set of divisors as the ones in \eqref{eq:T5divs} other than $G_2,G_3,G_4,G_5$.
Introduce the K\"ahler parameters $u_{ijk}^{a,b}$ dual to divisors $U_{ijk}^{a,b}$.
Then, the prepotential is
\begin{equation}
	\mathcal{F}= \frac{1}{6} J^3
\end{equation}
where $J$ is the K\"ahler form
\begin{equation}
	J=\sum_{i,j,k,a,b} u_{ijk}^{a,b}U_{ijk}^{a,b}.
\end{equation}
Here, the sum does not contain the divisors $G_2,G_3,G_4,G_5$.
In other parts of this paper, we use volumes of curves as K\"ahler parameters.
To relate them to the variables $u_{ijk}^{a,b}$ introduced here, one can compute the volume of the curve $C^{a,b}_{ijk}$ in the variables $u_{ijk}^{a,b}$ as
\begin{equation}
	\mathrm{vol}(C^{a,b}_{ijk}) = - \partial_{u_{ijk}^{a,b}}\partial_{u_{ijk}^{a+1,b}}\mathcal{F}.
\end{equation}

\clearpage

\bibliographystyle{JHEP}

\bibliography{compactCY}

\end{document}